\pdfoutput=1
\documentclass[12pt,a4paper]{article}

\usepackage{ifthen} 
\newboolean{pdflatex}
\setboolean{pdflatex}{true} 

\newboolean{articletitles}
\setboolean{articletitles}{true} 

\newboolean{uprightparticles}
\setboolean{uprightparticles}{false} 

\usepackage{rotating}
\usepackage{dashrule}
\usepackage{array} 
\usepackage{dcolumn}
\usepackage{comment}
\usepackage{xcolor}
\usepackage{graphpap} 
\usepackage{rotating} 
\usepackage{tikz}
\usepackage{xcolor}
\usepackage{longtable} 
\usepackage{lscape}
\usepackage{multirow} 
\usepackage{mathrsfs}
\usepackage{mathtools} 
\usetikzlibrary{patterns}
\usepackage{nicefrac} 
\usepackage{transparent}
\usepackage{afterpage}

\def\paperauthors{I/Belyaev, G.Carboni, N.Harnew, C.Matteuzzi, F.Teubert} 
\def\paperasciititle{The history of LHCb} 
\def\papertitle{The history of LHCb} 
\def\paperkeywords{{High Energy Physics}, {LHCb}} 
\def\papercopyright{\the\year\ CERN for the benefit of the LHCb collaboration} 

\def\paperlicenceurl{https://creativecommons.org/licenses/by/4.0/}

\newcommand{\plum}[1]{10^{#1} \; \mathrm{cm^{-2} s^{-1}}}

\makeatletter
\g@addto@macro\bfseries{\boldmath}
\makeatother

\usepackage[top=1in, bottom=1.25in, left=1in, right=1in]{geometry}

\usepackage{microtype}
\usepackage{lineno}  
\usepackage{xspace} 
\usepackage{caption} 

\usepackage{graphicx}  
\usepackage{color}
\usepackage{colortbl}
\graphicspath{{./figures/}} 

\usepackage{amsmath} 
\usepackage{amssymb}
\usepackage{amsfonts}
\usepackage{upgreek} 

\newcommand*\patchAmsMathEnvironmentForLineno[1]{%
\expandafter\let\csname old#1\expandafter\endcsname\csname #1\endcsname
\expandafter\let\csname oldend#1\expandafter\endcsname\csname
end#1\endcsname
 \renewenvironment{#1}%
   {\linenomath\csname old#1\endcsname}%
   {\csname oldend#1\endcsname\endlinenomath}%
}
\newcommand*\patchBothAmsMathEnvironmentsForLineno[1]{%
  \patchAmsMathEnvironmentForLineno{#1}%
  \patchAmsMathEnvironmentForLineno{#1*}%
}
\AtBeginDocument{%
\patchBothAmsMathEnvironmentsForLineno{equation}%
\patchBothAmsMathEnvironmentsForLineno{align}%
\patchBothAmsMathEnvironmentsForLineno{flalign}%
\patchBothAmsMathEnvironmentsForLineno{alignat}%
\patchBothAmsMathEnvironmentsForLineno{gather}%
\patchBothAmsMathEnvironmentsForLineno{multline}%
\patchBothAmsMathEnvironmentsForLineno{eqnarray}%
}

\usepackage{hyperxmp}

\usepackage[pdftex,
            pdfauthor={\paperauthors},
            pdftitle={\paperasciititle},
            pdfkeywords={\paperkeywords},
            pdfcopyright={Copyright (C) \papercopyright},
            pdflicenseurl={\paperlicenceurl}]{hyperref}

\usepackage[colorinlistoftodos,textsize=scriptsize]{todonotes}

\usepackage[bottom,flushmargin,hang,multiple]{footmisc}

\usepackage[all]{hypcap} 

\usepackage{xspace} 
\usepackage{upgreek}

\def\lhcb   {\mbox{LHCb}\xspace}

\def\babar  {\mbox{BaBar}\xspace}
\def\belle  {\mbox{Belle}\xspace}

\def\cdf    {\mbox{CDF}\xspace}
\def\dzero  {\mbox{D0}\xspace}

\def\lhc    {\mbox{LHC}\xspace}

\def\tevatron {Tevatron\xspace}
\def\bfactories {\mbox{\B Factories}\xspace}
\def\bfactory   {\mbox{\B Factory}\xspace}

\def\velo   {VELO\xspace}
\def\rich   {RICH\xspace}
\def\richone {RICH1\xspace}

\def\ttracker {TT\xspace}
\def\intr   {IT\xspace}

\def\ot     {OT\xspace}

\def\spd    {SPD\xspace}

\def\ecal   {ECAL\xspace}
\def\hcal   {HCAL\xspace}
\def\MagUp {\mbox{\em Mag\kern -0.05em Up}\xspace}

\def\lone   {L0\xspace}
\def\hlt    {HLT\xspace}
\def\hltone {HLT1\xspace}
\def\hlttwo {HLT2\xspace}

\ifthenelse{\boolean{uprightparticles}}%
{
 
 \def\Pgamma      {\ensuremath{\upgamma}\xspace}

 \def\Peta        {\ensuremath{\upeta}\xspace}

 \def\Pmu         {\ensuremath{\upmu}\xspace}

 \def\Ppi         {\ensuremath{\uppi}\xspace}                 
                  
 \def\Prho        {\ensuremath{\uprho}\xspace}

 \def\Pphi        {\ensuremath{\upphi}\xspace}                 
                  
 \def\Pchi        {\ensuremath{\upchi}\xspace}                 
 \def\Ppsi        {\ensuremath{\uppsi}\xspace}

 \def\PDelta      {\ensuremath{\Delta}\xspace}                 
 \def\PXi         {\ensuremath{\Xi}\xspace}                 
 \def\PLambda     {\ensuremath{\Lambda}\xspace}                 
 \def\PSigma      {\ensuremath{\Sigma}\xspace}                 
 \def\POmega      {\ensuremath{\Omega}\xspace}                 
 \def\PUpsilon    {\ensuremath{\Upsilon}\xspace}

 \def\PB      {\ensuremath{\mathrm{B}}\xspace}                 
 \def\PC      {\ensuremath{\mathrm{C}}\xspace}                 
 \def\PD      {\ensuremath{\mathrm{D}}\xspace}

 \def\PJ      {\ensuremath{\mathrm{J}}\xspace}                 
 \def\PK      {\ensuremath{\mathrm{K}}\xspace}

 \def\PP      {\ensuremath{\mathrm{P}}\xspace}

 \def\PW      {\ensuremath{\mathrm{W}}\xspace}                 
 \def\PX      {\ensuremath{\mathrm{X}}\xspace}                 
                  
 \def\PZ      {\ensuremath{\mathrm{Z}}\xspace}                 
                  
 \def\Pb      {\ensuremath{\mathrm{b}}\xspace}                 
 \def\Pc      {\ensuremath{\mathrm{c}}\xspace}                 
 \def\Pd      {\ensuremath{\mathrm{d}}\xspace}                 
 \def\Pe      {\ensuremath{\mathrm{e}}\xspace}

 \def\Ph      {\ensuremath{\mathrm{h}}\xspace}                 
 \def\Pi      {\ensuremath{\mathrm{i}}\xspace}

 \def\Pp      {\ensuremath{\mathrm{p}}\xspace}

 \def\Ps      {\ensuremath{\mathrm{s}}\xspace}                 
                  
 \def\Pu      {\ensuremath{\mathrm{u}}\xspace}

 \def\thebaroffset{0.0em}
}
{
 
 \def\Pgamma      {\ensuremath{\gamma}\xspace}

 \def\Peta        {\ensuremath{\eta}\xspace}

 \def\Pmu         {\ensuremath{\mu}\xspace}

 \def\Ppi         {\ensuremath{\pi}\xspace}                 
                  
 \def\Prho        {\ensuremath{\rho}\xspace}

 \def\Pphi        {\ensuremath{\phi}\xspace}                 
                  
 \def\Pchi        {\ensuremath{\chi}\xspace}                 
 \def\Ppsi        {\ensuremath{\psi}\xspace}                 
                  
 \mathchardef\PDelta="7101
 \mathchardef\PXi="7104
 \mathchardef\PLambda="7103
 \mathchardef\PSigma="7106
 \mathchardef\POmega="710A
 \mathchardef\PUpsilon="7107
                  
 \def\PB      {\ensuremath{B}\xspace}                 
 \def\PC      {\ensuremath{C}\xspace}                 
 \def\PD      {\ensuremath{D}\xspace}

 \def\PJ      {\ensuremath{J}\xspace}                 
 \def\PK      {\ensuremath{K}\xspace}

 \def\PP      {\ensuremath{P}\xspace}

 \def\PW      {\ensuremath{W}\xspace}                 
 \def\PX      {\ensuremath{X}\xspace}                 
                  
 \def\PZ      {\ensuremath{Z}\xspace}                 
                  
 \def\Pb      {\ensuremath{b}\xspace}                 
 \def\Pc      {\ensuremath{c}\xspace}                 
 \def\Pd      {\ensuremath{d}\xspace}                 
 \def\Pe      {\ensuremath{e}\xspace}

 \def\Ph      {\ensuremath{h}\xspace}                 
 \def\Pi      {\ensuremath{i}\xspace}

 \def\Pp      {\ensuremath{p}\xspace}

 \def\Ps      {\ensuremath{s}\xspace}                 
                  
 \def\Pu      {\ensuremath{u}\xspace}

 \def\thebaroffset{0.18em}
}
\newcommand{\offsetoverline}[2][\thebaroffset]{\kern #1\overline{\kern -#1 #2}}%

\makeatletter
\ifcase \@ptsize \relax
  \newcommand{\miniscule}{\@setfontsize\miniscule{4}{5}}
\or
  \newcommand{\miniscule}{\@setfontsize\miniscule{5}{6}}
\or
  \newcommand{\miniscule}{\@setfontsize\miniscule{5}{6}}
\fi
\makeatother

\DeclareRobustCommand{\optbar}[1]{\shortstack{{\miniscule (\rule[.5ex]{1.25em}{.18mm})}
  \\ [-.7ex] $#1$}}

\def\en         {{\ensuremath{\Pe^-}}\xspace}   
\def\ep         {{\ensuremath{\Pe^+}}\xspace}

\def\epem       {{\ensuremath{\Pe^+\Pe^-}}\xspace}

\def\mup        {{\ensuremath{\Pmu^+}}\xspace}
\def\mun        {{\ensuremath{\Pmu^-}}\xspace} 

\def\mumu       {{\ensuremath{\Pmu^+\Pmu^-}}\xspace}

\def\ellell     {\ensuremath{\ell^+ \ell^-}\xspace}

\def\g      {{\ensuremath{\Pgamma}}\xspace}

\def\W      {{\ensuremath{\PW}}\xspace}

\def\Z      {{\ensuremath{\PZ}}\xspace}

\def\uquark    {{\ensuremath{\Pu}}\xspace}

\def\dquark    {{\ensuremath{\Pd}}\xspace}
\def\dquarkbar {{\ensuremath{\overline \dquark}}\xspace}

\def\squark    {{\ensuremath{\Ps}}\xspace}
\def\squarkbar {{\ensuremath{\overline \squark}}\xspace}

\def\cquark    {{\ensuremath{\Pc}}\xspace}
\def\cquarkbar {{\ensuremath{\overline \cquark}}\xspace}
\def\ccbar     {{\ensuremath{\cquark\cquarkbar}}\xspace}
\def\bquark    {{\ensuremath{\Pb}}\xspace}
\def\bquarkbar {{\ensuremath{\overline \bquark}}\xspace}
\def\bbbar     {{\ensuremath{\bquark\bquarkbar}}\xspace}

\def\pion   {{\ensuremath{\Ppi}}\xspace}
\def\piz    {{\ensuremath{\pion^0}}\xspace}
\def\pip    {{\ensuremath{\pion^+}}\xspace}
\def\pim    {{\ensuremath{\pion^-}}\xspace}

\def\kaon    {{\ensuremath{\PK}}\xspace}

\def\KorKbar {\kern \thebaroffset\optbar{\kern -\thebaroffset \PK}{}\xspace}
\def\Kz      {{\ensuremath{\kaon^0}}\xspace}

\def\Kp      {{\ensuremath{\kaon^+}}\xspace}
\def\Km      {{\ensuremath{\kaon^-}}\xspace}

\def\KS      {{\ensuremath{\kaon^0_{\mathrm{S}}}}\xspace}

\def\Kstarz  {{\ensuremath{\kaon^{*0}}}\xspace}

\def\Kstar   {{\ensuremath{\kaon^*}}\xspace}

\def\Dbar    {{\ensuremath{\offsetoverline{\PD}}}\xspace}
\def\D       {{\ensuremath{\PD}}\xspace}

\def\DorDbar {\kern \thebaroffset\optbar{\kern -\thebaroffset \PD}\xspace}
\def\Dz      {{\ensuremath{\D^0}}\xspace}
\def\Dzb     {{\ensuremath{\Dbar{}^0}}\xspace}
\def\Dp      {{\ensuremath{\D^+}}\xspace}
\def\Dm      {{\ensuremath{\D^-}}\xspace}

\def\DpDm    {\ensuremath{\Dp {\kern -0.16em \Dm}}\xspace}

\def\Dstarb  {{\ensuremath{\Dbar{}^*}}\xspace}
\def\Dstarz  {{\ensuremath{\D^{*0}}}\xspace}
\def\Dstarzb {{\ensuremath{\Dbar{}^{*0}}}\xspace}

\def\Dstarp  {{\ensuremath{\D^{*+}}}\xspace}

\def\Ds      {{\ensuremath{\D^+_\squark}}\xspace}
\def\Dsp     {{\ensuremath{\D^+_\squark}}\xspace}
\def\Dsm     {{\ensuremath{\D^-_\squark}}\xspace}

\def\B       {{\ensuremath{\PB}}\xspace}
\def\Bbar    {{\ensuremath{\offsetoverline{\PB}}}\xspace}

\def\BorBbar {\kern \thebaroffset\optbar{\kern -\thebaroffset \PB}\xspace}
\def\Bz      {{\ensuremath{\B^0}}\xspace}

\def\Bd      {{\ensuremath{\B^0}}\xspace}
\def\Bdb     {{\ensuremath{\Bbar{}^0}}\xspace}
\def\BdorBdbar {\kern \thebaroffset\optbar{\kern -\thebaroffset \Bd}\xspace}
\def\Bu      {{\ensuremath{\B^+}}\xspace}
\def\Bub     {{\ensuremath{\B^-}}\xspace}
\def\Bp      {{\ensuremath{\Bu}}\xspace}
\def\Bm      {{\ensuremath{\Bub}}\xspace}
\def\Bpm     {{\ensuremath{\B^\pm}}\xspace}

\def\Bs      {{\ensuremath{\B^0_\squark}}\xspace}

\def\BsorBsbar {\kern \thebaroffset\optbar{\kern -\thebaroffset \Bs}\xspace}
\def\Bc      {{\ensuremath{\B_\cquark^+}}\xspace}

\def\Bds     {{\ensuremath{\B_{(\squark)}^0}}\xspace}

\def\BdorBs  {\Bds\xspace}

\def\jpsi     {{\ensuremath{{\PJ\mskip -3mu/\mskip -2mu\Ppsi}}}\xspace}
\def\psitwos  {{\ensuremath{\Ppsi{(2S)}}}\xspace}

\def\chiczero {{\ensuremath{\Pchi_{\cquark 0}}}\xspace}
\def\chicone  {{\ensuremath{\Pchi_{\cquark 1}}}\xspace}

\def\Y#1S{\ensuremath{\PUpsilon{(#1S)}}\xspace}

\def\proton      {{\ensuremath{\Pp}}\xspace}
\def\antiproton  {{\ensuremath{\overline \proton}}\xspace}

\def\Lz          {{\ensuremath{\PLambda}}\xspace}

\def\LorLbar     {\kern \thebaroffset\optbar{\kern -\thebaroffset \PLambda}\xspace}

\def\Xires       {{\ensuremath{\PXi}}\xspace}

\def\Omegares    {{\ensuremath{\POmega}}\xspace}

\def\Lc          {{\ensuremath{\Lz^+_\cquark}}\xspace}

\def\Xic         {{\ensuremath{\Xires_\cquark}}\xspace}
\def\Xicz        {{\ensuremath{\Xires^0_\cquark}}\xspace}
\def\Xicp        {{\ensuremath{\Xires^+_\cquark}}\xspace}

\def\Xiccp       {{\ensuremath{\Xires^+_{\cquark\cquark}}}\xspace}
\def\Xiccpp      {{\ensuremath{\Xires^{++}_{\cquark\cquark}}}\xspace}

\def\Lb           {{\ensuremath{\Lz^0_\bquark}}\xspace}

\def\Xibz         {{\ensuremath{\Xires^0_\bquark}}\xspace}
\def\Xibm         {{\ensuremath{\Xires^-_\bquark}}\xspace}

\def\Omegab       {{\ensuremath{\Omegares^-_\bquark}}\xspace}

\def\BF         {{\ensuremath{\mathcal{B}}}\xspace}
\def\BR         {\BF}

\newcommand{\decay}[2]{\ensuremath{#1\!\to #2}\xspace} 

\def\to                 {\ensuremath{\rightarrow}\xspace}

\def\order   {{\ensuremath{\mathcal{O}}}\xspace}

\def\epsK  {{\ensuremath{\varepsilon_K}}\xspace}

\def\CP                {{\ensuremath{C\!P}}\xspace}

\newcommand{\DGs}{{\ensuremath{\Delta\Gamma_{\squark}}}\xspace}

\newcommand{\phis}{{\ensuremath{\phi_{\squark}}}\xspace}

\def\BdToKstmm    {\decay{\Bd}{\Kstarz\mup\mun}}

\def\BsToJPsiPhi  {\decay{\Bs}{\jpsi\phi}}

\def\BTohh        {\decay{\B}{\Ph^+ \Ph'^-}}

\def\bsll     {\decay{\bquark}{\squark \ell^+ \ell^-}}

\def\AT#1     {\ensuremath{A_{\mathrm{T}}^{#1}}\xspace}           

\def\Bsmm     {\decay{\Bs}{\mup\mun}}
\def\Bdmm     {\decay{\Bd}{\mup\mun}}

\def\C#1      {\ensuremath{\mathcal{C}_{#1}}\xspace}                       
\def\Cp#1     {\ensuremath{\mathcal{C}_{#1}^{'}}\xspace}                    
\def\Ceff#1   {\ensuremath{\mathcal{C}_{#1}^{\mathrm{(eff)}}}\xspace}        
\def\Cpeff#1  {\ensuremath{\mathcal{C}_{#1}^{'\mathrm{(eff)}}}\xspace}       
\def\Ope#1    {\ensuremath{\mathcal{O}_{#1}}\xspace}                       
\def\Opep#1   {\ensuremath{\mathcal{O}_{#1}^{'}}\xspace}                    

\newcommand{\aunit}[1]{\ensuremath{\text{\,#1}}}       

\newcommand{\tev}{\aunit{Te\kern -0.1em V}\xspace}
\newcommand{\gev}{\aunit{Ge\kern -0.1em V}\xspace}
\newcommand{\mev}{\aunit{Me\kern -0.1em V}\xspace}
\newcommand{\kev}{\aunit{ke\kern -0.1em V}\xspace}
\newcommand{\ev}{\aunit{e\kern -0.1em V}\xspace}
 
\newcommand{\mevc}{\ensuremath{\aunit{Me\kern -0.1em V\!/}c}\xspace}
\newcommand{\gevc}{\ensuremath{\aunit{Ge\kern -0.1em V\!/}c}\xspace}
\newcommand{\mevcc}{\ensuremath{\aunit{Me\kern -0.1em V\!/}c^2}\xspace}
\newcommand{\gevcc}{\ensuremath{\aunit{Ge\kern -0.1em V\!/}c^2}\xspace}

\def\m    {\aunit{m}\xspace}

\def\pb {\aunit{pb}\xspace}
\def\invpb {\ensuremath{\pb^{-1}}\xspace}
\def\fb   {\ensuremath{\aunit{fb}}\xspace}
\def\invfb   {\ensuremath{\fb^{-1}}\xspace}

\def\fs   {\aunit{fs}}

\newcommand{\stat}{\aunit{(stat)}\xspace}
\newcommand{\syst}{\aunit{(syst)}\xspace}

\def\order{{\ensuremath{\mathcal{O}}}\xspace}

\def\gsim{{~\raise.15em\hbox{$>$}\kern-.85em
          \lower.35em\hbox{$\sim$}~}\xspace}
\def\lsim{{~\raise.15em\hbox{$<$}\kern-.85em
          \lower.35em\hbox{$\sim$}~}\xspace}

\def\sqs   {\ensuremath{\protect\sqrt{s}}\xspace}

\def\pt         {\ensuremath{p_{\mathrm{T}}}\xspace}

\def\msq        {\ensuremath{m^2}\xspace}

\def\tell1  {TELL1\xspace}
\def\ukl1   {UKL1\xspace}

\newcommand{\eg}{\mbox{\itshape e.g.}\xspace}

\usepackage{cite} 
\usepackage{mciteplus}

\usepackage{longtable} 

\def\bbbar{b\overline{b}}

\def\Bbar{\overline{\rm B}}

\begin{document}

\numberwithin{equation}{section}
\numberwithin{table}{section}
\numberwithin{figure}{section}

\renewcommand{\thefootnote}{\fnsymbol{footnote}}
\setcounter{footnote}{1}


\begin{titlepage}
\pagenumbering{roman}

\vspace*{2.0cm}

{\normalfont\bfseries\boldmath\huge
\begin{center}
  \papertitle 
\end{center}
}

\vspace*{1.0cm}

\begin{center}
  I.~Belyaev$^1$,
  G.~Carboni$^2$, 
  N.~Harnew$^3$,
  C.~Matteuzzi$^4$ and F.~Teubert$^5$
  \\ $^1$\,NRC Kurchatov Institute/ITEP, Moscow,
  $^2$\,INFN and Universit\`a di Roma Tor Vergata,
  $^3$\,University of Oxford,
 $^4$\,INFN and Universit\`a Milano-Bicocca, $^5$CERN, Geneva
\end{center}

\vspace{2cm}

\begin{abstract}
  \noindent
In this paper we describe the history of the \lhcb experiment over the last  three decades, and its remarkable successes and achievements. \lhcb  
was conceived  primarily  as a \bquark-physics experiment,
dedicated to \CP violation studies and measurements of very rare $\bquark$  decays, however
the tremendous potential for
 \cquark-physics was also clear.
At first data taking, the versatility of the experiment as a general-purpose detector in the forward region also became evident, with  measurements achievable such as electroweak physics, jets and 
new particle searches in open states. These were facilitated by the excellent capability of the detector to identify muons
and to reconstruct decay vertices close to the primary
$\proton\proton$~interaction region.

\noindent By the end of the \lhc Run 2 in
2018, before the accelerator paused for its second long shut down,
\lhcb had measured the CKM quark mixing matrix elements and \CP violation parameters to world-leading precision in the
heavy-quark systems. The experiment had also measured many rare
decays of \bquark~and \cquark~quark mesons and baryons to below their Standard Model expectations, some down to branching ratios of
 order 10$^{-9}$.
In addition, world knowledge of $\bquark$ and $\cquark$ spectroscopy had  improved significantly through discoveries of many new resonances already anticipated in the quark model, and also adding new exotic four and five quark states.

\noindent
The paper describes the evolution of the \lhcb detector, from conception to its operation at the present time.  The authors' subjective summary of the experiment's important contributions is then presented, demonstrating the wide domain of successful physics measurements that have been achieved over the years.

\end{abstract}

\vspace*{\fill}

\begin{center}
  To appear in EPJH
\end{center}

\end{titlepage}


\newpage
\setcounter{page}{2}
\mbox{~}
%
%
%
%


\renewcommand{\thefootnote}{\arabic{footnote}}
\setcounter{footnote}{0}

\tableofcontents
\cleardoublepage


\pagestyle{plain} 
\setcounter{page}{1}
\pagenumbering{arabic}



\section{Introduction}

\lhcb is an experiment at the CERN \lhc, dedicated to the study of heavy flavours with large statistics. 
The resulting high precision makes possible the observation of tiny deviations from the predictions of the Standard Model  (SM)
in \CP violation and rare
phenomena, variations which could hint at New Physics (NP) processes.
\lhcb started taking data in 2010. The so-called Run~1 
commenced at an initial centre of mass energy of  $\sqrt{s}=7$ TeV which was then increased to $\sqrt{s}=8$ TeV,
collecting an integrated luminosity of $3.23\,\invfb$ until the end of 2012.
After a two-year shutdown, \lhc operation continued from 2015 to 2018 (Run~2), when the experiment
took data at $\sqrt{s}=13$ TeV, recording an integrated luminosity of $\sim 6 \,\invfb$. Throughout the running periods,
\lhcb collected and analysed an unprecedented number of $\bquark$ decays, and also enlarged its
scope to include charm physics, $W$ and $Z$ measurements, jets and nuclear collisions. 

In this paper the motivation for the \lhcb
experiment is described,  recalling how its design was
developed and evolved. The experiment's major achievements in terms of physics results are then summarised. 
Finally we discuss the plans for the future upgrade, in the period when the Super-KEKB collider will also operate. 

The layout of this paper is as follows.
The Introduction (Sect. 1) describes the status of the $\bquark$ physics programme at the time when the \lhcb detector was conceived, and provides an evolution of its design. 
In Sect. 2 
the basic elements of the detector are described,  optimised for the requirements of heavy-flavour physics measurements, together 
with a description of the triggers.
The measurements which were originally the major aims of the experiment,
i.e. the CKM matrix measurements and \CP violation 
 and  very rare decays  
of the $b$-quark, are reviewed in Sects. 3 and 4, respectively.
In Sect. 5,
a summary of the wide-ranging results in $\bquark$- and $\cquark$- spectroscopy is presented.
A review is given of the many {\it {non planned}} physics areas in Sect. 6, such as results on jets,  electro-weak (EW) physics, and searches for new particles not associated with heavy flavours.
In all these domains, \lhcb proves to be an extremely versatile detector, providing complementary measurements to those of the \lhc General Purpose Detectors (GPDs).
The paper concludes in Sect. 8 with a short description of the 
upgrade plans, which will ensure \lhcb operation beyond 2030.


The paper presents the authors' subjective summary of \lhcb's
many major physics results, however the review 
inevitably omits a substantial number of important measurements. 
To this end, additional information 
can  be found in the paper's exhaustive bibliography.

\subsection{$\bquark$ physics at the end of the XXth century}

In 1970 Glashow, Iliopoulos and Maiani postulated the existence of a fourth
quark, {\em charm},  necessary to explain the smallness
of $\Kz$ oscillations in the framework of the 
Standard Model of Weak Interactions~\cite{Glashow:1970gm}. 
The so-called Glashow-Iliopoulos-Maiani (GIM) mechanism generalized
Cabibbo's idea of rotated  weak currents to the quark model with two doublets, 
introducing a $2 \times 2$ mixing matrix written in terms of the Cabibbo angle. 
An experimental confirmation of the fourth quark hypothesis came in 1974 with the discovery of $\jpsi$~\cite{Augustin:1974xw, Aubert:1974js}, soon
identified as a $\ccbar$ bound state, and followed by the discovery of open charm. 
In 1973 Kobayashi and Maskawa \cite{Kobayashi:1973fv}
proposed a third heavy-quark doublet 
in order to describe  \CP violation in the framework of the SM, thus 
generalizing the Cabibbo matrix to the $3 \times 3$ CKM (Cabibbo, Kobayashi, Maskawa) matrix. 
The  discovery of the $\PUpsilon$ in 1975, followed by the charged and neutral $\B$-mesons in 1983 
\cite{Behrends:1983er} 
 proved the validity of their idea and held the  
prospect of understanding quantitatively \CP violation.

The level of \CP violation  in
$\bquark$ quark decays  was expected to be orders of magnitude larger than in the neutral $\kaon$ system, but
unfortunately the relevant decay channels had only tiny branching fractions, so the  lack of intense
``$\bquark$-quark sources''  slowed the progress  of beauty physics: in particular, in 1986 the PDG  only listed five decay modes of $\Bz$ and $\Bpm$.
However in 1987 a new impetus came from the discovery at ARGUS of
$\B^0 - {\overline{\B}^0}$  oscillations
 \cite{Schroder:1987vf}.
It was clear that the forthcoming LEP machine,
designed for an entirely different purpose, 
and the symmetrical CESR collider, could not 
yield an exhaustive answer to all the questions related to the CKM hypothesis,
despite their valuable contributions to many facets of $\bquark$-physics
\cite{Abbaneo:2000ej,doi:10.1146/annurev-nucl-010609-115108}. 
When in 1989 P. Oddone proposed an asymmetric $\epem$ collider \cite{Oddone:1989ir} operating at the $\PUpsilon(4S)$ energy with a
luminosity above  $ \plum{33}$, an intense period of accelerator studies ensued. This gave birth to the PEP-II and KEKB $\PB$ Factories, which were approved in 1994, 
and  started operating in 1998, soon reaching and passing  their 
design luminosity. 

Around this time, proponents pursued the idea of exploiting hadron beams to attack
the problem of detecting \CP violation in the $\bquark$ sector. 
The idea behind this was that the large hadronic $\bquark$ production cross-section plus the high-intensity hadron beams at the already existing and planned proton accelerators
would  produce a large number of $\bbbar$ pairs, sufficient to gather evidence for \CP violation
at least in the so-called ``golden channel'' $\Bz \to \jpsi \KS$.
To achieve an adequate background
rejection, the experimental difficulties were formidable because of the small ratio of the $\bquark$ cross-section 
to the total hadronic cross-section at the $\sqrt{s}$ values available in fixed-target and collider experiments. 

\begin{figure}[!htbp]
 \begin{center}
\includegraphics[height=7cm]{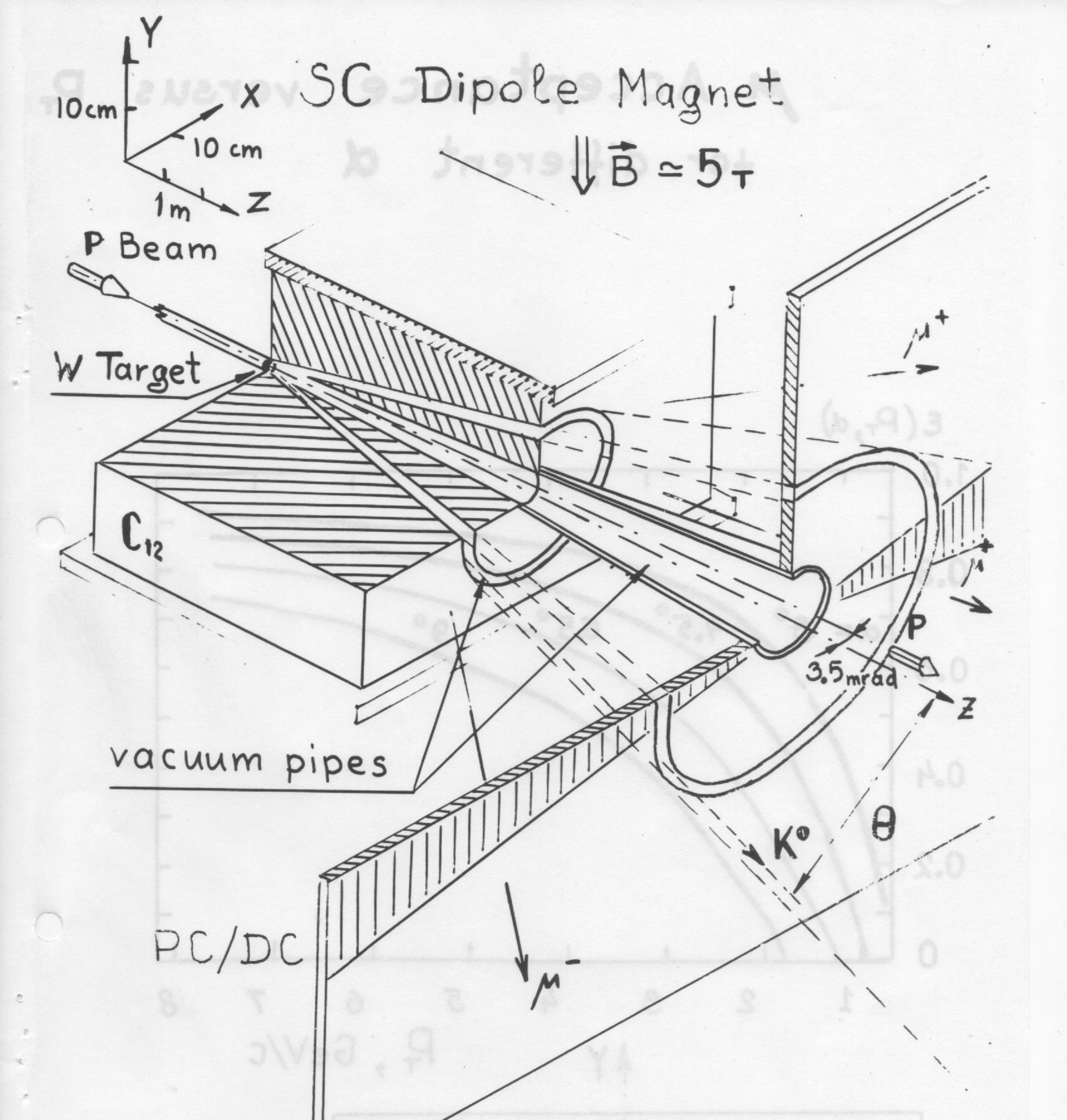}
\end{center}
\caption{A sketch of a proposed beam-dump style experiment designed
  to detect  $\Bz \to \jpsi \KS$ decays. The 
  $\KS$ would traverse the conical slit before decaying 
  \cite{Fidecaro:1990py}. }
 \label{fig:kekel_det}
 \end{figure}

 In 1985, the fixed-target WA75 hybrid experiment  at the CERN SPS \cite{Albanese:1985wk}
 observed in emulsions the first partially reconstructed $\bbbar$ pair produced by an extracted pion beam 
 of 350 GeV/c, confirming that the production cross-section at such low energies was very small. 
 To circumvent this problem, simple experiments were proposed 
 \cite{Fidecaro:1990py} 
 for the CERN SPS and for the planned UNK machine \cite{Myznikov:1985bj} at Serpukhov, 
 with a brute-force approach based on a high-intensity extracted beam and a minimalist detector designed 
 to reconstruct the  $\Bz \to \jpsi \KS$ decay and to provide flavour tagging (Fig.~\ref{fig:kekel_det}).
 There was no  
 time-dependent analysis of the decay since the beam-dump character of the experiments made the use
 of a micro vertex detector impossible. There were exploratory
 fixed-target experiments, at CERN (WA92) \cite{Adamovich:1996ih} 
 and at Fermilab (E653, E672, E771, E789)  \cite{Mishra:1993hn}, which tried to observe and measure beauty events,  albeit without (or very limited) success.

 In 1989 P. Schlein proposed a dedicated Beauty experiment 
 exploiting the large $\bquark$ cross-section 
 expected at the CERN SPS proton-antiproton
 collider ($\sqrt{s} = 630$ GeV) \cite{P238:aa}.
The authors of the proposal (P238) remarked that the bulk of
 $\bbbar$ production occurred at very small angles with respect to the beams, therefore
 making a compact experiment practical. The heart of the detector was a Silicon
 Microvertex Detector operating very close to the beam (1.5 mm) coupled
 to a fast readout and track-reconstruction electronics. 
 The Microvertex Detector provided the trigger by requiring that
 accepted events had to be inconsistent with a single vertex. 
 P238 was not approved but the CERN R\&D Committee, established to support
 new detector developments in view of the LHC,
 approved in 1991 a test of the Microvertex Detector \cite{Ellett:291228} 
 in the SPS Collider. This proved very successful \cite{Ellett:1991bh}
 and paved the way towards  the future COBEX experimental proposal.

 At the time when the $\epem$ colliders were approved,
the HERA-B experiment had been conceived at DESY \cite{Krizan:1994ts}. 
Approved in 1994, HERA-B exploited the 920 GeV HERA proton beam
on a fixed target made of metallic wires, placed  inside Roman Pots in the vacuum pipe, and immersed in the beam halo. 
HERA-B was approved to take data in 1998, one year before PEP-II and KEKB. 
The sophisticated apparatus consisted of a single-arm spectrometer, including a RICH,
a large microvertex silicon detector, a high-resolution tracker, 
plus an electromagnetic calorimeter.
HERA-B was designed primarily for the detection of the  
$\Bz \to \jpsi \KS$ decay and its trigger was based on  $\jpsi$ reconstruction at the first level. 

At $\sqrt{s}=40$ GeV, the $\bbbar$
cross-section is about $10^{-6}$ of the total hadronic cross-section, hence HERA-B had to achieve
a background rejection around $10^{-11}$ for the 
$\Bz \to \jpsi \KS$ decay. 
 Data taking conditions were similar to those of the current \lhcb experiment (a 40 MHz interaction rate), as were the requirements of radiation resistance. 
 HERA-B started data taking in 2000 but it soon emerged that 
 the detector did not have sufficient rejection power against background and the track reconstruction 
 was not as efficient as expected.
 The large number of 
 detector stations and their total thickness  in terms of radiation lengths made secondary interactions 
 an important 
 issue for event reconstruction. Eventually  HERA-B could not observe $\bquark$ events efficiently, but taught several valuable lessons  for any future experiment working in a crowded hadron-collision 
 environment.
 These were a need of a robust and efficient tracking and a flexible trigger systems able to adapt to 
 harsher environments than may have been expected, as well as the 
 need to design the thinnest and lightest detector (in terms of radiation and interaction lengths).

\subsection{Towards the \lhc}
Over the same period, the planned \lhc and SSC machines, with their
large energies,  promised spectacular increases
of the $\bbbar$ cross-section, thus making the task of
background rejection much simpler. 
This was particularly true for operation in collider mode,
but even in fixed-target mode ($\sqrt{s} \approx \mathcal{O}(130)$~GeV) the $\bquark$ cross-section was
expected to be a respectable $1 ~\mu$b at the \lhc and $2 ~\mu$b at
the SSC \cite{Berger:1993yw}. 
The large cross-section and corresponding good
background rejection would facilitate a hadron \bfactory, competitive and complementary with
$\epem$  colliders, which could focus primarily on the measurement 
of \CP violation and also allow  the study of the spectrum
of all $\bquark$ particles.  Given the intrinsically 
``democratic'' nature of hadronic production, the new hadron machines would also give access
to large samples of $\Bs$ and of $b$ baryons, something
not possible at $\epem$ colliders operating at the $\PUpsilon$ resonances. 
  
Two schools of thought soon emerged: one pursuing a fixed-target (FT) strategy
and the other based on a collider mode. The more favourable ratio of
the $\bbbar$ to the total hadronic cross-section, about two orders
of magnitude larger in collider mode,  gave this a competitive advantage.

There was, however, a strong reason in favour of the FT concept:
a  collider $\PB$ experiment could not operate at the design luminosity of the machine
($10^{33-34}\, {\rm cm}^2{\rm s}^{-1})$ because of the significant number of 
overlapping interactions (pile-up) with multiple vertices. This would have  
required dedicated low-luminosity 
running, creating a potential conflict with the major experiments
and considerably
reducing the data-taking time. Later it was ascertained that 
the individual experiment luminosities could be tuned
over a broad range with an appropriate design of beam optics in the
interaction regions, hence this would become a moot point, but at the time it was a serious one.    

  Moreover, for the advocates of the FT approach, the  
  advantage of the larger cross-section in collider mode was partially
offset by the higher event multiplicity
and by the shorter flight path of beauty particles.
In addition, while the $\pt$ of the
$\bquark$ decay products would to a good approximation be the same in the
two modes, the $\pt$ of the other collision products would be smaller in FT mode,
thus making the trigger simpler. 
Active silicon targets were also possible with an extracted beam, where
the track of a charged
$\bquark$-hadron would be directly measured. 

  Finally, $\bbbar$ production kinematics is forward peaked
in the centre-of-mass (CM) system,
but the Lorentz boost of the centre-of-mass in the FT mode ($\beta \gamma > 60$) 
concentrates the event at smaller angles
than in collider mode. It was therefore possible, in principle,  to build a more
compact detector,  achieving larger angular acceptance at lower cost. There was also the
possibility of recycling components (in particular dipole magnets)
from existing detectors. 
The cost was an important consideration, since a
dedicated $\PB$ experiment at the \lhc (or SSC) was generally considered  to have secondary importance
with respect to the general-purpose experiments.

\begin{figure}[!htbp]
\begin{center}
\includegraphics[height=6cm]{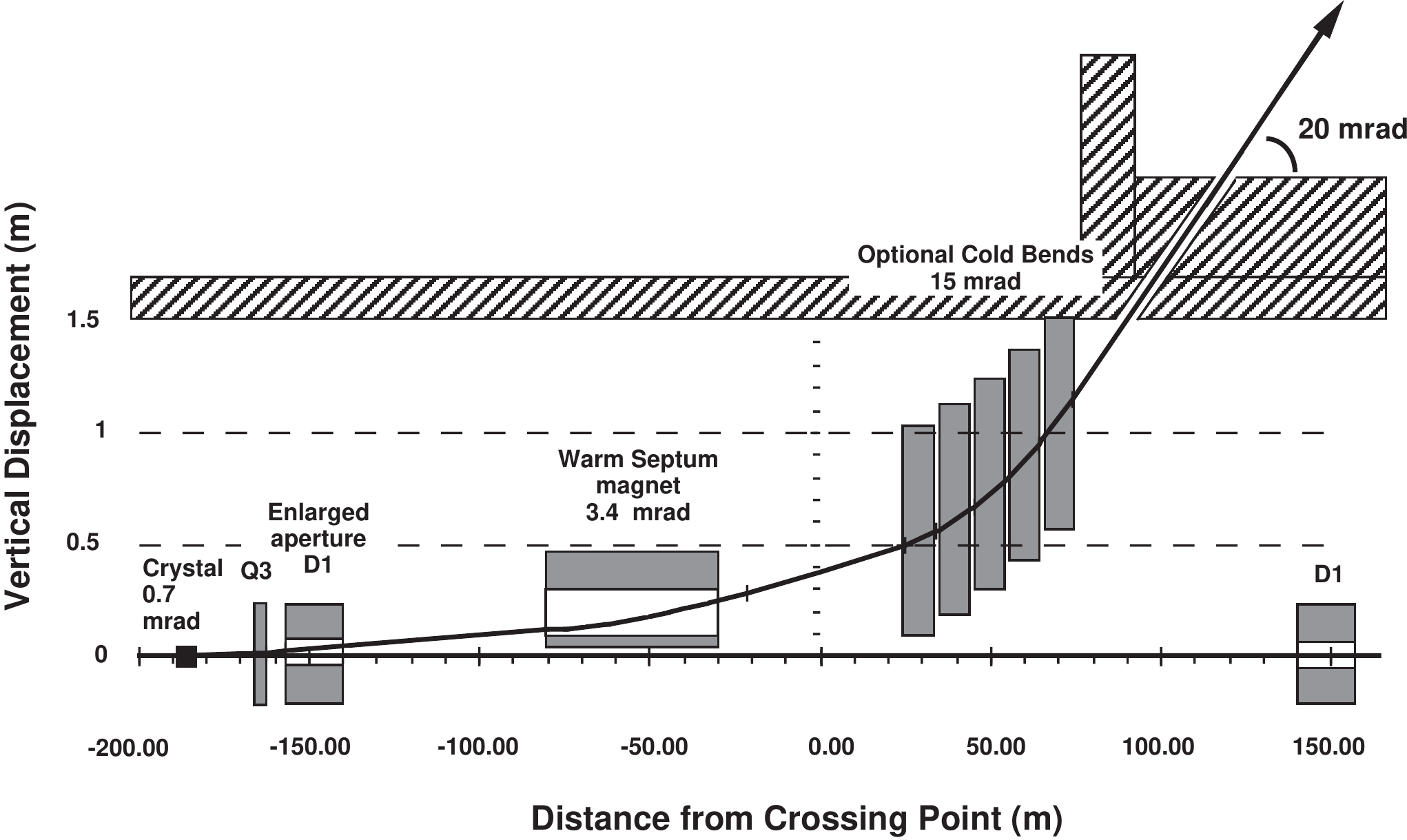}
\end{center}
\caption{Proposed LHC beam extraction scheme  based on crystal channeling.
The channeled beam, deflected by 0.7 mrad in a bent silicon crystal,
was guided towards the beam transport tunnel 
by several conventional magnets. The cost of the extracted beam was
a non-negligible fraction of the detector cost.}
\label{fig:extraction}
\end{figure}

The anticipated  demise of the SSC  led three groups to study and propose dedicated
$\bquark$ experiments at the \lhc. The three Letters of Intent (LoI)  were presented in 1993. 
COBEX \cite{Erhan:1994tt}, an acronym for Collider Beauty Experiment,
with P. Schlein as spokesperson, was a collider experiment with
a backward-forward geometry. The other two proposals,
Gajet \cite{Dauncey:1994vn} (spokesperson T. Nakada) and LHB
\cite{Waldi:1994jx} (spokesperson G. Carboni)
were fixed target experiments, the first,
as the name suggests, using a gas jet target,
the latter exploiting an extracted beam. 
Since no traditional beam extraction was foreseen for the \lhc, LHB (Large Hadron Beautyfactory) 
used a parasitic extraction technique, based on channeling in a bent
silicon crystal placed close to one of the circulating beams
(Fig.~\ref{fig:extraction}).

\noindent A dedicated R\&D experiment, RD22, approved by the CERN DRDC \cite{Jensen:291236} 
to test the feasibility of this idea at the SPS, demonstrated
that high-efficiency beam extraction (larger than 10\%) was possible \cite{Akbari:1993mb}.
The three proposed experiments 
were presented in their final form in 1994 at the Beauty'94 Conference~\cite{Schlein:1994kqa}. 
It should also be noted that, by then, CDF had already 
contributed important $\bquark$-physics results, offering a glimpse of what would later prove to be the extraordinary
success of $b$-physics at hadronic machines. 
 
Following the submission of the LoIs,
the \lhc Committee (LHCC) considered the proposals in June 1994.
One of its concerns was the small size of the three collaborations: the total number
of physicists involved was barely one hundred. In addition the
  Committee remarked that beam extraction by channeling could not be guaranteed at that stage,
  because accelerator experts feared possible interference with  normal \lhc operation.
  Finally the following recommendation was issued:
\textit{``The collider model approach has the greater potential in view of the
very high rate of $\bquark$ production, the much better
signal/background ratio and the possibility of exploring other
physics in the forward direction at 14 TeV''.}

\noindent The LHCC  encouraged the three collaborations to join together and design a new experiment,
incorporating attributes of each, and operating in collider mode. The LHCC noted that 
its ambitious request was justified by the fact that, at the startup of the \lhc, the  experiment
would not be an exploratory one, since \CP violation would already have been observed 
``at HERA-B, FNAL or \bfactories''.
The LHCC also issued guidelines requiring a number of issues to be
addressed and solved. The three experiments combined into the new
LHC-B Collaboration (then named), which published its Letter of Intent in 1995~\cite{Dijkstra:1995cha}.

  The new experiment derived several of its characteristics from the parent proposals: notably
  the concept of a silicon vertex detector in retractable pots, the calorimeter design,
  and the high $\pt$ first-level trigger.

\begin{figure}[!htbp]
\begin{center}
\includegraphics[width=0.9\textwidth]{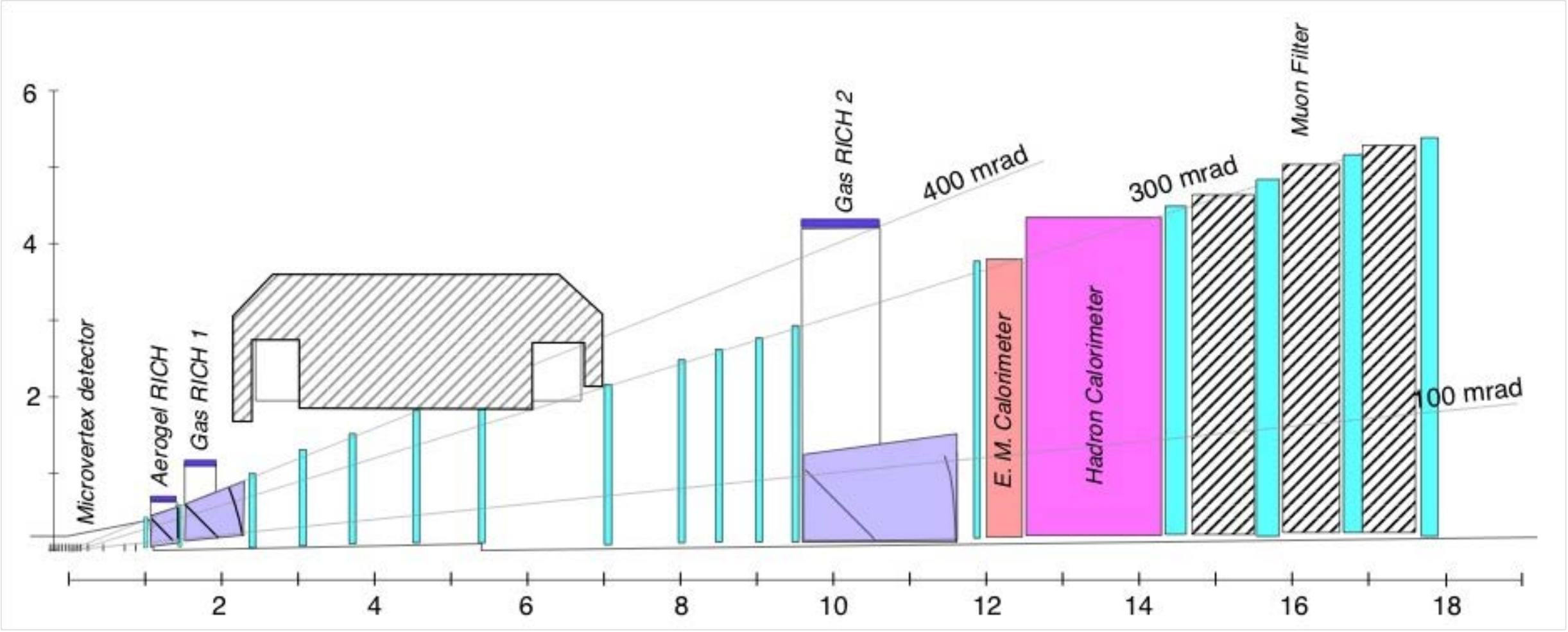}
\end{center}
\caption{The LHC-B detector as it was proposed in 1995. All the basic
  components shown would be part of the final detector,
  albeit with  many refinements and optimisations that will be
  described in the following Section. 
}  
\label{fig:LHC-B}
\end{figure}

  In the transition to LHC-B, a number of proponents from the former
  three collaborations decided not to continue. This 
  unfortunately included
P. Schlein and his UCLA colleagues who had pushed very strongly for the collider-mode idea  with COBEX, and who had been a driving force up to that point.
The LHCC decision sent a strong signal to the high-energy physics 
community: that CERN were prepared to give their strong support to one   dedicated $\PB$ experiment. 
This encouraged many  physicists from institutions around 
the world to join the new collaboration over the subsequent years.  
T. Nakada, who was an instigator of the Gajet proposal, was elected LHC-B spokesperson.
The 1995 Letter of Intent of LHC-B established the basis for the new detector design, which was
refined in the following years, until its approval in 1998. By that time, the name
had changed  from LHC-B to simply \lhcb. Fig.~\ref{fig:LHC-B} shows the detector as it appeared in the
Letter of Intent.

Soon afterwards, a competing $\B$-physics  
experiment, BTeV~\cite{Kulyavtsev:2000zz}, was proposed to run at the Fermilab Tevatron,
incorporating  a single magnet,
a double arm spectrometer and a vertex trigger at the first level,
in order to recover the reduced $\bquark$  production cross section.
Following the decision to shut down the Tevatron, BTeV was not approved,
however several of the experiment's innovative ideas were carried through
to the future \lhcb experiment.

\clearpage

\section{ The \lhcb Detector}
The basic mechanism for heavy-quark production at the \lhc is via gluon-gluon fusion. The angular distribution of $\ccbar$ 
or $\bbbar$ pairs is peaked at small angles with respect to the 
beam-line, with high correlation between the constituents of the pair. 
This allows the detection with good acceptance of the
resulting hadrons in a rather limited solid angle. QCD calculations give cross-section values of
$\sigma_{\ccbar} \simeq 1.5$ mb and $\sigma_{\bbbar} \simeq 0.5$ mb, respectively. 
The large event multiplicity requires a high granularity of the detector, together with 
minimal thickness in terms of radiation and interaction lengths to reduce secondary interactions. 

The  LHC-B LoI~\cite{Dijkstra:1995cha} presented a forward spectrometer with 400 mrad acceptance and a single 
large dipole magnet. The 
apparatus would be located inside the former DELPHI cavern at LEP with little modifications to the 
existing infrastructure but, for reasons of cost, 
the detector sacrificed half of the solid angle by only being a {\em single} arm spectrometer. The LoI design 
 inherited important features from the three ancestor experiments and from the contemporary
HERA-B, which had rate and radiation issues similar to those expected at the \lhc.  In contrast to the
latter experiment,  LHC-B had in addition a hadron calorimeter and a second (upstream) RICH with two radiators. 
Initially it was thought that an efficient tracking system in a harsh environment would require a large number 
of tracking stations, so, paralleling HERA-B,  LHC-B had  twelve  tracking stations in the large-angle region. This
number was reduced to ten in the Technical Proposal presented in 1998~\cite{Amato:1998xt2}, which by then had changed its name to \lhcb. 

The disappointing performance of HERA-B was largely 
ascribed to the 
large amount of material in the detector, which prompted the \lhcb 
collaboration to perform a thorough review of the 
apparatus, with the aim to reduce material without sacrificing performance. 
A Technical Design Report submitted in 2003 presented the \lhcb ``Reoptimized'' Detector~\cite{LHCb:2003ab}. This is the basis on which the experiment was eventually built and is described in the following
subsections.

\subsection{Overview}
\label{sez:detector}

\begin{figure}[!htbp]
\begin{center}
\includegraphics[height=7cm]{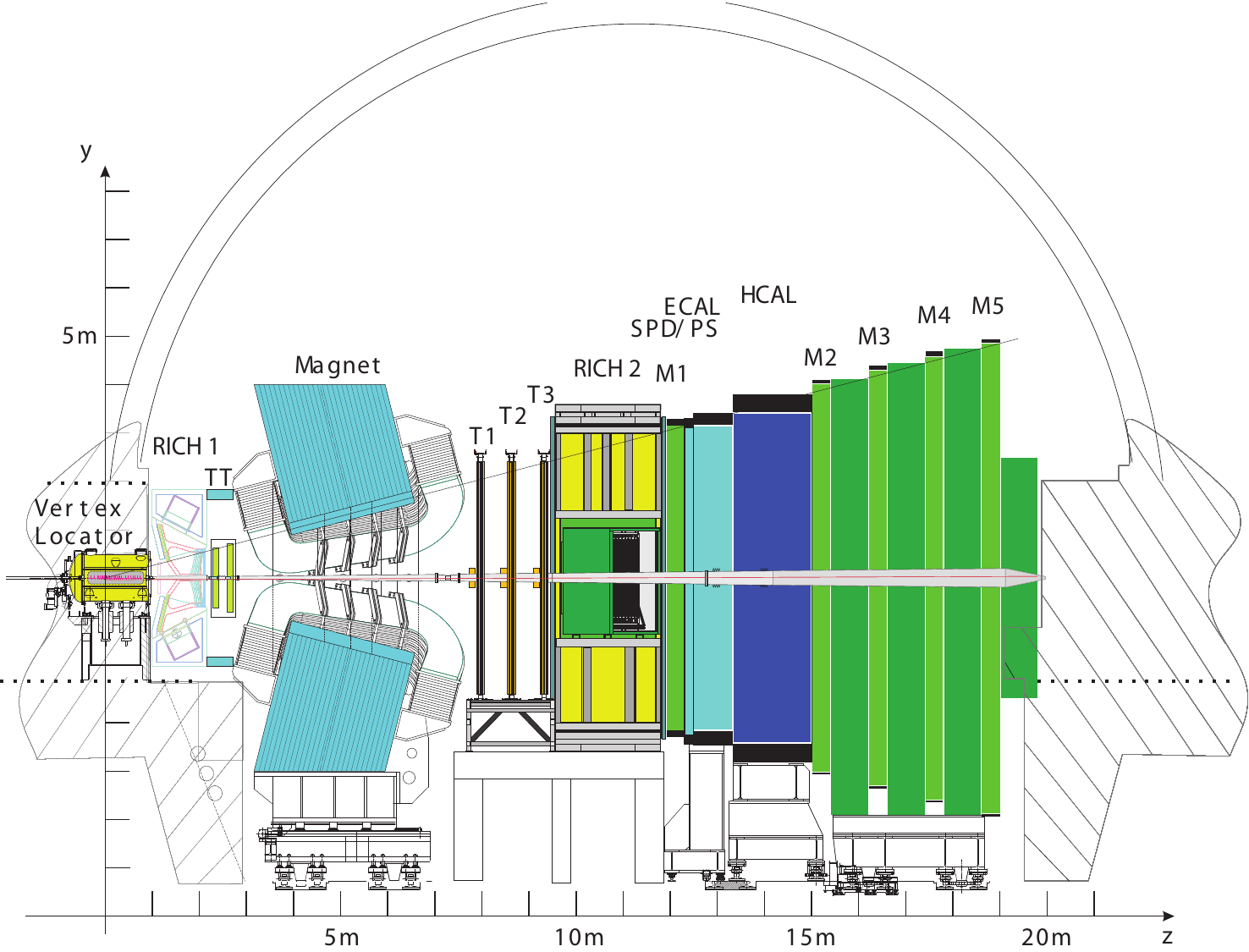}
\end{center}
\caption{The LHCb detector: side view}\label{fig:lhcb_dect}
\end{figure}

\begin{figure}[!htbp]
\begin{center}
\includegraphics[height=7cm]{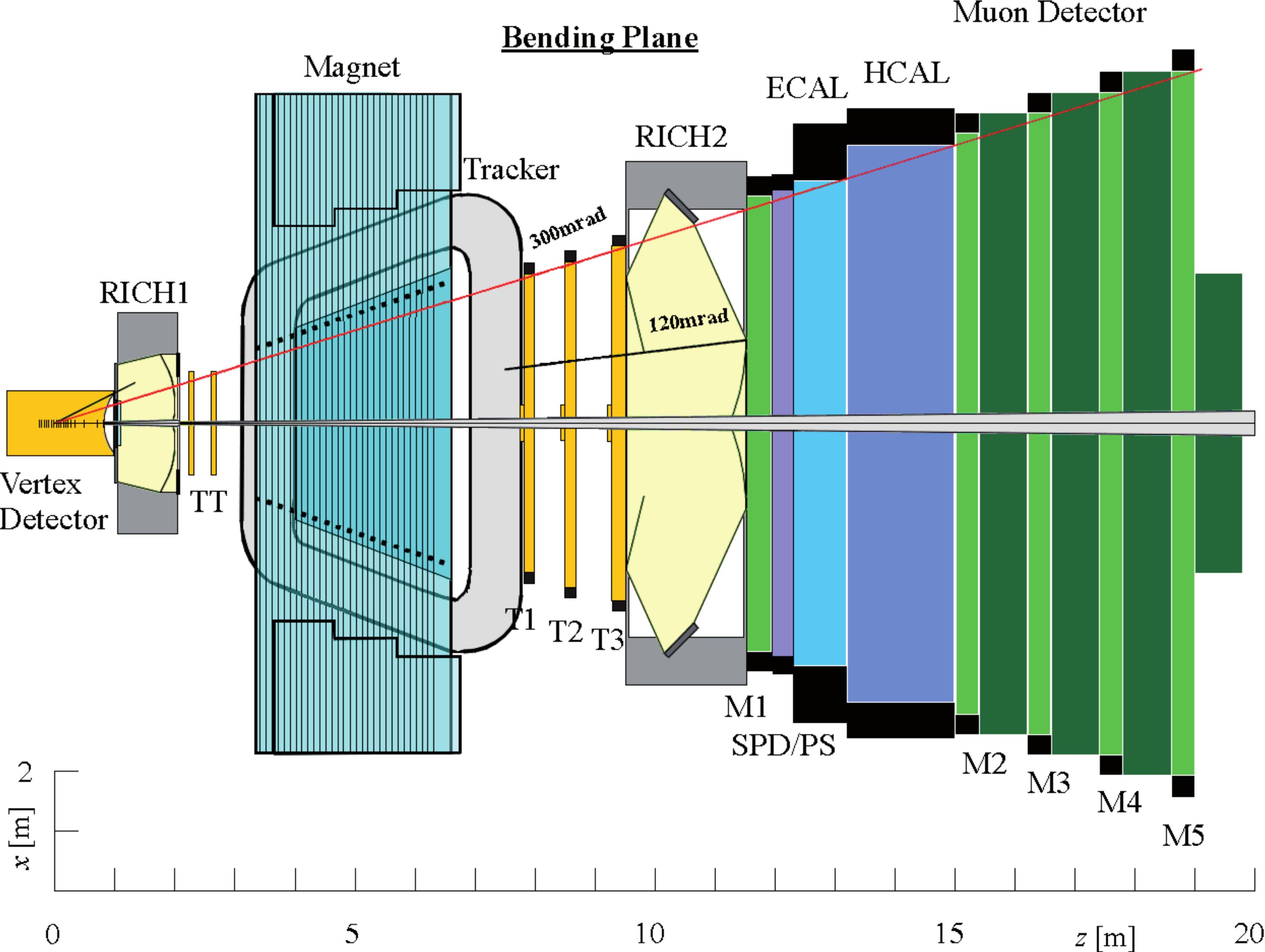}
\end{center}
\caption{The LHCb detector: top view}\label{fig:bending}
\end{figure}

The \lhcb detector~\cite{LHCb-DP-2008-001}  is a forward spectrometer, shown in Fig.~\ref{fig:lhcb_dect}, and is 
installed at Intersection Point 8 of the \lhc.
A modification to the \lhc optics,  shifting the  interaction point
by about 11~m from the centre, allowed maximum use of the cavern space. This results in a detector
length of approximately 20~m, and with maximum transverse dimensions about 
$6 \times 5 ~\rm{m}^{3}$. 
The angular acceptance ranges from approximately 
10~mrad to 300~mrad (Fig.~\ref{fig:bending}) in the horizontal magnetic-bending plane,  and 
from 10~mrad to 250~mrad in the vertical  plane 
(Fig.~\ref{fig:lhcb_dect}). With this geometry the detector is able to reconstruct
approximately 20\% of all $\bbbar$ pairs produced.

To measure the momenta of charged particles, a dipole magnet producing a vertical magnetic field is used. 
It is a warm magnet providing an integrated field of 4 Tm, with saddle-shaped coils in a window-frame yoke, and with sloping poles in
order to match the required detector acceptance. The 
design of the magnet allows for a level of fringe 
field inside the upstream Ring Imaging Cherenkov detector (RICH\,1, see Sec. \ref{sec:RICH}) of less than 2 mT whilst providing a residual field 
in the regions between
the upstream tracking stations.

The tracking system consists of a silicon VErtex LOcator detector (VELO)~\cite{LHCb-DP-2014-001},
surrounding the interaction region, and four planar tracking stations, the TT tracker
upstream of the dipole magnet and three tracking stations T1-T3 downstream of the
magnet~\cite{LHCb-DP-2013-003,LHCb-DP-2017-001}. The
T1-T3 stations consist of an Inner Tracker (IT), located at the centre of the stations and surrounding the beam-pipe, and an Outer Tracker (OT) for the outer regions. A minimum momentum of around 1.5\gevc is
required for a track to reach the downstream  stations~\cite{LHCb-DP-2014-002}.\\

\newpage

Particle identification (PID) is a fundamental to the goals of
the \lhcb experiment by separating pions, kaons and protons produced in heavy-flavour decays. Accurate  reconstruction of electrons and muons
is crucial for flavour tagging. All aspects of PID are accomplished by a set of specialized detectors.  

The PID system is based around two RICH detectors,  designed to cover almost the full momentum
range of tracks in \lhcb. The upstream detector, RICH\,1, covers the low momentum region from 
$\simeq 2$ to $60$ \gevc using aerogel (in Run~1 only) and $\hbox{C}_{4}\hbox{F}_{10}$  radiators, 
whilst the downstream detector, RICH\,2, covers the
high momentum range from 15 \gevc up to and beyond 100 \gevc using a $\hbox{C}\hbox{F}_{4}$ radiator. 

Two calorimeters, one electromagnetic (ECAL) and the other hadronic (HCAL), supplemented by a 
Preshower Detector (SPD/PS) \cite{LHCb-DP-2013-004},  provide identification of electrons, photons and hadrons
and a measurement of their energy. This measurement is used at the trigger level to select candidates on the basis of their
transverse energy. Muons play a crucial role in many of \lhcb's measurements because of the cleanliness of the 
signature. Their identification is achieved by five muon stations (M1 --  M5), interspersed with iron filters. The muon
system also supplies measurements of muon transverse momenta for the trigger. 


\subsection{The Vertex Locator}
\label{sec:velo}

The role of the VELO is to measure the impact parameters of
all tracks relative to the primary vertex (PV), to reconstruct the production points and decay vertices of hadrons containing $\bquark$- and $\cquark$-quarks and
to allow precision measurements of their mean lifetimes.
The subdetector accepts particles with pseudorapidities in the range
$1.6~<~\eta~<~4.9$ and which have PVs within $|z|~<~10.6$~cm from the nominal collision point along the beam direction. 
The VELO is split into two halves surrounding the beam-pipe, each containing 21 modules. Each module is then made up of
two silicon half discs of 300~$\mu$m thickness, one with strips in  the radial, $r$ coordinate, the other in the polar, $\phi$
coordinate. 
This cylindrical geometry allows 
a fast track- and vertex-reconstruction to be made at the second stage of trigger. The strip segmentation is such to
limit the highest occupancy of the strips to less than 1.1 \%. 

The VELO is positioned, with an accuracy better than 4 $\mu$m, at the closest distance possible from the beam, about ~7 mm during data taking, 
The sensors operate within a so-called \textit{Roman pot}  configuration, 
located inside a secondary vacuum of less than $2 \cdot 10^{-7}$~mbar
pressure, separated from the primary \lhc vacuum.
The sensors are retracted during beam injection and are then quickly moved in 
for physics operation when the \lhc
beams are stable.

The vessel containing the silicon discs and the front-end electronics (RF-box) 
has aluminium walls of 300~$\mu$m thickness to minimize multiple scattering. The average material budget
of the detector for tracks in the \lhcb acceptance is 0.22 $X_0$.
In order to minimize radiation damage and to dissipate the produced heat, a cooling
system keeps the temperature range between
-10 to 0$^\circ$C.

\begin{figure}[!htbp]
\begin{center}
\includegraphics[width=7cm]{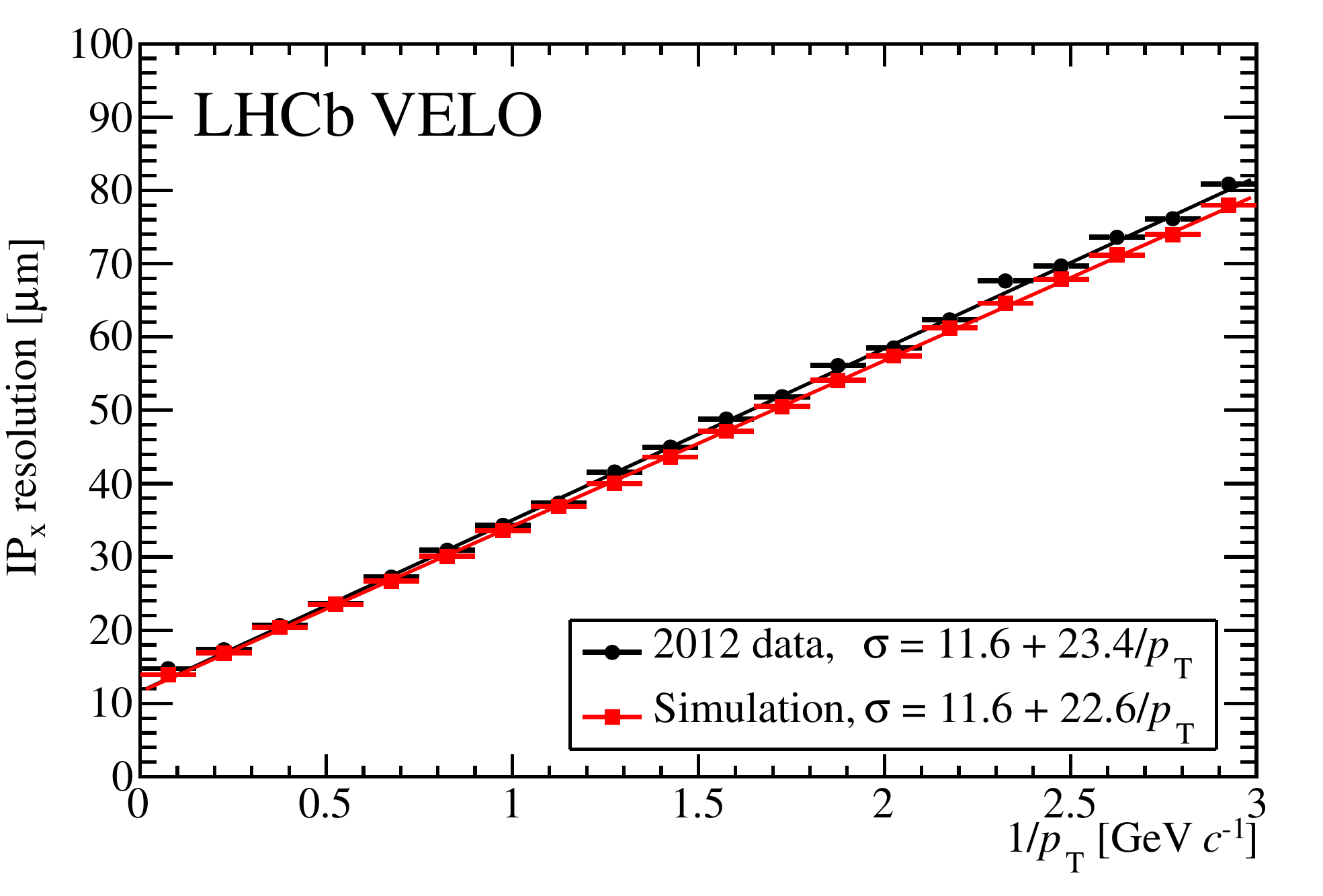}
\includegraphics[width=7cm]{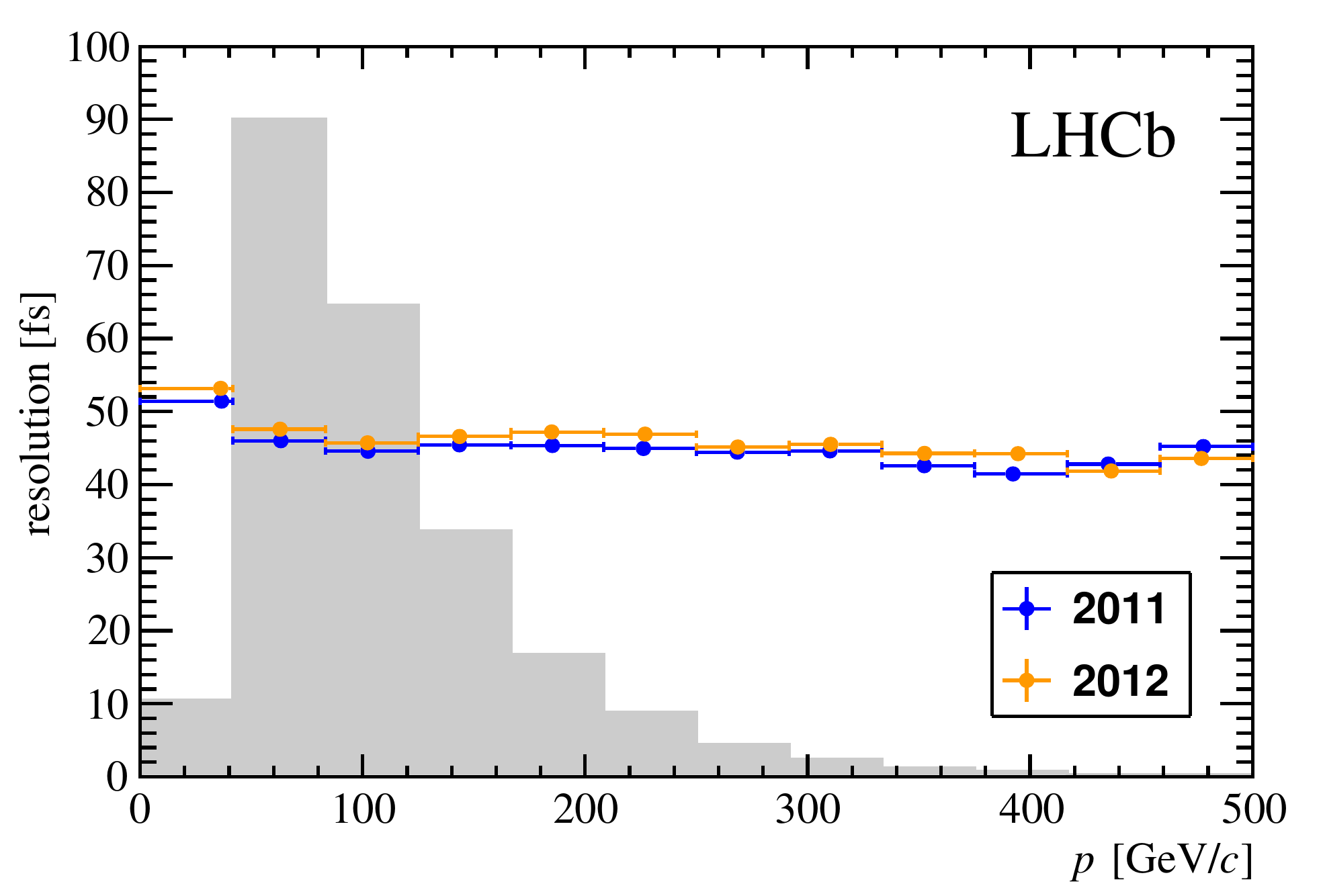}
\end{center}
\caption{Left: Projected impact parameter resolution as a function of $1/p_T$.
Right: Decay time resolution (points) as a function of momentum for $\BsToJPsiPhi$  decays. 
The superimposed histogram shows the distribution of momentum for the decay.}
\label{fig:velo_dz}
\end{figure}

Fig.~\ref{fig:velo_dz} summarizes the VELO performance in terms of impact parameter and decay time resolution \cite{LHCb-DP-2014-001}.  

\vspace{6mm}

\subsection{The TT and Downstream Tracking System}
Following the VELO, the tracking system  is composed of the TT station, located between RICH\,1 and the magnet, and three stations (T1,T2,T3) downstream of the magnet.
The TT is composed of four stations grouped in pairs, called TTa and TTb, spaced by 30 cm. Each station 
consists of silicon microstrip planar modules covering a rectangular area of 150~cm $\times$ 130~cm (width times height), 
covering the \lhcb acceptance of
300~mrad in the horizontal plane and 250~mrad in the vertical. The strips of the first and the fourth stations
are vertical  and measure the bending $x$ coordinate, whilst the
 second and  third planes have stereo angles of $\pm 5^\circ$, respectively.

\begin{figure}[!htbp]
\begin{center}
\includegraphics[height=6cm,width=16cm]{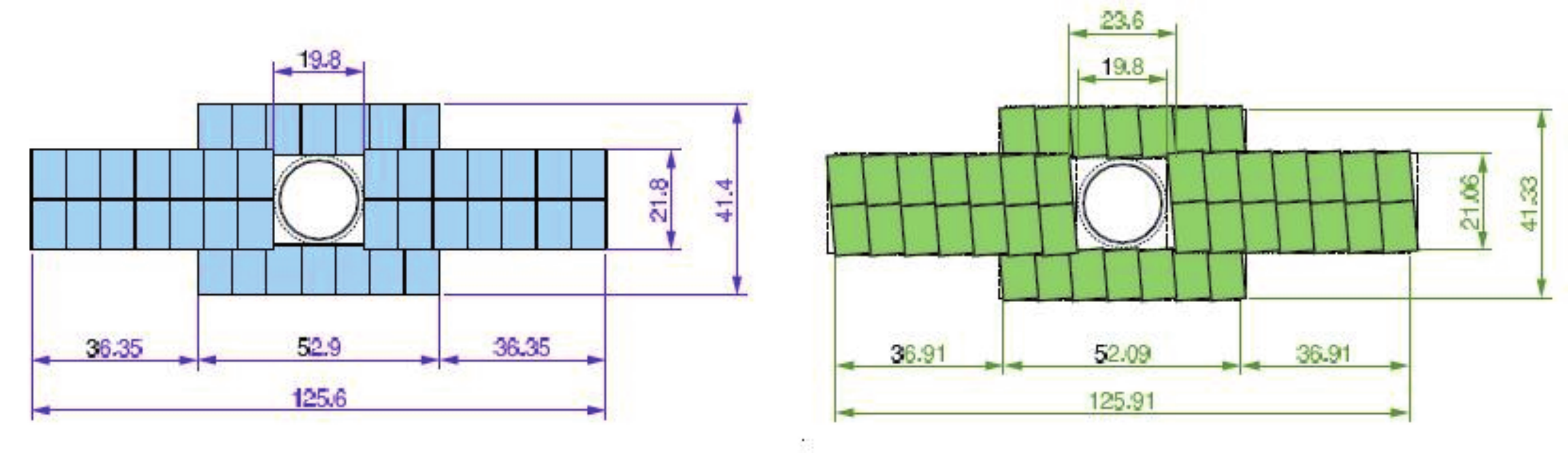}
\end{center}
\caption{The \lhcb inner detector, (left) a vertically-aligned layer, (right) a stereo layer.
 }
\label{fig:it}
\end{figure}

Tracking stations T1 - T3 
each consist  of an inner part (IT) surrounding the beam pipe, 
and an outer part (OT) beyond. 
 Each IT station consists
of four overlapping silicon layers, two rotated by a stereo angle of $\pm 5$ deg  and
two aligned to the $y$ (vertical) axis. Each layer is made up of four independent
modules placed around the beam pipe, covering about a $120 \times 40$~cm$^{2}$ area, as shown in Fig. \ref{fig:it}.

\vspace{6mm}

The spatial resolutions of both the TT and IT are approximately 50~$\mu$m per hit, 
with strip pitches of about 200~$\mu$m.
The hit occupancies 
vary between 1.9\% for the inner sectors to 0.2\% for the outermost modules.
To minimize radiation damage, the sensors operate at 5$^\circ$C temperature.

\begin{figure}[!htbp]
\begin{center}
\includegraphics[height=6cm]{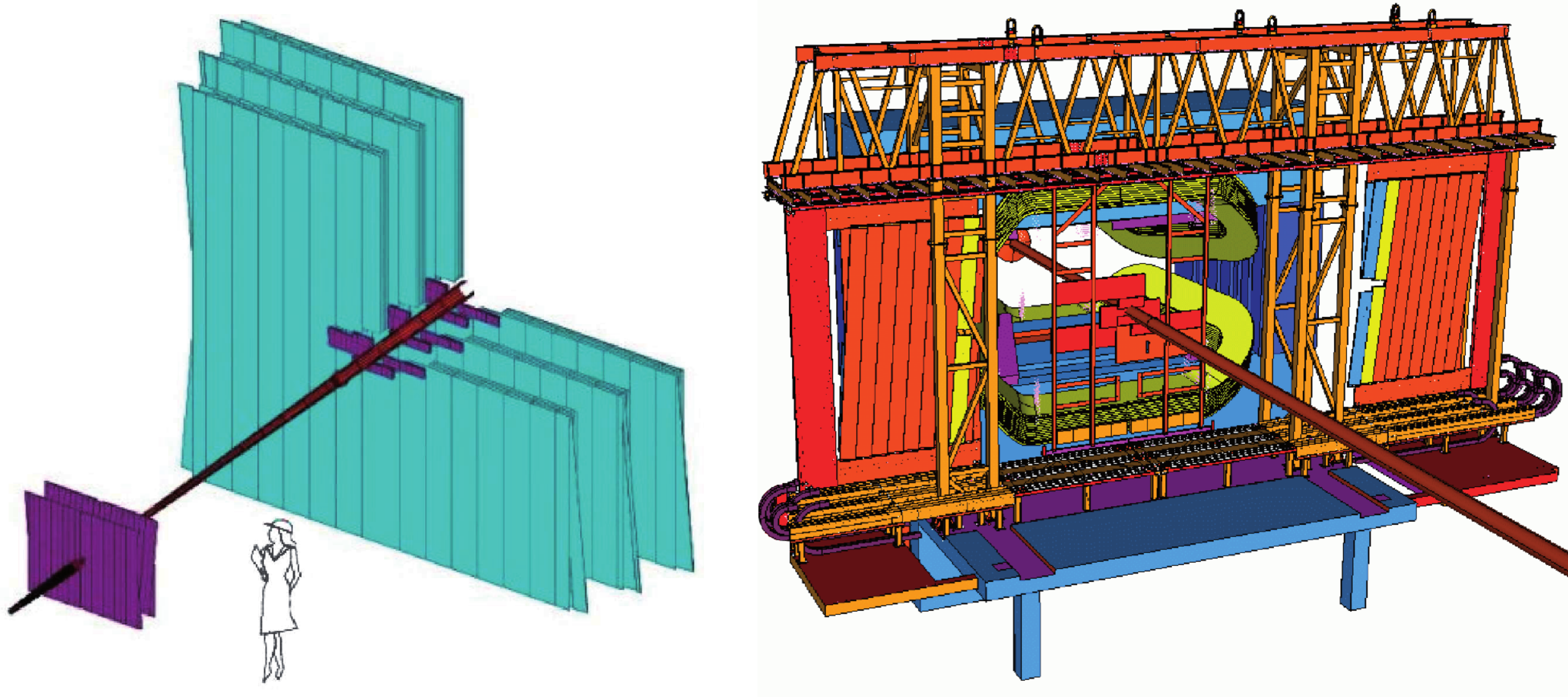}
\end{center}
\caption{Schematic views of the \lhcb outer detector.
 }
\label{fig:ot}
\end{figure}

\begin{figure}[!htbp]
\begin{center}
\includegraphics[height=8cm]{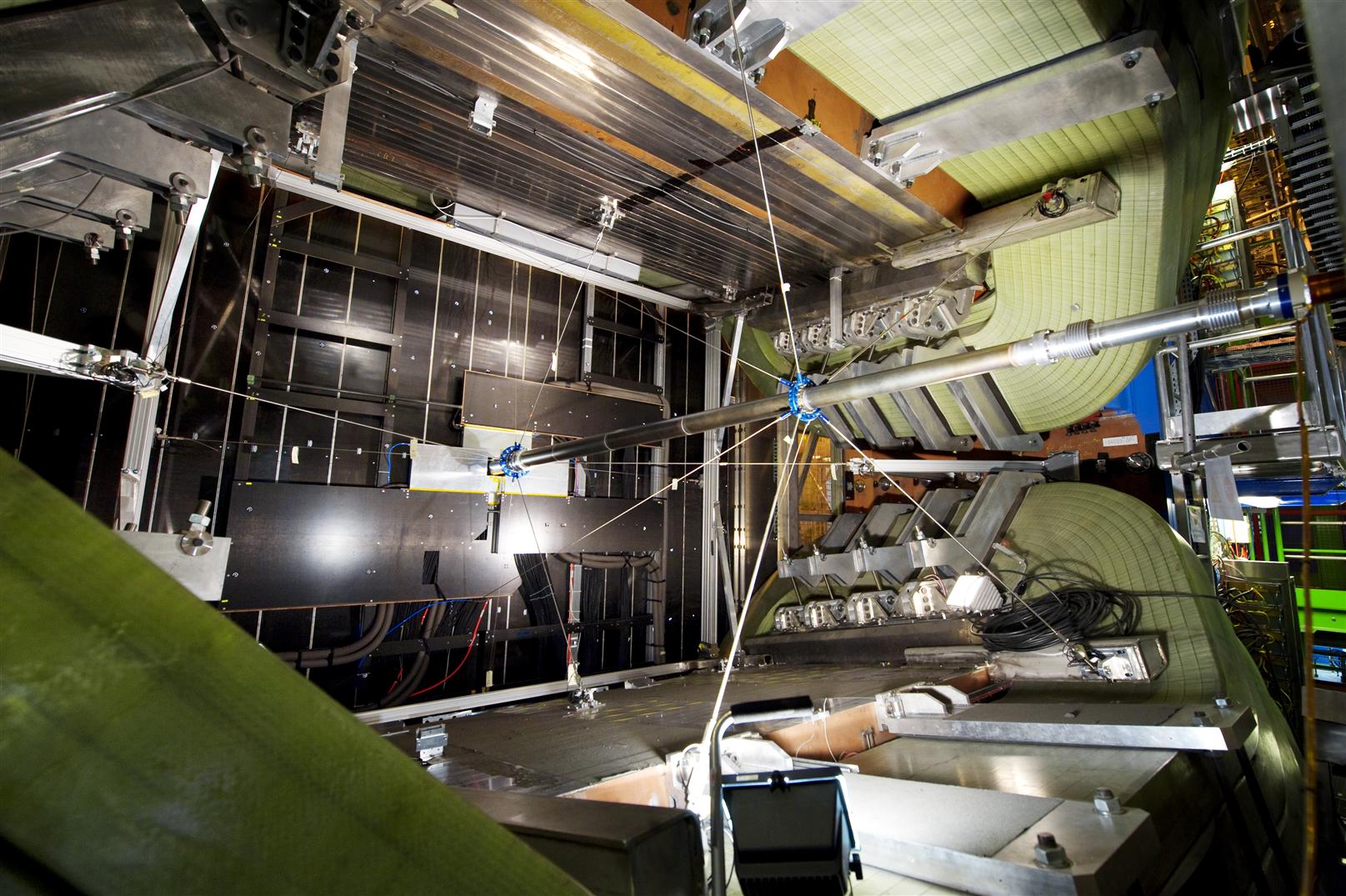}
\end{center}
\caption {The LHCb Outer Detector in place on the beam line. Well visible is the  beam pipe.
 }
\label{fig:picture-ot}
\end{figure}

The OT is a drift detector~\cite{LHCb-DP-2013-003}
consisting of straw tubes with internal diameters of 4.9 mm, each filled with an Ar/CO$_{2}$ gas mixture in a ratio $70-30$\%. The straws
provide a 35 ns maximum drift time and 205~$\mu$m spatial resolution with 17\% maximum straw occupancy.  Each of the three stations is  made of four modules, shown schematically in Fig.~\ref{fig:ot}. 
A picture of the assembled OD is shown in Fig.~\ref{fig:picture-ot}.
In the first and third modules the straw tubes are aligned to the vertical axis while  the third and fourth modules have stereo angles of $\pm 5$ deg.
The total active area is about
$ 5.97 \times 4.85 \; \msq$, 
covering the full \lhcb acceptance.

The overall tracking efficiency for ``long'' tracks (i.e. those tracks measured in all the tracking detectors
including the VELO) is greater than 96 \% for $5 < p < 200 \; \gevc$. The momentum resolution $dp/p$ is 0.5\% at low
momentum, increasing to 1.1\% at $240~\gevc$. The mass resolution is $14.3~\mevcc$ for the $\jpsi$ resonance.

\vspace{6mm}

\vspace{6mm}

\subsection{The RICH system}
\label{sec:RICH}

The role of the RICH system \cite{LHCb-DP-2012-003} is to provide $\pi/K/p$ discrimination for \lhcb, which is essential for most \CP-violation studies, background rejection and flavour tagging.
The momentum range which contains 90\% of kaons, pions and protons from $\B$ meson decay is between 2 and 150 \gevc and, to achieve this  separation,
two Cherenkov detectors, RICH\,1 and RICH\,2, are employed. 

The RICH\,1 detector differentiates particles with low
and intermediate momenta, from 1 to $\sim$60 \gevc. It  
is located close to
the interaction region, upstream of the magnet, and covers the acceptance
from $\pm 25$ mrad to
$\pm 300$ mrad (horizontal plane) and to $\pm 250$ mrad (vertical plane).
RICH\,1 initially contained two
different radiator materials: an aerogel layer 5~cm thick with refractive index
$n=1.03$ and a $\hbox{C}_{4}\hbox{F}_{10}$ gas layer of 
length 85~cm  with refractive index $n=1.0014$.
Aerogel has the power to provide $\pi/K$
discrimination from about 1 up to 10 \gevc, however it was removed for
Run~2 due to occupancy problems. The $\hbox{C}_{4}\hbox{F}_{10}$ radiator extends
the positive $\pi/K$ identification from about 10 \gevc to   60~GeV/c, however $\pi/K$ discrimination below 10 \gevc is still possible by operating the RICH in kaon veto mode.

RICH\,2 has a smaller angular acceptance of $\pm15$ mrad to  $\pm 120$ mrad (horizontal plane) and to $\pm 100$ mrad (vertical plane) and covers the region where high momentum particles are most abundant. It is located downstream of the magnet,   between T3
and the first muon station M1.
RICH\,2  uses a $\hbox{CF}_{4}$ gas radiator with refractive index $n=1.00046$. 

In both detectors, Cherenkov photons are detected by a combined system of plane and spherical mirrors  
to focus photons onto a pair of photo-detector planes, where Hybrid Photon Detectors are employed to detect the Cherenkov rings.
The photo-detectors
are located outside the detector acceptance, in regions of low magnetic field and relatively low radiation.

\begin{figure}[!htbp]
\begin{center}
\includegraphics[height=5cm]{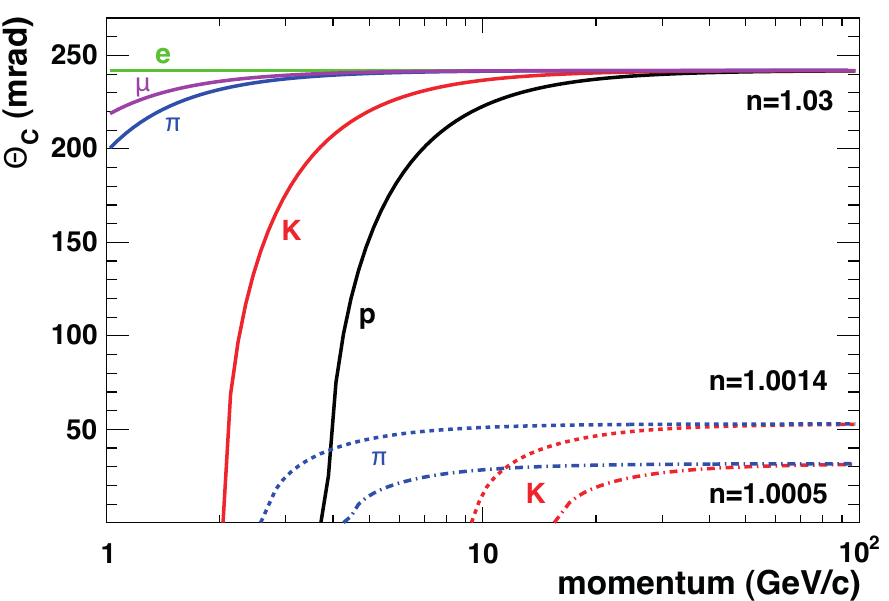}
\includegraphics[height=5.5cm]{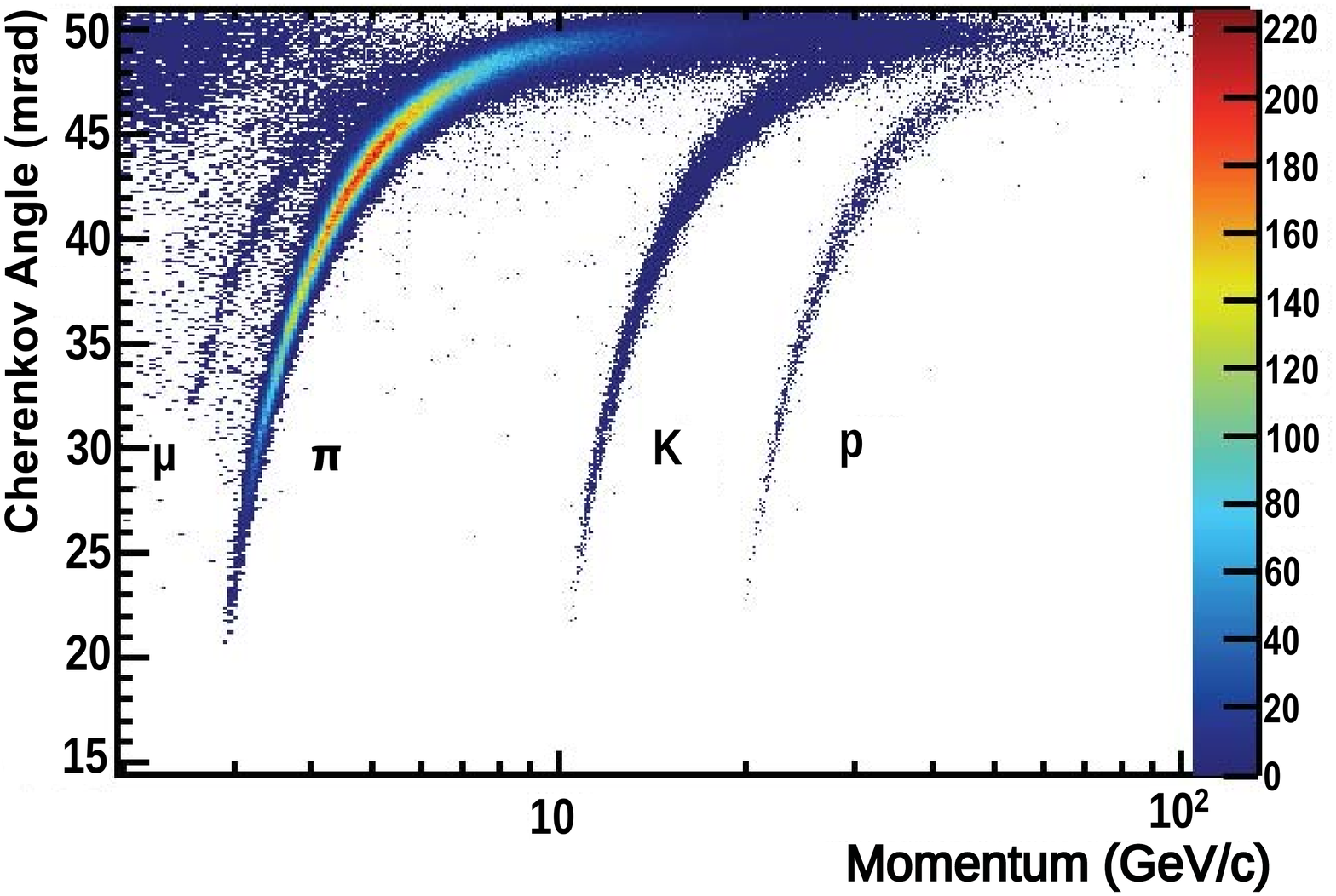}
\end{center}
\caption{(Left) Values of Cherenkov angle as a function of momentum for different particles
for the three RICH radiators of refractive index $n$. (Right) Measured Cherenkov angles in \lhcb data \cite{LHCb-DP-2012-003}.
}
\label{fig:pid_rich_2}
\end{figure}

The Cherenkov angles for the three different RICH radiators and for different particles as a
function of  momentum are shown in Fig.~\ref{fig:pid_rich_2} (left).
A measurement of RICH performance in \lhcb data is shown for the two gaseous radiators In Fig.~\ref{fig:pid_rich_2} (right) \cite{LHCb-DP-2012-003}.
\vspace{6mm}

\subsection{Calorimeters}

\begin{figure}[!htbp]
\begin{center}
\includegraphics[height=6cm,width=16cm]{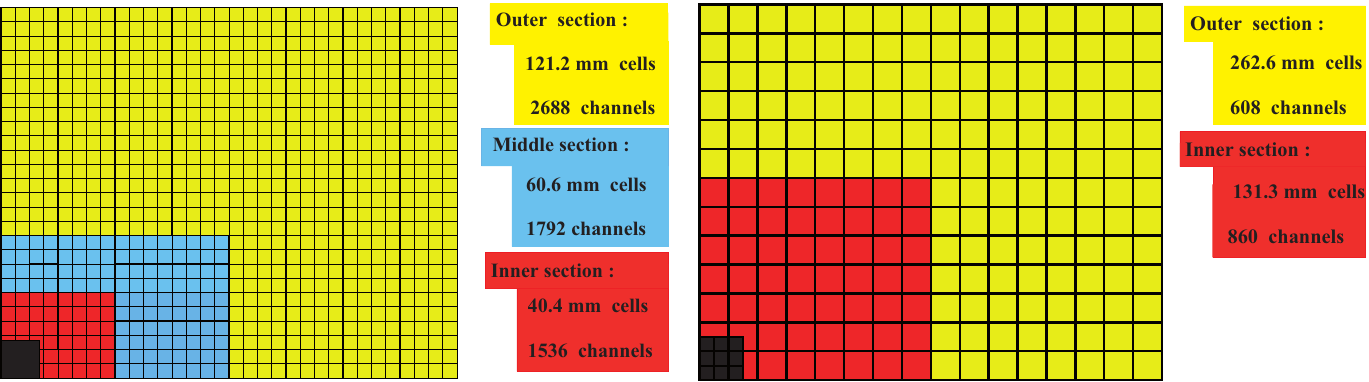}
\end{center}
\caption{The  (left) \lhcb electromagnetic calorimeter
and (right) hadronic calorimeter. The bottom left regions indicate the areas occupied by the beam-pipe.
 }
\label{fig:cal}
\end{figure}

The calorimeter system identifies hadrons, electrons and photons, and also measures their energies
and positions for the Level-0 trigger.
The  system is composed of four sub-detectors:
the scintillator Pad Detector (SPD),
 the PreShower detector (PS),
 the Electromagnetic Calorimeter (ECAL), and the
Hadron Calorimeter (HCAL).
The SPD and PS are located just upstream of the ECAL.  The ECAL, PS and SPD are segmented into three sections in the $xy$ plane, with active pads growing from the inner to the outer regions. The
HCAL is similarly divided in two sections. 
The corresponding granularities are outlined in Fig.~\ref{fig:cal}.

\vspace{6mm}

The SPD and PS are used at the
trigger level and offline, in association with the ECAL, to indicate the presence of electrons, photons
and neutral pions.
The detectors have two plastic scintillator layers separated
by a 15~mm thick lead plate where electrons and photons can radiate; the downstream scintillator then
samples the radiated energy. 
The light from the scintillators
is sent to photomultipliers by wavelength-shifter (WLS) optical fibers.

\begin{figure}[!htbp]
\begin{center}
\includegraphics[height=6cm]{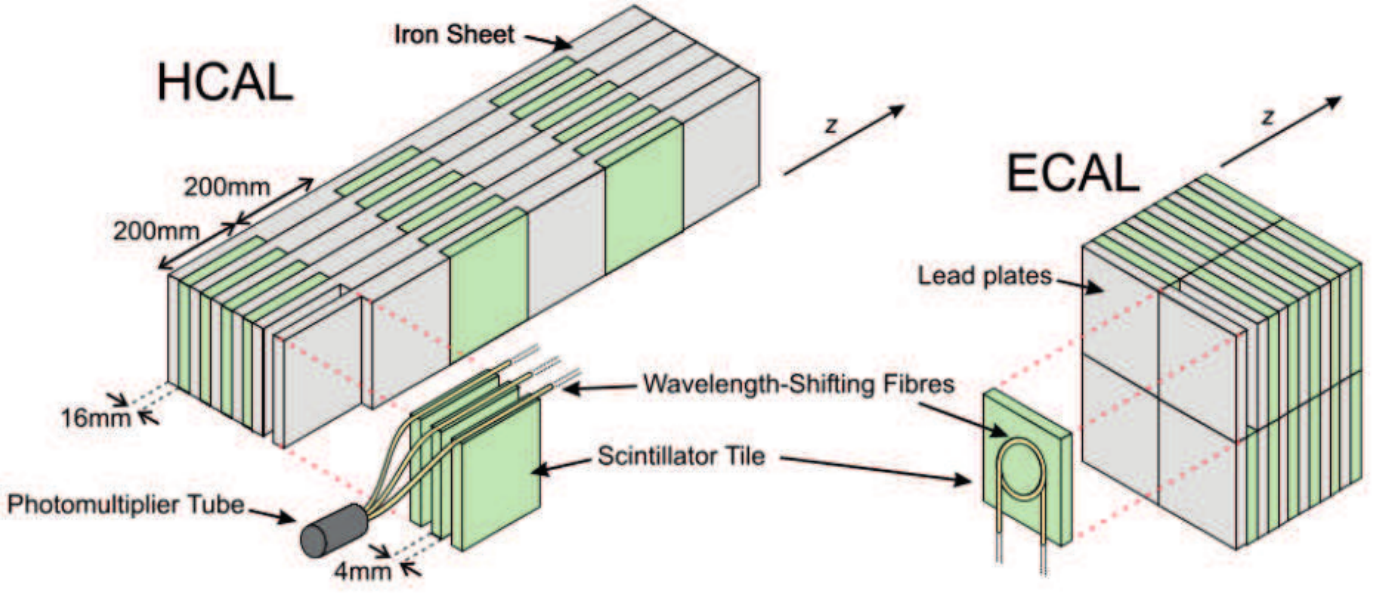}
\end{center}
\caption{A schematic showing the layout and segmentation of the \lhcb ECAL and HCAL calorimeters.
  }
\label{fig:ecal-hcal}
\end{figure}

The ECAL employs the \textit{Shashlik} technology, where independent modules,
constructed from scintillating tiles and lead plates, are alternated (see Fig.~\ref{fig:ecal-hcal}). The ECAL has  $66$ layers of such modules consisting of 2~mm of lead
followed by 4~mm of scintillator material. 
The ECAL also uses WLS optical fibers to guide the light from the detector to photomultipliers,
placed on the back face of each module. The energy resolution achieved \cite{LHCb-DP-2013-004} is
\begin{equation}
\frac{\sigma\left(E\right)}{E}=\frac{10\%}{\sqrt{E}}\oplus 1\%
\end{equation}
where $E$ is the electron energy expressed in \gev.

The HCAL is also a sampling calorimeter.  It is constituted of iron absorber with scintillating tiles as the  active material. The innovative feature of this sampling
structure is the orientation of the scintillating material: the tiles run parallel to the beam axis.
In the lateral direction, tiles are spaced with 1~cm iron, while longitudinally
 the length of the tiles and iron spacers correspond to the hadron interaction length $\lambda_{I} \simeq 20$ cm in steel.
Light is collected by WLS optical fibres running along the detector towards the back
side where the photomultiplier tubes are located (see Fig.~\ref{fig:ecal-hcal}).

\vspace{6mm}

The HCAL is used to measure the hadronic shower transverse energy for the Level-0 trigger and
to improve the high momentum electron/hadron separation.
The energy resolution achieved is
\begin{equation}
\frac{\sigma\left(E\right)}{E}=\frac{(69 \pm 5) \% }{\sqrt{E}}\oplus(9 \pm 2)\% 
\end{equation}
where $E$ is the hadron energy in \gev.

\subsection{The Muon System}
\label{sez:rivelatore_mu}

The Muon System  consists of five stations, 
M1-M5, of rectangular shape. The complete system is made up by 1368 Multi Wire Proportional Chambers 
supplemented by 12 Triple GEM Chambers in the inner region of the first   station, to cope with the 
very high particle rate. The chambers employ a variety of readouts, optimized for a precise
$p_T$ measurement for the trigger. 
The complete system has an acceptance in the bending plane from 20~mrad to
306~mrad, and in the non-bending plane from 16~mrad to 258~mrad. This results in a total
acceptance of about 20\% for muons from semileptonic inclusive $\bquark$ decays.

The  M1 station is located in front of the calorimeters and is used in order to improve the $p_{t}$ measurement
for the trigger. The geometry of the five stations is projective;
all the transverse dimensions scale as the distance from the interaction point. Stations M2-M5 are placed downstream of the calorimeters and are
interleaved with 80~cm thick iron absorbers. The total absorption thickness, calorimeters included,
is about 20 interaction lengths. In this way the minimum momentum for muons
crossing the five stations is about 6 \gevc. 

Each muon station is designed to achieve an efficiency above 99\% in a 20 ns time window
with a noise rate below 1 kHz per physical channel, as
described in~\cite{LHCb-DP-2012-002}. To reach such an efficiency, four chamber layers per station are used
in M2-M5 (two layers in M1).
The time resolution is achieved by a fast gas mixture
Ar/CO$_{2}$/CF$_{4}$ in the ratio 40:55:5. A ratio 45:15:40 is employed in the Triple GEM chambers.

\subsection{The trigger}
\label{sez:Trigger}

Even with the relatively large $\bbbar$ cross-section at \lhc energies, 
only approximately 1\% of visible \proton\proton interactions result in a $\bbbar$ event.
Moreover, only about 15\% of those events will produce at least one \bquark-hadron
with all  decay products passing within the acceptance of the spectrometer. The
branching fractions of decays used to study \CP violation are typically less than
$10^{-3}$. Further reductions are unavoidable in the offline selection, where stringent cuts must be applied
to enhance signal over background. Therefore 
the purpose of the \lhcb trigger is to achieve the highest efficiency for the events later selected
in the offline analysis while rejecting drastically most of the  uninteresting background events.
To achieve this goal, the  trigger uses information from all \lhcb sub-detectors. 

\begin{figure}[!htbp]
\begin{center}
\includegraphics[height=8cm]{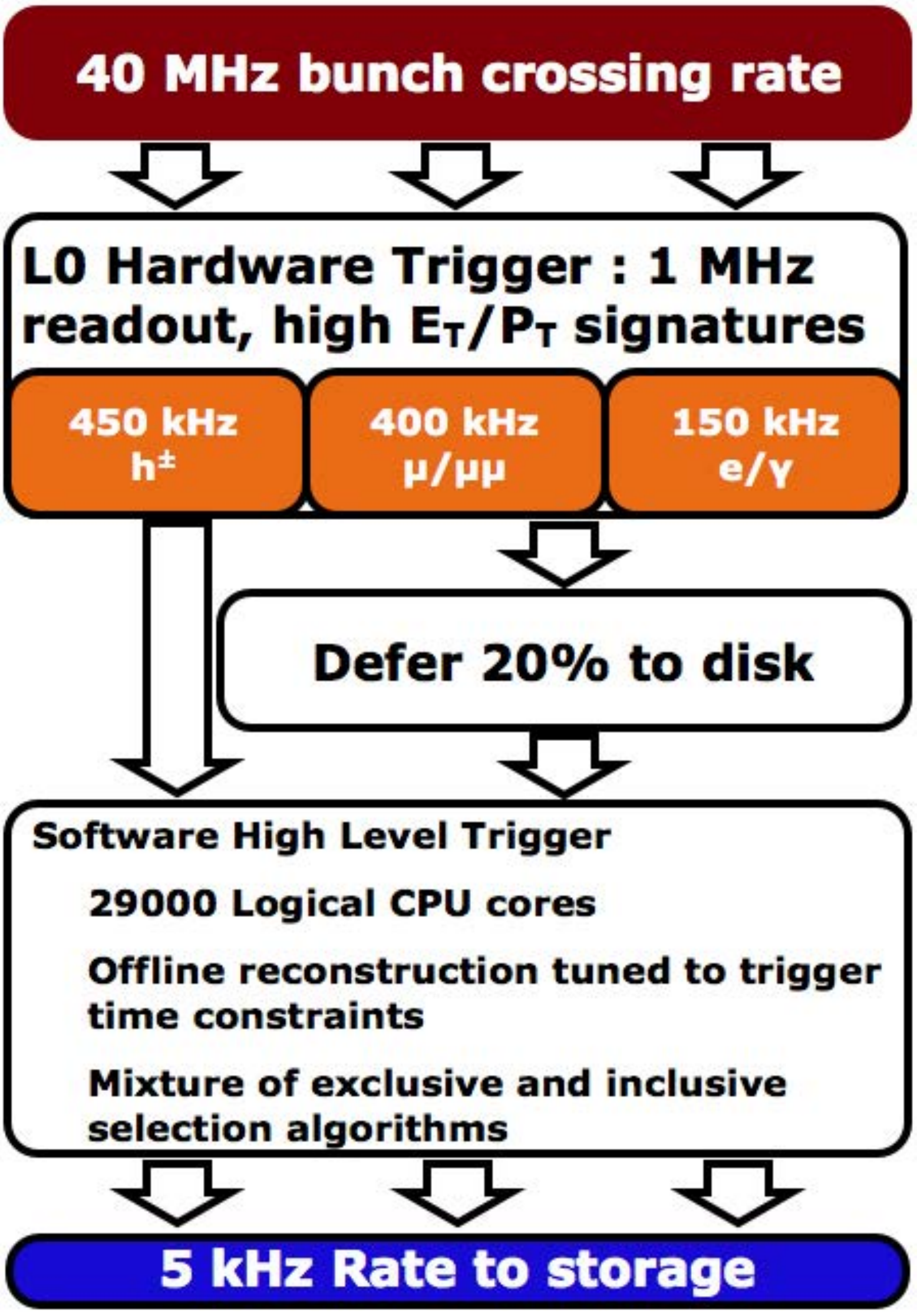}
\includegraphics[height=8cm]{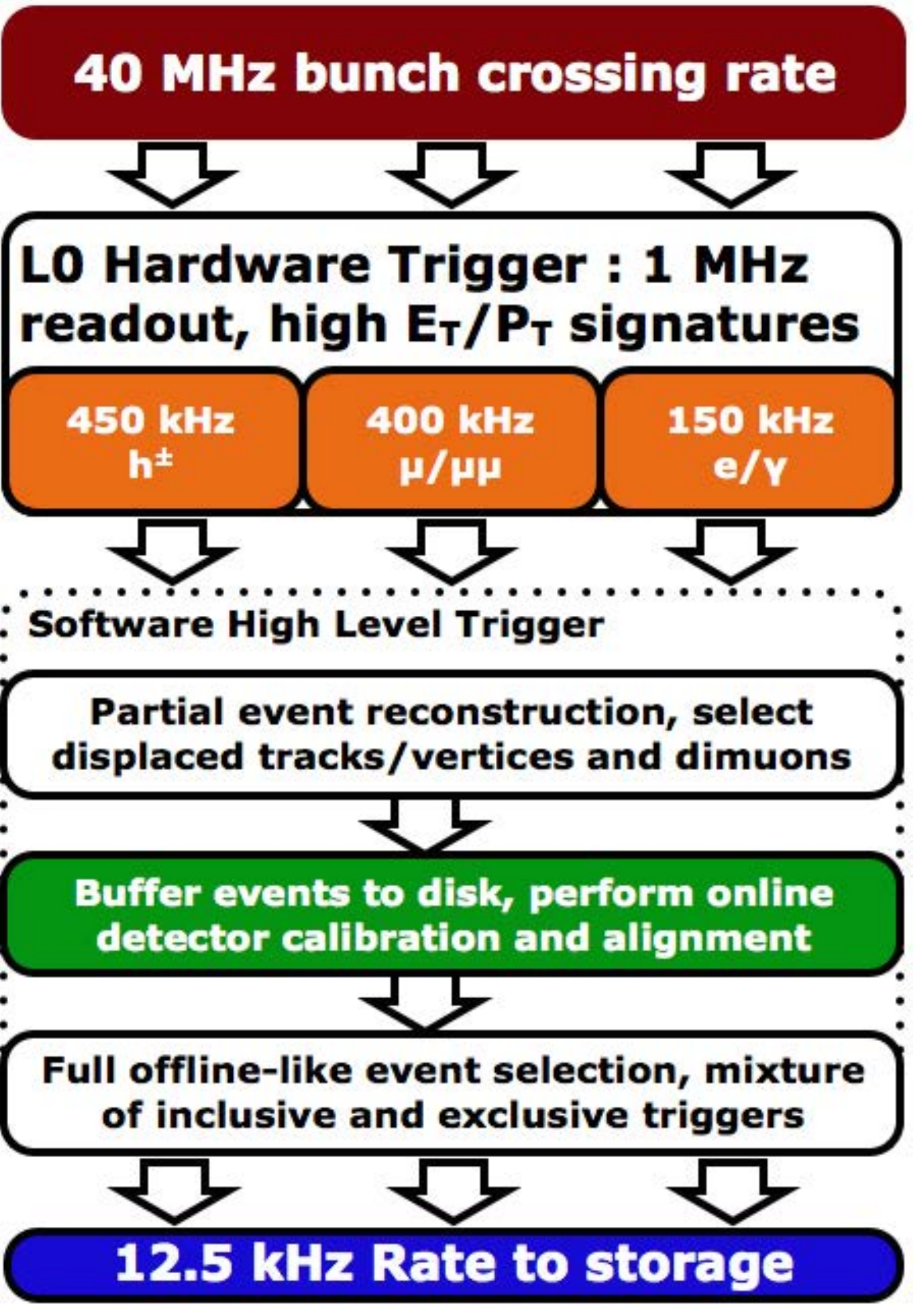}

\end{center}
\caption{Trigger overview in Run~1 (left) and Run~2 (right)} \label{fig:LHCb_Trigger_RunI}
\end{figure}

The trigger is organised in two different levels: the Level-0 (\lone) trigger
based on custom electronic boards, and the High-Level Trigger (\hlt), implemented in a computer farm.
Level-0 uses the information from the calorimeter and muon systems, performing 
a selection in order to reduce the event rate from 40 MHz to below 1 MHz, which is the maximum
frequency allowed to read out the entire detector. 
The \hlt is a software application running
on a processor farm that further reduces the rate of events in the kHz range for storage
(see Fig. \ref{fig:LHCb_Trigger_RunI}).

The \hlt has significantly evolved over time  from the original design in the \lhcb Technical Proposal (TP)~\cite{Amato:1998xt} in 1998, 
to the trigger design in the Technical Design Report (TDR)~\cite{LHCb-TDR-010} in 2003, to the Run~1 (2010-2012) actual implementation~\cite{LHCb-DP-2012-004} and  finally to the additional features introduced during Run~2 (2015-2018)~\cite{LHCb-DP-2019-001}.
In the  TP it was assumed that a first \hlt trigger level (L1) would reduce the 1~MHz input rate to a
40~kHz output rate with a variable latency of less than 256~$\mu$s, using coarse information from the vertex detector to 
reconstruct vertices and tracks with no momentum information (the \velo r-$\phi$ geometry was designed for this purpose). 
A second \hlt trigger level (L2) was fashioned to extrapolate  \velo tracks into the magnetic field to the 
tracking stations downstream of the magnet and reduce the output rate to 5~kHz with
an average latency of 10~ms. Finally  a third level (L3) would implement the full event reconstruction and a set of exclusive 
selections to bring down the rate to 200~Hz. 

By the time of the trigger TDR in 2003, it became clear that the \lhc was not going to start before 
the end of the decade when much more powerful processing units would become available. In addition a series of test-beam 
and detailed simulation studies convinced the collaboration of the need to have momentum information at the first stage of 
the \hlt. Therefore a new tracking station just upstream of the magnet was introduced (the \ttracker station). In addition, a  shield which had been protecting \richone from stray magnetic fields 
was removed to allow for a rough estimation for the momentum of tracks reconstructed between the \velo and \ttracker  stations. 
The software trigger 
then had two levels: Level-1 able to reduce the output rate to 40~kHz using \lone, \velo and \ttracker information with an average latency of 1~ms, 
and \hlt to reduce the output rate to 200~Hz with a combination of inclusive and exclusive selections. Between the time of the trigger TDR 
and the first physics run (Run~1), the interest in having a more performing \hlt for charm physics ($\ccbar$ with a factor 20 larger production 
cross section than $\bbbar$) and a much more robust system, convinced the collaboration to push for much more inclusive selections in the 
final trigger stage and a much larger trigger output rate (3-5 kHz). This implied a complete redefinition of the offline data processing model.

Furthermore,
it had been assumed  that the \lhc would operate with a 25 ns bunch separation, limiting the
number of overlapping events 
to a mean number of $\mu \simeq 0.4$ per bunch crossing at a luminosity of
$2 \times 10^{32}$ cm$^{-2}$s$^{-1}$. When, from 2011, a separation of 50 ns
was adopted for early \lhc operation, the experiment decided to run at $\mu \approx 1.4$ to compensate for the
lower number of bunches. Therefore the \hlt had to adapt to running conditions rather different than first assumed.
This was made possible by the highly flexible design of the \hlt. 

After the success of the \lhcb trigger performance in Run~1, the good 
understanding of the trigger reconstruction allowed the introduction of the  "real-time analysis" concept during Run~2. After the first \hlt
trigger level (\hltone), events are buffered to disk storage in the online system. This is done for two purposes, firstly events can be processed further 
during inter-fill periods, and secondly the detector can be calibrated and aligned run-by-run before the \hlttwo stage. Once the detector is 
calibrated and aligned, events are passed to \hlttwo, where a full event reconstruction of ``offline quality'' is performed. This allows for a wide 
range of inclusive and exclusive final states to trigger and obviates the need for further offline processing. In addition, new techniques to reduce the amount of information saved per event~\cite{LHCb-DP-2019-002} allowed to increase significantly the output rate to 10-15 kHz, as in Fig.~\ref{fig:LHCb_Trigger_RunI}, while the output of \hltone could be increased to $\order$(110 kHz). The decrease in requests for 
offline reconstruction also helped to mitigate the pressure on the offline computing model.

\subsubsection{Level-0 hardware trigger}
The \lone trigger is divided into three independent components: the \lone-Calorimeter trigger, the
\lone-Muon trigger and the \lone-PileUp trigger. The latter is used
to reject multiple visible interactions in a  bunch crossing by means of the
\hbox{``ad hoc''} \textit{Pile-Up System} detector housed in the \velo.
The first two components are briefly described below.

\paragraph{}
The \lone-Calorimeter part of the trigger obtains informations from the \spd, PS, \ecal and \hcal subdetectors
and computes the transverse energy deposited by incident particles: $E_{T} = E_{0} \cos \theta$, where $E_{0}$
is the energy of the particle and $\theta$ is the polar angle given by the cell hit in the detector.
Together with energy information, the total number of hits in the \spd (\spd multiplicity)
is also determined in order to veto large multiplicity events that would take too large a fraction of the
available processing time in the \hlt. From the calorimeter information, three types of
candidates are built and selected according to specific $E_{T}$ criteria: i) Hadron candidate (L0Hadron); ii) Photon candidate (L0Photon); and iii) Electron candidate (L0Electron).

\paragraph{}
The \lone-Muon part of the trigger  requires a muon candidate to have a hit in all five muon stations.
The \lone muon processor boards select the two highest $p_{T}$ muon tracks in each quadrant
of the muon system with a maximum of eight candidates. The trigger sets a single threshold
either on the largest muon $p_{T}$ (\lone muon trigger) or on the product of the largest and the 2nd largest
(\lone dimuon trigger).
Events with \spd multiplicity $>$ 600 are excluded in the \lone muon trigger in order to minimize the track multiplicity.
This limit is raised to 900 in the \lone dimuon trigger at the expense of  a small increase in rate.

The total output rate of the \lone trigger is limited to 1 MHz, which is the maximum rate
 accepted by the \hltone. Such an output rate consists of about
400 kHz of muon triggers, about 450 kHz of hadron triggers and about 150 kHz of electron and photon
triggers (the individual triggers have an overlap of about $10\%$).

\subsubsection{High Level Trigger}

Data from \lone are sent to the Event Filter computer Farm (EFF) which runs the \hlt algorithms.
The \hlt is a software application whose $29500$ instances run
on the EFF. Each instance is made up of independently operating trigger lines; each line
consists of selection parameters for a specific class of events.

The \hlt is divided into two stages. The first stage (\hltone) processes the full \lone rate and uses
partial event reconstruction to reduce the rate to about 110 kHz. The second stage (\hlttwo) reduces the rate to
about 12.5 kHz, performimg a more complete event reconstruction \cite{LHCb-DP-2019-001}.

\paragraph{}
{\bf \hltone} reconstructs the trajectories of charged particles traversing the full \lhcb tracking system which have a \pt larger than 500 MeV. 
The hits in the \velo are combined to form straight-line tracks loosely pointing towards the beam line. Next, at least three hits in the \ttracker 
are required in a small region around a straight-line extrapolation from the \velo. The \ttracker is located in the fringe 
field of the \lhcb dipole magnet, which allows the momentum to be determined with a relative resolution of about $20\%$, and  this estimate is  used to reject low \pt tracks. Tracks are then extrapolated to the T-stations downstream of the magnet. The search window in the 
\intr and \ot is defined by the maximum possible deflection of charged particles with \pt larger than 500 MeV. The search is also 
restricted to one side of the straight-line extrapolation by the charge estimate of the track. Subsequently, all tracks are fitted with a 
Kalman filter to obtain the optimal parameter estimate using a simplified geometry description of the \lhcb detector. The set of fitted 
\velo tracks is re-used to determine the positions of the PVs. 

Tight timing constraints in \hltone mean that most particle-identification algorithms 
cannot be executed. The exception is muon identification due to its clean signature. Hits in the muon stations are searched for in 
momentum-dependent regions of interest around the track extrapolations. Tracks with $p < 3$ GeV cannot be identified as muons, as they 
would not be able to reach the muon detectors.

\hltone has two inclusive trigger lines which select events containing a particle whose decay vertex is displaced from the PV: a line 
which selects a single displaced track with high \pt, and a line which selects a displaced two-track vertex with high \pt.
Both lines start by selecting good quality tracks that are inconsistent with originating from the PV. The single-track trigger then 
selects events based on a hyperbolic requirement in the 2D plane of the track displacement and \pt. The two-track displaced vertex 
trigger selects events based on a multivariate discriminant whose input variables are the vertex-fit quality, the vertex displacement, 
the scalar sum of the \pt of the two tracks and the displacement of the tracks making up the vertex. The two-track line is more 
efficient at low \pt, whereas the single track line performs better at high \pt, such that in combination they provide high efficiency 
over the full \pt range.

The \hltone muon lines select muonic decays of $\bquark$ and $\cquark$ hadrons, as well muons originating from decays of $\W$ and $\Z$ bosons. 
There are four main lines: one line that selects a single displaced muon with high $p_{T}$, a second single muon line that selects 
very high $p_{T}$ muons without displacement for electroweak physics, a third line that selects a dimuon pair compatible with 
originating from a decay of a charmonium or bottonium resonance or from Drell-Yan production, and a fourth line that selects displaced 
dimuons with no requirement on the dimuon mass. During Run~2, typically about 80~kHz were allocated to the inclusive \hltone lines, 
while about 20 kHz to the muon lines. The rest of the \hltone output is dedicated to special low multiplicity triggers and calibration trigger  lines.

\paragraph{}
{\bf \hlttwo} can perform the full event reconstruction since the output of \hltone is buffered. The full event reconstruction 
consists of three major steps: the track reconstruction of charged particles, the reconstruction of neutral 
particles and particle identification. The \hlttwo track reconstruction exploits the full information from the 
tracking sub-detectors, performing additional steps of the pattern recognition which are not possible in \hltone.
Tracks with a \pt larger than 80 MeV are reconstructed in \hlttwo, without the requirement to have hits 
in the \ttracker station. This is to avoid inefficiencies due to the \ttracker acceptance, which is crucial for part of the charm and kaon physics 
programme. In addition, tracks produced by long-lived resonances that decay outside the \velo are reconstructed using 
T-station segments that are extrapolated backwards through the magnetic field and combined with hits in the \ttracker.
Similarly, the most precise neutral cluster reconstruction algorithms are executed. Finally, in addition to 
the muon identification available in \hltone, \hlttwo exploits the full particle identification from the \rich 
detectors and calorimeter system. 

The \hlttwo inclusive $\bquark$-hadron trigger lines look for a two-, three-, or four-track vertex with sizeable \pt, 
significant displacement from the PV, and a topology compatible with the decay of a $\bquark$-hadron, using a multivariate 
discriminant. Whenever one or more tracks are identified as muons, the requirements on the discriminant are relaxed 
to increase the efficiency. As in the case of \hltone, several muon lines are used to select muonic decays of $\bquark$ 
and $\cquark$ hadrons and of $W$ and $Z$ bosons. However in \hlttwo, the muon reconstruction is identical to the 
offline procedure, having access to exactly the same information. During Run~2, typically about 3~kHz of the trigger 
rate is from the inclusive $\bquark$-hadron trigger while the muon lines take about 1~kHz. A large fraction of the 
trigger bandwidth (2-4 kHz) is allocated to exclusive selection of charm decays, where a reduced amount of information 
is saved per event. The rest of the trigger bandwidth is due to other special triggers and calibration trigger lines.

\clearpage

\newpage

\clearpage

\section{\lhcb contributions to CKM measurements and \CP violation }
\label{sec:CPviolation}

The violation of the combined operation of charge conjugation and parity, \CP, 
was first observed in 1964 in  
decays of neutral kaons~\cite{Christenson:1964fg}. 
The \babar~\cite{Aubert:2001nu} 
and \belle~\cite{Abe:2001xe} \bfactory experiments and
the \cdf experiment~\cite{Affolder:1999gg}
established \CP violation in the decays of neutral 
\Bd mesons.  \lhcb now extends measurements to much greater precision, 
and also probes the  $\B_s$ system, which is vital
to explore the full range of \CP violation measurements. 

In the Standard Model,  
the  Cabibbo-Kobayashi-Maskawa (CKM) unitary matrix~\cite{Cabibbo:1963yz,Kobayashi:1973fv},
$\textrm{V}_{\textrm{CKM}}$, 
describes the electroweak coupling strength $V_{ij}$ of the \W boson to quarks $i$ and $j$:

\begin{equation}\label{equ:CKM}
  \textrm{V}_{\textrm{CKM}} =   
\left( \begin{array}{ccc} 
V_{ud} & V_{us} & V_{ub}\\
V_{cd} & V_{cs} & V_{cb}\\
V_{td} & V_{ts} & V_{tb}\\
\end{array} \right).
\end{equation}

\CP is violated in the Standard Model if any element of the CKM matrix is complex.
The parametrisation of the CKM matrix due to Wolfenstein~\cite{Wolfenstein:1983yz} 
is given  by  
\begin{equation}\label{equ:CKMwolf}
\textrm{V}_{\textrm{CKM}} =   
\left( \begin{array}{ccc} 
1 - \frac{1}{2}\lambda^2 - \frac{1}{8}\lambda^4 & \lambda & A \lambda^3(\rho - i\eta)\\
-\lambda + \frac{1}{2}A^2\lambda^5[1-2(\rho+i\eta)] & 1 - \frac{1}{2}\lambda^2-\frac{1}{8}\lambda^4(1+4A^2) & A\lambda^2 \\
A\lambda^3[1-(1-\frac{1}{2}\lambda^2)(\rho+i\eta)]  & - A\lambda^2 + \frac{1}{2}A\lambda^4[1-2(\rho+i\eta)] &  1-\frac{1}{2}A^2\lambda^4\\
\end{array} \right)  \\
\end{equation}
for the four Standard Model parameters $(\lambda, A, \rho, \eta)$.
The expansion parameter, $\lambda$, equal to the sine of the Cabibbo angle,
has a value $|V_{us}| = 0.22$~\cite{PDG2019}, and in Equ.\,\ref{equ:CKMwolf} the 
expansion is given for terms up to order $\lambda^5$. 

The unitarity of the CKM matrix leads to six orthogonality conditions between any pair of columns 
or any pairs of rows of the matrix.  The orthogonality  means the six conditions can be represented as six 
triangles in the complex plane. 
The interesting relations for \CP violation are those given by:
\begin{equation}\label{equ:uniOne}
V_{ud}V^*_{ub}+V_{cd}V^*_{cb}+V_{td}V^*_{tb}=0  ~~~: {\mathrm{the}}  ~{\mathrm{ unitarity ~ triangle}}, 
\end{equation}
\begin{equation}\label{equ:uniTwo}  
V_{us}V^*_{ub}+V_{cs}V^*_{cb}+V_{ts}V^*_{tb}=0   ~~~: {\mathrm{the}}~B_s  ~ {\mathrm{triangle, ~  and }} 
\end{equation}
\begin{equation}\label{equ:uniThree}  
V_{cd}V^*_{ud}+V_{cb}V^*_{ub}+V_{cs}V^*_{us}=0   ~~~: {\mathrm{the~charm~triangle}}. 
\end{equation}
The unitarity triangle has sides with lengths that are the same order in $\lambda$, namely $\order(\lambda^3)$, 
which implies large \CP asymmetries in \Bd and \Bpm decays. The  $\B_s$ triangle has two sides of 
$\order(\lambda^2)$ and the third of $\order(\lambda^4)$. Hence \CP violation in $\B_s$  mixing is significantly smaller
than in the $\Bd$ system.
Moreover, the charm triangle has two sides of 
$\order(\lambda)$ and the third of $\order(\lambda^5)$, hence \CP violation in the charm system is expected to be 
extremely small.
Note that all three triangles have equal  area~\cite{Jarlskog:1989bm}.

To study \CP violation, the $\B-$physics experiments measure the
complex phases of the CKM elements  and 
measure the lengths of the sides of the triangles to check for a self-consistent picture. \CP violation is
predicted in many (often very rare) $\B$ hadron decays, hence \lhcb
utilises large samples of $\B$, $\B_s$, $\B_c$ mesons and $\B-$baryons. 
New physics can be discovered and studied when new particles appear in, for example, 
virtual loop processes of rare $\B$ decays, 
leading to observable deviations from Standard Model expectations, both in branching ratios 
and \CP observables. Hence the \lhcb strategy is to determine with high precision the CKM elements and to
compare  measurements of the same parameters, especially those where one is sensitive to new physics 
and the other to Standard Model processes.


\subsection{The status of the unitarity triangle before \lhcb}
\label{sec:beforeLHCb}

\begin{figure}
\begin{center}
\includegraphics[width=10cm]{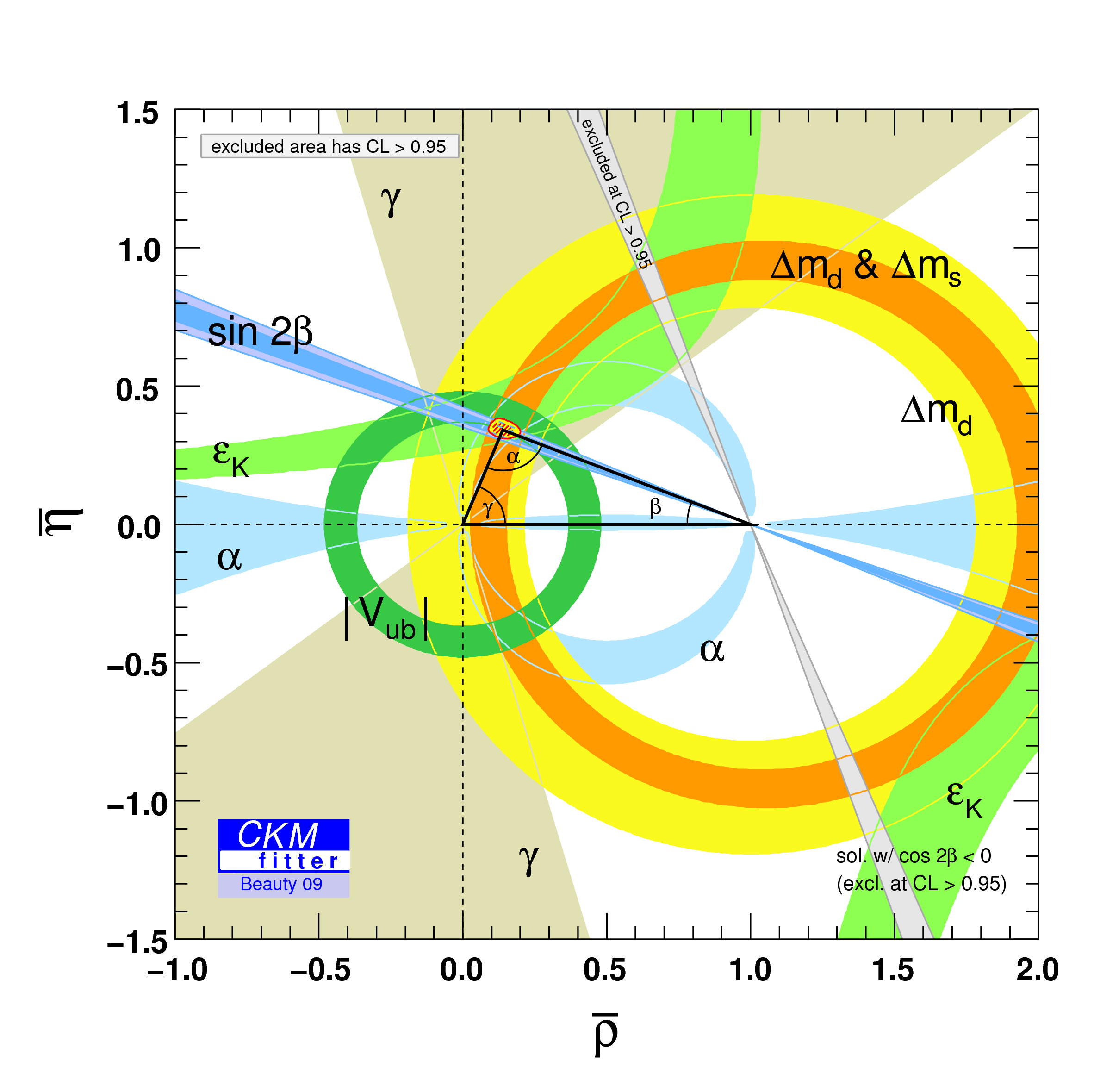}
\caption{The knowledge of the unitarity triangle as of autumn 2009~\cite{CKMfitter2005}.}
\label{fig:CKM-fitter-2009}
\end{center}
\end{figure}

The first generation \bfactory experiments to study
\CP violation in the $\B-$system, \babar and \belle,
made huge in-roads into testing the Standard Model description of \CP violation; the status was summarised extensively at the Beauty 2009 Conference \cite{Beauty2009}. \cdf and \dzero extended these studies at the \tevatron, and make first explorations in the $B_s$ sector.
Fig.~\ref{fig:CKM-fitter-2009} shows the status of the unitarity triangle measurements
compiled by the CKM-Fitter Group~\cite{CKMfitter2005} in 2009, when the \bfactories had been running for around ten years.
Here graphical results are displayed in the $\rho-\eta$ plane and the best fit to the apex of the
triangle (Equ.~\ref{equ:uniOne}) to the 95\% confidence level is shown. The fit
to the CKM parametrisations include measurements of the sides of the triangle through measurements of the CKM elements and the angles, 
information from rare $K$ and $\B$ meson decays, and $\Bs-\overline{B^0_{s}}$  mixing.

Before 2009, when the \lhc turned on, the \bfactory experiments, \cdf and,  to a lesser degree \dzero, measured the parameters of the unitarity triangle with varying degrees of precision:
\begin{itemize}
\item 
The quantity $\textrm{sin\,} 2 \beta$  was
measured in all channels, including the ``gold plated'' channel $\Bd \rightarrow {J}/\psi {K}^0_{S}$, to a precision of around $\sim$0.03;
\item
The sides $|V_{td}/V_{ts}|$ and $|V_{ub}/V_{cb}|$
were known from $\Bs-\overline{B^0_{s}}$  mixing 
and from $\bquark\to \,\uquark$ decays, respectively each to $\sim$10\%, but
limited by theory. The $B_s$ mixing phase ($\phis$) was unmeasured; 
\item 
The angle $\alpha$
was measured in the channels $\B \to \pi\pi$, $\rho\pi$ and $\rho\rho$ with a statistical precision of $\sim 5^\circ$;
\item 
There was a statistics-limited measurement in  
$\B \to \,\D \kaon$ modes of the angle $\gamma$ 
to around $20-25^\circ$. A measurement of $\gamma$ from $\B_s$ modes such as 
$\Bs \to \Ds \kaon^-$ had been completely unexplored.
\item The parameter $\epsK$, measured in kaon decays, provided a very loose constraint on the triangle vertex;
\item 
The $B_s$ sector had been
largely unexplored by the first generation experiments, as had $\bquark-$baryons. Running  at the $\Upsilon(4S)$, the \bfactories produced predominantly $\B_{u,d}$ meson pairs, however Belle record a significant sample of $\Upsilon(5S)$ data, allowing some interesting measurements of $\B_s$ pairs. At the \lhc, $\Bd$, $\Bpm$, $B_s$, $B_c$ and 
$\bquark-$baryons are  produced in approximately in the ratios 
\mbox{$\sim 40: 40 :10 :0.1 :10 \%$};
\item The \bfactories were statistics limited for very rare processes
with branching ratios $\lesssim 1 \times 10^{-6}$, such as $\bquark \to \,\squark$ flavour-changing neutral current (FCNC) transitions, e.g. $b\to s\gamma$ and $b\to s l^+l^-$. Super-rare transitions such as $\B_{(s,d)}\to \mu^+\mu^-$ were also unobserved.
\end{itemize}

\begin{figure}[ht]
\begin{center}
\includegraphics[width=12cm]{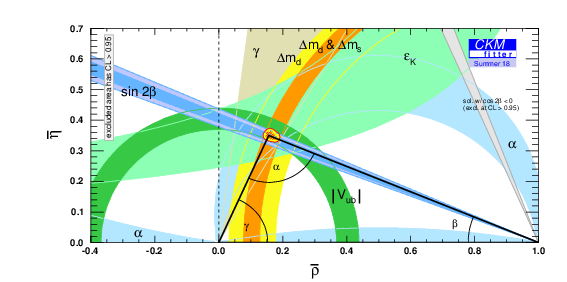}
\caption{The current knowledge of the unitarity triangle (as of Summer 2019)~\cite{CKMfitter2005}.}
\label{fig:CKM-fitter}
\end{center}
\end{figure}

In contrast, Fig.~\ref{fig:CKM-fitter} shows the  status of the unitarity triangle measurements today ~\cite{CKMfitter2005}.

\subsection{Heavy quark mixing measurements}
\label{sec:mixing}

\begin{figure}[ht]
\begin{center}
\includegraphics[width=14cm]{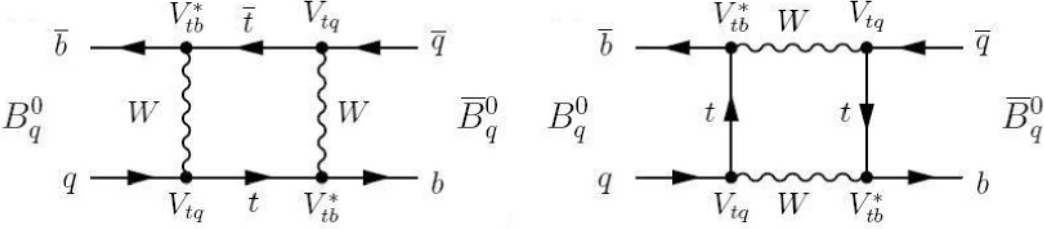}
\caption{The box diagrams representing mixing between $\B^0_q$ and $\overline{\B}^0_q$ mesons. }
\label{fig:boxDiag}
\end{center}
\end{figure}

Since flavour is not conserved in the weak interaction, mixing between $\B^0_q$ and 
$\overline{\B}^0_q$ mesons (where $q = d$ or $s$) is possible via the box diagrams shown in Fig.~\ref{fig:boxDiag}. 
The probability for finding a $\overline{\B}_q$ (or a $\B_q$), 
given the initial state was a $\B_q$ (or a $\overline{\B}_q$) at time $\Delta t$ after production,
is given by:
\begin{equation}\label{eqn:probBbar}
P_{mix}(\Delta t)  =  \exp{-\dfrac{\Delta t}{\tau_{B^0}}} \left( 1 - \cos(\Delta m_q \Delta t)\right) .
\end{equation}
Here  $\Delta m_q$ is the mass difference $m_H - m_L$, 
where $m_{H,L}$ are the masses of the heavy and light mass eigenstates
and $\tau_{B^0}$ is the  $\B$ lifetime. 
At the \lhc, the two neutral $\B$ mesons produced can oscillate independently at any time after production.

Any \CP measurement from a time-dependent analysis of neutral $\B$ decays needs the determination of the $\B$ flavour ($b$ or 
${\overline b}$) at production. This requires $b-$quark ``tagging'', and several  algorithms have been developed by  \lhcb involving the combination of  so-called opposite side~\cite{LHCb-PAPER-2011-027}
and same side taggers~\cite{LHCb-PAPER-2016-039}.

\begin{itemize}
\item {\bf $\B^0$ mixing:}
 $\B^0$ oscillations are measured in the channel  $\B^0 \to D^- \mu^+ \nu_\mu X$ and the charge conjugate modes. The  oscillation  measurements rely on tagging, and are
 shown for 3 fb$^{-1}$ of  data  in 
Fig.~\ref{fig:BOscillations} (left). 
The \lhcb $\Delta m_{d}$ measurement, which is the world's best,  is
$\Delta m_{d}=(505.0 \pm 2.1 \pm 1.0)\, \rm{ns}^{-1}$~\cite{LHCb-PAPER-2015-031}.

\item {\bf $\B_s$ mixing:}
 $\B^0_s - {\overline{\B}^0_s}$ oscillations were first observed by 
  \cdf~\cite{Abulencia:2006mq}. 
$\B_s$ oscillations are measured at \lhcb in the  mode $\Bs \to \Dsm \pip$ and its charge conjugate state.
The oscillations are shown for 1 fb$^{-1}$ of  data  in 
Fig.~\ref{fig:BOscillations} (right). 
The plot shows the proper-time distribution of  $\Bs \to \Dsm \pip$ candidates, in five different $\Bs$ decay channels,
that have been flavour-tagged as not having  oscillated. 
The \lhcb $\Delta m_{s}$ measurement, which is again the world's best,  is
$\Delta m_{s}=(17.768 \pm 0.023 \pm 0.006)\, {\rm ps}^{-1}$~\cite{LHCb-PAPER-2013-006}.

\begin{figure}[ht]
\begin{center}
\includegraphics[width=7.5cm]{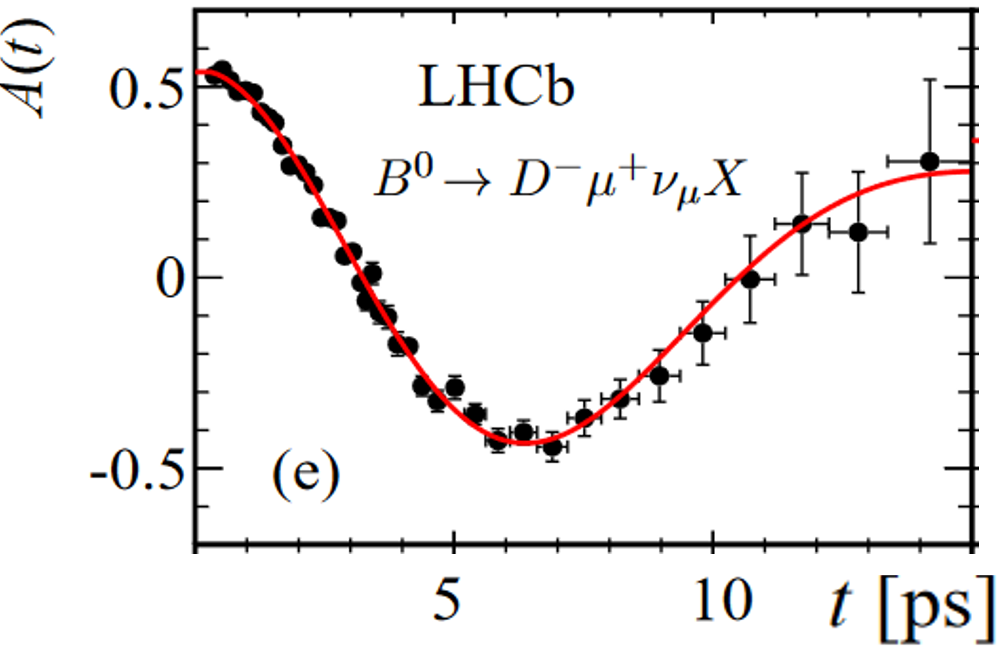}
\includegraphics[width=8cm]{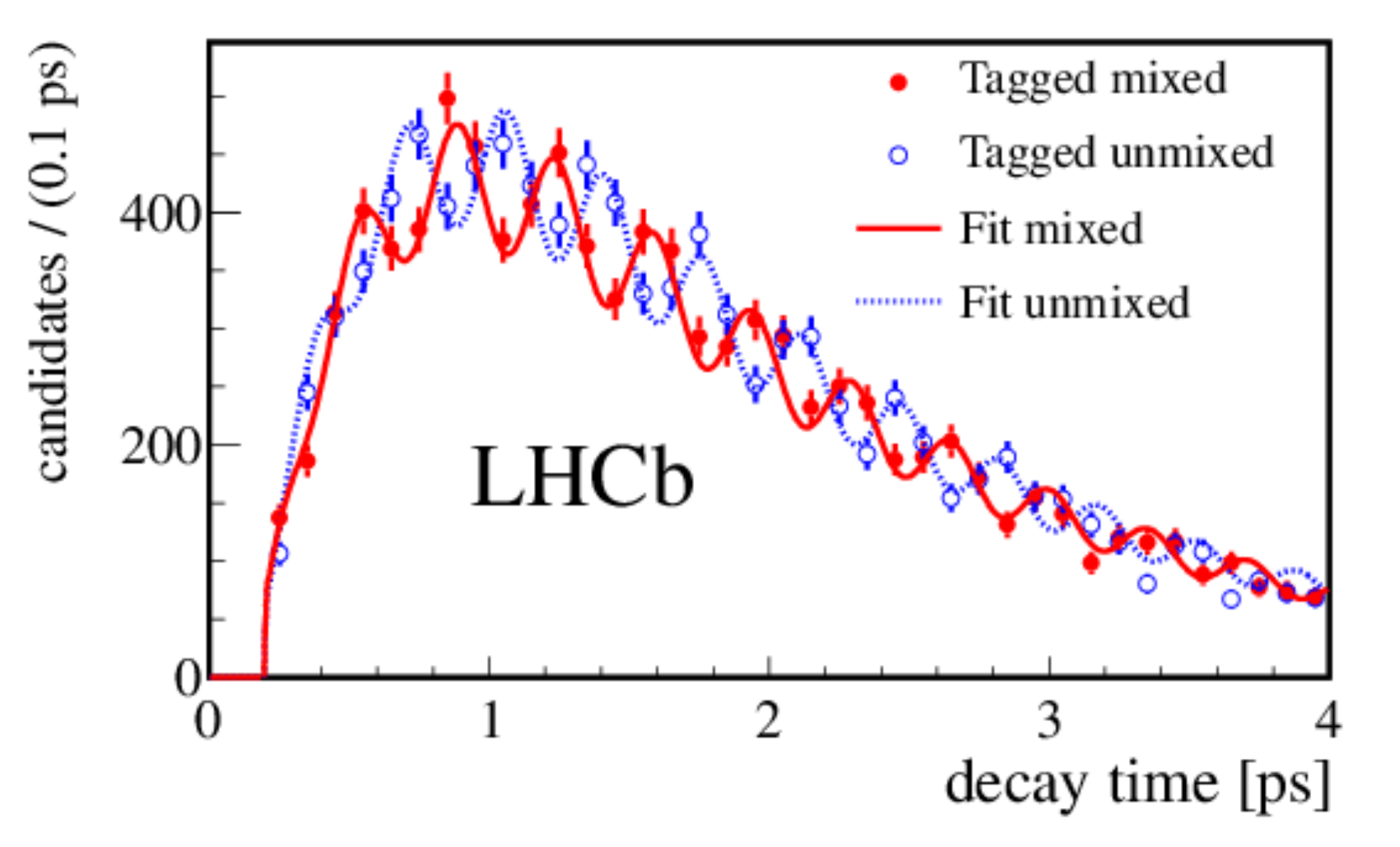}
\includegraphics[width=7cm]{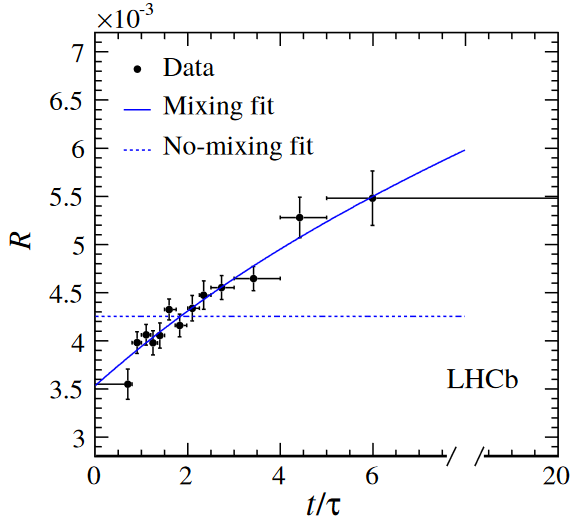}
\caption{Proper-time distribution of (left) $\B^0 - {\overline{\B}^0}$ 
oscillations in $\B^0 \rightarrow D^- \mu^+ \nu_\mu X$ decays,
 (right) $\B^0_s - {\overline{\B}^0_s}$ oscillations in $\Bs \to \Dsm \pip$ decays, 
and (lower) 
$\D^0 - {\overline{\D}^0}$ oscillations in $\D^{0} \rightarrow K^{+} \pi^{-}$ decays.
All distributions rely on flavour tagging, and the curves correspond to the fitted oscillations.}
\label{fig:BOscillations}
\end{center}
\end{figure}
 
\item {\bf Charm meson mixing:}
 
$\D^0 - {\overline{\D}^0}$ mixing was established in a single experiment in 2013 by \lhcb in $\D^{0} \rightarrow K^{+} \pi^{-}$ decays,
although it was confirmed earlier by combining several \bfactory results.
The $\D^0$ mixing parameters were measured in \lhcb in a decay-time-dependent fit to 
the ratio {\large $\frac{N_{\D^{0} \to K^{+} \pi^{-}}}{N_{\D^{0} \to K^{-} \pi^{+}}}$}. 
The time-dependent fit is shown in  Fig.\ref{fig:BOscillations} (lower) for 1 fb$^{-1}$ of data.  The no-mixing scenario is excluded at $9.1 \sigma $ ~\cite{LHCb-PAPER-2012-038}.

\end{itemize}

\subsection{ \lhcb measurements of the unitarity triangle parameters}
\label{sec:measurements}

The \lhcb experiment performs a high-statistics study of \CP violation with 
unprecedented precision in many different
and complimentary channels, providing a sensitive test
of the Standard Model and physics beyond it. 

\subsubsection{Measurements of the CKM angle $\beta$}

The time-dependent decay asymmetry of the channel $\Bd \to {J}/\psi {K}^0_{S}$ allows a
measurement of the angle $\beta$. 
This is known as the ``golden'' decay mode because the channel is virtually free of penguin pollution (which
enters with the same overall phase), resulting in very small theoretical uncertainty, of order 1\% ~\cite{Fleischer:2004xw}.
\CP violation in this channel  occurs in the interference between mixing and decay, where the 
mixing process introduces a relative \CP-violating weak phase of 2$\beta$.
Experimentally the \CP asymmetry is measured from the ratio of the numbers of $\overline{\B}$ and $\B$ mesons,
$N_{\overline{\B} \to f}$ and $N_{{\B} \to f}$, decaying into final state $f$: 
\begin{eqnarray}\label{equ:asymm-BdJpsi}
\mathcal{A_{CP}}  = {{N_{\overline{\B} \to f} - N_{{\B} \to f}}\over
                  {N_{\overline{\B} \to f} + N_{{\B} \to f}}}  
				  = \textrm{sin\,}(\Delta m_d t) ~\textrm{sin\,} (2 \beta) -  \textrm{cos\,}(\Delta m_d t) ~\textrm{cos\,} (2 \beta) . 
\end{eqnarray}

\begin{figure}[ht]
\begin{center}
\includegraphics[width=7cm]{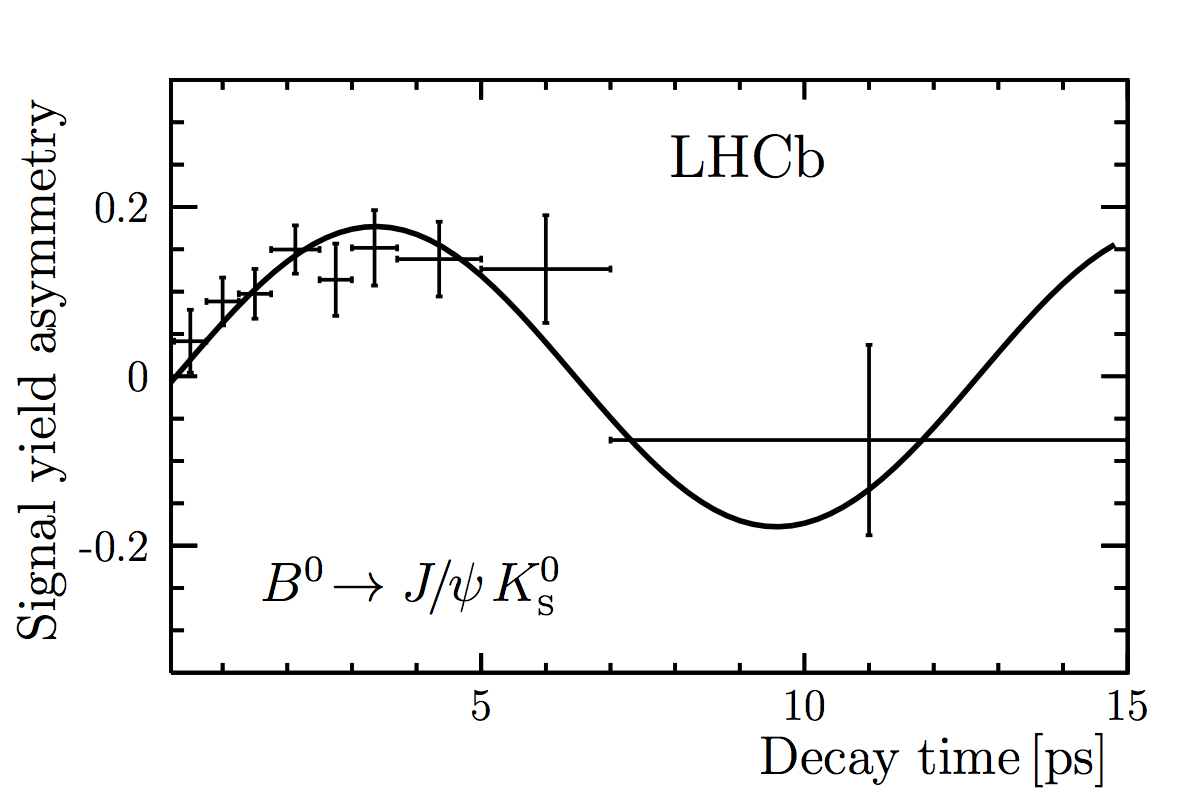}
\includegraphics[width=7cm]{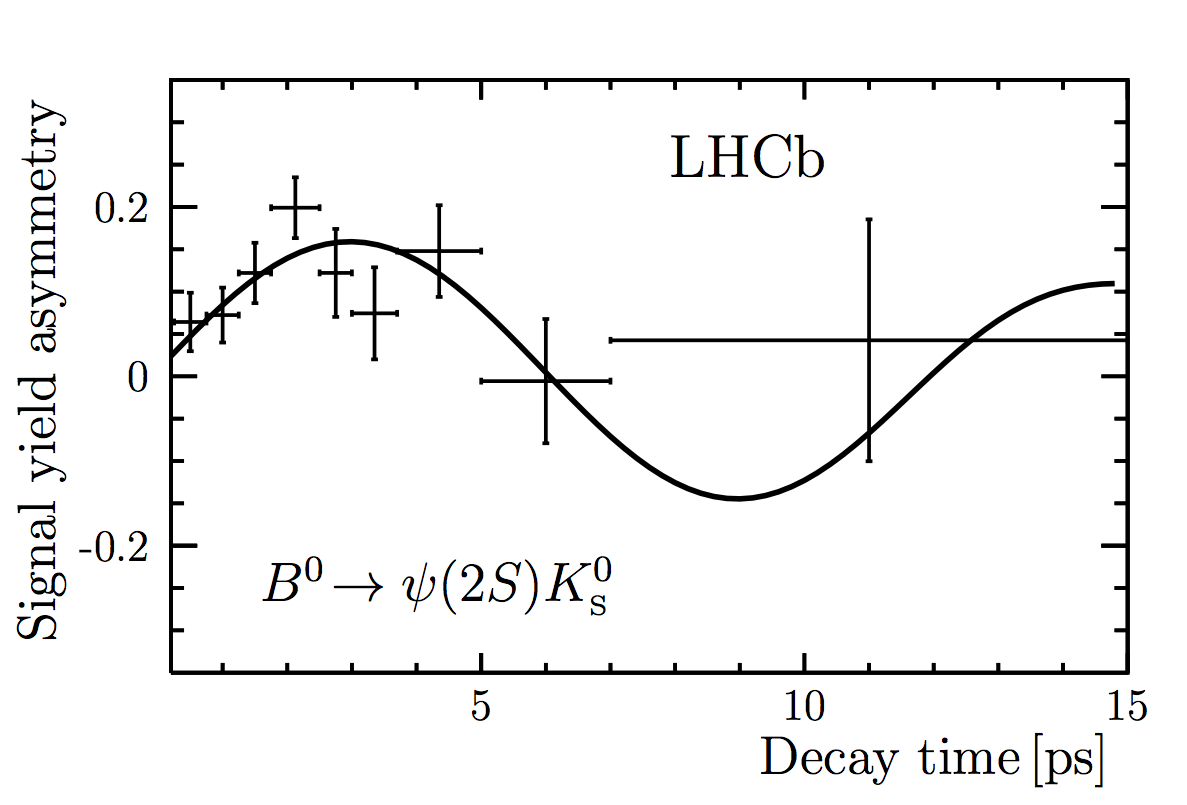}
\caption{The \lhcb measured and fitted asymmetry in $\Bd \to {J}/\psi {K}^0_{S}$ decays 
for (left) the $J/\psi(1S)$ and (right) the $\psi(2S)$.}
\label{fig:BtoJKs_asymmetry}
\end{center}
\end{figure}

The \lhcb measured and fitted asymmetries for the $J/\psi$ $(1S)$ and $(2S)$ states
are shown in  
Fig.~\ref{fig:BtoJKs_asymmetry} for 3 fb$^{-1}$ of data at  7 and  8 TeV~\cite{LHCb-PAPER-2017-029}.  
These measurements are
$\textrm{cos\,}(\Delta m_d t) = -0.017 \pm 0.029$ and 
$\textrm{sin\,}(\Delta m_d t) = 0.760 \pm 0.034$, where
an observation of  a 
direct \CP-violation contribution proportional 
to $\textrm{cos\,}(\Delta m_dt)$ would be an indication of new physics~\cite{Fleischer:2004xw}.
The \lhcb measurement is now competitive with \babar and \belle
measurements; the current world 
average of $\textrm{sin\,} 2 \beta = 0.695 \pm 0.019$~\cite{HFLAV16}
 is dominated by \lhcb together with  the
 \bfactory measurements in the complementary channels
$\Bd \to {J}/\psi {K}^0_{S}$ and  
$\Bd \to {J}/\psi {K}^0_{L}$.  
The measurement by \lhcb of $\textrm{sin\,} 2 \beta$ in gluonic penguins will further contribute to this study. 

\subsubsection{Measurements of the CKM angle $\alpha$}

The primary method at \lhcb for the measurement of $\alpha$ is through an amplitude analysis via the $\B \to \rho \pi$ decay modes~\cite{Snyder:1993mx},
however these channels are difficult at \lhcb due to the need to efficiently reconstruct $\pi^0$s.
Penguin pollution is present and must be constrained, with the additional application of isospin symmetry.
The precision on $\alpha$ at \lhcb  is expected to be dominated by systematic uncertainties, and any
measurement is not expected to improve on a combination of the \bfactory measurements,
$\alpha=\left(86.4^{+4.5}_{-4,3}\right)^\circ$~\cite{CKMfitter2005}.

\subsubsection{Measurements of the CKM angle $\gamma$}
\label{Sec:gamma}

A precise measurement of the angle $\gamma$ is  key to understanding the
closure (or otherwise) of the 
unitarity triangle.
Constraints on the unitarity-triangle apex largely come from loop decay measurements
which  are very sensitive to the presence of new physics.
$\gamma$ is the only angle accessible at tree level and hence forms a SM benchmark
to which the loop measurements can be compared (assuming no significant new physics in tree decays).
The $\gamma$  measurement also relies on theoretical input which is very well
understood~\cite{Brod:2013sga,Brod:2014bfa}.

Determination of $\gamma$ from a combined fit to all measured parameters of the unitarity triangle 
currently gives a value   
$\gamma=\left(65.8^{+1.0}_{-1.7}\right)^\circ$ ~\cite{CKMfitter2005}.
Conversely the measurement of $\gamma$ alone from a combination of 
all direct measurements from tree decays gives
\mbox{$\gamma=\left(72.1^{+5.4}_{-5.7}\right)^\circ$}. 
Hence reaching degree-level precision from direct $\gamma$ measurements is crucial.

\lhcb makes measurements of $\gamma$ by a variety of methods, where complementary is vital.  Examples of the most sensitive \lhcb measurements are outlined below.

\begin{itemize} 

\item {\bf $\gamma$ in the ``time integrated''   $\Bpm \to \D^0\kaon^\pm$  modes}

 The measurement of $\gamma$ is made in  direct CP-violation via $\Bpm \to \D^0\kaon^\pm$ by three different methods: 
 the GLW method (decay into a \CP eigenstate)~\cite{Gronau:1990ra,Gronau:1991dp}, the   ADS method (decay into a flavour-specific mode)~\cite{Atwood:1996ci},
 and the
GGSZ method (Dalitz analysis)~\cite{Giri:2003ty}. These all
access $\gamma$ through interference between the $\Bpm \to \D^0\kaon^\pm$ and $\Bpm \to \overline{\D}^0\kaon^\pm$ 
decay paths, where the $\D^0$ and $\overline{\D}^0$ decay to the same final state. 
When using these methods, the decay modes are self-tagging. In addition 
time-dependent analyses are not necessary. For the ALD and GLW modes,  the charge-conjugate event yields are simply counted to 
determine the CP asymmetries (i.e. effectively a ``counting experiment'').


Figure~\ref{fig:ADS} shows an example of a \CP asymmetry in the $\Bpm \to \D^0\kaon^\pm$ ADS mode~\cite{LHCb-PAPER-2016-003}.  
Here  the $\D^0$ is produced in a Cabibbo favoured mode ($V_{cb}$) 
but decays via a suppressed mode ($V_{cd}$) into   $K^+\pi^-$. 
This interferes with the $\overline{\D}^0$ charge-conjugate state which is produced in a suppressed mode
($V_{ub}$)  but decays  to the same final state   $K^+\pi^-$ via a favoured mode
($V_{cs}$). The branching fraction for the favoured $\B$ decay is only $\sim 10^{-4}$, so these 
measurements require high statistics.  
The asymmetry observed in Fig.~\ref{fig:ADS} has a magnitude of around 40\% and has a significance of 7$\sigma$.

\begin{figure}[ht]
\begin{center}
\includegraphics[width=15cm]{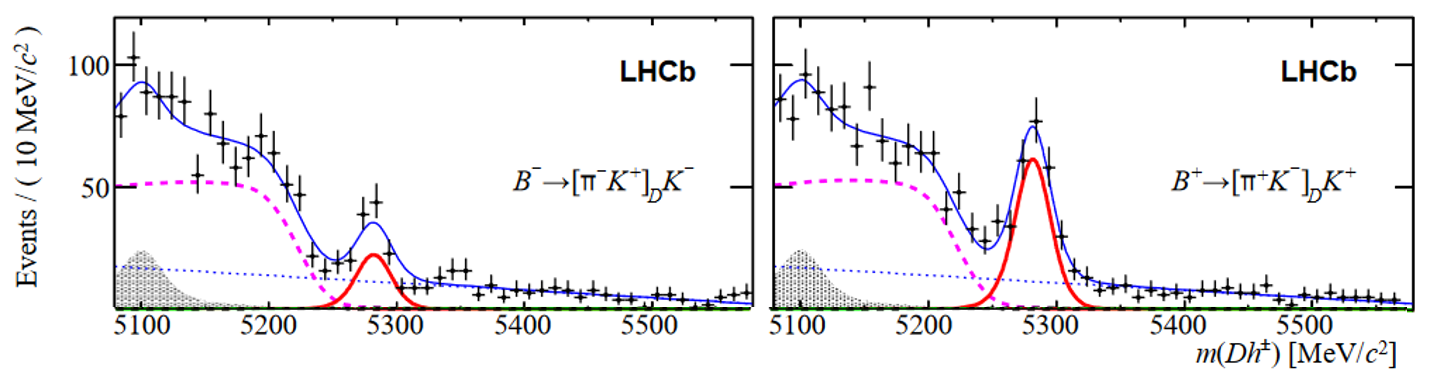}
\caption{Example of a \CP asymmetry in the ADS mode $\Bpm \to \D^0\kaon^\pm$, where the
$\D^0 $ and $\overline{\D}^0$ both decay into $K^+\pi^-$. The event yield is shown as a function of $\D \kaon$ mass.}
\label{fig:ADS}
\end{center}
\end{figure}

A specific example of an \lhcb analysis using the GGSZ Dalitz method is in the decay
$\Bpm \to \D^0\kaon^\pm$ where
$\D^0 \to K_S^0\pi^+\pi^-$ or $\D^0 \to K_S^0 K^+ K^-$~\cite{LHCb-PAPER-2018-017}.
There is a rich Dalitz plot structure with the presence of large interference effects. 
The Dalitz space is divided up  into 
symmetric bins, chosen to optimise sensitivity.
An amplitude analysis can then be used to extract $\gamma$.

In all $\Bpm \to \D^0\kaon^\pm$ modes and decays listed above, $\gamma$ can also be extracted from the corresponding $\Bpm \to \D^0\kaon^{\star\pm}$ modes, albeit with reduced $\gamma$ sensitivity. In addition $\Bd \to \D^{(*)}\kaon^{(*)}$ GGSZ modes are also included in the global fit to extract the $\gamma$ average value.

\item {\bf  $\gamma$ from the ``time-dependent'' $\Bs \to \Dsm\Kp$   mode} 

The channel $\Bs \to \Dsm\Kp$, and its charge conjugate states,
provide  a theoretically clean measurement of the angle  
($\gamma + \phis$) 
where  $\phis$ is the (small valued) $\B_s$ mixing phase, with no significant penguin contribution expected~\cite{Aleksan:1991nh}.
Here both $\Bs$, and via the mixing diagram $\overline{\B}^0_s$ , can decay  to the same 
final state $\Dsm\Kp$, resulting in interference which is sensitive to $\gamma$. 
The same is true  for decay into the charge conjugate state $\Dsp\Km$. 
%
Hence four time-dependent decay rates are measured:
$\Bs \to \Dsm\Kp$, ~$\Bs \to \Dsp\Km$, 
~$\overline{\B}^0_s \to \Dsm\Kp$ and $\overline{\B}^0_s \to \Dsp\Km$.
The method is then to fit two asymmetries of the form
\begin{eqnarray}\label{eqn:eq_asymmBstoDSK}
\mathcal{A_{CP}}  =
\dfrac{{N_{\overline{\B} \to f} - N_{{\B} \to f}}}
      {N_{\overline{\B} \to f} + N_{{\B} \to f}} 
\,.
\end{eqnarray}
These measurements
 yield values for  the strong phase difference $\delta_{QCD}$ between the amplitudes 
${\B} \to f$ and $\overline{\B} \to f$, the  amplitude ratio, and ($\gamma +  \phis$). 

The current measurement by \lhcb in 1~fb$^{-1}$ of data  yields a value 
$\gamma=\left(115^{+28}_{-43} 
\right)^\circ$~\cite{LHCb-PAPER-2014-038}.  
This complements the measurements 
in $\Bpm \to \D^0\kaon^\pm$, although with less statistical precision.




\item {\bf The $\gamma$ combination}

The  \lhcb measurement  of $\gamma$ averaged  over all the above methods, which includes all $\B^0, B^\pm$ and $B_s$ modes, is
$\gamma=\left(74.0^{+5.0}_{-5.8} 
\right)^\circ$~\cite{LHCb-CONF-2018-002}. 
This measurement dominates the  current world
average. The confidence limits as a function of $\gamma$  for the combination
is shown in Fig.~\ref{fig:gamma-combination} for the various measurement channels. 
The agreement between $\B_s$ and $\B^\pm$ initial states is currently at the  2$\sigma$ level.

\begin{figure}[ht]
\begin{center}
\includegraphics[width=10cm]{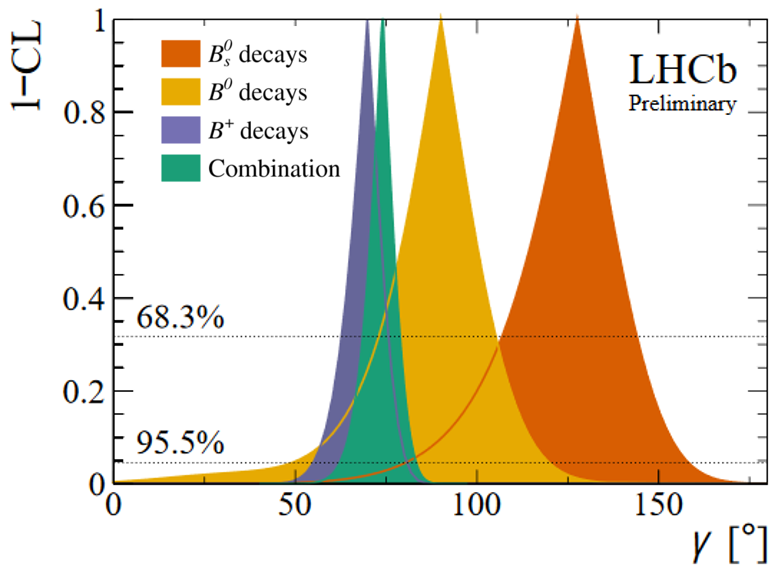}
\caption{Confidence limits as a function of $\gamma$ in the \lhcb combination for the various measurement channels.}
\label{fig:gamma-combination}
\end{center}
\end{figure}

\end{itemize}

\subsubsection{The sides of the triangle}

\begin{itemize}
\item {\bf The side opposite to $\beta$ }

Currently the closure test of the unitarity triangle is   limited mainly by the side 
opposite to $\beta$ which has a length proportional to $|V_{ub}| / |V_{cb}|$ in the Standard Model. This limitation is a consequence of 
tension between \bfactory inclusive and exclusive $|V_{ub}|$ measurements which differ
by $\sim 3.5 \sigma$~\cite{HFLAV16}.
$|V_{ub}|^2$  is directly proportional to the decay rate $\B^0\to X_u \mu^- \nu_\mu$, where $X_u$ is a 
meson containing a $u$ quark. Theoretical input from Heavy Quark Effective Theory and 
lattice calculations are also necessary to calculate $|V_{ub}|$, although several of the theoretical uncertainties cancel in the ratio to calculate the side.

$|V_{ub}| / |V_{cb}|$ is a very difficult measurement at \lhcb due to presence of a neutrino, the identification
of which was never in \lhcb's original plans. Although the \bfactory favoured channel
 $\B^0\to \pi^+ \mu^- \nu_\mu$  cannot currently be identified at \lhcb, 
the equivalent baryonic channel $\Lambda_b \to p \mu^- \nu_\mu$ has been measured. The 
signal is separated from the lower-mass backgrounds, shown in Fig.~\ref{fig:Vub}.
Using form factors from lattice calculations,
\lhcb measures~\cite{LHCb-PAPER-2015-013}
\begin{equation*}
\left|V_{ub}\right|  = (3.27 \pm 0.15\,{\rm (exp)} \pm 0.17\,{\rm (theory)} 
\pm  0.06\,( \left|V_{cb}\right|)) \times 10^{-3}
\end{equation*}
This is to be compared to the world average of
\mbox{$\left|V_{ub}\right|  = (3.94 \pm 0.36) \times 10^{-3}$}~\cite{HFLAV16}.

\begin{figure}[ht]
\begin{center}
\includegraphics[width=8cm]{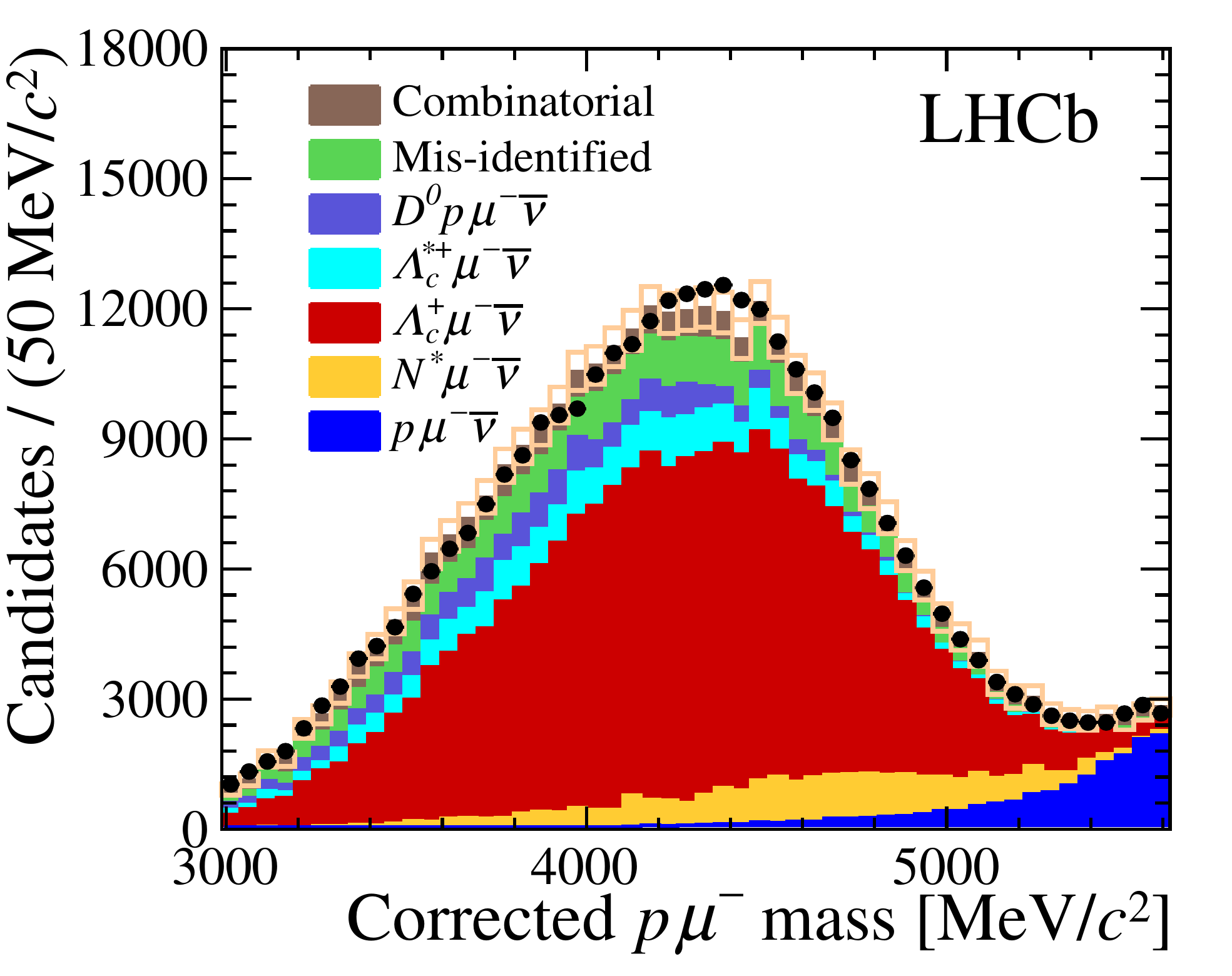} 
\caption{The $(p \mu^-)$ mass distribution in the measurement of $|V_{ub}|$,  showing the contribution from 
$\Lambda_b \to p \mu^- \nu_\mu$ decays and the various backgrounds. }
\label{fig:Vub}
\end{center}
\end{figure}

\item{ \bf The side opposite to $\alpha$ } 

The  mass difference $\Delta{m_s}$ measured in $\B^0_s$ mixing (Fig.~\ref{fig:boxDiag}),
which is dominated by the top-quark loop, 
provides a measurement of the third side of the 
triangle, {\large ${{V_{td}}\over{\lambda V_{ts}}}$}. This is proportional to the
ratio of mixing frequencies  {\large $\sqrt{\frac{\Delta m_{d}}{\Delta m_{s}}}$}. 
Corrections are calculated from the lattice with a theoretical error of $\sim$5-10\%
and systematic errors largely cancel in the ratio.

Following the measurement by \lhcb of the mixing parameters presented 
above, the ratio  $\left|V_{t d} / V_{t s}\right|$ is $0.210 \pm 0.001 \pm 0.008$~\cite{PDG2019}.
Systematic errors can be reduced in the future by improved  lattice QCD calculations.
\end{itemize}

\subsection{Other \CP violation measurements}

\subsubsection{$\B_s$ weak mixing phase $\phis$ in $\Bs \to J/\psi \phi$}

It can be seen in Fig.~\ref{fig:boxDiag} 
that $V_{ts}$ appears twice in the $\Bs-\overline{\B}^0_s$ mixing process, introducing a relative  
``weak mixing phase'', of $\phis$  to fourth order in $\lambda$. 
The $\B_s$ mixing phase can be  measured in the channel $\Bs \to J/\psi \phi$, which  
 is governed by a single tree-level diagram
with a negligible  penguin contribution. 
Hence this mode is the strange-quark analogue 
of the golden mode $\Bd \to {J}/\psi {K}^0_{S}$ in the $\B^0$ system. In the $\B_s$ system  \CP asymmetry arises from the interference of the $\Bs \to J/\psi \phi$  with the mixed process
$\Bs \to \overline{\B}^0_s \to J/\psi \phi$.
In the Standard Model, $\phis$  is expected to be very small, $\sim 0.036 \pm 0.002$~rad~\cite{CKMfitter2005},  
hence this channel is a very sensitive probe for new physics.


\lhcb reconstructs  
\mbox{$\Bs \to J/\psi \phi$} 
events in the 
decay modes \mbox{$J/\psi \to \mu^+\mu^-$}, 
and
\mbox{$\phi \to K^+K^-$}~\cite{LHCb-PAPER-2019-013}.
This $\Bs$ final state is an admixture of \CP-even and odd contributions, therefore an angular analysis of decay products is required.
Good tagging performance of $\Bs$ and $\overline{\B}^0_s$ is important, with a total tagging power 
in this analysis of  $4.73 \pm 0.34$. 

\begin{figure}[ht]
\begin{center}
\includegraphics[width=11cm]{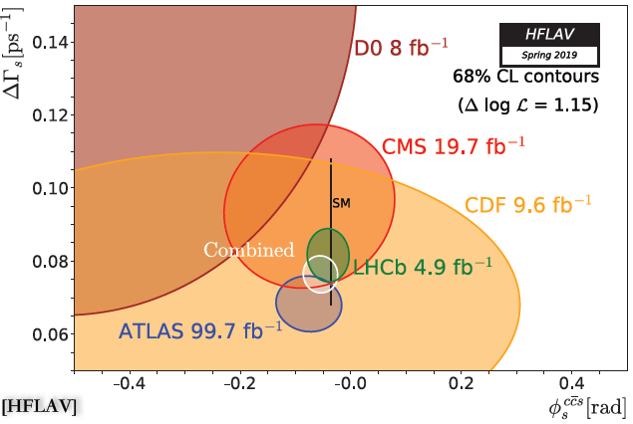} 
\caption{The  measurement of the $\Bs$ weak mixing phase in the 
$\left(\phis, \DGs\right)$ plane, showing the \lhcb result in combination
with those from other experiments, the Standard Model expectation, and the  $1\sigma$ 
contour of the combination~\cite{HFLAV16}. }
\label{fig:HFLAV-phiS}
\end{center}
\end{figure}

The fitted value of $\phis$   is correlated with $\DGs$, the width difference of the $\Bs$
mass eigenstates. 
The decay $\Bs \to J/\psi \pi^{+} \pi^{-}$ is also added  to 
improve the sensitivity~\cite{LHCb-PAPER-2019-003}. Contours in the $\left(\phis, \DGs\right)$ plane are plotted in Fig.~\ref{fig:HFLAV-phiS}. 
The \lhcb measurements are  $\DGs=0.0816 \pm 0.0048~ \mathrm{ps}^{-1}$ 
 with the  \CP-violating phase   $\phis=-0.041 \pm 0.025 ~\mathrm{rad}$.
 Figure~\ref{fig:HFLAV-phiS} also shows the world-averaged  measurement of
 $ \phis ~\mathrm{ versus }~ \DGs$ showing the
 \lhcb result in combination with those from other experiments and the Standard Model expectation~\cite{HFLAV16}.

\subsubsection{\CP violation in charm}

The Standard Model prediction of \CP violation  in the charm system is expected 
to be very small $\order\left(10^{-4}\right) \to \order\left(10^{-3}\right)$, where
\CP violation can arise in Cabibbo-suppressed (CS) decays in the interference between tree and penguin amplitudes.
In particular \lhcb has measured asymmetries in the direct \CP-violating
channels $\D^{0} (\overline{\D}^{0}) \to \pi^+\pi^- $
and $\D^{0} (\overline{\D}^{0}) \to K^+K^-$~\cite{LHCb-PAPER-2019-006}.

In the \lhcb analysis, $\D^{0}$ and $\overline{\D}^{0}$ decays are identified via  two self-tagging 
decay paths. ``Prompt''  decays ($\D$ decays originating  from the primary vertex) 
are  characterised by the presence of a ``soft'' low-momentum
pion from a $\D^*$ i.e. $\D^{*+} \to \D^{0} \pi^{+}_{\rm soft}$ and the charge-conjugate mode. ``Semileptonic'' decays  are secondary $\D$'s which originate from prompt $\B$ decays, 
i.e. $\B^{+} \to \overline{\D}^{0} \mu^{+} X$ and its charge-conjugate state.

The raw asymmetry $(\mathcal{A})$ for
$\D^{0} \rightarrow h^{+}h^{-}$ decays $(h = K$ or $\pi)$ is defined as

\begin{equation}
\mathcal{A}(D \rightarrow f)=\frac{N(D \rightarrow f)-N(\overline{D} \rightarrow \overline{f})}{N(D \rightarrow f)+N(\overline{D} \rightarrow f)}
\end{equation}

\noindent which includes both physics and detector terms: $\mathcal{A}=\mathcal{A}_{\CP}+\mathcal{A}_{D}+\mathcal{A}_{P}$.
Detection asymmetry arises from small charge differences associated with the $\pi^{\pm}_{{\rm soft}}$  or $\mu^{\pm}$.
 Production asymmetry arises from different production rates of $\D^{*}$ and $\B$ in $pp$ collisions.
To eliminate these two contributions and cancel associated systematics, 
the $\Delta \mathcal{A}_{\CP}$ parameter is measured in \lhcb :
\begin{equation}
\Delta \mathcal{A}_{\CP}=\mathcal{A}\left(K^+K^-\right)-\mathcal{A}\left(\pi^+\pi^-\right)=\mathcal{A}_{\CP}\left(K^+K^-\right)-\mathcal{A}_{\CP}\left(\pi^+\pi^-\right) .
\end{equation}

\noindent The raw symmetries are obtained from mass fits, and then by simply counting the 
numbers of $\D$'s decaying to $\pi^+\pi^-$ and  $K^+K^-$, respectively.

A measurement performed with  Run~1 and Run~2 \lhcb  data combined gives 
 \mbox{$\Delta \mathcal{A}_{\CP}=(-15.4 \pm 2.9) \times 10^{-4}$}. 
 This is a 5.3$\sigma$ measurement of 
\CP violation in the charm system and opens a new window for the study of \CP violation in the future.

\subsubsection{\CP violation in beauty baryons}

\CP violation has been observed in $\B$, $K$, and $\D$ decays, but not yet in baryon decays.
A search for \CP  violation 
in the multi-body mode $\Lambda_{b}^{0} \to p^{+} \pi^{-} \pi^{+} \pi^{-}$ decays
was  performed on \lhcb Run 1 data~\cite{LHCb-PAPER-2016-030}.
This decay proceeds via tree and loop diagrams with similar
contributions and through  numerous intermediate resonances, enhancing the possibility
for \CP violation,  although in areas where re-scattering effects can play a role. 
A 3.3$\sigma$  deviation from \CP
symmetry was  observed, however introducing 6.6 fb$^{-1}$ of Run 2 data has not confirmed
this result~\cite{LHCb-PAPER-2019-026}.
Hence this measurement awaits further statistics, and will be improved when cleaner
2-body $\B$-baryon decays can be added to the study.

\subsubsection{\CP violation in charmless 3-body $\Bpm$ decays}

\begin{figure}[ht]
\begin{center}
\includegraphics[width=11cm]{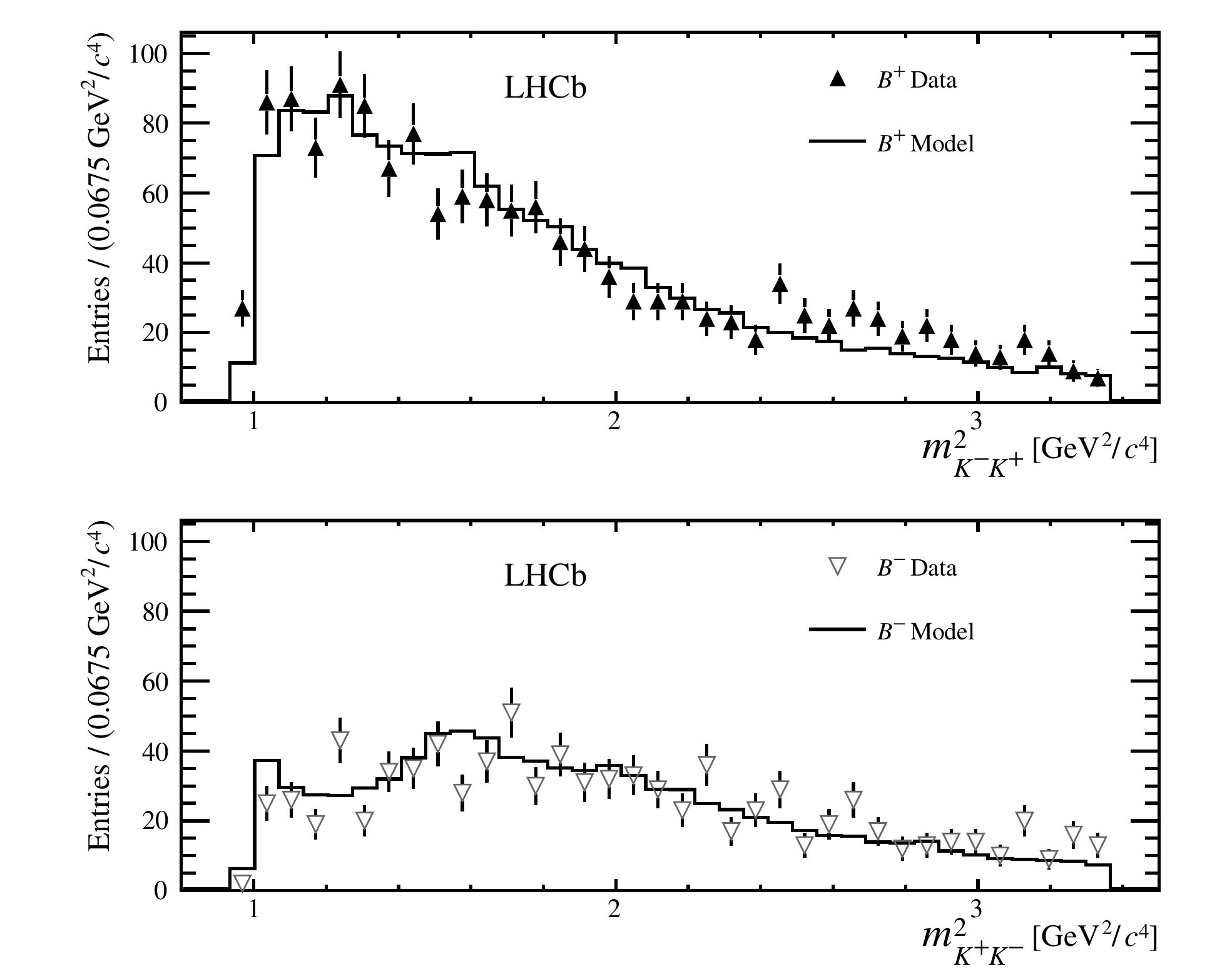}
\caption{Yields of (top) $\Bp \to \pi^+ K^+ K^-$
and (bottom) $\Bm \to \pi^- K^- K^+$ decays as a function of $K^+K^-$ mass squared. }
\label{fig:3-body}
\end{center}
\end{figure}

Yields of $\Bp \to \pi^+ K^+ K^- $
and $\Bm \to \pi^- K^- K^+$ 
decays show striking asymmetries
in the region of phase space dominated by re-scattering effects~\cite{LHCb-PAPER-2018-051}.  
Similarly huge \CP-violating effects  between $\Bp \to \pi^+ \pi^+ \pi^- $ 
and $\Bm \to \pi^- \pi^- \pi^+$ decays 
are observed in a region of phase space including the $\rho^0(770)$ and $f_2(1270)$ 
resonances~\cite{LHCb-PAPER-2019-018,LHCb-PAPER-2019-017}.  
Figure~\ref{fig:3-body} shows an example of the spectacular asymmetries observed in $\B \to \pi K K$ decays, which exceed  50\% at low values of $K^+K^-$ mass.

\clearpage
%
%


\section{ Rare decays }

Within the SM the interplay of weak and Higgs interactions implies that Flavour Changing
Neutral Currents (FCNCs) can occur only at higher orders in the electroweak interactions and are 
strongly suppressed by the GIM mechanism. This strong suppression
makes FCNC processes natural candidates to search for physics beyond the SM. If the new 
degrees of freedom do not have the same flavour structure of the quarks/leptons-Higgs
interactions present in the SM, then they could contribute to FCNCs at a comparable (or even larger) level 
to the SM amplitudes.
 
\begin{figure}[ht]
\begin{center}
\includegraphics[width=0.45\linewidth]{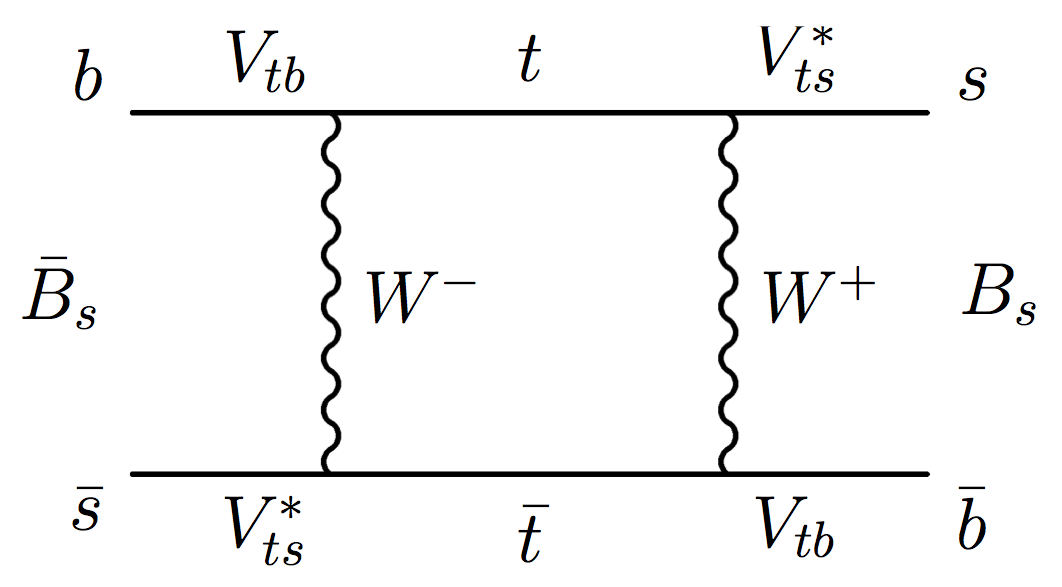}
\includegraphics[width=0.45\linewidth]{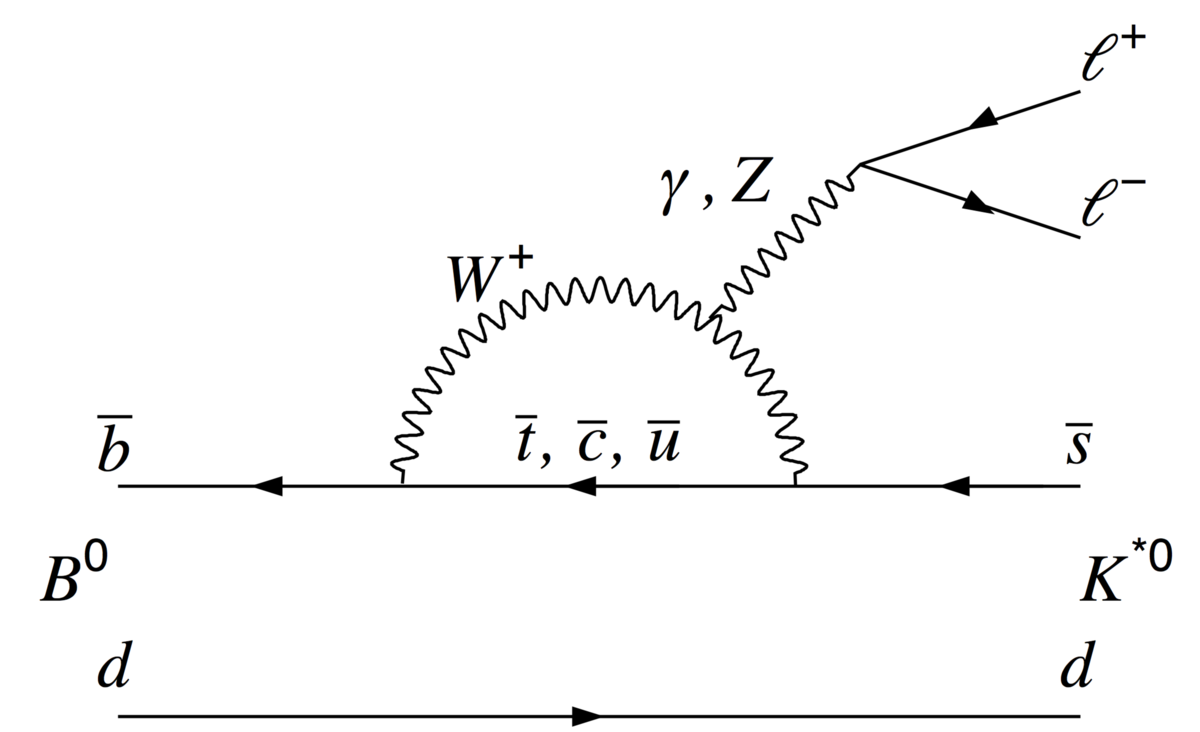}
\caption{Examples of loop processes within the SM that allow  FCNC
$\bquark \rightarrow \squark$ quark transition. On the left is an
example of a box diagram and on the right  an example 
of a penguin diagram.}
\label{FeynmanRareDecays}
\end{center}
\end{figure}

In \B-meson decays, experimenters have measured $\bquark \rightarrow \squark$  and 
$\bquark \rightarrow \dquark$ quark transitions, while $\cquark \rightarrow \uquark$ and
$\squark \rightarrow \dquark$ transitions have been measured in \D-meson and \kaon-meson decays, respectively.
At first order, these transitions can occur through two kinds of Feynman diagram shown in
Fig.~\ref{FeynmanRareDecays}. The first corresponds to the so-called ``box'' diagram and describes the mixing between neutral mesons, discussed in Section \ref{sec:CPviolation},
the example of Fig.~\ref{FeynmanRareDecays} shows $\Bs$ mixing. The second kind 
of diagram, the so-called "penguin" diagram, is responsible for a large variety of FCNC rare
decays. The example shown in Fig.~\ref{FeynmanRareDecays} is 
that of a \bsll
transition. In particular,
if the  radiated bosons are of electroweak type  (\Z, \W or \g-like), the uncertainties in the 
calculation of the SM predictions due to 
non-perturbative QCD effects are
drastically reduced as compared with the case where a gluon is radiated.
These "electroweak penguins" are 
the subject of the discussions in this section.

Before the first physics run of the \lhc accelerator in 2010, the main contributors to the study of rare \B and \D-meson decays were 
the \bfactory experiments (\babar and \belle) and the \tevatron experiments (\cdf and \dzero). However the production rate of 
\bquark\bquarkbar pairs in the \ep\en \bfactories was typically five orders of magnitude smaller than at the \lhc. In addition, the lower 
\proton\proton collision energy of the \tevatron   (with a correspondingly reduced \bquark\bquarkbar cross section which  is proportional to the collision energy), and the
detector's reduced trigger acceptance  for rare \B and \D-meson decays, 
implied that \lhcb would already be the most sensitive experiment even after 1 $\rm{fb}^{-1}$ accumulated in 2011.
For example, prior to the \lhc, the rarest \B-meson decay ever measured was 
$\BR(\Bu \rightarrow \Kp \mup \mun) \sim 5\times10^{-7}$ with a $20\%$ precision; \belle and \babar had analyzed $\order (200)$ $\Bd \rightarrow \Kstarz \ellell$ events and \cdf had reached a sensitivity of \BR(\Bsmm) $< 3 \times 10^{-8}$ at the $95\%$ C.L. As will be clear from the  next sections, \lhcb has already  reached after Run~1 of the \lhc an order of magnitude larger statistical power.

\subsection{$B^0_{\rm{(s)}} \rightarrow \mu^+ \mu^-$}
\label{Sec:bsmm}

The pure leptonic decays of \kaon,~\D and \B mesons are a particular interesting case
of electroweak penguins, where the final quark leg in Fig.~\ref{FeynmanRareDecays} (right)
needs to be swapped 
to the initial state.
The helicity configuration of the purely leptonic final state suppresses the
vector and axial-vector contributions by a factor in terms of masses  proportional to
$\left[ \tfrac {m_\mu}  {m_{\kaon,\D,\B}} \right]^2$. Therefore, these decays 
are particularly sensitive to new (pseudo-) scalar interactions.
In the case of \Bd and \Bs-meson decays, the
contribution of the absorptive part can be safely neglected.
As a consequence, 
the rate is well predicted theoretically~\cite{Beneke_2019} :
\mbox{$\BR(\Bsmm)=(3.66\pm0.05)\times 10^{-9}$} and
\mbox{$\BR(\Bdmm)=(1.03\pm0.05)\times 10^{-10}$}.
In the \Bs case, this prediction corresponds to a
flavour-averaged time-integrated measurement, taking into account
the correction due to the non-negligible width difference.

The experimental signature is sufficiently clean to reach an expected signal over background ratio
$S/B \sim 3$ for the \Bs decay, assuming the SM branching fraction.
The main background in the  region 
around the \Bs invariant mass is due to combinations of uncorrelated muons
and can be estimated from the mass sidebands. The most important handle to reduce
this combinatorial background is the 
invariant mass resolution of the experiment, $\sigma \sim 23$~MeV, which is also crucial to differentiate
between  \Bd and \Bs decays ($\Delta m \sim 87$~MeV).
Moreover, the large fraction of 
\BTohh
decays is an important source of background
due to hadrons being misidentified as muons in the region around the 
\Bd mass (this background is very small in the \Bs mass 
region). 

\begin{figure}[ht]
\begin{center}
\includegraphics[width=0.90\linewidth]{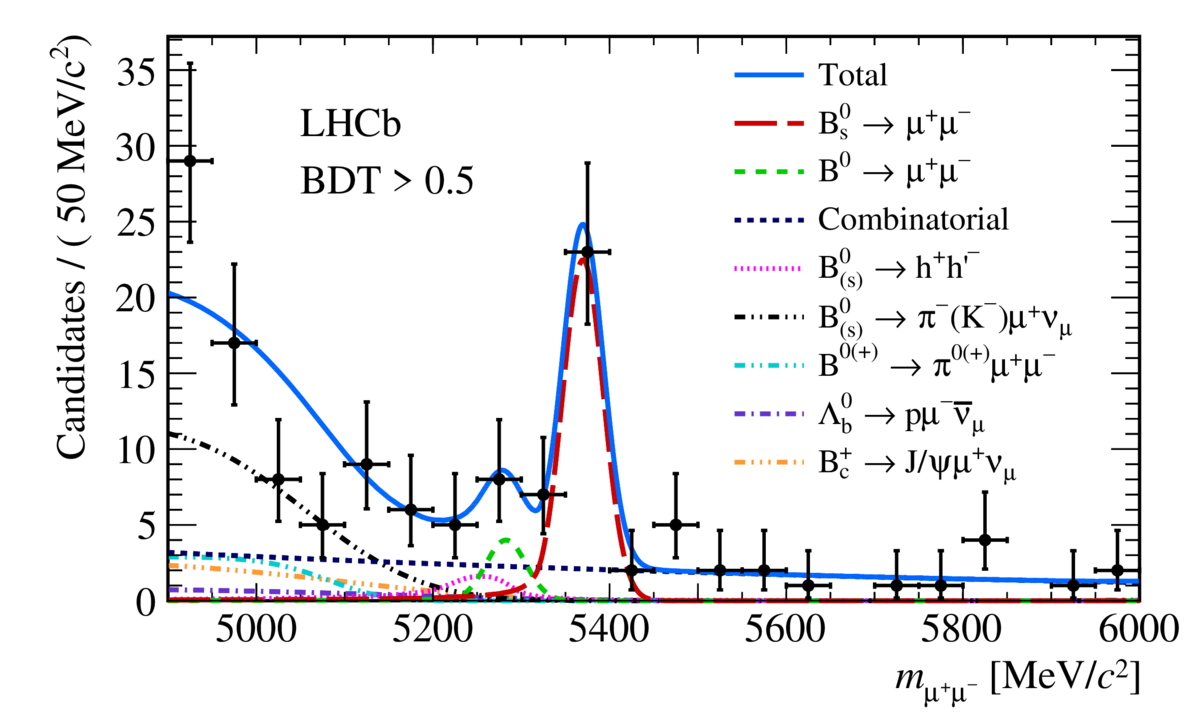}
\caption{Invariant mass distribution of  dimuons selected from Ref.~\cite{Aaij:2017vad}.
Superimposed on the data points in blue (solid line) is the combined fit, 
and its components as quoted in the insert.}
\label{LHCbBsmm}
\end{center}
\end{figure}

Given their experimental detector resolution and trigger acceptance
during Run~1 and Run~2, the CMS experiment with $61~\rm{fb}^{-1}$ of data has
similar sensitivity to the \lhcb experiment with 
$4.4~\rm{fb}^{-1}$ of data collected in the same period. Both experiments have
provided a clear observation of the decay \Bsmm
(CMS and \lhcb observe about 90 and 30 
\Bs signal candidates with a background of about 40 and
10 events respectively within the invariant mass interval expected to
contain $95\%$ of the signal). Neither experiment has yet reached the 
sensitivity to observe the decay \Bdmm.
The invariant mass 
distribution obtained by the LHCb experiment is shown 
in Fig.~\ref{LHCbBsmm} and the LHCb~\cite{Aaij:2017vad} and
CMS~\cite{CMS-PAS-BPH-16-004} results are 
\mbox{$\BR(\Bsmm) =(3.0\pm0.6_{-0.2}^{+0.3})\times 10^{-9}$} and
\mbox{$\BR(\Bsmm) =(2.9_{-0.6}^{+0.7} \pm 0.2)\times 10^{-9}$} respectively,
while for the \Bd  decay only limits can be quoted.
The ATLAS collaboration has also published after a 
measurement\cite{Aaboud_2019} using $51~\rm{fb}^{-1}$ of data,
\mbox{$\BR(\Bsmm) =(2.8_{-0.7}^{+0.8})\times 10^{-9}$} in agreement with the previous results.


\begin{figure}[ht]
\begin{center}
\includegraphics[width=0.90\linewidth]{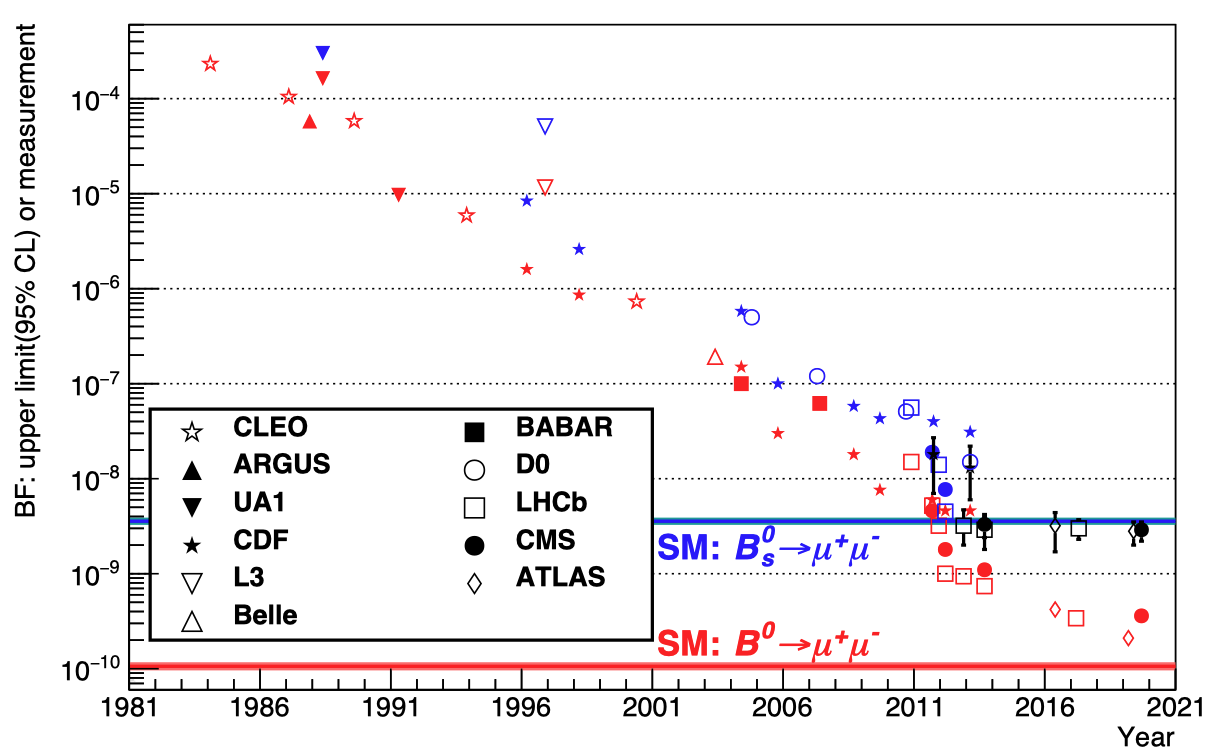}
\caption{Upper limits (\Bdmm) and eventual measurements of
the \Bsmm branching ratio  by several
experiments over the last 40 years.}
\label{LHCbBsmmHistory}
\end{center}
\end{figure}

The $\Bsmm$ very rare decay has been searched for ever since the discovery of \B mesons,
around 40 years ago. Thanks to the ingenuity and persistence of the
experimenters, it has been eventually measured at the \lhc and found to be 
in agreement with the SM within current uncertainties, as shown in
Fig.~\ref{LHCbBsmmHistory}. Over the next decade it will be extremely interesting to
see how the measurement of \BR(\Bdmm) evolves, 
for which only upper limits are currently available.

\subsection{\BdToKstmm}

The language of effective field theory is used to parameterise NP contributions
in terms of a sum of local four-fermion operators ($Q_i$) which depend only on SM fermions
modulated by Wilson coefficients ($C_i$) 
which in turn depend on the heavy degrees of freedom i.e. NP particles.
The decay \BdToKstmm is the so-called ``golden mode'' to test new
vector (axial-vector) couplings, i.e. the $C_9$ and $C_{10}$  Wilson coefficients contributing to the 
$\bquark \rightarrow \squark$ transition.
The \BdToKstmm channel also complements the $\bquark \rightarrow \squark \g$
decay which is mostly sensitive to NP dipole 
operators (i.e. $C_7$). and also the  \Bsmm decay which is mostly
sensitive to NP (pseudo-) scalar operators (i.e. $C_S$ and $C_P$). The charge
of the pion in the decay $\Kstar \rightarrow K\pi$ defines 
the flavour of the \B meson and an angular analysis can be performed unambiguously
to test the helicity structure of the electroweak penguin.

The above system is completely defined by four variables: $q^2$, the square of
the invariant mass of the dimuon system, $\theta_l$, the angle between the positive
lepton and the direction opposite to the \B-meson in the dimuon 
rest frame, $\theta_K$, the equivalent angle of the $K^+$ in the $K^*$ rest frame
and $\phi$ the angle between the two planes defined by $(K,\pi)$ and $(\mu^+,\mu^-)$ in
the \B-meson rest frame. The four-fold differential distribution 
contains a total of eleven angular terms that can be written in terms of seven $q^2$-dependent
complex decay amplitudes. These amplitudes can be expressed in terms of five complex Wilson
coefficients ($C_S, C_P, C_7, C_9$ 
and $C_{10}$), their five helicity counterparts and six form-factors, which play a role of
nuisance parameters in the fit.

The \lhcb experiment, with $3~\rm{fb}^{-1}$ of data collected in Run~1, has triggered and selected about 2400 \BdToKstmm candidates in the range $0.1 < q^2 < 19$ GeV$^2$ with 
signal over background $(S/B) >5$.
This is about one order of magnitude 
larger than the samples available at previous experiments (\babar, \belle and \cdf) and similar to
the samples collected by ATLAS and CMS with ten times the luminosity, however with significantly worse $S/B$.

\begin{figure}[ht]
\begin{center}
\includegraphics[width=0.45\linewidth]{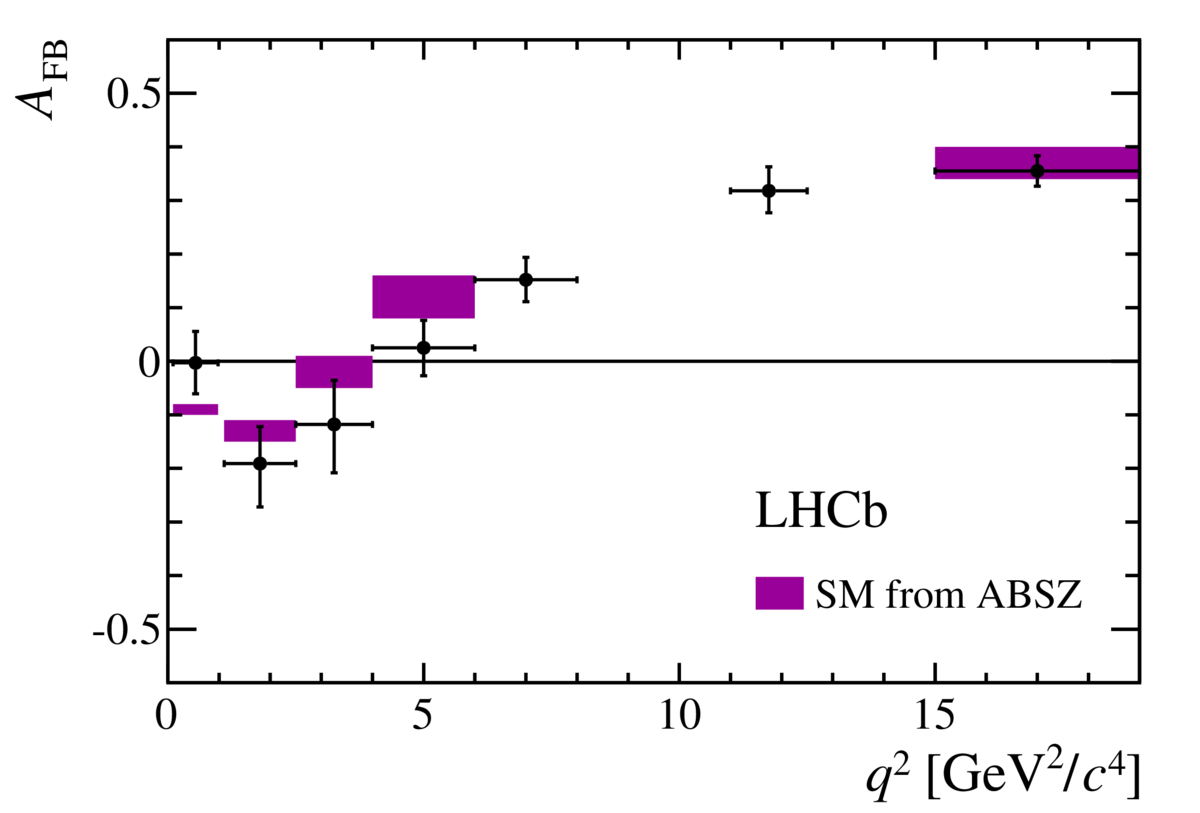}
\includegraphics[width=0.45\linewidth]{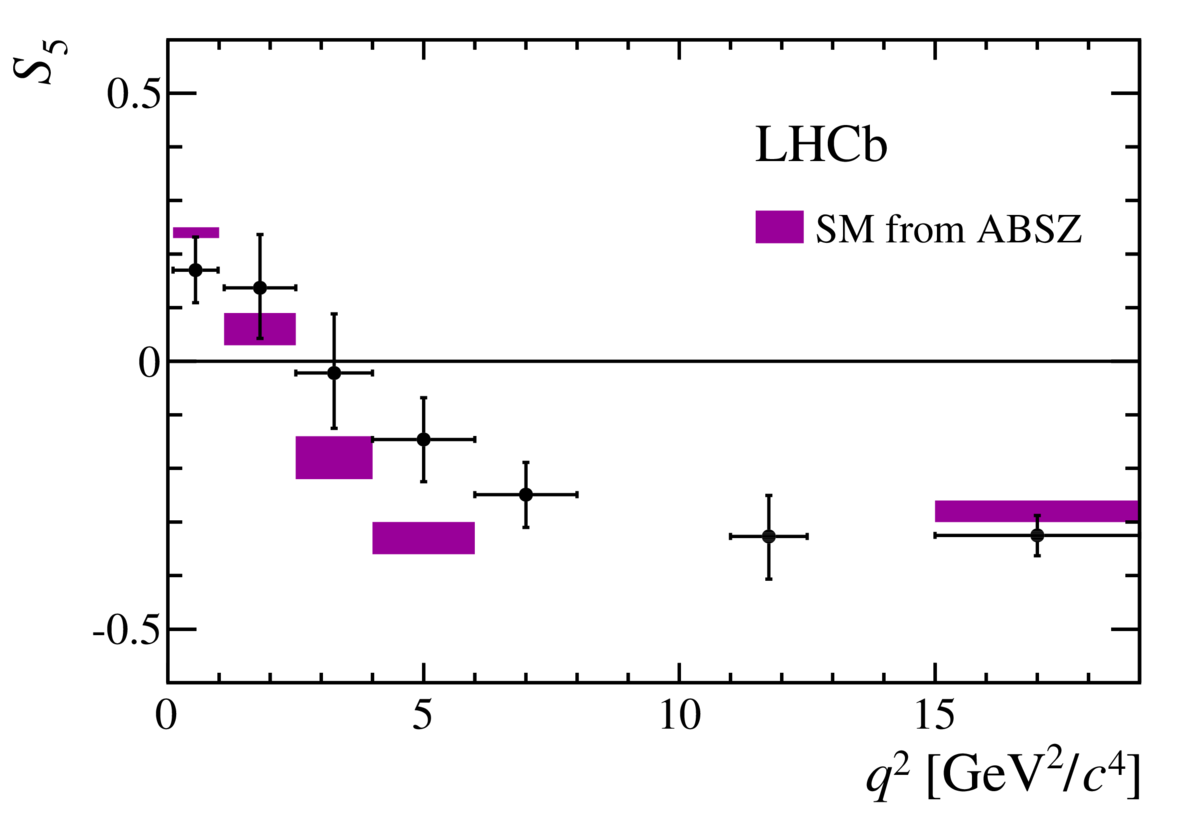}
\caption{Two examples of the \CP-averaged coefficients in
the \BdToKstmm angular terms as a function of $q^2$.
The shaded boxes show the SM predictions taken
from Ref.~\cite{Altmannshofer_2015}.}
\label{AngularAnalysis}
\end{center}
\end{figure}

The statistics and the quality of the data accumulated by the LHCb experiment allows for
a full angular analysis of \BdToKstmm decays to be performed for the first time.
The results~\cite{LHCb-PAPER-2015-051} of this ``tour de force'' analysis  
mostly agree  with SM predictions, however with some hints of disagreement for some
specific distributions. In Fig.\ref{AngularAnalysis}, two examples of the \CP-averaged angular coefficients
(i.e. the average of the coefficients measured with 
\Bd and \Bdb  decays)     are shown as a
function of $q^2$. For these two examples, $A_{\rm{FB}}$ (modulating
the $\sin^2\theta_K \times \cos\theta_l$ angular term) and 
$S_5$ (modulating the $\sin(2\theta_K) \times \sin\theta_l \times \cos\phi$
angular term) seem to agree less well with  SM predictions.
However these are early days, and more data will be required (the Run~2 data analysis will be released soon).
Also a careful reassessment of the SM uncertainties are
needed before drawing definitive conclusions.

\begin{figure}[ht]
\begin{center}
\includegraphics[width=0.60\linewidth]{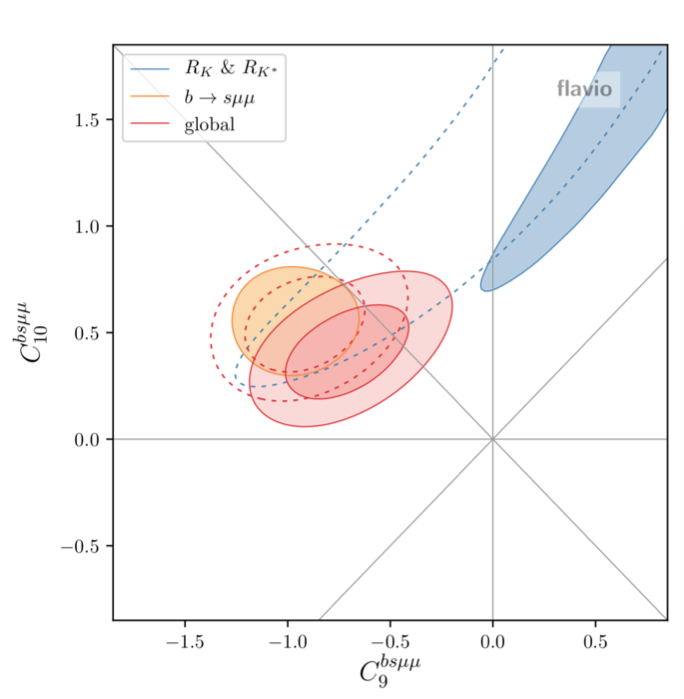}
\caption{Constrains on the contribution of NP to the real parts
of $C_9$ and $C_{10}$ at the $1\sigma$ and  $2\sigma$ level
taken from Ref.~\cite{aebischer2019bdecay}, corresponding to
different measurements as indicated in the insert. The 
dotted lines correspond to the status before the latest
updates for the 2019 physics winter conferences.}
\label{WILSON}
\end{center}
\end{figure}

Several authors have already attempted to see if the overall
pattern of the angular measurements is consistent with a given value of the relevant
Wilson coefficients. As previously discussed, the inclusive 
$\bquark \rightarrow \squark \gamma$ measurements strongly constrain non-SM values
for $C_7$. The scalar $C_S$ and pseudo-scalar $C_P$ coefficients are constrained,
for example, by the measurement of the 
branching fraction of the very rare decay \Bsmm.
Therefore, the small disagreements observed in the   angular analysis of
the decay \BdToKstmm and other decays, seem to 
be consistent with a non-SM  value of the $C_9$ Wilson coefficient,
as can be seen in Fig.~\ref{WILSON} taken from Ref.~\cite{aebischer2019bdecay}.

\subsection{Lepton Universality}

In the SM, the electroweak couplings of leptons are flavour independent, or lepton ``universal''.
However, this may not necessarily be the case for new particles beyond the SM. In particular,
if the hints described in the 
previous section are an indication of new particles modifying the penguin diagram
in Fig.~\ref{FeynmanRareDecays}, it is interesting to measure the ratios of
branching fractions in decays of different lepton families.
For example, the ratios 
\begin{equation}
R_{X} = \int \frac{ d\Gamma(B^0 \rightarrow X \mu \mu) }{dq^2} dq^2 \Bigg/
\int \frac{ d\Gamma(B^0 \rightarrow X e e) }{dq^2} dq^2 
\end{equation}
between \B decays to final states with muons and electrons,
where $X$ is a hadron containing an \squark-quark or  \dquark-quark, are predicted
to be very close to unity in the SM~\cite{Hiller:2003js,Bobeth:2007dw,Bouchard:2013mia}.
The uncertainties from QED corrections are found to be at the percent level~\cite{Bordone:2016gaq}. 

\begin{figure}[ht]
\begin{center}
\includegraphics[width=0.60\linewidth]{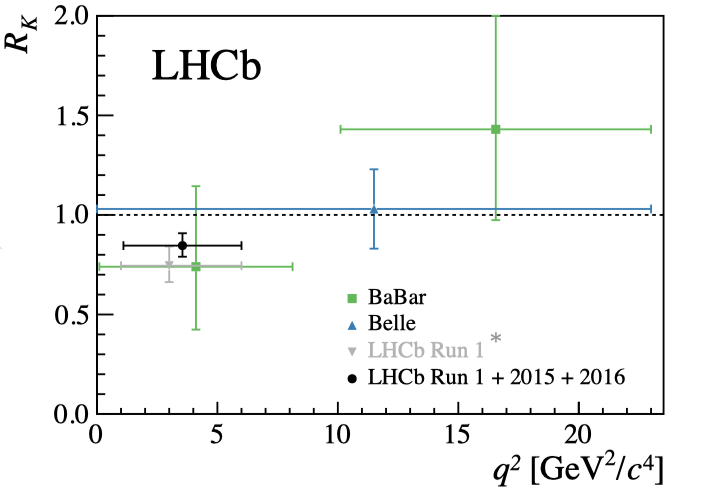}
\caption{ Measurement of $R_K$ as a function of the $q^2$ bins by different experiments.}
\label{RK}
\end{center}
\end{figure}

\lhcb has measured the above ratio in several channels. Using Run~1 and part of
 Run~2 data ($4.4~\rm{fb}^{-1}$),  \lhcb measures~\cite{LHCb-PAPER-2019-009}
$R_K = 0.846^{+0.060}_{-0.054}\rm{(stat)}^{+0.014}_{-0.016}\rm{(syst)}$ in the range 
$1.1 < q^2 < 6$ GeV$^2$, about $2.5 \sigma$ below the SM prediction,
and using only Run~1 data ($3~\rm{fb}^{-1}$) \lhcb measures~\cite{LHCb-PAPER-2017-013}
$R_\Kstar = 0.69^{+0.11}_{-0.07}\rm{(stat) }\pm 0.05 \rm{(syst)}$ in the same $q^2$
range, with a similar level of disagreement with the SM prediction. The latest
$R_K$ results from LHCb  in bins of $q^2$ can be seen in Fig.~\ref{RK}, compared
with previous results from the \babar and \belle collaborations. The consistency
between different experiments, and different channels, although with very different
precision, has motivated many theoretical studies that relate these hints of lepton
non-universality with the discrepancies described in the previous section.
Figure~\ref{WILSON} shows the status of the compatibility of both
sets of measurements when assuming that only the $\bquark\squark\mu\mu$ Wilson coefficients
are modified (and $\bquark\squark e e$ coefficients are as predicted by the SM).
Whilst the initial results showed remarkable consistency between different
sets of measurements as shown by the dotted lines in Fig.~\ref{WILSON}, the
2019 latest  updates show a less clear picture. As of today, it is difficult
to draw reliable conclusions,  and more data is eagerly awaited.

\clearpage
\section{Spectroscopy}

A deep understanding of  quantum chromodynamics, 
the~theory of strong interactions, is vital 
for~precision tests of the~Standard Model  
and in searches for new physics beyond.
QCD is intensively tested 
in deep\nobreakdash-inelastic\nobreakdash-scattering processes
and  heavy vector boson production, 
however in the low\nobreakdash-energy regime, 
there is a~lack of precise QCD predictions.
QCD, being a~non\nobreakdash-pertubative theory,
does not calculate hadron properties, namely masses
and decay widths from first principles. 
Alternative theoretical approaches
are developed, such as 
heavy quark effective theory,
heavy quark expansion or lattice calculations.
These approaches require verification with experiment in 
various regimes, \eg 
testing the~agreement with data 
for hadrons with different quark content and quantum numbers. 
Spectroscopic measurements of hadron  masses and widths or 
lifetimes provide a~wide variety of 
tests for QCD models.

Huge production cross sections of charm and beauty in 
high\nobreakdash-energy $\proton\proton$~collisions in the~forward 
region at \lhcb~\cite{LHCb-PAPER-2010-002, 
LHCb-PAPER-2011-003, 
LHCb-PAPER-2011-036,
LHCb-PAPER-2011-043,
LHCb-PAPER-2013-004,
LHCb-PAPER-2013-016,
LHCb-PAPER-2013-066,
LHCb-PAPER-2014-029,
LHCb-PAPER-2015-037,
LHCb-PAPER-2015-041,
LHCb-PAPER-2015-045,
LHCb-PAPER-2016-031,
LHCb-PAPER-2016-042,
LHCb-PAPER-2017-037,
LHCb-PAPER-2018-002,
LHCb-PAPER-2018-049,
LHCb-PAPER-2019-024},
together 
with a good reconstruction efficiency, 
versatile trigger scheme and 
an excellent momentum and mass resolution, 
opens up
exciting opportunities for spectroscopy measurements.
The employment of \lhcb's~powerful  hadron identification system~\cite{LHCb-PROC-2011-008,LHCb-DP-2012-003,LHCb-DP-2014-002} 
enables a substantial reduction 
in the combinatorial background  
  specific to~high\nobreakdash-energy hadron\nobreakdash-hadron 
collisions. The~unique hadron identification           
is also especially important for spectroscopy measurements  
involving charged kaons and/or protons in the~final state.
The~excellent momentum and vertex resolutions provided by 
the~\lhcb tracking system allows unprecedented precision on mass and width measurements: indeed  
 the~most precise  
measurements of mass for all open beauty particles 
and lifetimes of all open heavy flavour particles, 
currently result from the~\lhcb experiment~\cite{PDG2018}.
Additional good control over the~momentum 
scale and the detector alignment~\cite{LHCb-PAPER-2012-048,
LHCb-PAPER-2013-011} also 
allows  the~natural widths
of  hadronic resonances to be probed with 
world-leading  sub\nobreakdash-$\!\mev$~precision~\cite{LHCb-PAPER-2012-012,
LHCb-PAPER-2014-061,
LHCb-PAPER-2016-010,
LHCb-PAPER-2017-036,
LHCb-PAPER-2019-045,
LHCb-PAPER-2020-008,
LHCb-PAPER-2020-009}.

Almost 30 new hadrons have been 
discovered~\cite{Koppenburg:2020uix} using 
the~Run~1\&2 data-set, among them 
several  open charm mesons and baryons, 
double\nobreakdash-charm baryon, a~charmonium state, 
several beauty mesons, 
fifteen beauty baryons, and 
pentaquark\nobreakdash-like
$P_{\cquark}(4312)^+$,
$P_{\cquark}(4400)^+$ and 
$P_{\cquark}(4457)^+$~states.  
The~quantum numbers of many hadrons, especially 
for charm mesons and charmonia\nobreakdash-like exotic candidates,  
have been determined using  amplitude\nobreakdash-analysis techniques. 
In the~case of exotic particles such as 
pentaquark and tetraquark candidates, the determination 
of quantum numbers is vital for 
the~understanding of their nature. 
Large statistics and low levels of background in decays of beauty hadrons with 
\mbox{$\decay{\jpsi}{\mup\mun}$}
or \mbox{$\decay{\psitwos}{\mup\mun}$}~final states 
allow the study of the~spectroscopy of light unflavoured or strange  hadrons~\cite{
LHCb-PAPER-2011-002,
LHCb-PAPER-2011-026,
LHCb-PAPER-2012-022,
LHCb-PAPER-2012-040,
LHCb-PAPER-2012-045,
LHCb-PAPER-2012-053,
LHCb-PAPER-2013-023,
LHCb-PAPER-2013-047,
LHCb-PAPER-2013-055,
LHCb-PAPER-2013-069,
LHCb-PAPER-2014-009,
LHCb-PAPER-2014-012,
LHCb-PAPER-2014-014,
LHCb-PAPER-2014-016,
LHCb-PAPER-2015-056,
LHCb-PAPER-2015-038,
LHCb-PAPER-2016-009,
LHCb-PAPER-2016-018,
LHCb-PAPER-2016-019,
LHCb-PAPER-2016-040}.
Selected spectroscopy highlights are discussed in this chapter.

\subsection{Charm hadrons}\label{sec:spec:charm}

Two  complementary methods for the study of spectroscopy of charm hadrons 
have been exploited in \lhcb: 
\begin{itemize} 
\item the study of promptly produced charm hadrons, 
\item the study of charm particles produced in weak decays of beauty hadrons, 
  \eg~from exclusive
  $\decay{\B}{\D^{(\ast)}}\Ppi\Ppi$~decays. 
\end{itemize}
The~first technique allows the most efficient exploitation of the~huge 
prompt $\cquark\cquarkbar$~production cross section in high energy 
hadron collisions,  but this is usually affected by a~large background   
from light hadrons produced at the~$\proton\proton$~collision vertex. 
The~second  technique exploits a~full amplitude analysis of the~exclusive 
decays of beauty hadrons and therefore involves much lower statistics,  however  the method often allows the determination of quantum numbers 
of the~charm hadrons. 
The study of $D \pi$ final states enables a~search for natural 
spin\nobreakdash-parity resonances,\,($P=(-1)^J$, labelled as $\D^*$) 
whilst the study of $\D^*\pi$ final states provides the possibility 
of studying both natural and unnatural spin\nobreakdash-parity states, 
except for the $J^P=0^+$ case, which is forbidden because of angular momentum and parity conservation.
In~inclusive $D^{(*)}\pi$ production, the~production of any $J^P$ state is permitted.
An amplitude analysis of \B~decays allows a~full spin\nobreakdash-parity analysis of 
the~charmed mesons present in the decay. 

Both the above approaches are  complementary, and   have resulted in 
discoveries of several new charm hadrons, amongst them  the excited charm mesons. 
Many previous meson and baryon states,  discovered earlier by other experiments,
have been confirmed with high 
statistical significance, and their masses and widths have been measured
with high precision. For many, the quantum 
numbers were either measured or constrained. 

\paragraph{Excited charm mesons in the
  $\Dstarp\pim$, 
  $\Dp\pim$ and 
  $\Dz\pip$ spectra}
have been studied in \lhcb
using a 1\invfb data-set collected at $\sqs=7\tev$~\cite{LHCb-PAPER-2013-026}.
Four resonances labelled   
$\D_0(2550)$, 
$\D^{\ast}_J(2600)$, 
$\D(2740)$
and 
$\D_3^{\ast}(2750)$ have been observed.
The~$\D_0(2550)$, 
and 
$\D(2740)$
decay angular distributions 
are consistent with an~unnatural 
spin\nobreakdash-parity, 
whilst the 
$\D^{\ast}_J(2600)$, 
and 
$\D_3^{\ast}(2750)$ 
states are assigned natural parities.
For~the~$\D_0(2550)$, 
meson, angular distributions 
are consistent with a $J^P=0^-$ assignment, 
however for the other states, no definite assignment exists.

The~excited charm mesons were also studied in \B-decay amplitude analyses 
of
\mbox{$\decay{\Bz}{\Dzb\pip\pim}$}
and 
\mbox{$\decay{\Bm}{\Dp \pim \pip}$}~decays
using the~Run~1 data-set~\cite{LHCb-PAPER-2014-070,LHCb-PAPER-2016-026}
and 
\mbox{$\decay{\Bm}{\Dstarp \pim \pip}$}~\cite{LHCb-PAPER-2019-027}
using a~data-set corresponding to 4.7\invfb, 
collected at
\mbox{$\sqs=7$}, 8 and~13\tev.
The~$\D^{\ast}_3(2760)^0$ and $\D^{\ast}_2(3000)^0$~states were observed
for the~first time~\cite{LHCb-PAPER-2016-026}, 
and the~most precise determination of masses, 
widths and quantum numbers has been performed for the
$\D^{\ast}_0(2400)^-$, 
$\D^{\ast}_2(2660)^-$, 
$\D^{\ast}_3(2760)^-$~\cite{LHCb-PAPER-2014-070},
$\D^{\ast}_2(2660)^0$,
$\D^{\ast}_1(2680)^0$,
$\D^{\ast}_3(2760)^0$,
$\D^{\ast}_2(3000)^0$~\cite{LHCb-PAPER-2016-026}
and for the
$\D_1(2420)^0$, 
$\D_1(2430)^0$, 
$\D_0(2550)^0$, 
$\D^{\ast}_1(2600)^0$,
$\D_2(2740)^0$,
$\D^{\ast}_3(2750)^0$~\cite{LHCb-PAPER-2019-027}~states.

\paragraph{Excited $\Ds$~mesons}were studied both in inclusive production  
in the $\Dp\KS$ and $\Dz\Kp$~spectra using 
a~data-set of 1\invfb collected
at $\sqs=7\tev$~\cite{LHCb-PAPER-2012-016},  and 
through the~amplitude analysis of 
$\decay{\Bs}{\Dzb\Km\pip}$~decays
using  Run~1 data~\cite{LHCb-PAPER-2014-035}.
It~has been shown that the previously reported 
$\D_{\squark J}^{\ast}(2860)^+$~state~\cite{Aubert:2006mh,Aubert:2009ah,LHCb-PAPER-2012-016}
is an admixture of spin\nobreakdash-1 and spin\nobreakdash-3 resonances.
The~masses and width of new states, 
dubbed  $\D_{\squark 1}^{\ast}(2860)^+$ 
and $\D_{\squark 3}^{\ast}(2860)^+$,
are precisely measured,
as well as the masses and widths 
of the~$\D_{\squark 2}^{\ast}(2573)^+$ and 
$\D_{\squark 1}^{\ast}(2700)^+$~\cite{LHCb-PAPER-2012-016,LHCb-PAPER-2014-035}.

\paragraph{Excited \Lc~baryons}were studied in their decays 
to the $\Dz\proton$~final state  via the~amplitude analysis of  \Lb~baryon decays 
using an~integrated luminosity sample of~3\invfb 
collected at $\sqs=7$ and 8\tev~\cite{LHCb-PAPER-2016-061}. 
The~analysis uses a~sample of $11\,212\pm126$ signal~\mbox{$\Lb\to\Dz\proton\pim$}~decays,
where the \Dz~mesons are reconstructed in the$\Km\pip$  final state.
The~amplitude fit is performed in the~four-phase space regions in the~Dalitz plot. 
For the near\nobreakdash-threshold $m_{\Dz\proton}$~region, an enhancement in the~$\Dz\proton$~
amplitude is studied. The~enhancement is
consistent with being a~resonant state, dubbed the $\PLambda_{\cquark}(2860)^+$, 
with quantum numbers \mbox{$J^P=\tfrac{3}{2}^+$}, and
with the~parity measured relative to that of 
the~$\PLambda_{\cquark}(2880)^+$~state. 
The~other quantum numbers are excluded with a~significance greater
than 6~standard deviations. The~phase motion of the $\tfrac{3}{2}^+$ component with respect 
to the~non-resonant amplitudes is obtained in a~model\nobreakdash-independent way 
and is consistent with resonant behavior.
The mass of the~$\PLambda_{\cquark}(2860)^+$~state is consistent with 
predictions for an~orbital D\nobreakdash-wave \Lc-excitation 
with quantum numbers $\tfrac{3}{2}^+$, based on the~nonrelativistic 
heavy quark\nobreakdash-light diquark model 
and from QCD sum rules in the~HQET framework. 
Also the~fit allowed the~most precise determination of~the masses and widths of 
the~known resonances $\PLambda_{\cquark}(2880)^+$ and $\PLambda_{\cquark}(2940)^+$, 
as well as constraining their quantum numbers. 

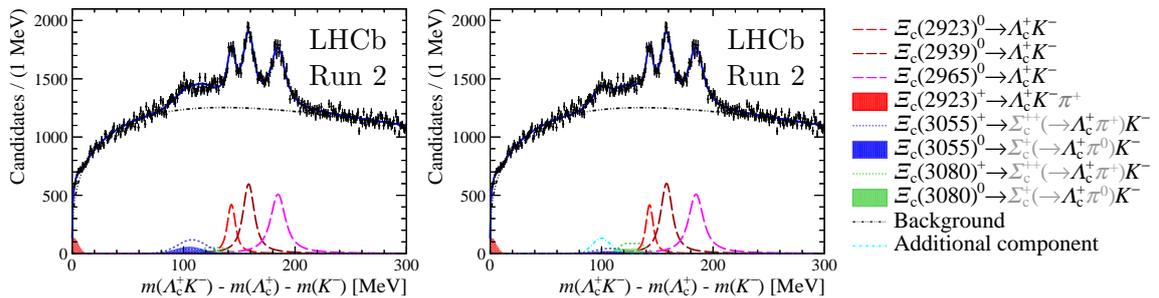
\begin{figure}[htb]
  \centering
  \setlength{\unitlength}{1mm}
  \begin{picture}(150,40)
    \put( 0, 0){ 
      \includegraphics*[width=55mm,
     ]{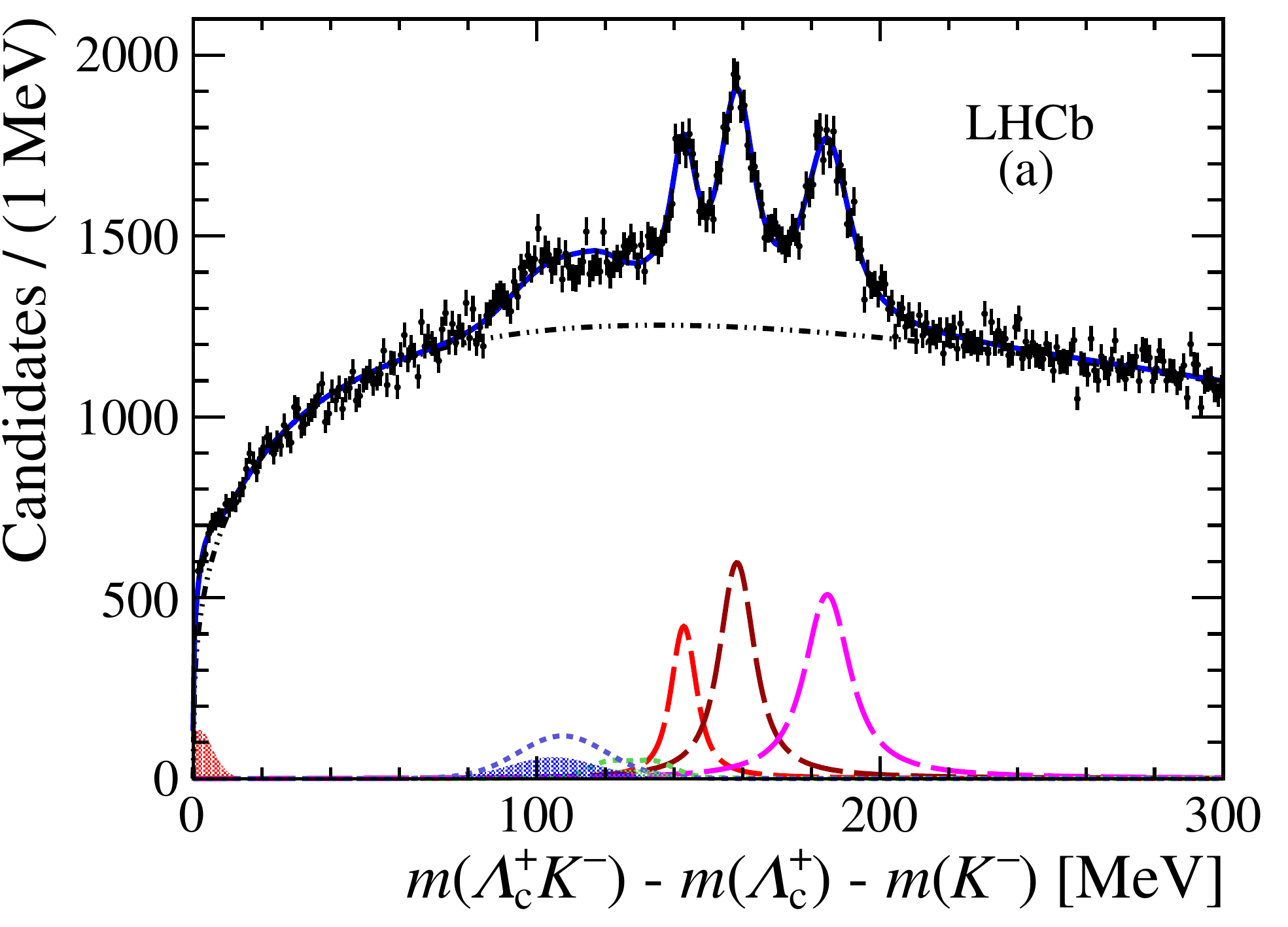}
    }
    \put(55, 0){ 
      \includegraphics*[width=55mm,
     ]{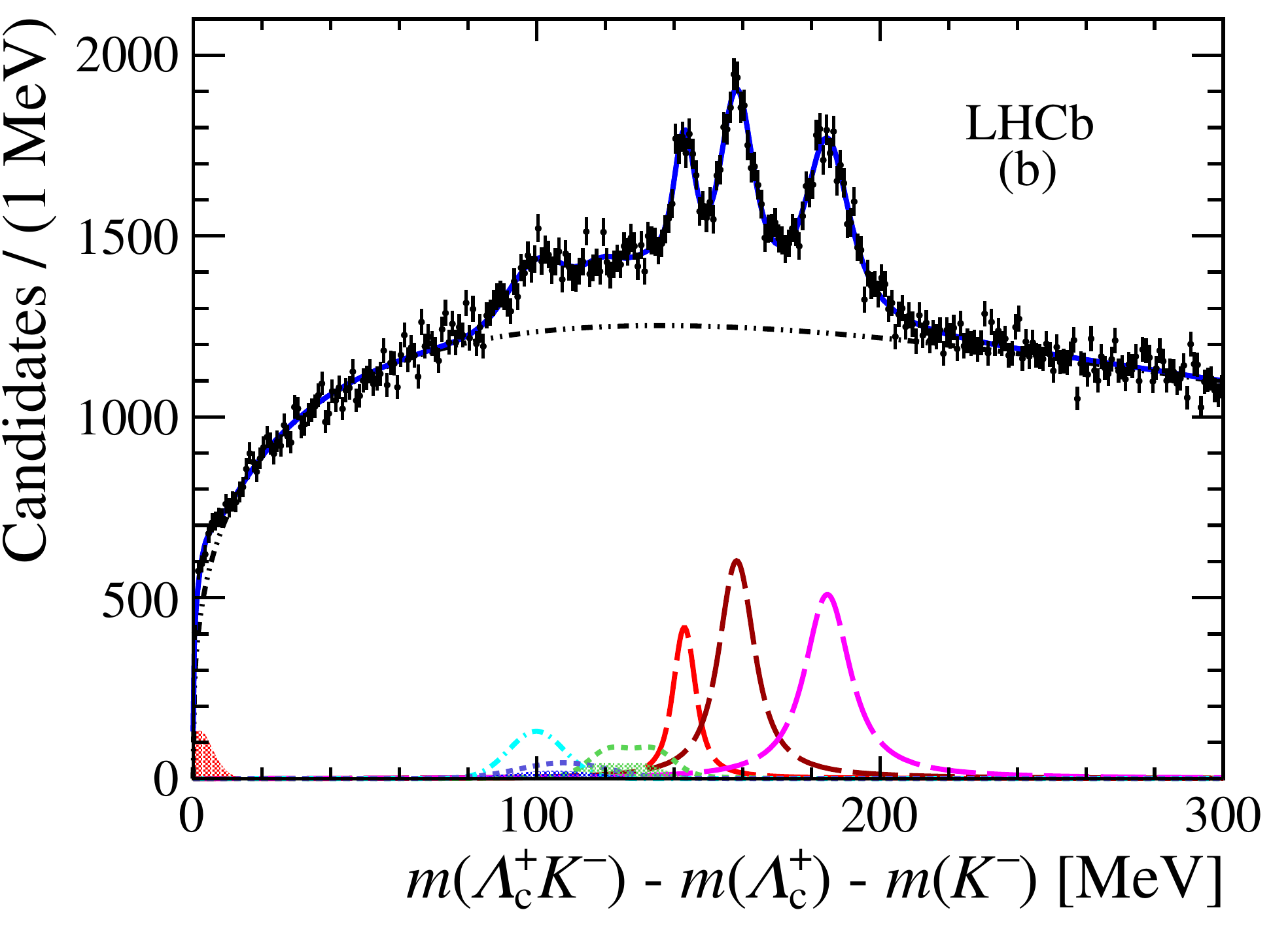}
    }
    \put(110,0){ 
      \includegraphics*[width=40mm,
     ]{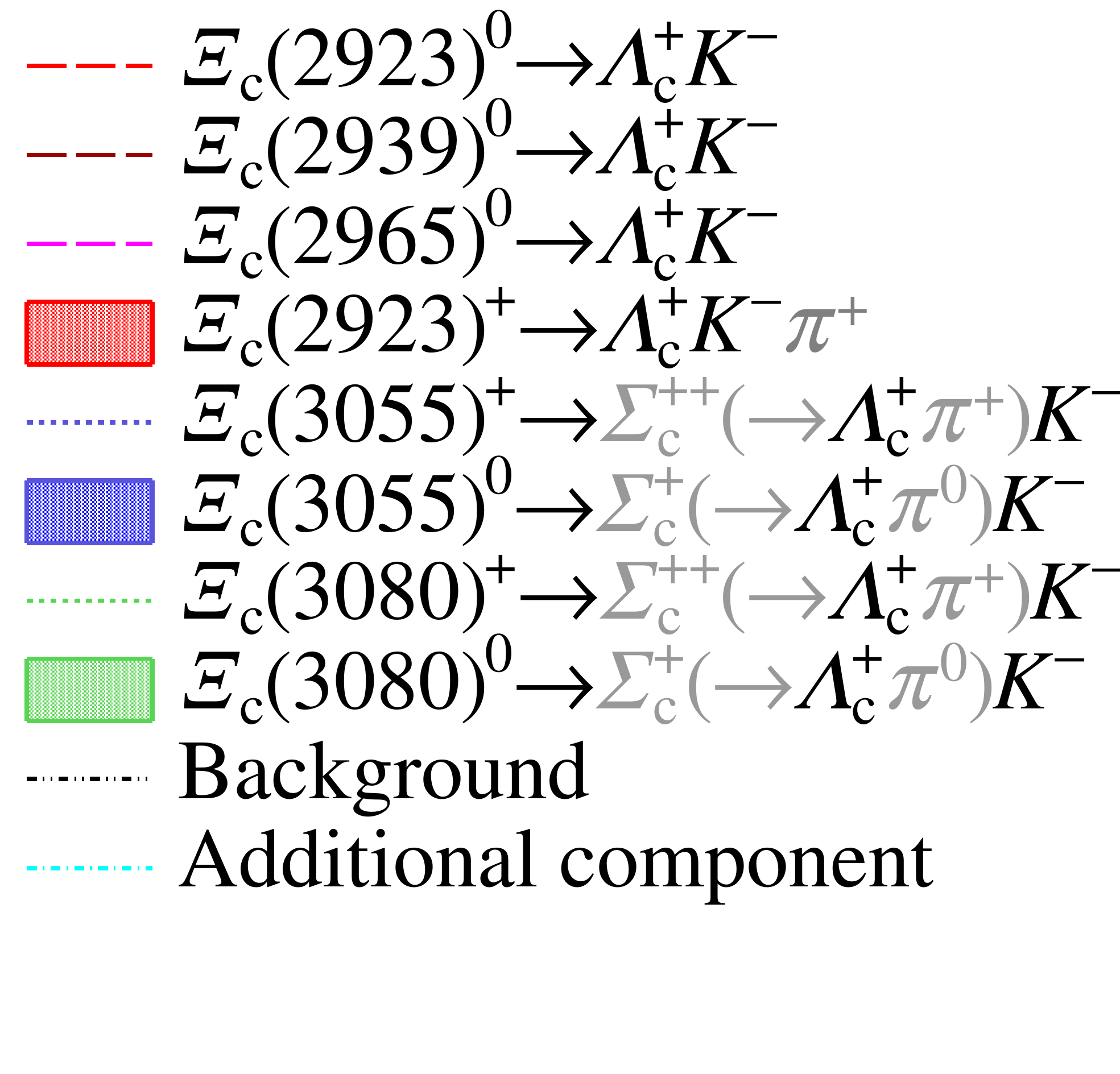}
    }
    \put( 39,28){\textcolor{white}{\rule{14mm}{10mm}}}
    \put( 94,28){\textcolor{white}{\rule{14mm}{10mm}}}
    \put( 39,31){\small$\begin{array}{c}\lhcb\\ \text{Run~2}\end{array}$}
    \put( 94,31){\small$\begin{array}{c}\lhcb\\ \text{Run~2}\end{array}$}
  \end{picture}
  \caption { \small
    Mass difference 
    $\Delta m = m_{\Lc\Km} - m_{\Lc} - m_{\Km}$ 
    distributions~\cite{LHCb-PAPER-2020-004}.
    Fits accounting for (left)~three and (right)~four excited 
    $\PXi_{\cquark}^{\ast0}$ states 
    are superimposed.  
  }
  \label{fig:spectra:lxic}
\end{figure}

\paragraph{Three excited $\PXi_{\cquark}^{\ast0}$~baryons} have been observed 
in the $\Lc\Km$~mass spectrum using the~Run~2 data-set~\cite{LHCb-PAPER-2020-004}. 
The~mass difference \mbox{$\delta m \equiv m(\Lc\Km)-m(\Lc)-m(\Km)$}~spectrum for selected 
$\Lc\Km$~combinations is shown in Fig.~\ref{fig:spectra:lxic}.
Three narrow structures, denoted  
$\PXi_{\cquark}(2923)^0$,
$\PXi_{\cquark}(2939)^0$ and  
$\PXi_{\cquark}(2965)^0$\, are clearly visible with a significance 
exceeding $20\sigma$ for each signal. 
The~data and fit show the~least compatibility 
in the~region 
\mbox{$\delta m \approx 110\mev$}, 
that could be  evidence for  a fourth 
$\PXi_{\cquark}^{\ast0}$~state.
Figure~\ref{fig:spectra:lxic}(right) shows 
the~$\delta m$~distribution for the~signal samples, 
where  a~structure in this region is  added into 
the~fit.
A~large improvement in the~fit quality is achieved. 

\begin{figure}[htb]
  \centering
  \setlength{\unitlength}{1mm}
  \begin{picture}(150,60)
    \put( 0, 0){ 
      \includegraphics*[width=75mm,height=60mm,%
      ]{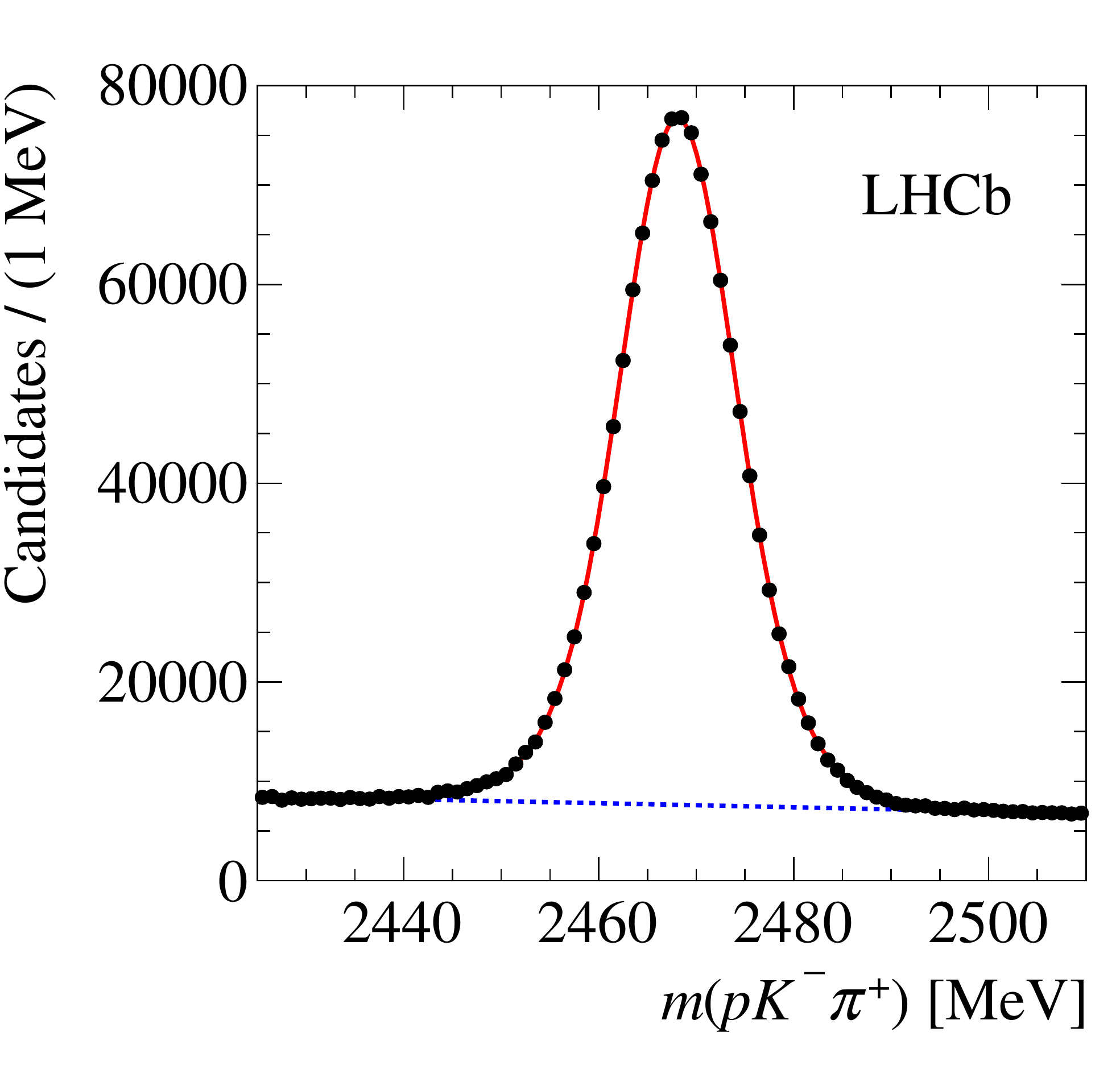}
    }
    \put(75, 0){ 
      \includegraphics*[width=75mm,height=60mm,%
      ]{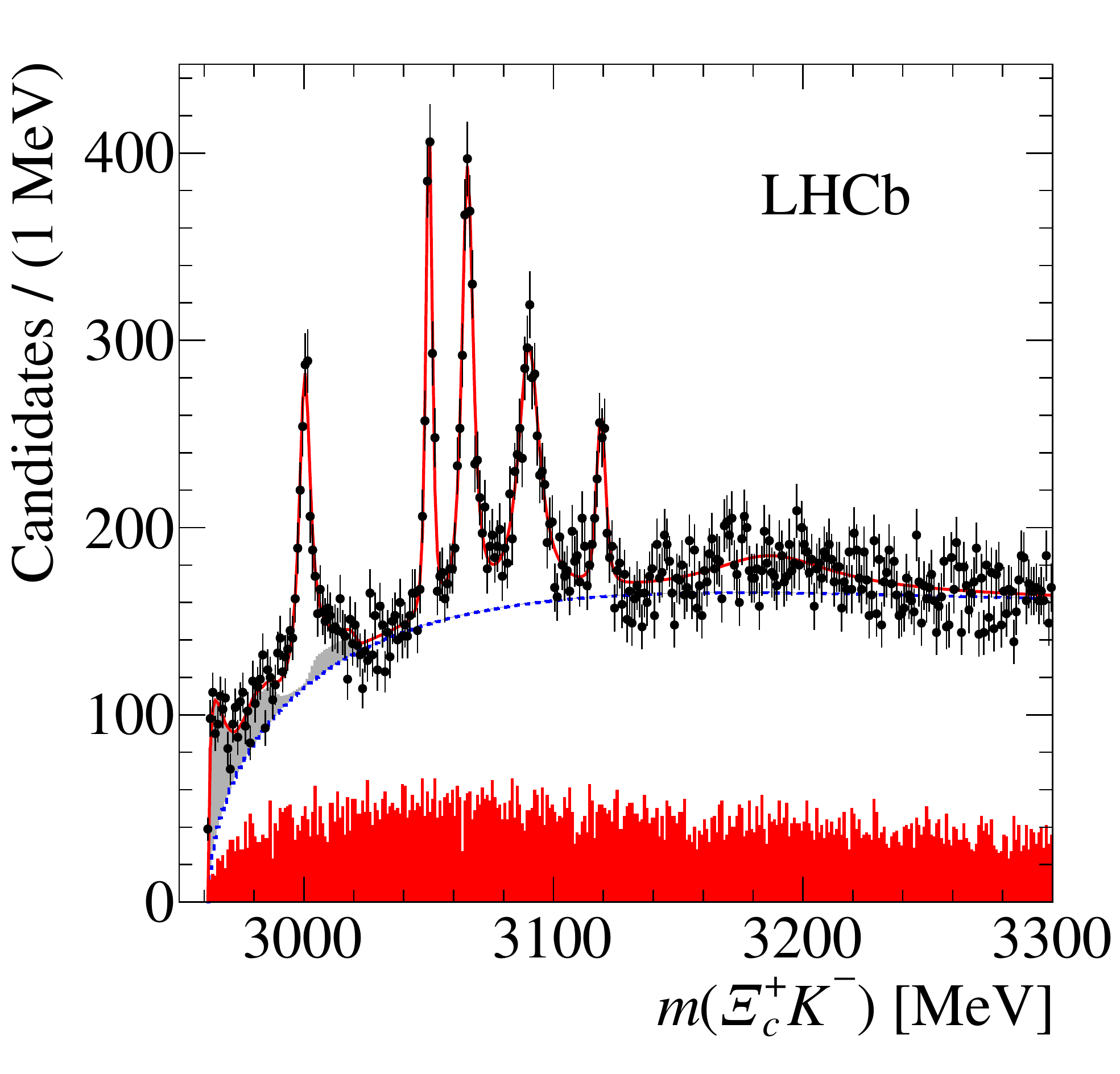}
    }
    \put(20,48){\small$\Xicp\to\proton\Km\pip$}
    \put( 55,44){\textcolor{white}{\rule{14mm}{10mm}}}
    \put(127,44){\textcolor{white}{\rule{14mm}{10mm}}}
    \put( 55,47){$\begin{array}{c}\lhcb\\ \text{Run~1}\end{array}$}
    \put(127,47){$\begin{array}{c}\lhcb\\ \text{Run~1}\end{array}$}
  \end{picture}
  \caption { \small
    (left)~Distribution of the~reconstructed invariant mass $m_{\proton\Km\pip}$ for all \Xicp~candidates.
    The~solid\,(red) curve shows the~result of the~fit, and the~dashed\,(blue) line indicates the~fitted background.
    (right)~Distribution of the reconstructed invariant mass of $\Xicp \Km$~combinations; 
    the~solid\,(red) curve shows the~result of the~fit, 
    and the~dashed\,(blue) line indicates the fitted 
    background~\cite{LHCb-PAPER-2017-002}.
    The~shaded\,(red) histogram shows the~corresponding mass spectrum from 
    the~\Xicp side-bands and the~shaded\,(light grey) distributions indicate the~feed-down 
    from partially reconstructed $\Omegares_\cquark(X)^0$ resonances. 
  }
  \label{fig:spectra:omegac}
\end{figure}
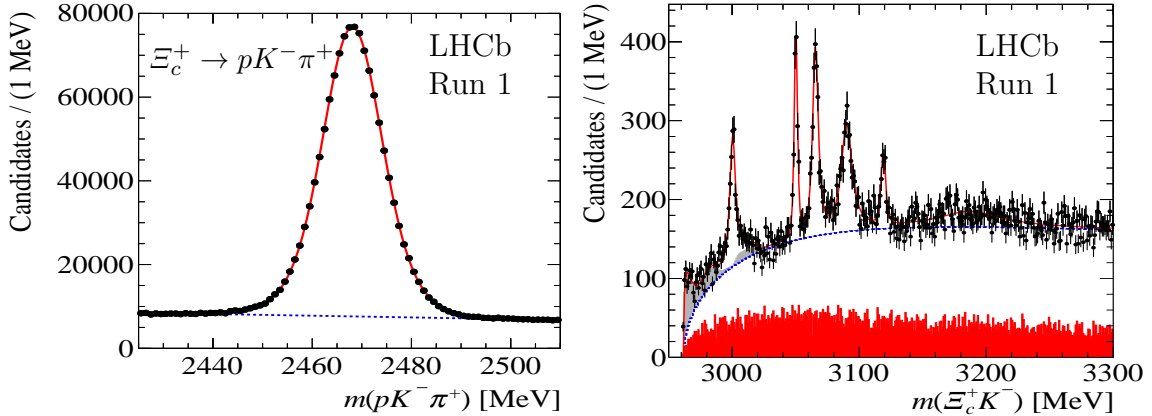

\paragraph{Five narrow excited $\POmega_{\cquark}^{\ast}$~baryons} have been observed 
in the $\Xicp\Km$~mass spectrum using the~Run~1 data-set~\cite{LHCb-PAPER-2017-002}.
A~large sample of \Xicp~candidates
were reconstructed  in their Cabibbo\nobreakdash-suppressed 
mode~\mbox{$\decay{\Xicp}{\proton\Km\pip}$}. 
In~total around $1.05\times 10^6$ 
\mbox{$\decay{\Xicp}{\proton\Km\pip}$} candidates with 
with a purity of 83\%~were selected, shown in Fig.~\ref{fig:spectra:omegac}(left).
The~mass distribution for $\Xicp\Km$~combinations is shown
in Fig.~\ref{fig:spectra:omegac}(right), and
five narrow peaks are clearly visible. 
The~natural widths of the~peaks are found to be 
between 0.8 and~$8.7\mev$, and two of them, 
named $\POmega_{\cquark}(3050)^0$ and 
$\POmega_{\cquark}(3119)^0$, are found to be extremely narrow,
with 95\%CL limits of 1.2 and~2.8\mev, respectively.
It~is found that the fit improves if an~additional broad 
Breit$-$Wigner function is included in the~3188\mev mass region.
This~broad structure may represent a~single resonance, 
 be  the~superposition of several resonances, 
be a feed\nobreakdash-down from higher states, 
or  some combination of the~above. 
The~interpretation of the~narrow states is still an~open question.
The naive~quark model expects five states in the region, 
but some  have to be relatively broad.
The~molecular model predicts two states with $J^P=\frac{1}{2}^+$
and two with $J^P=\frac{1}{2}^+$, and three of 
the observed states are
in remarkable agreement,  
both in mass and width, with this hypothesis~\cite{Debastiani:2017ewu}.

\subsection{Double-charm~baryons}

Three weakly decaying  states with charm number $C=2$ are expected in the quark model:
one isospin doublet $\PXi_{\cquark\cquark}$ and one  
isospin singlet $\POmega_{\cquark\cquark\squark}$,
each with spin\nobreakdash-parity $J^P =\tfrac{1}{2}^+$.
The~properties of these baryons have been calculated 
with a~variety of theoretical models. 
In~most cases, the~masses of the~$\PXi_{\cquark\cquark}$ states are
predicted to lie in the~range 3500 to 
3700\mevcc~\cite{Gershtein:1998un}.
The~masses of the~\Xiccpp and \Xiccp states are expected 
to differ by only a~few~\mevcc due to approximate isospin
symmetry. 
Most predictions for the~lifetime of the~\Xiccp baryon are in the~range 50 to 250\fs, 
and the~lifetime of the~\Xiccpp baryon is expected to be three to four times longer at 
200 to 700\fs,
While both are expected to be produced at hadron colliders
the~longer lifetime of the~\Xiccpp baryon should make it significantly easier
to experimentally observe  than the~\Xiccp baryon.

Experimentally, there is a~longstanding puzzle in the~$\PXi_{\cquark\cquark}$ system.
Observations of the~\Xiccp baryon 
in the $\Lc \Km \pip$\, final state at a~mass of $3519 \pm 2$\mevcc 
with signal yields of 15.9~events over $6.1 \pm 0.5$ events background
($6.3\sigma$~significance), 
and 5.62~events over $1.38 \pm 0.13$ events  background
in the~final state~$\proton \Dp \Km$\,($4.8\sigma$~significance), 
were reported by the~SELEX collaboration~\cite{Mattson:2002vu,Ocherashvili:2004hi}.
The SELEX results included a~number of unexpected features,
notably a~short lifetime and a~large production rate relative to that of the~singly charmed \Lc~baryon. 
The~lifetime was reported to be shorter than $33\fs$ at  the 90\% confidence level, and SELEX concluded that 
20\% of all $\Lc$~baryons observed by the~experiment originated
from $\Xiccp$~decays, implying a~relative $\PXi_{\cquark\cquark}$~production rate several
orders of magnitude larger than theoretical expectations. 
Searches from the~FOCUS~\cite{Ratti:2003ez}, \babar~\cite{Aubert:2006qw}, and \belle~\cite{Chistov:2006zj} 
experiments did not find evidence for a~state with the~properties reported by  
the~SELEX collaboration,
and neither did a~search at \lhcb with data corresponding to an~integrated luminosity 
of $0.65$\invfb~\cite{LHCb-PAPER-2013-049}. 
However, because the~production environments 
at all the above experiments differ from that of SELEX, which studied collisions of a~hyperon beam
on fixed nuclear targets, these null results do not exclude the~original observations.

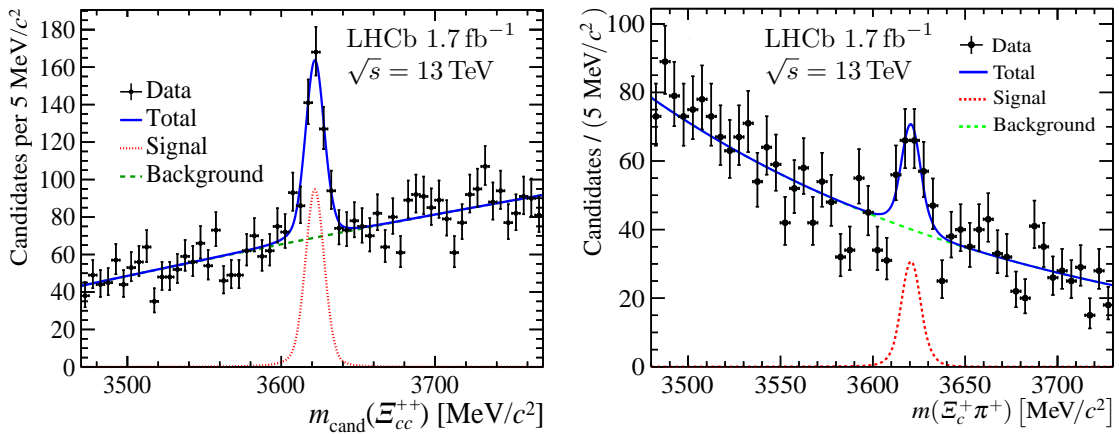
\begin{figure}[htb]
  \centering
  \setlength{\unitlength}{1mm}
  \begin{picture}(150,60)
    \put(0 , 0){ 
      \includegraphics*[width=75mm,height=60mm,%
      ]{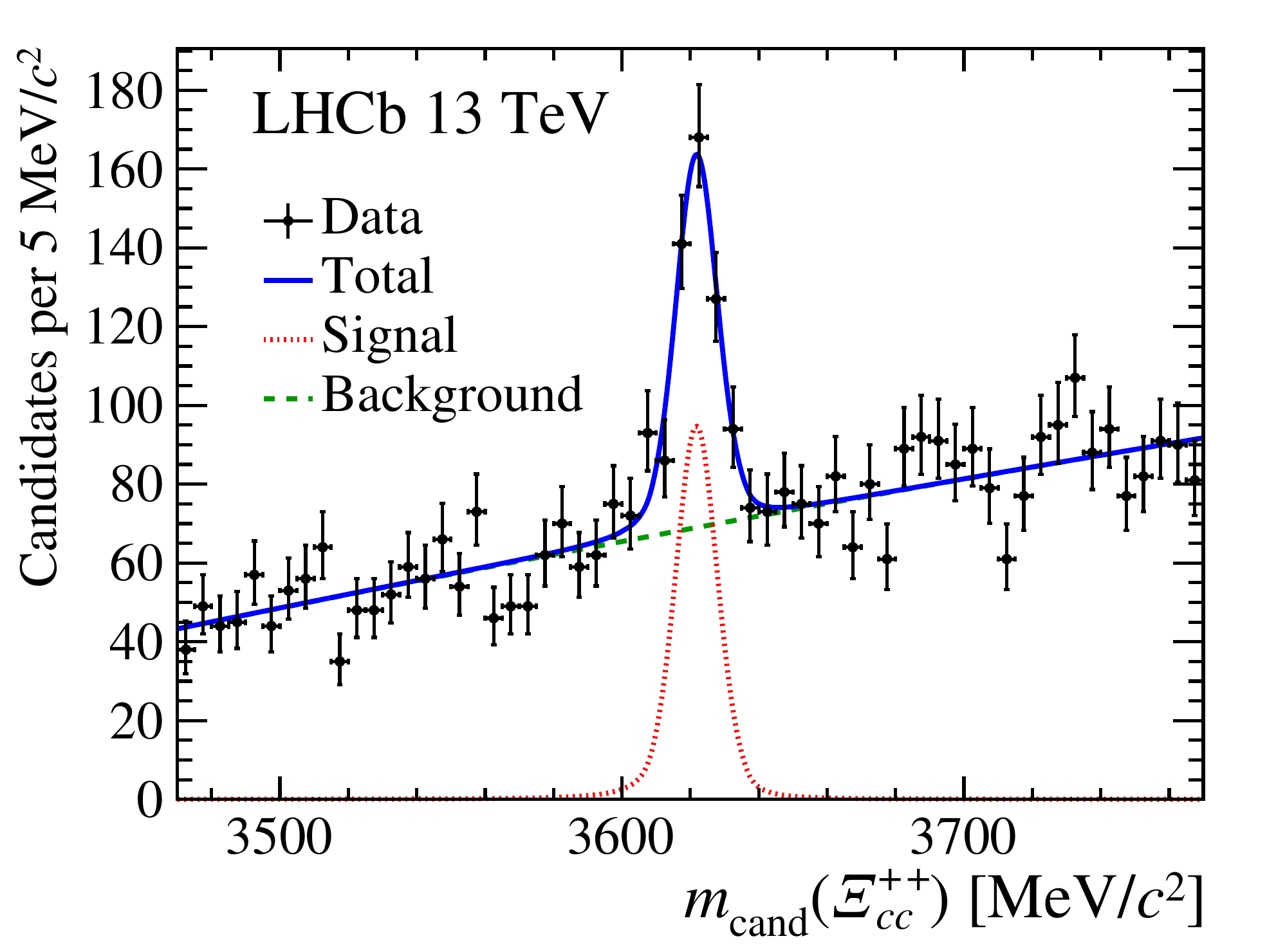}
    }
    \put(75, 0){ 
      \includegraphics*[width=75mm,height=63mm,%
      ]{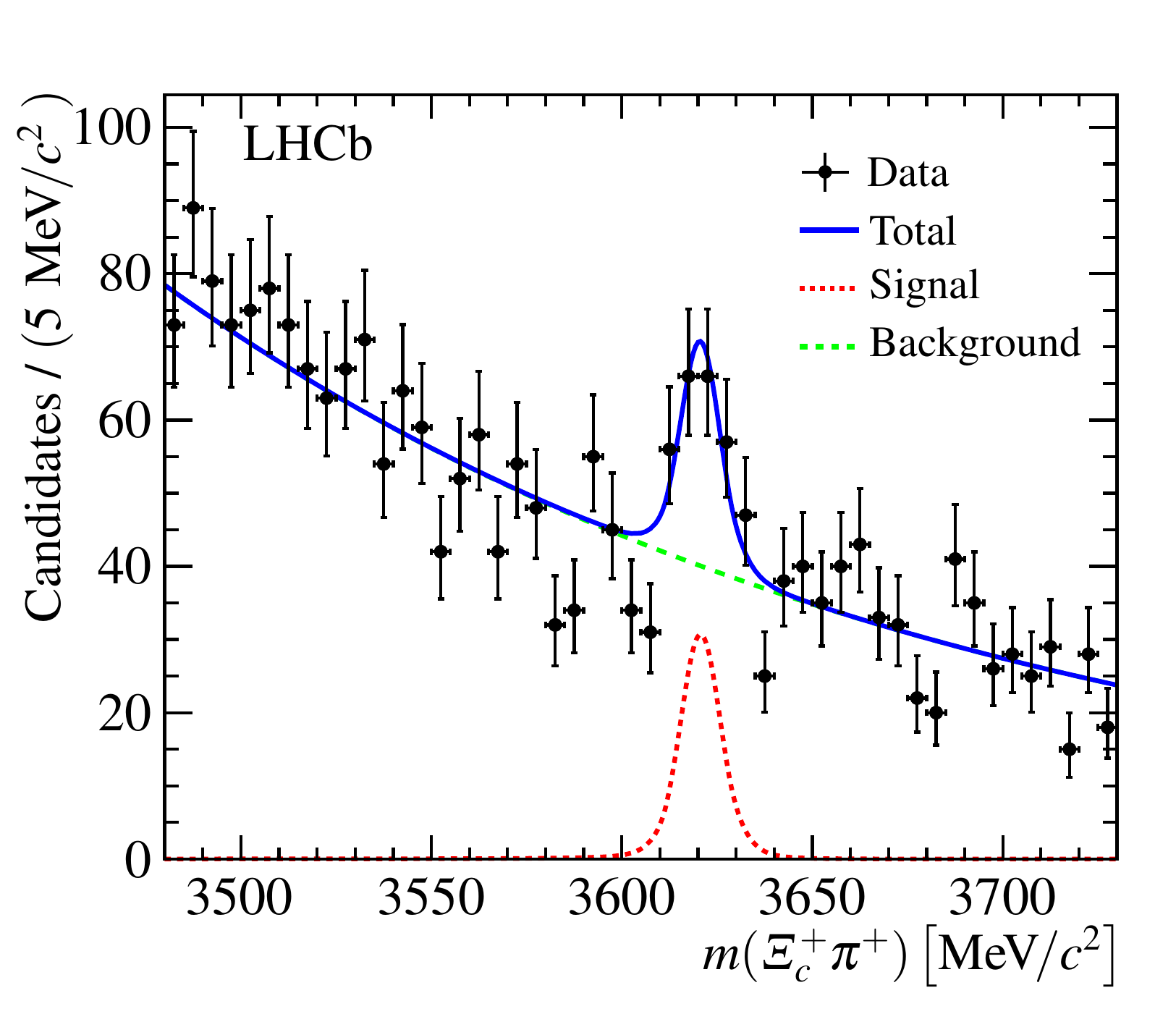}
    }
    \put( 15,50){\textcolor{white}{\rule{25mm}{6mm}}}
    \put( 45,50){\footnotesize{$\begin{array}{l}\lhcb~1.7\invfb\\ \sqs=13\tev\end{array}$}}
    \put( 92,50){\textcolor{white}{\rule{25mm}{6mm}}}
    \put(100,50){\footnotesize{$\begin{array}{l}\lhcb~1.7\invfb\\ \sqs=13\tev\end{array}$}}
  \end{picture}
  \caption{\small 
    (Left)~Mass distribution of  $\Lc\Km\pip\pip$ candidates with fit projections overlaid~\cite{LHCb-PAPER-2017-018}.
    (Right)~Mass distribution of  $\Xicp\pip$~candidates~\cite{LHCb-PAPER-2018-026}. 
  }
  \label{fig:spectra:xicc}
\end{figure}

\lhcb has searched for the $\Xiccpp$  decaying into  $\Lc\Km\pip\pip$
using  
a~sample of $\proton\proton$~collision data 
at 
13\tev, 
corresponding to an~integrated luminosity of 1.7\invfb.
A~highly significant structure is observed in 
the~mass spectrum, where the~\Lc~baryon is reconstructed
into $\proton\Km\pip$, shown in Fig.~\ref{fig:spectra:xicc}(left).
The~structure is consistent with originating from a~weakly decaying particle,
identified as the~doubly charmed baryon~\Xiccpp.
The observation of the~state is confirmed using 
an~additional sample of data collected at~8\tev.
Soon after, the~observation was further confirmed  by observing 
the~same state in the~decay ~\mbox{$\Xiccpp\to\Xicp\pip$}, 
shown in Fig.~\ref{fig:spectra:xicc}(right). 
The~mass of the~\Xiccpp~state 
was measured to be
in very good agreement with the~value measured in 
the~\mbox{$\Xiccpp\to\Lc\Km\pip\pip$}~decay channel~\cite{LHCb-PAPER-2018-026}.
The~lifetime and mass of the ~$\Xiccpp$~baryon were also precisely 
measured~\mbox{\cite{LHCb-PAPER-2018-019,LHCb-PAPER-2019-037}}, 
where the 
  lifetime favours smaller values
in the range of the theoretical 
predictions. 

\subsection{Beauty hadrons}

\begin{figure}[htb]
  \centering
  \setlength{\unitlength}{1mm}
  \begin{picture}(150,60)
    \put( 0, 0){ 
      \includegraphics*[width=75mm,height=60mm,%
      ]{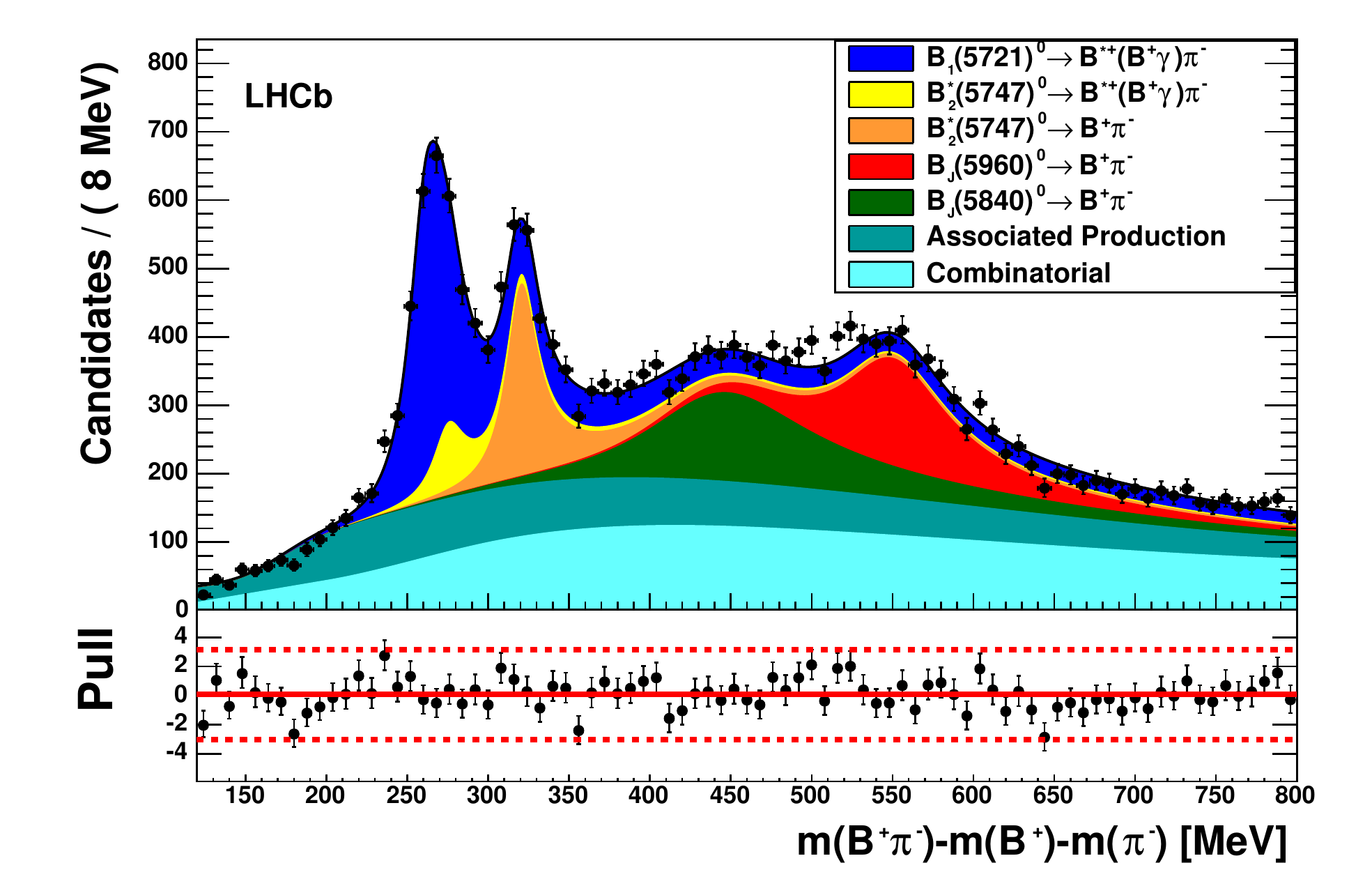}
    }
    \put(75, 0){ 
      \includegraphics*[width=75mm,height=60mm,%
      ]{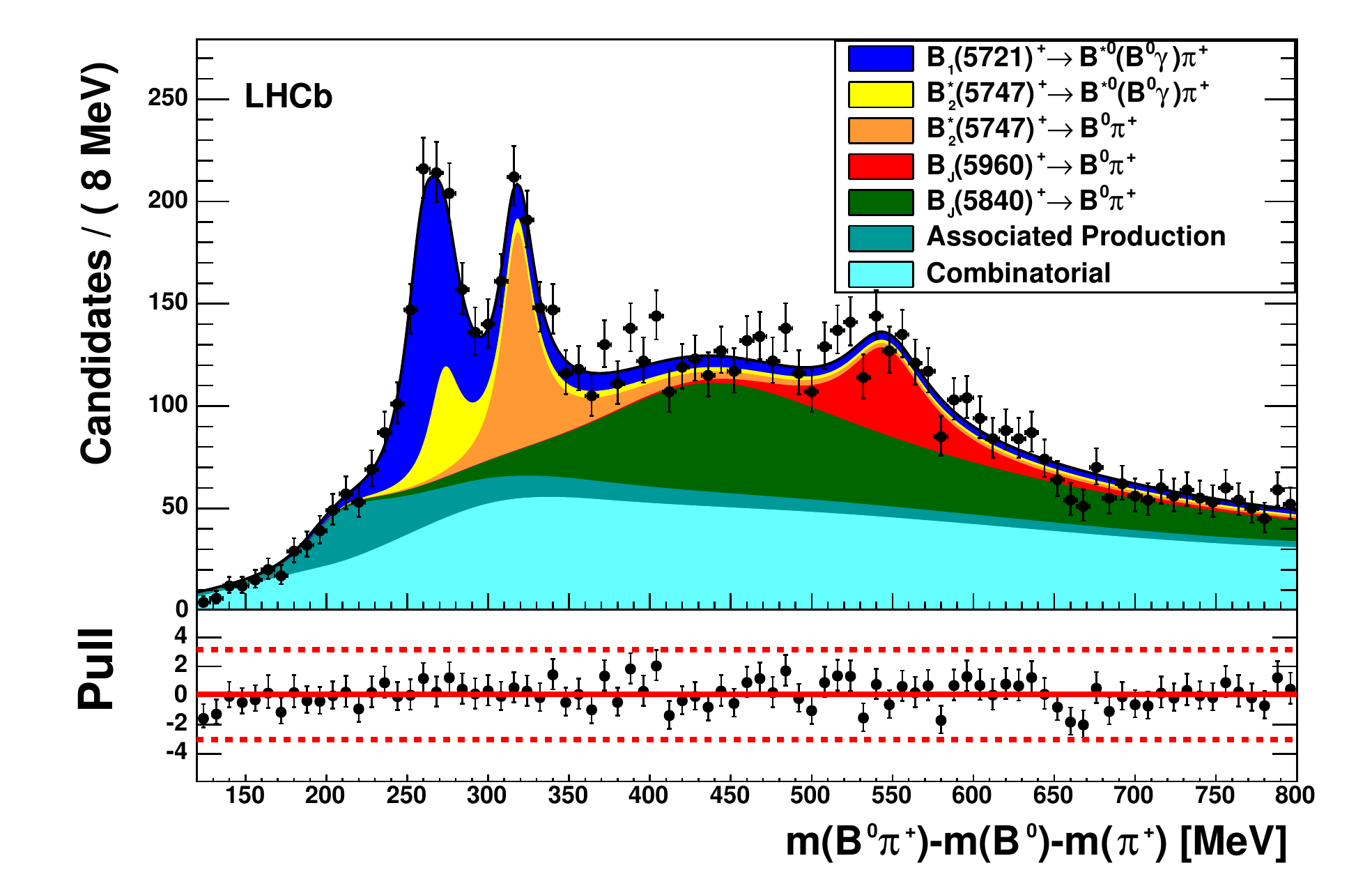}
    }
    \put( 14,52){\textcolor{white}{\rule{8mm}{4mm}}}
    \put( 89,52){\textcolor{white}{\rule{8mm}{4mm}}}
    \put( 15,52){\footnotesize{\lhcb~Run~1}}
    \put( 90,52){\footnotesize{\lhcb~Run~1}}
  \end{picture}
  \caption { \small  
    Spectra of 
    \mbox{$\Delta m \equiv m_{\B\Ppi} - m_{\B} - m_{\Ppi}$}
    for (left)~$\Bp\pim$ and 
    (right) $\Bz\pip$ candidates
    with $p_{\mathrm{T}}(\Ppi)>2\gevc$~\cite{LHCb-PAPER-2014-067}.
  }
  \label{fig:spectra:bmesons}
\end{figure}

\paragraph{Excited \Bu and \Bd~mesons} have been investigated in the
mass distributions of $\Bp\pim$ and $\Bz\pip$ combinations using  a $3\invfb$ data sample 
at 7 and $8 \tev$. 
The \Bu and \Bz candidates were reconstructed through the
\mbox{$\Bu\to \Dzb\pi^+$},
\mbox{$\Bu\to \Dzb\pi^+\pi^+\pi^-$},
\mbox{$\Bu\to J/\psi K^+$},
\mbox{$\Bz\to \Dm \pi^+$},
\mbox{$\Bz\to \Dm \pi^+\pi^+\pi^-$} and
\mbox{$\Bz\to J/\psi \Kstarz$}~decay chains.
Samples of about 1.2~million \Bz and 2.5~million \Bu~candidates have been obtained 
with purity depending on decay mode, but always better than 80\%.
The~$\Bp\pim$ and $\Bz\pip$~mass spectra with   requirements that $\pt>2\gevc$
are shown in Fig.~\ref{fig:spectra:bmesons}, where 
ten peaking structures are    
reconstructed. Out of these,
six narrow low\nobreakdash-mass structures correspond to the~decays of 
the~four $\B_1(5721)^{0,+}$  and $\B_2^*(5747)^{0,+}$~states 
observed by the~CDF and D0~collaborations~\cite{Aaltonen:2008aa,Abazov:2007vq,Aaltonen:2013atp}: 
$\B_1(5721)\to\B^{\ast}\Ppi$ and  $\B_2(5747)\to\B^{(\ast)}\Ppi$.
The~high statistics of \lhcb has allowed the~most precise measurements 
of the~masses and widths of the~$\B_1(5721)^{0,+}$  and $\B_2^*(5747)^{0,+}$~states to be  made. 

In addition to the~six low\nobreakdash-mass structures, 
 four wider high\nobreakdash-mass structures are observed, 
particularly prominent at high pion transverse momentum. 
These  structures are consistent with the~presence 
of four new excited \B~mesons, 
labeled $\B_J(5840)^{0,+}$ and $\B_J(5960)^{0,+}$, 
whose masses and widths are obtained under different hypotheses of their quantum numbers~\cite{LHCb-PAPER-2014-067}.

\begin{figure}[htb]
  \centering
  \setlength{\unitlength}{1mm}
  \begin{picture}(150,57)
    \put( 0,  0){ 
      \includegraphics*[width=75mm,height=57mm,%
      ]{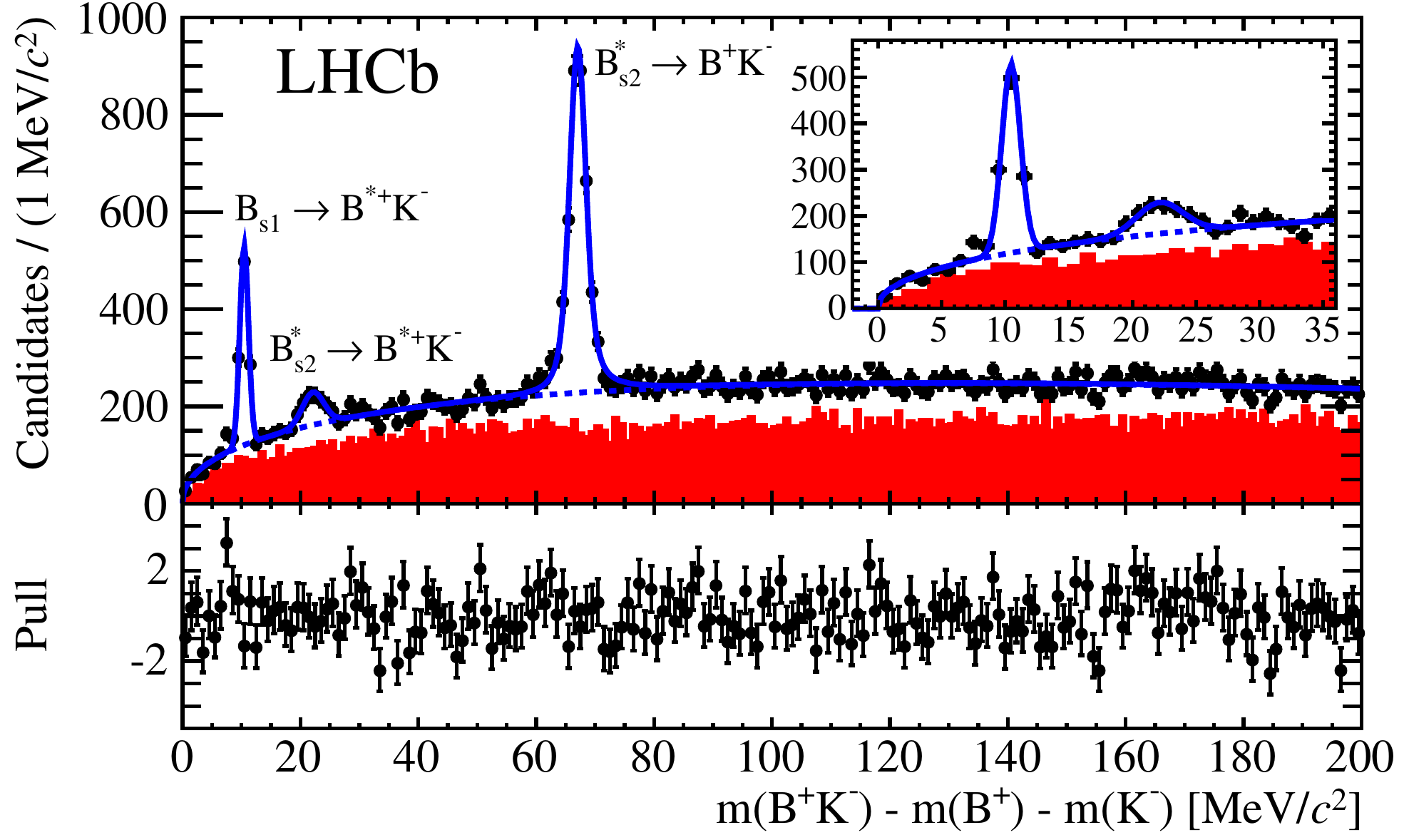}
    }
    \put(75, -2){ 
      \includegraphics*[width=75mm,height=60mm,%
      ]{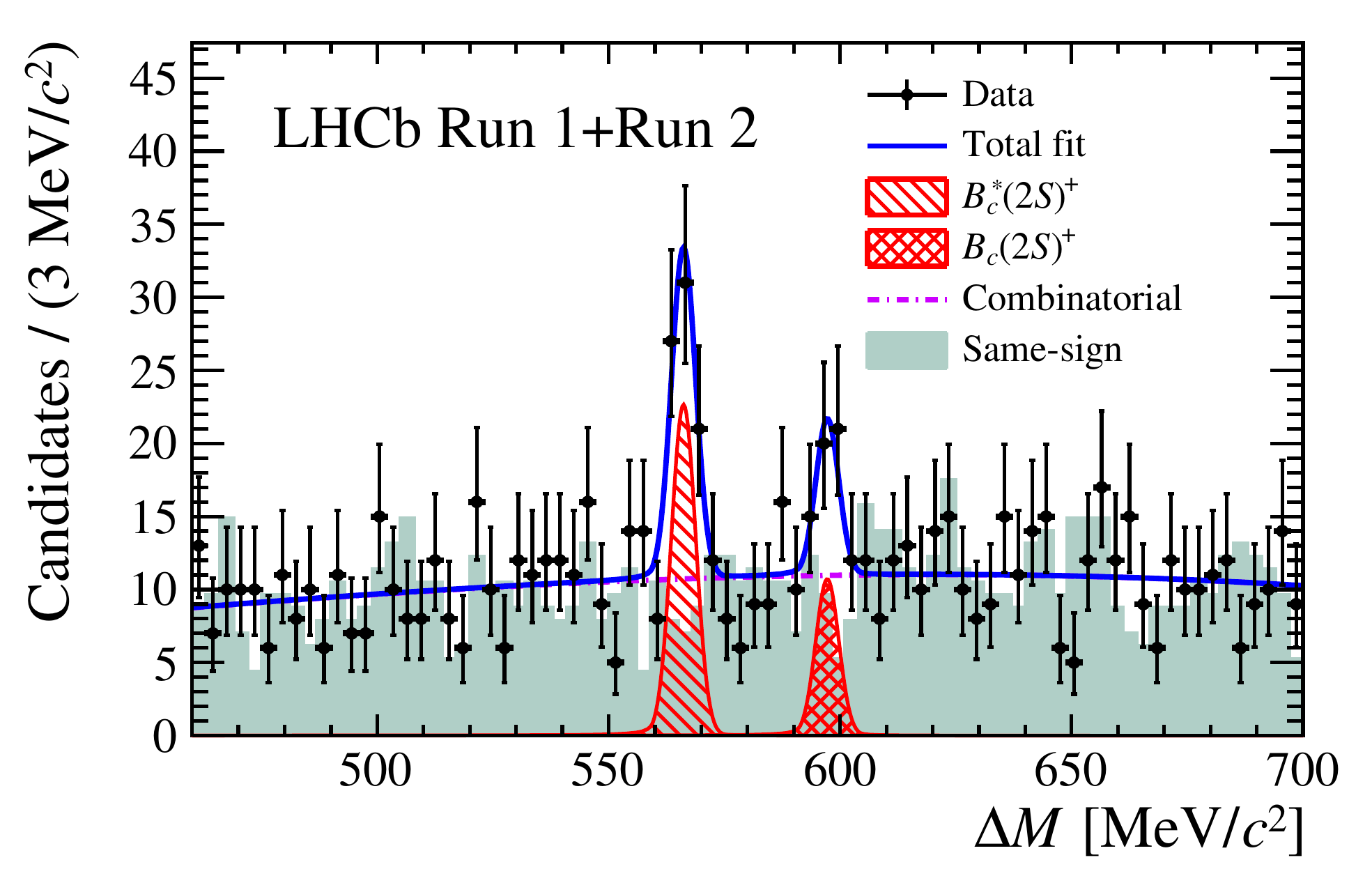}
    }
    \put( 14,48){\textcolor{white}{\rule{11mm}{6mm}}}
    \put( 12,50){\scriptsize{$\begin{array}{l}\lhcb~1\invfb\\ \sqs=7\tev\end{array}$}}
    \put( 90,48){\textcolor{white}{\rule{30mm}{6mm}}}
    \put( 90,50){\footnotesize{\lhcb Run~1\&2}}
    \put( 90,42){\scriptsize{$\begin{array}{l}\B_{\cquark}(2S)^{(\ast)+} \\ \rightarrow \B_{\cquark}^{(\ast)+}\pip\pim \end{array}$}}
  \end{picture}
  \caption { \small 
    (left)~The mass difference distribution $m_{\Bu\Km}-m_{\Bu}-m_{\Km}$ 
    in the $\Bu\Km$~mass spectra. 
    The~three peaks are identified as (left-to-right)~\mbox{$\B_{\squark1}(5830)^0\to \B^{\ast+}\Km$}, 
    \mbox{$\B_{\squark2}^{\ast}(5840)^0\to \B^{\ast+} \Km$}, 
    and \mbox{$\B_{\squark2}^{\ast}(5840)^0\to \Bu \Km$}. 
    The~total fit function is shown as a~solid blue line, while
    the~shaded red region is the~spectrum of like-charge $\Bu\Kp$ combinations. 
    The~inset shows an expanded view of 
    the~\mbox{$\B_{\squark1}(5830)^0/\B_{\squark2}^{\ast}(5840)^0\to \B^{\ast+}\Km$} region.
    (right)~Distribution of \mbox{$\Delta M \equiv m_{\Bc\pip\pim} - m_{\Bc}$} in the $\Bc\pip\pim$ mass spectra with  fit results overlaid.
    The~same-sign distribution has been normalised to the~data in the~$\B_{\cquark}(2S)$~side band region.
  }
  \label{fig:spectra:hbmesons}
\end{figure}

\paragraph{Orbitally excited \Bs~mesons} have been  studied using only 1\invfb of data, collected 
at~\mbox{$\sqs=7\tev$}. 
The~$\Bu\Km$~mass spectra were investigated with  the \Bu~mesons 
being reconstructed in  four   decay modes.
Previously, two~narrow peaks had been observed in the $\Bu\Km$~mass distribution by 
the~CDF 
and D0~collaborations~\cite{Aaltonen:2007ah,Abazov:2007af}, 
named the $\B_{\squark1}^{*}(5830)^0$ and $\B_{\squark2}^{*}(5840)^0$. They are 
putatively identified as members of $j_q=\tfrac{3}{2}$ HQET doublet~\cite{DiPierro:2001dwf}.
The~two states are also visible in \lhcb data,  here  as three narrow peaks shown in 
Fig.~\ref{fig:spectra:hbmesons}(left),
corresponding to 
the~decays~\mbox{$\B_{\squark1}(5830)^0\to \B^{\ast+}\Km$}, 
\mbox{$\B_{\squark2}^{\ast}(5840)^0\to \B^{\ast+} \Km$}, 
and \mbox{$\B_{\squark2}^{\ast}(5840)^0\to \Bu \Km$}, where 
a~soft photon from $\B^{\ast+}\to\Bu\g$~is  undetected.
This~is the~first observation of 
the~\mbox{$\B_{\squark2}^{\ast}(5840)^0\to \B^{\ast+} \Km$}~decay mode
and a $J^P = 2^+$~assignment is favoured for this state.
Large statistics, low background and \lhcb's excellent mass resolution has
allowed the first determination of the~$\B_{\squark2}^{\ast}(5840)^0$~width
as well as the most precise mass measurements of    both states.
Due~to the small   energy  release, 
the~position and the~shape of the~peaks depends 
on the~mass of the $\B^{\ast+}$~states, allowing 
the~most precise  determination of the 
$m_{\B^{\ast+}}$ mass, as well as the~mass 
difference~\mbox{$m_{\B^{\ast+}}-m_{\Bu}$}. 


\paragraph{Excited \Bc~mesons} have been searched for via 
their decays into the $\Bc\pip\pim$~final state.  
A wide peak, interpreted  as the $\B_{\cquark}(2S)^+$, 
was observed by 
the~ATLAS collaboration  using a sample of about 300 reconstructed 
\mbox{$\decay{\Bc}{\jpsi\pip}$}~candidates~\cite{Aad:2014laa}, 
with a large
relative production rate with respect to the base \Bc~state.
\lhcb has searched for this state using 
2\invfb 
of data collected at $\sqs=8\tev$, observing
a sample   of $3325\pm73$ reconstructed  
\mbox{$\decay{\Bc}{\jpsi\pip}$}~decays.  
No signal is observed and
an upper limit on the relative production rate 
has been obtained~\cite{LHCb-PAPER-2017-042}.
This upper limit is smaller than the relative production 
rate reported by ATLAS. 

In~2018, the~CMS collaboration,  
using a~huge data set corresponding to 143\invfb collected 
at $\sqs=13\tev$ and containing of $7629\pm225$ signal
\mbox{$\decay{\Bc}{\jpsi\pip}$}~decays,
reported observation of a doublet of 
two narrow states, interpreted as 
spin-triplet $\B_{\cquark}(2S)^{\ast+}$
and spin-singlet $\B_{\cquark}(2S)^+$~states~\cite{Sirunyan:2019osb}.
These observations were confirmed by 
\lhcb using 
a    $8.5\invfb$ data-set collected at 
$\sqs=7,8$ and 13\tev~\cite{LHCb-PAPER-2019-007},
with $3785\pm73$ signal \mbox{$\decay{\Bc}{\jpsi\pip}$}~decays.
Two~narrow peaks with a width compatible with the~detector
resolution are seen in 
the~mass\nobreakdash-difference 
\mbox{$m_{\Bc\pip\pim} - m_{\Bc}$}~spectrum, 
shown in Fig.~\ref{fig:spectra:hbmesons}(right).
The~local\,(global) significances 
of the two peaks  are estimated to be 
$6.8\sigma$\,($6.3\sigma$) and 
$3.2\sigma$\,($2.2\sigma$) for the
low\nobreakdash-mass and 
high\nobreakdash-mass states, respectively.
The~low-mass signal is interpreted as the
spin\nobreakdash-triplet state 
$\B_{\cquark}(2S)^{\ast+}$,
decaying into $\B_{\cquark}^{\ast+}\pip\pim$ 
with the subsequent decay of the
$\B_{\cquark}^{\ast+}$ into $\Bc\g$. The
high\nobreakdash-mass peak is attributed to
the decay of the spin\nobreakdash-singlet 
$\B_{\cquark}(2S)^{+}$~state into the 
$\Bc\pip\pim$~final state.

\begin{figure}[ht]
  \centering
  \setlength{\unitlength}{1mm}
  \begin{picture}(150,67)
    \put(  0, 0){ 
      \includegraphics*[height=67mm,width=75mm%
      ]{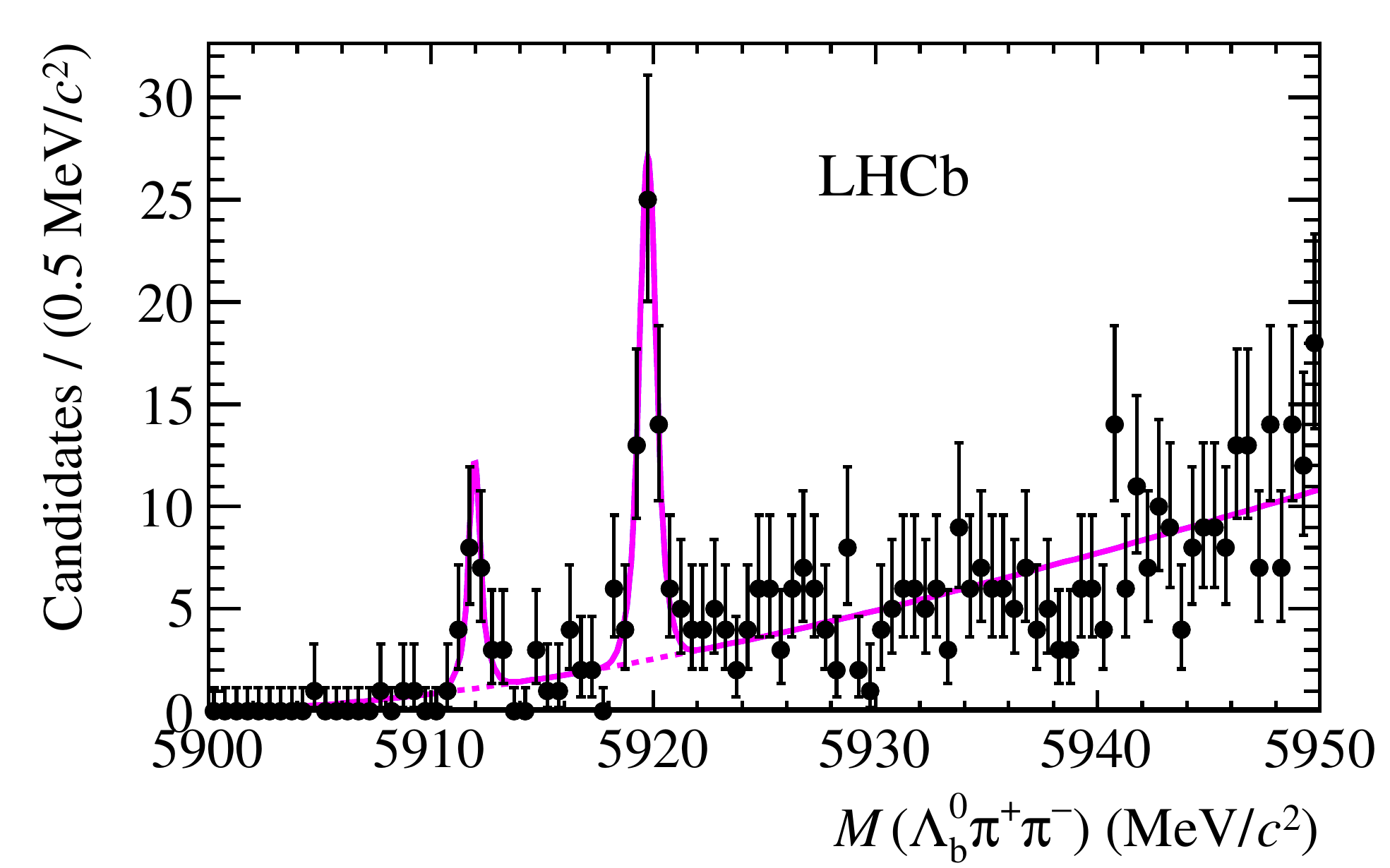}
    }
    \put( 75, 2){ 
      \includegraphics*[height=63mm,width=75mm%
      ]{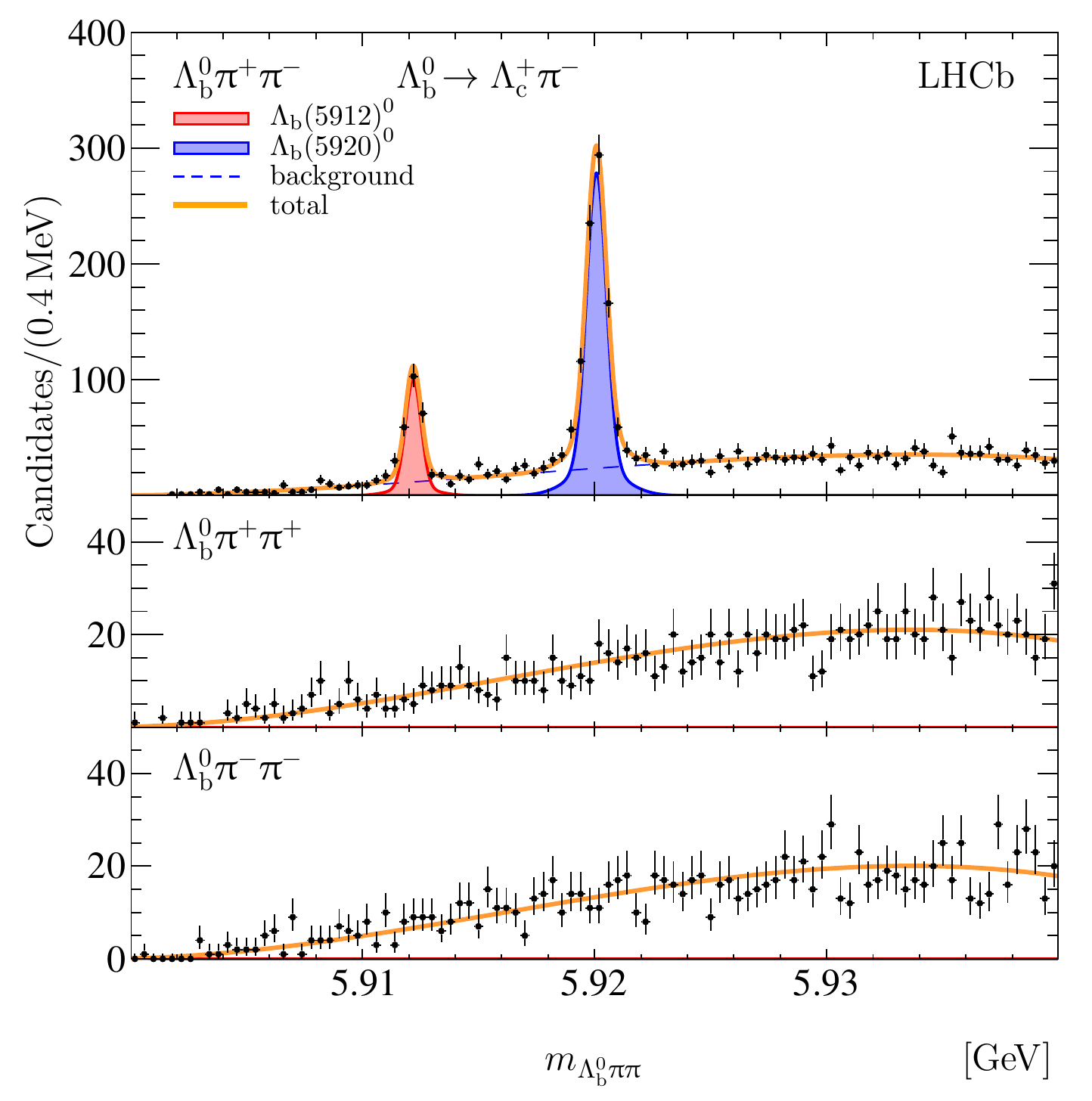}
    }
    \put( 45,50){\textcolor{white}{\rule{20mm}{10mm}}}
    \put( 50,55){\footnotesize{$\begin{array}{l}\lhcb~1\invfb\\ \sqs=7\tev\end{array}$}}
    \put(125,52){\textcolor{white}{\rule{22mm}{10mm}}}
    \put(122,57){\footnotesize{\lhcb~Run~1\&2}}
    \put( 95,0){\textcolor{white}{\rule{55mm}{7mm}}}
    \put(105,2){\small$m(\Lb\Ppi\Ppi)$}
    \put(135,2){\small$\left[\!\gevcc\right]$}
  \end{picture}
  \caption { \small
    (Left)~Mass spectrum of $\Lb\pip\pim$~combinations~\cite{LHCb-PAPER-2012-012}.
    The~points with error bars are the~data, 
    the~solid line is the result of a fit, and 
    the~dashed line is the~background contribution. 
    (Right)~Mass spectrum of (top)\,$\Lb\pip\pim$,
    (middle)\,$\Lb\pip\pip$ and (bottom)\,$\Lb\pim\pim$~combinations
    in the~full Run~1 and 2 data-sets~\cite{LHCb-PAPER-2019-045}.
  }
  \label{fig:spectra:lambdab}
\end{figure}
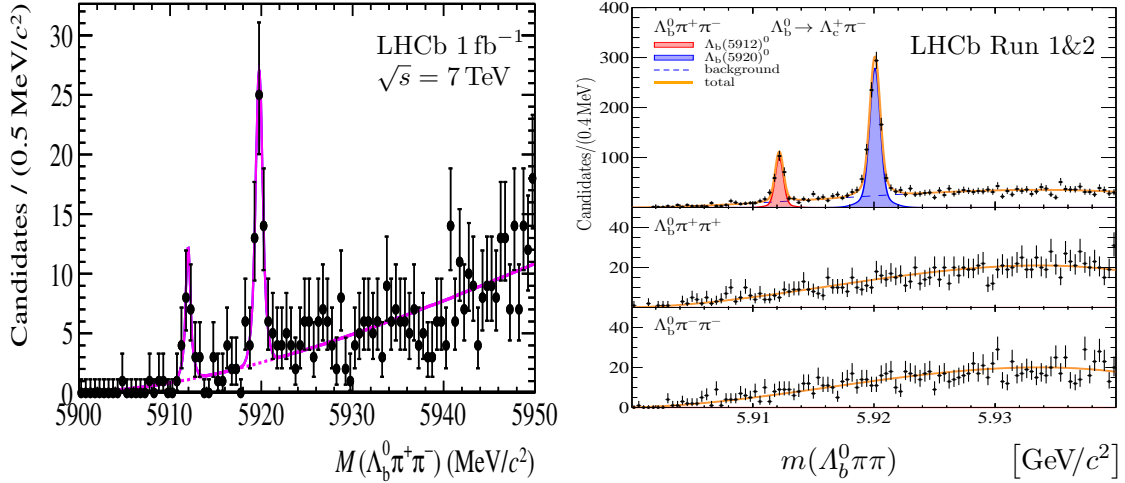

\paragraph{Excited $\Lb$~baryons} were discovered 
in the $\Lb\pip\pim$~mass spectrum using only 1\invfb~of \lhcb data, accumulated
at $\sqs=7\tev$ in 2011~\cite{LHCb-PAPER-2012-012}.
The~\Lb~candidates were reconstructed via 
\mbox{$\decay{\Lb}{\Lc\pim}$} followed by 
\mbox{$\decay{\Lc}{\proton\Km\pip}$}.
In~total $(70.54\pm 0.33)\times 10^3$ signal $\Lb$~decays 
were selected and then  combined with $\pip\pim$~pairs.
The~$\Lb\pip\pim$~spectrum
in the region $5.90\le m_{\Lb\pip\pip}\le5.95\gevcc$
is shown in Fig.~\ref{fig:spectra:lambdab}(left), 
where two narrow peaks, consistent with detector resolution, 
are visible.  The~significances of the~observations  
are~5.2 and 10.2~standard deviations 
for the low\nobreakdash-mass and high\nobreakdash-mass peaks,  respectively.
The~observed states are interpreted as the~doublet of orbitally-excited $\PLambda_{\bquark}(1P)^0$~states  
with quantum numbers $J^P=\tfrac{1}{2}^+$ and $\tfrac{3}{2}^+$.
In 2020 the analysis was updated using the~full Run~1 and 2 data-sets,
shown in Fig.~\ref{fig:spectra:lambdab}(right).
The masses of these states are measured with 
unprecedented precision~\cite{LHCb-PAPER-2019-045}.

\begin{figure}[htb]
  \centering
  \setlength{\unitlength}{1mm}
  \begin{picture}(150,75)
    \put(0, 0){ 
      \includegraphics*[width=75mm,height=75mm,%
      ]{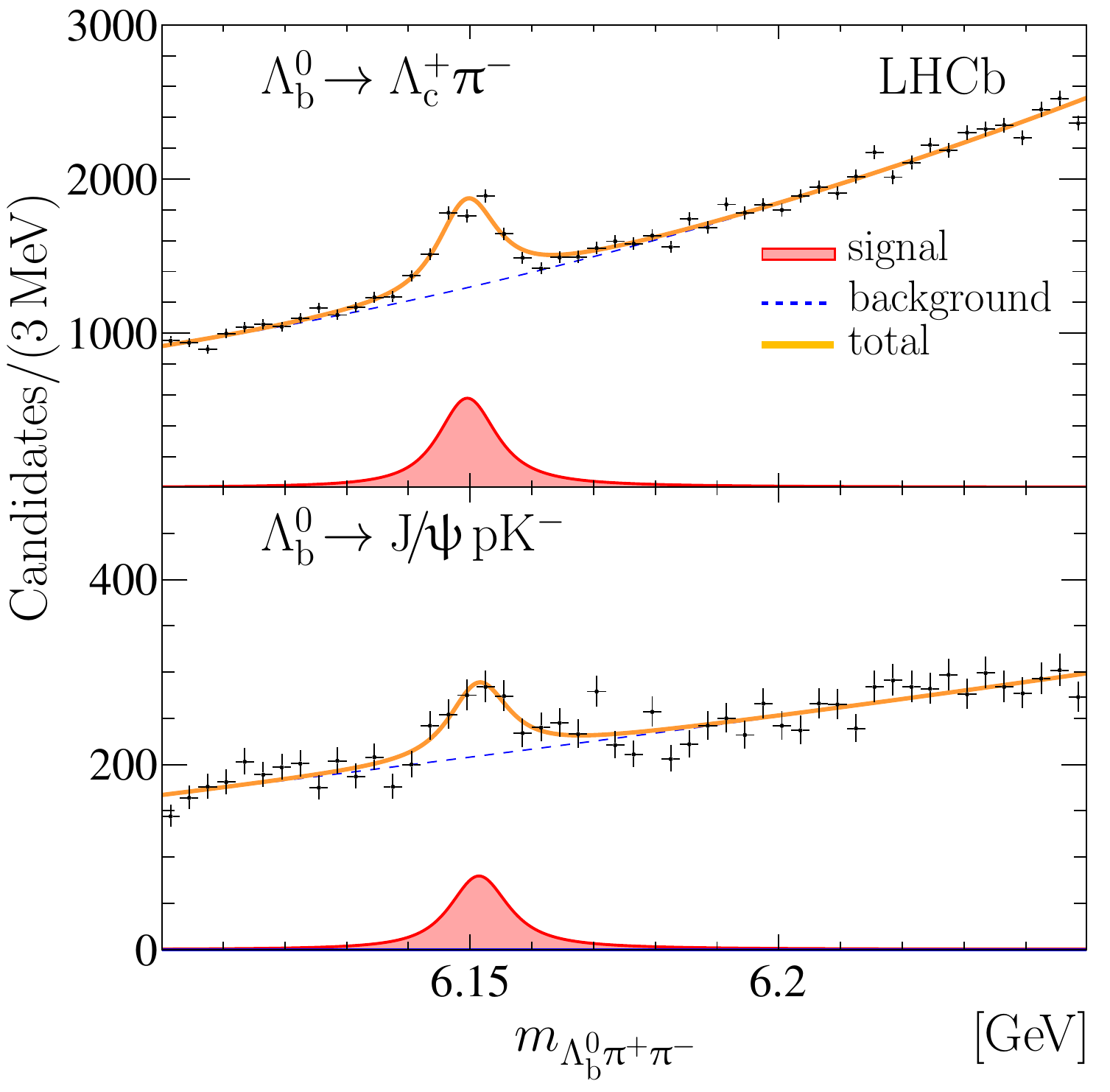}
    }
    \put(75, 0){ 
      \includegraphics*[width=75mm,height=75mm,%
      ]{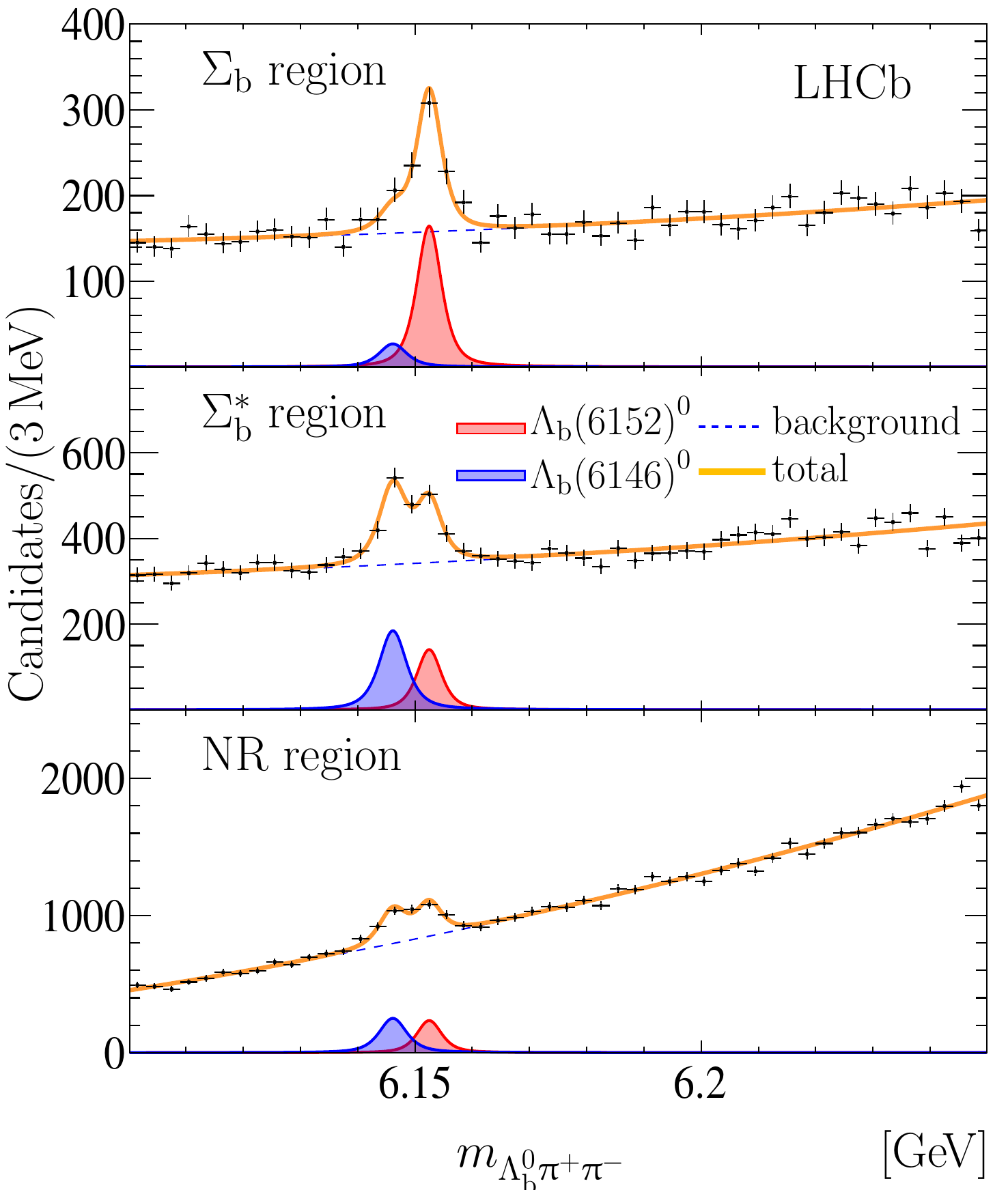}
    }
    \put( 45,68){\textcolor{white}{\rule{25mm}{4mm}}}
    \put( 43,68){\footnotesize{\lhcb~Run~1\&2}}
    \put(120,68){\textcolor{white}{\rule{25mm}{4mm}}}
    \put(118,68){\footnotesize{\lhcb~Run~1\&2}}
    \put( 95,-1){\textcolor{white}{\rule{55mm}{6mm}}}
    \put(110,1){\small$m(\Lb\pip\pim)$}
    \put(135,1){\small$\left[\!\gevcc\right]$}
    \put( 20,-1){\textcolor{white}{\rule{55mm}{6mm}}}
    \put( 35,1){\small$m(\Lb\pip\pim)$}
    \put( 60,1){\small$\left[\!\gevcc\right]$}
  \end{picture}
  \caption { \small
    (left)~The mass distribution of selected $\Lb\pip\pim$~candidates
    for ~(top)~the \mbox{$\decay{\Lb}{\Lc\pim}$}
    and (bottom)~the \mbox{$\decay{\Lb}{\jpsi\proton\Km}$} decay modes.
    (right)~Mass distributions of selected $\Lb\pip\pim$~candidates
    for the~three regions in $\Lb\Ppi^{\pm}$~mass:
    (top)~$\PSigma_{\bquark}$,
    (middle)~$\PSigma_{\bquark}^{\ast}$
    and (bottom)~the nonresonant\,(NR) region~\cite{LHCb-PAPER-2019-025}.
  }
  \label{fig:spectra:lambdabd}
\end{figure}

Using the~full Run~1 and 2 data-sets, 
the~mass spectrum of $\Lb\pip\pip$~combinations,  where 
\mbox{$\decay{\Lb}{\Lc\pim}$} and  
\mbox{$\decay{\Lb}{\jpsi\proton\Km}$}, was explored for higher 
masses of $\Lb\pip\pim$~combinations.
A  significant broad 
structure is found at $m~\approx 6.150\gevcc$ with 
a~width around 10\mev, shown in Fig.~\ref{fig:spectra:lambdabd}(left)~\cite{LHCb-PAPER-2019-025}.
The~mass  and width agree well when measured in the two decay modes \mbox{$\decay{\Lb}{\Lc\pim}$} and  
\mbox{$\decay{\Lb}{\jpsi\proton\Km}$}, where the~significance exceeds
26~and 9~standards deviations, respectively.
Since the~mass of the~new structure is above 
the~$\PSigma_{\bquark}^{(\ast)\pm}\Ppi^{\mp}$~kinematic thresholds, 
the~$\Lb\pip\pim$~mass spectrum is investigated 
in the $\Lb\Ppi^{\pm}$~mass regions
populated by the~$\PSigma_{\bquark}^{(\ast)\pm}$~resonances.

The~data are split into three 
non overlapping regions:
candidates with a~$\Lb\Ppi^{\pm}$ mass 
within the~natural width
of the~known $\PSigma_{\bquark}^{\pm}$~mass,
candidates with a~$\Lb\Ppi^{\pm}$ mass 
within the~natural width 
of the~known $\PSigma_{\bquark}^{\ast\pm}$~mass,
and the~remaining nonresonant\,(NR) region.
The~$\Lb\pip\pim$~mass spectra in these three regions 
are shown in Fig.~\ref{fig:spectra:lambdabd}(right).
The~spectra in the~$\PSigma_{\bquark}$ and 
$\PSigma_{\bquark}^{\ast}$~regions
look different and suggest 
the~presence of two narrow peaks 
with very similar widths.
The~two\nobreakdash-signal hypothesis 
is favoured over 
the~single\nobreakdash-signal hypothesis 
with a~statistical significance exceeding 
seven~standard deviations. 
The masses of the two states measured 
are consistent with
 predictions for the~doublet of $\PLambda_{\bquark}(1D)^0$~states 
with quantum numbers $J^P=\tfrac{3}{2}^+$ and $\tfrac{5}{2}^+$. 

\begin{figure}[htb]
  \centering
  \setlength{\unitlength}{1mm}
  \begin{picture}(150,75)
    \put(0, 0){ 
      \includegraphics*[width=75mm,height=75mm,%
      ]{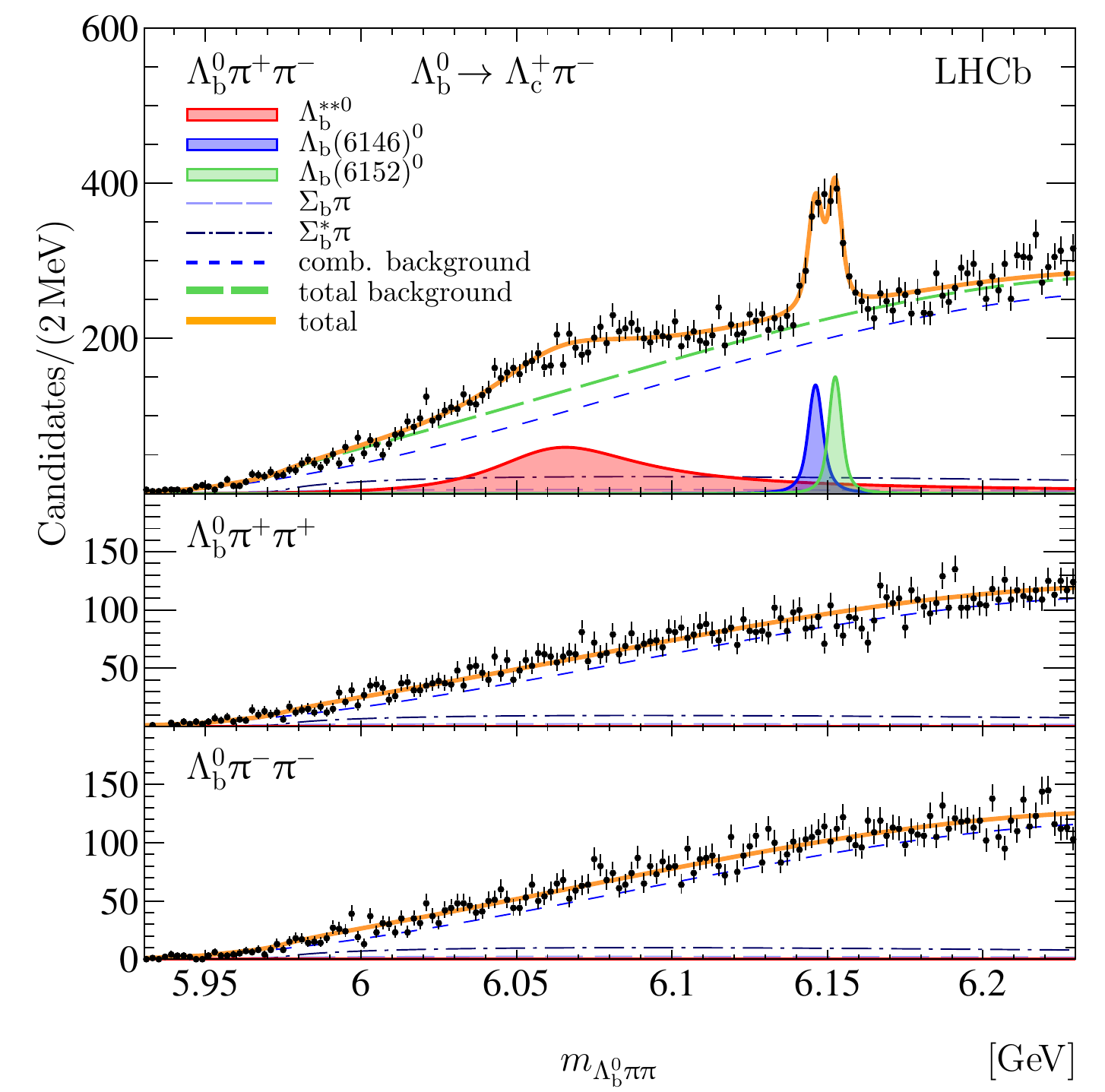}
    }
    \put(75, 0){ 
      \includegraphics*[width=75mm,height=75mm,%
      ]{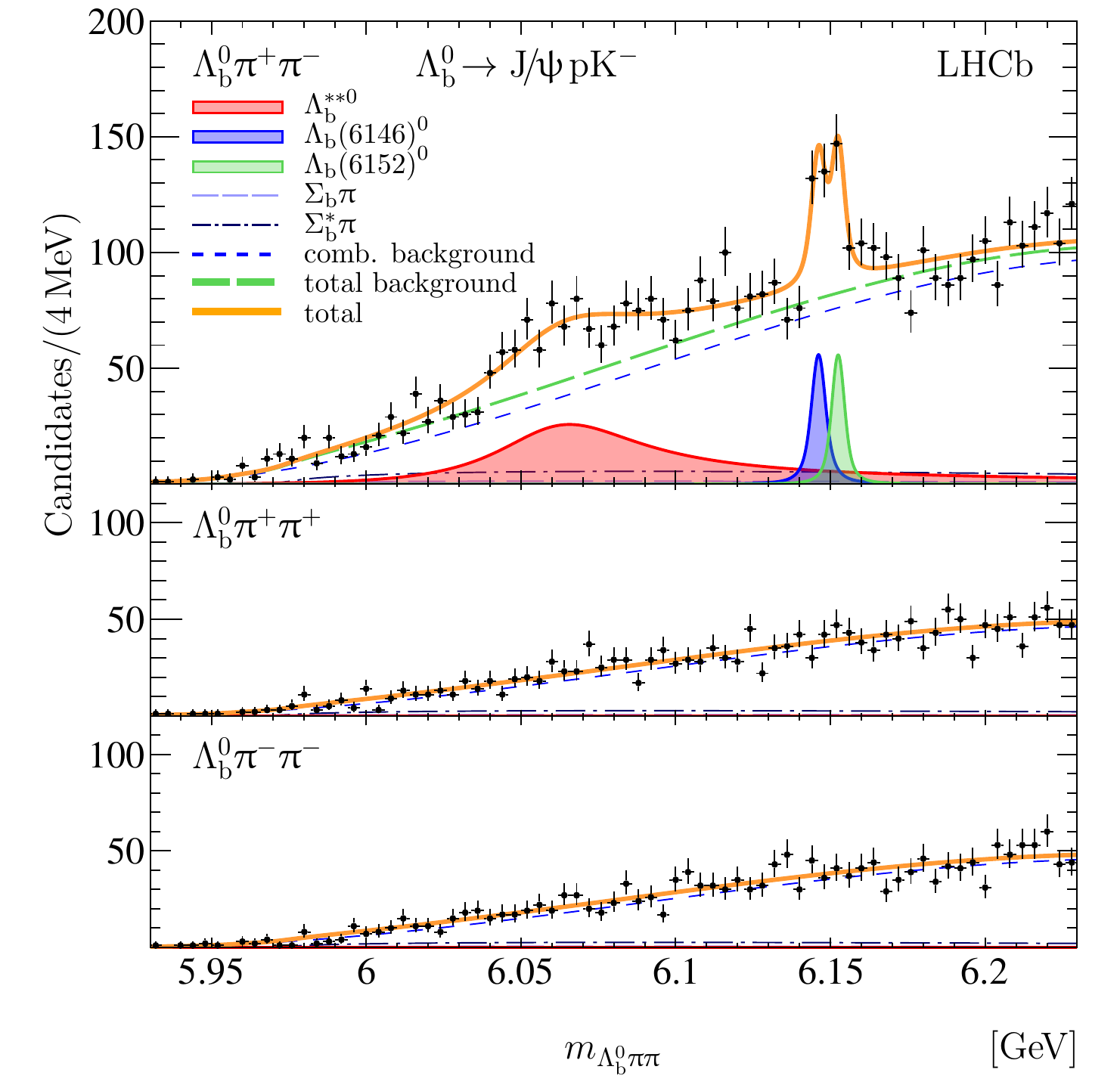}
    }
    \put( 48,68){\textcolor{white}{\rule{25mm}{4mm}}}
    \put( 48,68){\footnotesize{\lhcb~Run~1\&2}}
    \put(123,68){\textcolor{white}{\rule{25mm}{4mm}}}
    \put(123,68){\footnotesize{\lhcb~Run~1\&2}}
    \put( 95,0){\textcolor{white}{\rule{55mm}{6mm}}}
    \put(110,2){\small$m(\Lb\Ppi\Ppi)$}
    \put(135,2){\small$\left[\!\gevcc\right]$}
    \put( 20,0){\textcolor{white}{\rule{55mm}{6mm}}}
    \put( 35,2){\small$m(\Lb\Ppi\Ppi)$}
    \put( 60,2){\small$\left[\!\gevcc\right]$}
  \end{picture}
  \caption { \small
    Mass distribution of selected 
    (top)~$\Lb\pip\pim$,
    (middle)~$\Lb\pip\pip$ and 
    (bottom)~$\Lb\pim\pim$
    candidates
    for the~(left)~\mbox{$\decay{\Lb}{\Lc\pim}$}
    and (right)~\mbox{$\decay{\Lb}{\jpsi\proton\Km}$} decay modes~\cite{LHCb-PAPER-2019-045}.
  }
  \label{fig:spectra:lambdabtwos}
\end{figure}

In 2020, the fifth excited $\PLambda_{\bquark}$~state was observed  
in the $\Lb\pip\pim$~mass spectra, 
using the~full \lhcb Run~1 and 2 data-sets.
Two decay modes of the~$\Lb$~baryon were used,
\mbox{$\decay{\Lb}{\Lc\pim}$} and 
\mbox{$\decay{\Lb}{\jpsi\proton\Km}$}, 
and the significance of the new state, 
denoted  $\PLambda_{\bquark}^{\ast\ast0}$, 
is in excess of 
14 and 7 standard deviations in the two decay modes, respectively. This is    
shown in  Fig.~\ref{fig:spectra:lambdabtwos}.
Unlike the previously observed four narrow $\PLambda_{\bquark}$ states, 
$\PLambda_{\bquark}(5912)^0$,
$\PLambda_{\bquark}(5920)^0$,
$\PLambda_{\bquark}(6146)^0$ and 
$\PLambda_{\bquark}(6152)^0$,
the new state is rather broad, \mbox{$\Gamma=72\pm11\pm2\mev$}.
The~measured mass and width agree with the interpretation of 
this state being the~first 
radial excitation, the $\PLambda_{\bquark}(2S)^0$~resonance~\cite{Capstick:1986bm}.
This resonance is also consistent with 
a~broad excess of events in the~$\Lb\pip\pim$~mass spectrum,
previously reported by 
the~CMS collaboration~\cite{Sirunyan:2020gtz}.

\begin{figure}[htb]
  \centering
  \setlength{\unitlength}{1mm}
  \begin{picture}(150,60)
    \put(0 , 0){ 
      \includegraphics*[width=75mm,height=60mm,%
      ]{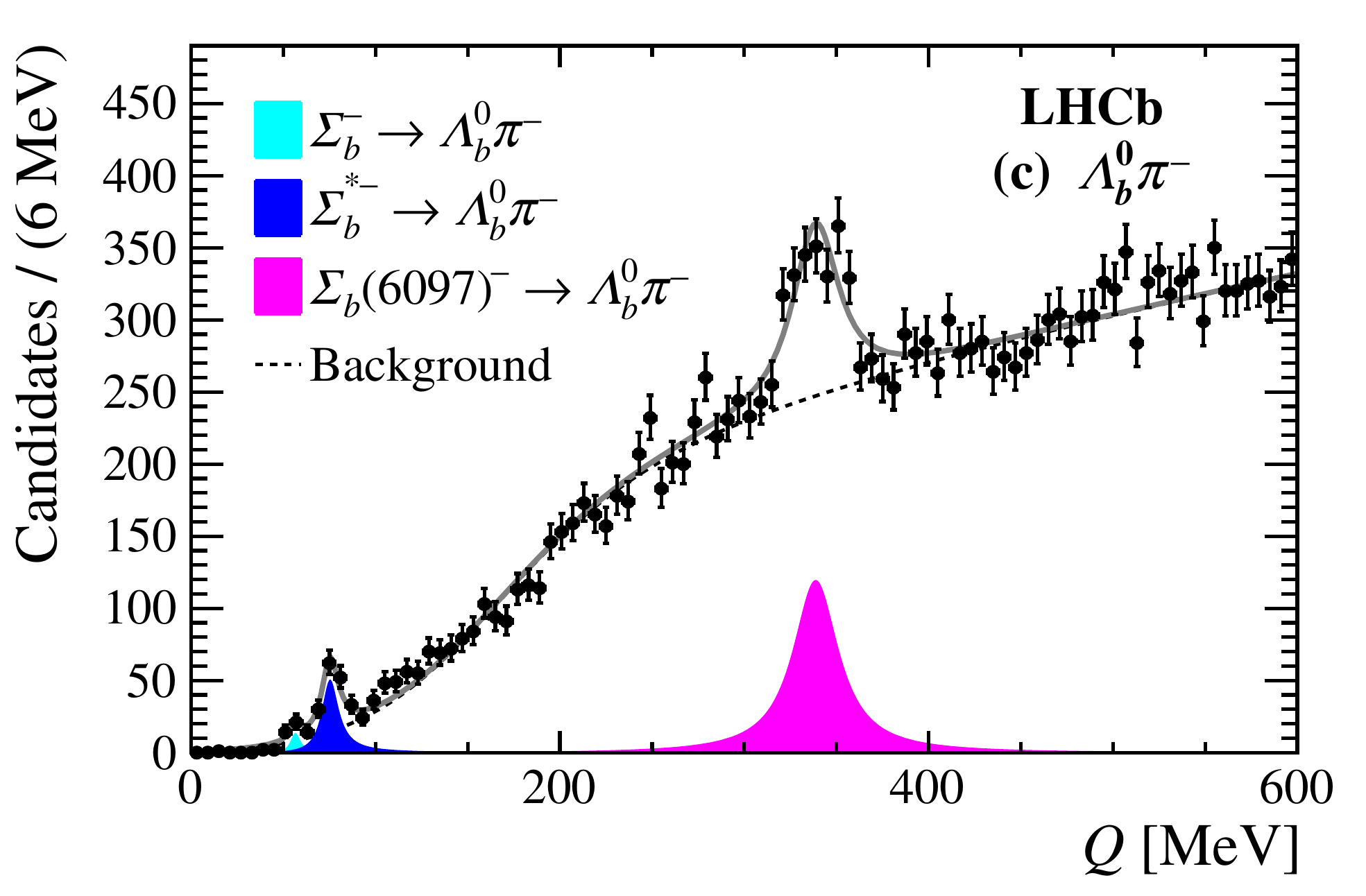}
    }
    \put(75, 0){ 
      \includegraphics*[width=75mm,height=60mm,%
      ]{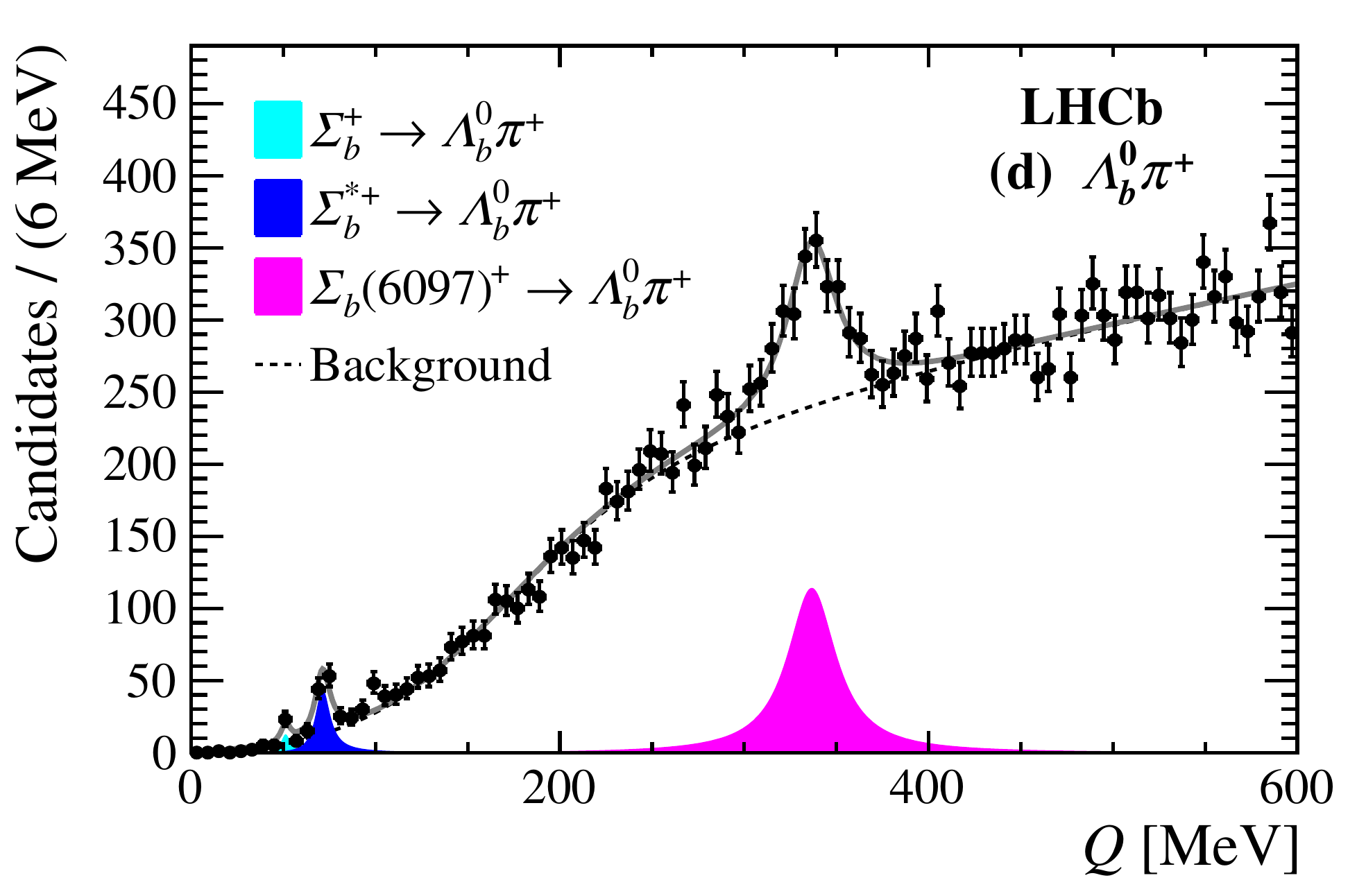}
    }
    \put( 58,48){\color{white}{\circle*{10}}}
    \put(133,48){\color{white}{\circle*{10}}}
    \put( 45,51){\textcolor{white}{\rule{25mm}{5mm}}}
    \put( 48,51){\footnotesize{\lhcb~Run~1}}
    \put(120,51){\textcolor{white}{\rule{25mm}{5mm}}}
    \put(123,51){\footnotesize{\lhcb~Run~1}}
  \end{picture}
  \caption{\small
    The~distributions of energy release 
    $Q\equiv m_{\Lb\Ppi^{\pm}}-m_{\Lb}-m_{\Ppi}$
    for selected $\Lb\Ppi^{\pm}$~candidates~\cite{LHCb-PAPER-2018-032}. 
    The~points show experimental data.
    The~left\,(right) plot shows $\Lb\pip$\,($\Lb\pim$)~combinations.
  }
  \label{fig:spectra:sigmab}
\end{figure}
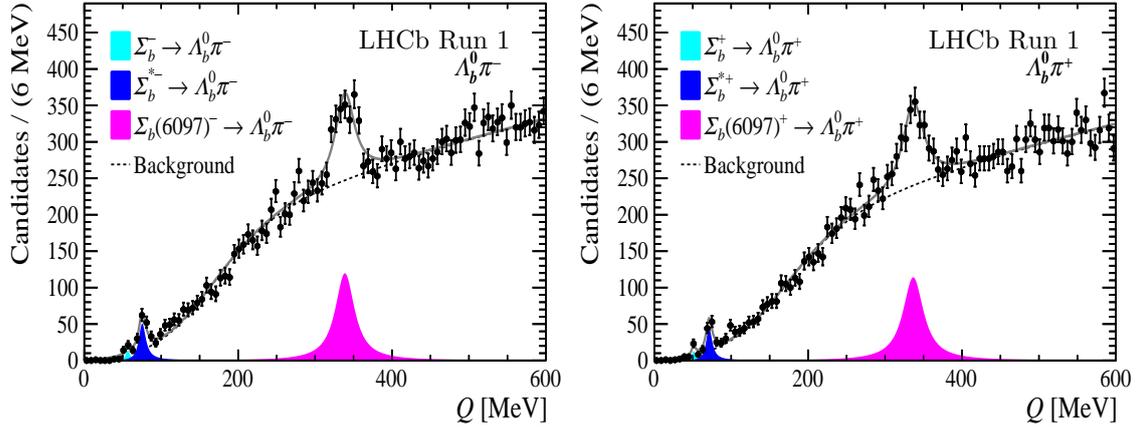

\paragraph{Excited $\PSigma_{\bquark}^{\pm}$~baryons} have been studied
in the $\Lb\Ppi^{\pm}$~mass spectra 
using the~Run~1 \lhcb data-set~\cite{LHCb-PAPER-2018-032}.
In~total $(234.27\pm0.90)\times 10^3$~signal 
$\Lb$~baryons  were reconstructed in the~decay 
mode~\mbox{$\decay{\Lb}{\Lc\pim}$}. 
Distributions of the~energy release in the decay, 
$Q \equiv m_{\Lb\Ppi^{\pm}}-m_{\Lb}-m_{\Ppi}$, are shown 
in Fig.~\ref{fig:spectra:sigmab}. 
At~low values of~$Q$, there are 
previously-known signals from   $\PSigma_{\bquark}^{(\ast)\pm}$~states, 
observed and characterised by 
the~CDF 
collaboration~\cite{PhysRevLett.99.202001,CDF:2011ac}.
New peaks  in  the  $\Lb\pim$\,($\Lb\pip$)~spectra
are visible at 
\mbox{$Q=338.8\pm1.7\mev\,(336.6\pm1.7\mev)$}, 
with a~local significance 
of $12.7\sigma$\,($12.6\sigma$),
based on the differences in
log\nobreakdash-likelihoods between the
 fits with zero signal and the nominal fit.
In~the~heavy\nobreakdash-quark limit, 
five $\PSigma_{\bquark}(1P)$ states are expected, 
and several predictions of their masses have been 
made.
Since the~expected density of baryon states is high, it
cannot be excluded that the~new observed structures are 
the~superposition of more than one
(near-)degenerate state. 
Taking into account that
the~predicted mass and width depend
on the~as\nobreakdash-yet\nobreakdash-unknown 
spin and parity, 
the~newly observed structures are compatible
with being $\PSigma_{\bquark}(1P)^{\pm}$~excitations. 
Other interpretations, such as molecular states, may also be
possible. 

\begin{figure}[htb]
  \centering
  \setlength{\unitlength}{1mm}
  \begin{picture}(150,60)
    \put(0 , 0){ 
      \includegraphics*[width=75mm,height=60mm,%
      ]{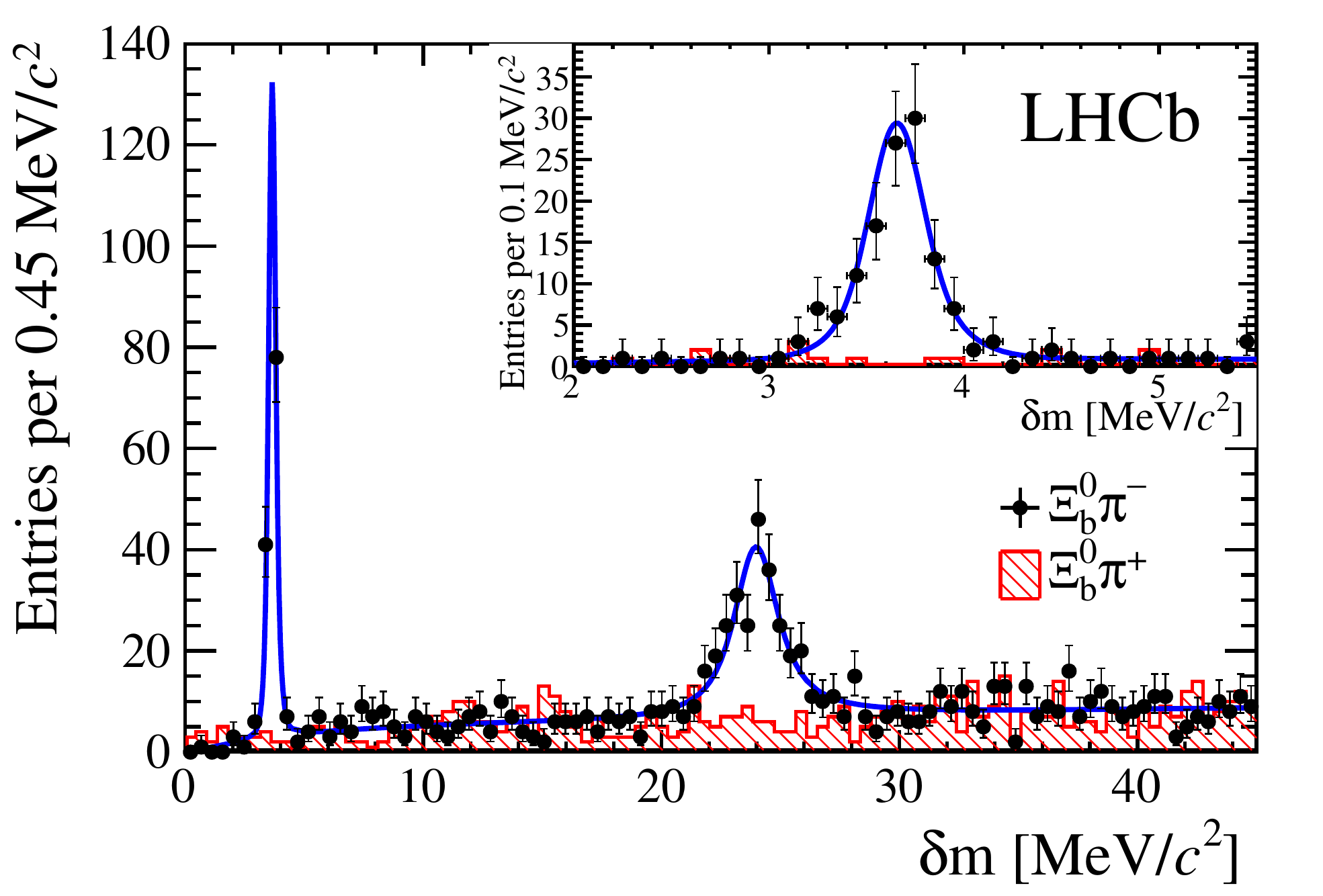}
    }
    \put(75, 0){ 
      \includegraphics*[width=75mm,height=60mm,%
      ]{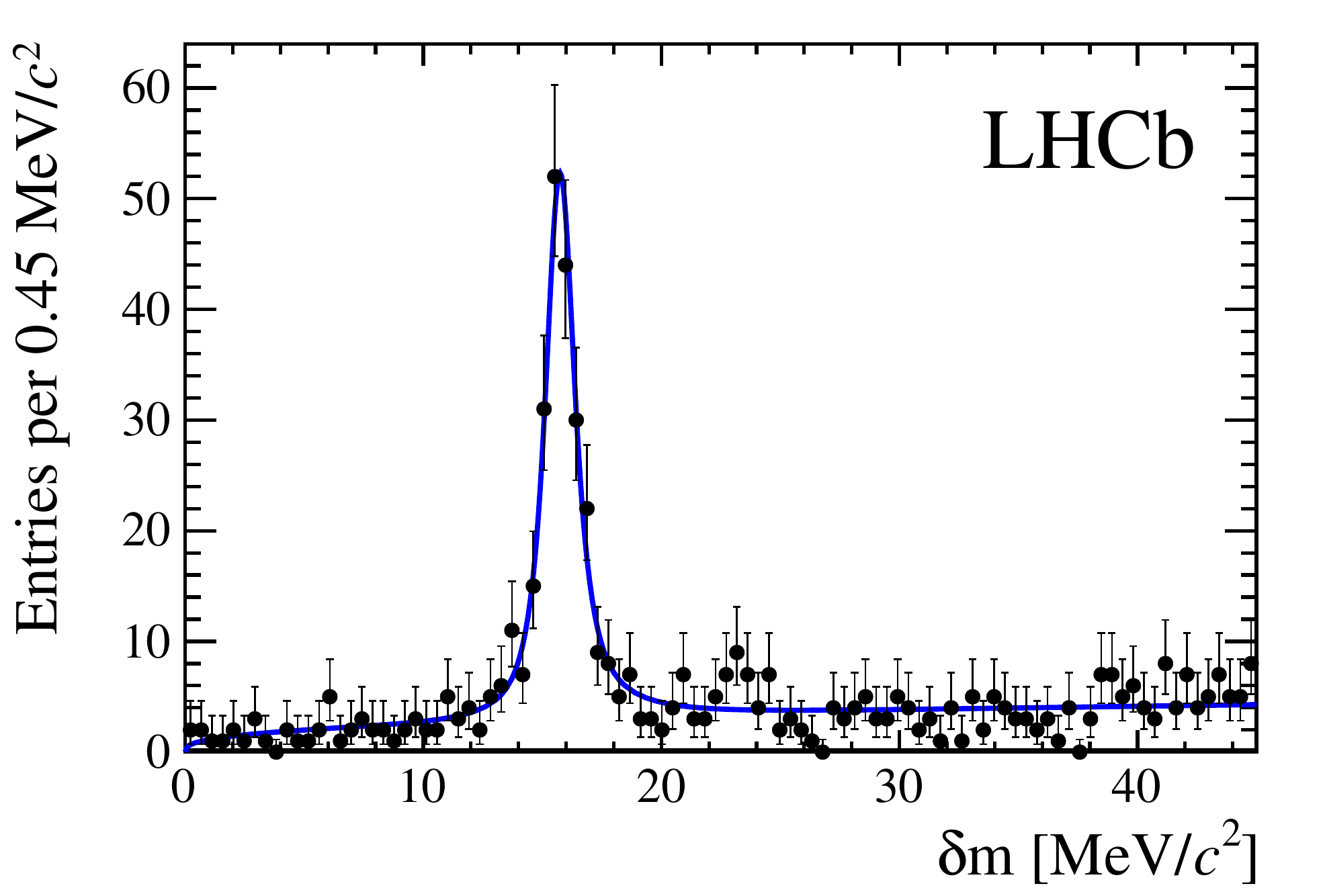}
    } 
    \put( 55,50){\textcolor{white}{\rule{15mm}{6mm}}}
    \put( 60,50){\footnotesize{$\begin{array}{l}\lhcb\\ \text{Run~1}\end{array}$}}
    \put(130,48){\textcolor{white}{\rule{15mm}{8mm}}}
    \put(132,50){\footnotesize{$\begin{array}{l}\lhcb\\ \text{Run~1}\end{array}$}}
    \put ( 35,29){\small$\decay{\PXi_{\bquark}^{\ast-}}{\Xibz\pim}$} 
    \put ( 36,53){\scriptsize$\decay{\PXi_{\bquark}^{\prime-}}{\Xibz\pim}$} 
    \put (123,40){\small$\decay{\PXi_{\bquark}^{\ast0}}{\Xibm\pip}$} 
 \end{picture}
  \caption{\small 
    (left)~The distribution of the~mass difference, \mbox{$\delta m \equiv m_{\Xibz\pim} - m_{\Xibz} - m_{\pim}$}, 
    for~\mbox{$\Xibz \pim$}~candidates~\cite{LHCb-PAPER-2014-061}.
    The~points with error bars show
    right-sign candidates in the~\Xibz mass signal region,
    and the~hatched histogram shows
    wrong-sign candidates with the~same selection.
    The~curve shows the nominal fit to the~right-sign candidates.
    Inset: detail of the~region $2.0$--$5.5$\mevcc.
    (right)~Distribution of $\delta m$ and the~fit
    for~\mbox{$\Xibm \pip$}~candidates~\cite{LHCb-PAPER-2016-010}.
  }
  \label{fig:spectra:xibl}
\end{figure}
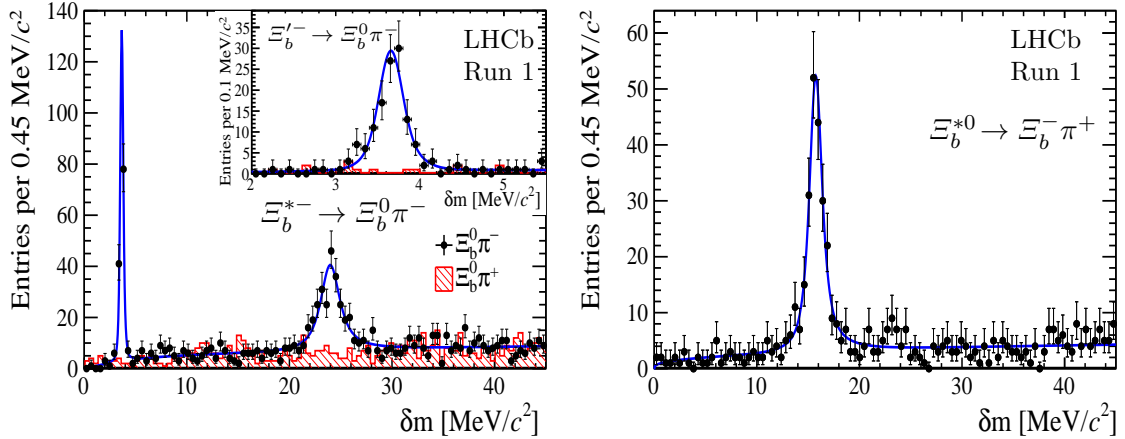

\paragraph{$\PXi_{\bquark}^{\prime-}$ and $\PXi_{\bquark}^{\ast-}$~baryons} 
have been observed in the $\Xibz\pim$~mass spectrum
using the \lhcb Run~1 data-set~\cite{LHCb-PAPER-2014-061}.
Signal $\Xibz$ candidates were reconstructed in 
the~final 
state $\Xicp\pim$ with \mbox{$\decay{\Xic}{\proton\Km\pip}$}.
Two~peaks are
clearly visible in the
\mbox{$\delta m \equiv m_{\Xibz\pim} - m_{\Xibz} - m_{\pim}$} spectrum, shown in Fig.~\ref{fig:spectra:xibl}(left),
a~narrow state at $\delta m \approx 3.7\mevcc$  
and a~broader state at~$\delta m \approx 24\mevcc$.
No~structure is observed in 
the~wrong\nobreakdash-sign sample, 
nor in the~\Xibz~mass side bands.
The~fitted natural width of 
the~lower\nobreakdash-mass state
is found to be consistent 
with zero. The~fitted yields of the~lower and
higher\nobreakdash-mass peaks are $121\pm12$ and 
$237\pm24$ events, respectively, with statistical significance in
excess of $10$~standard deviations. 
The~non-zero value of the~natural 
width of the~higher\nobreakdash-mass state,
\mbox{$\Gamma=1.65\pm0.31\pm0.10\mev$} 
is   significantly different from zero.
The~signals are interpreted as  $\PXi_{\bquark}^{\prime-}$
and $\PXi_{\bquark}^{\ast-}$~baryons.

\paragraph{The~$\PXi_{\bquark}^{\ast0}$ baryon} was first observed 
at CMS~\cite{Chatrchyan:2012ni}, 
and later studied in detail 
by the~\lhcb collaboration using 
the~Run~1 data-set~\cite{LHCb-PAPER-2016-010}.
The~$\PXi_{\bquark}^{\ast0}$~candidates have been reconstructed 
in the decay \mbox{$\decay{\PXi_{\bquark}^{\ast0}}{\Xibm\pip}$},
with \mbox{$\decay{\Xibm}{\Xicz\pip}$} and 
\mbox{$\decay{\Xicz}{\proton\Km\Km\pip}$}.
The~$\delta m$~distribution,  defined as 
\mbox{$m_{\Xibm\pip} - m_{\Xibm} - m_{\pip}$},
is shown in Fig.~\ref{fig:spectra:xibl}(right).
A~narrow peak is clearly visible with a fitted signal yield of $232\pm19$ events. 
The~non-zero value of the~natural width of the~peak, $\Gamma = 0.90\pm0.16\mev$,
is also highly significant; the change in log\nobreakdash-likelihood when 
the~width is  fixed to zero exceeds 30 units. 
No~other statistically-significant structures 
are seen. 
The~peak position and the~width are
consistent with, and about a~factor of ten more precise 
than, the CMS measurements~\cite{Chatrchyan:2012ni}.
The~measured width of the~state is in line with theory expectations; 
a~calculation based on lattice QCD predicts a~width of 
$0.51\pm0.16\mev$~\cite{Detmold:2012ge}, 
and another using the~${}^3P_0$~model 
obtains a~value of 0.85\mev~\cite{Chen:2007xf}.
The~measured production ratio with respect to 
the~\Xibm~state is measured to be 
$(28\pm3\pm1)\%$, suggesting that 
in high energy $\proton\proton$~collisions 
at~7~and 8\tev, a~large fraction of~\Xibm 
baryons are produced through feed\nobreakdash-down 
from higher\nobreakdash-mass states.

\paragraph{A high-mass excited $\PXi_{\bquark}^-$~baryon} has been observed
in the~$\Lb\Km$ and $\Xibz\pim$~mass spectra, 
using a~3.5\invfb \lhcb data-set at 
$\sqs=7$, 8 and 13\tev~\cite{LHCb-PAPER-2018-013}.
The~$\Lb$~baryons were reconstructed via~\mbox{$\decay{\Lb}{\Lc\pim}$}
and~\mbox{$\decay{\Lb}{\Lc \mun X}$}, with the 
$\Xibz$ decaying to~\mbox{$\decay{\Xibz}{\Xicp\mun X}$},
with~\mbox{$\decay{\Lc,\Xic}{\proton\Km\pip}$}.
Full~reconstruction of the \Lb~baryon allows the~excellent resolution 
of the~$\Lb\Km$~mass spectra to be reached. In addition, partial reconstruction of 
\Lb and \Xibz~baryons in their semileptonic modes allows a significant 
increase in the~sample of   \Lb and \Xibz~baryons, where 
the~missing neutrinos do not prevent  a~peaking structure in the spectra of
mass differences $(m_{"\Lb"\kaon}-m_{"\Lb"})$ 
and 
$(m_{"\Xibz"\pim}-m_{"\Xibz})$. 
The~resolution is improved by applying 
the~ 4-vector constraint~\mbox{$(p_{H_{\cquark}^+}+p_{\mun}+p_{\rm miss})^2=m_{H_{\bquark}^0}^2$}, 
where $H_{\cquark}^+$ stands for~$\Lc$ and $\Xicp$ and 
$H_{\bquark}^0$~stands for~$\Lb$ and~$\Xibz$.

\begin{figure}[htb]
  \centering
  \setlength{\unitlength}{1mm}
  \begin{picture}(150,80)
    \put(  0,40){ 
      \includegraphics*[width=50mm,height=40mm,%
      ]{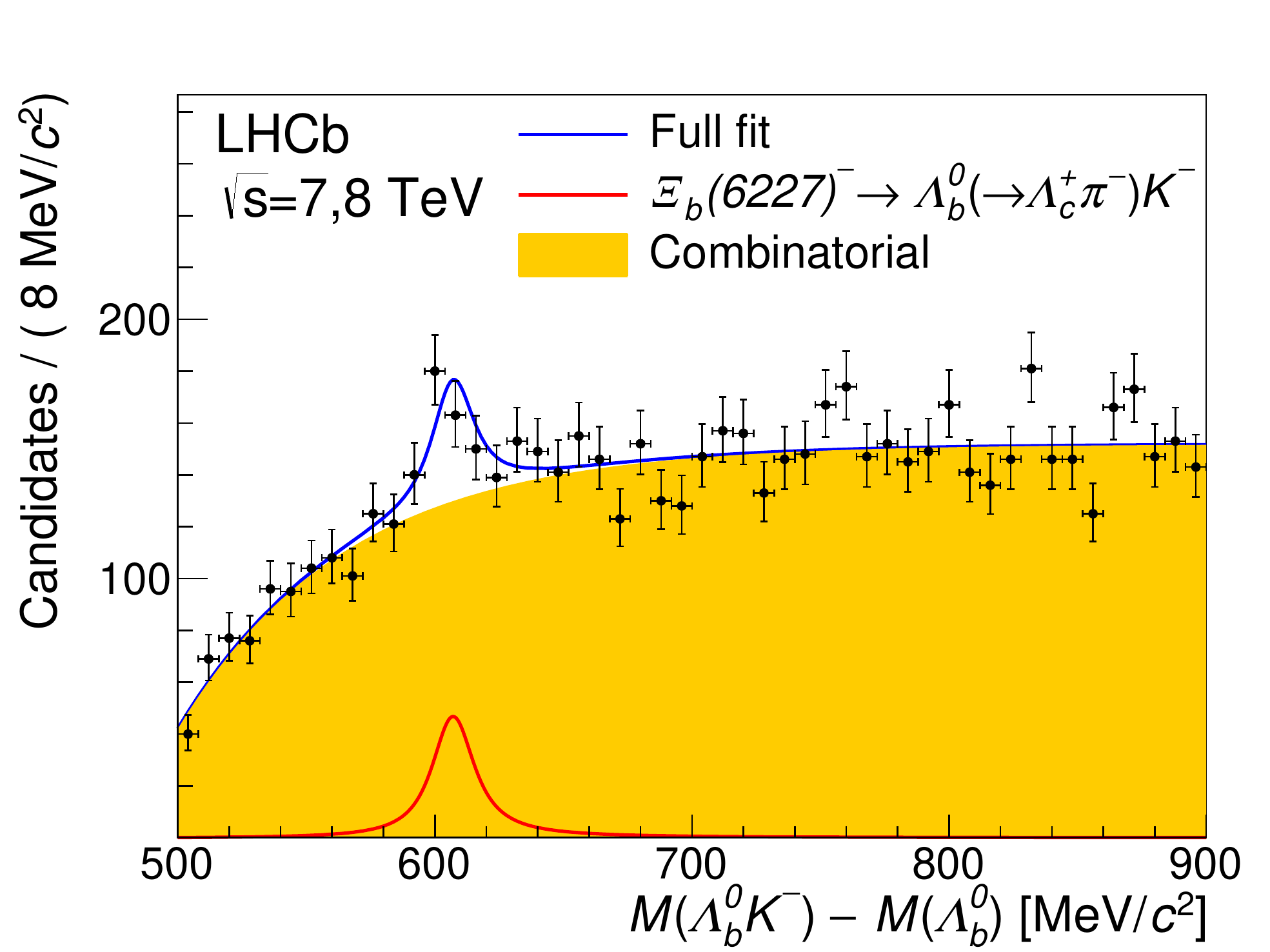}
    }
    \put( 50,40){ 
      \includegraphics*[width=50mm,height=40mm,%
      ]{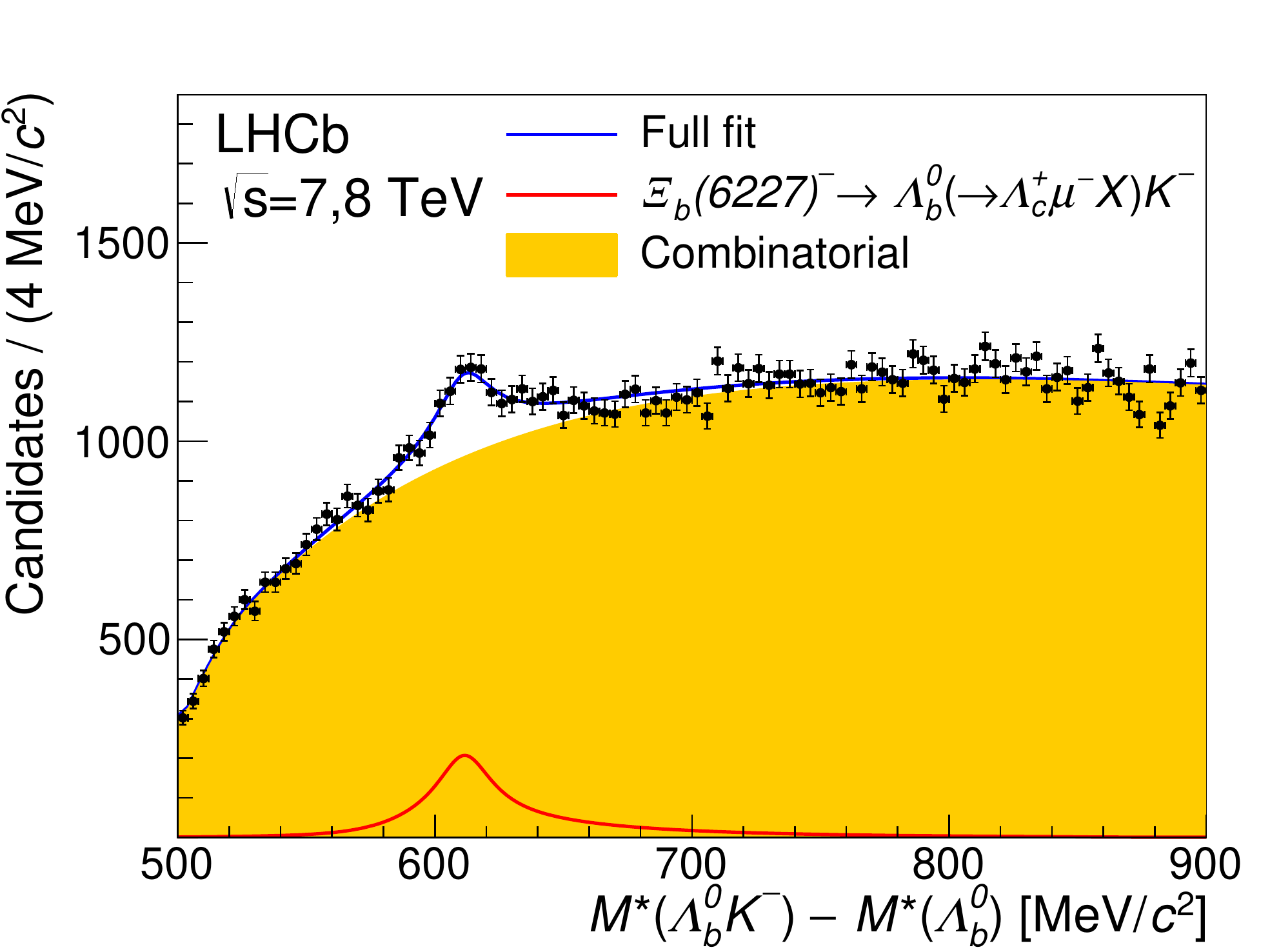}
    }
    \put(100,40){ 
      \includegraphics*[width=50mm,height=40mm,%
      ]{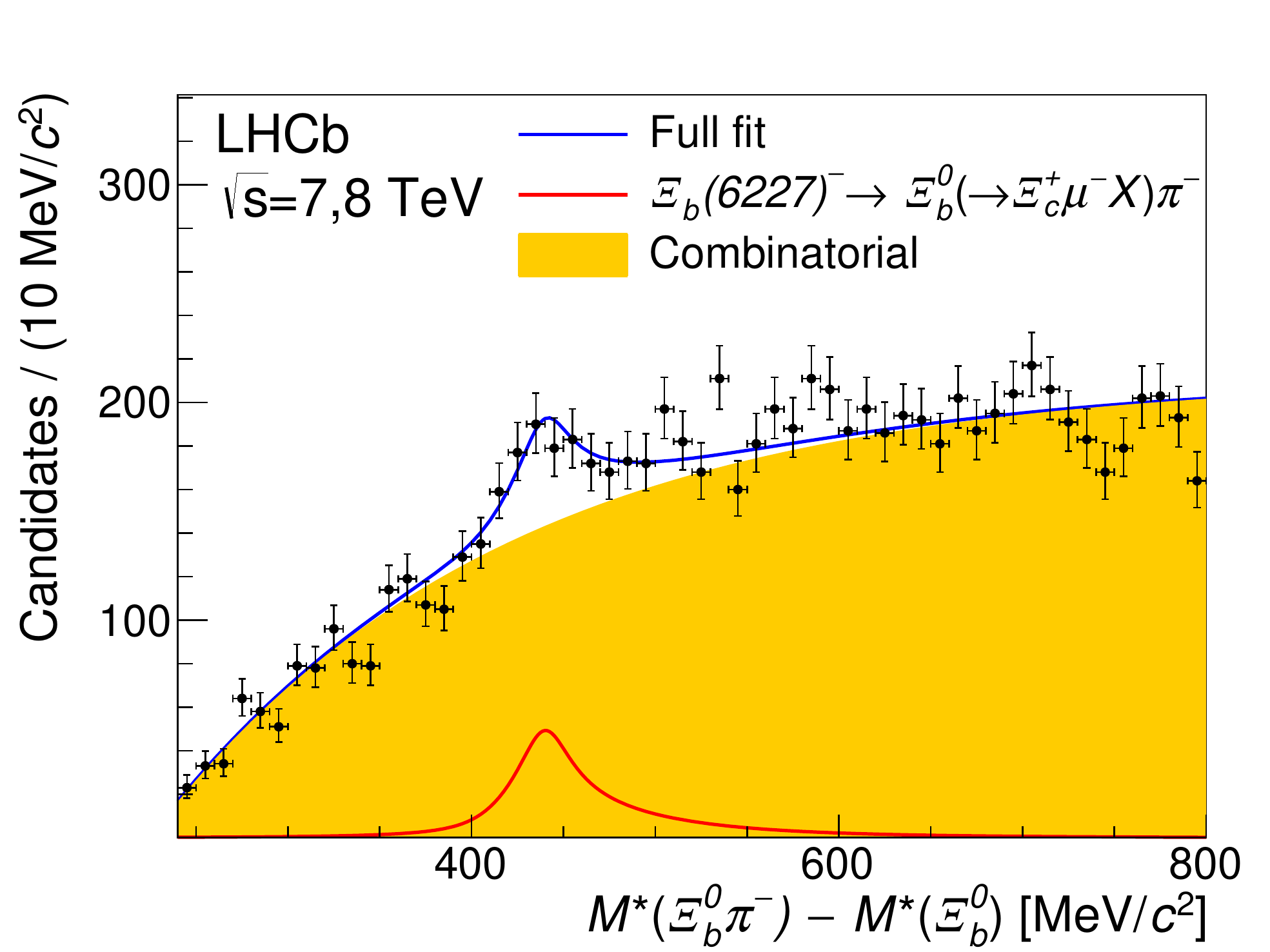}
    }
    \put(  0,0){ 
      \includegraphics*[width=50mm,height=40mm,%
      ]{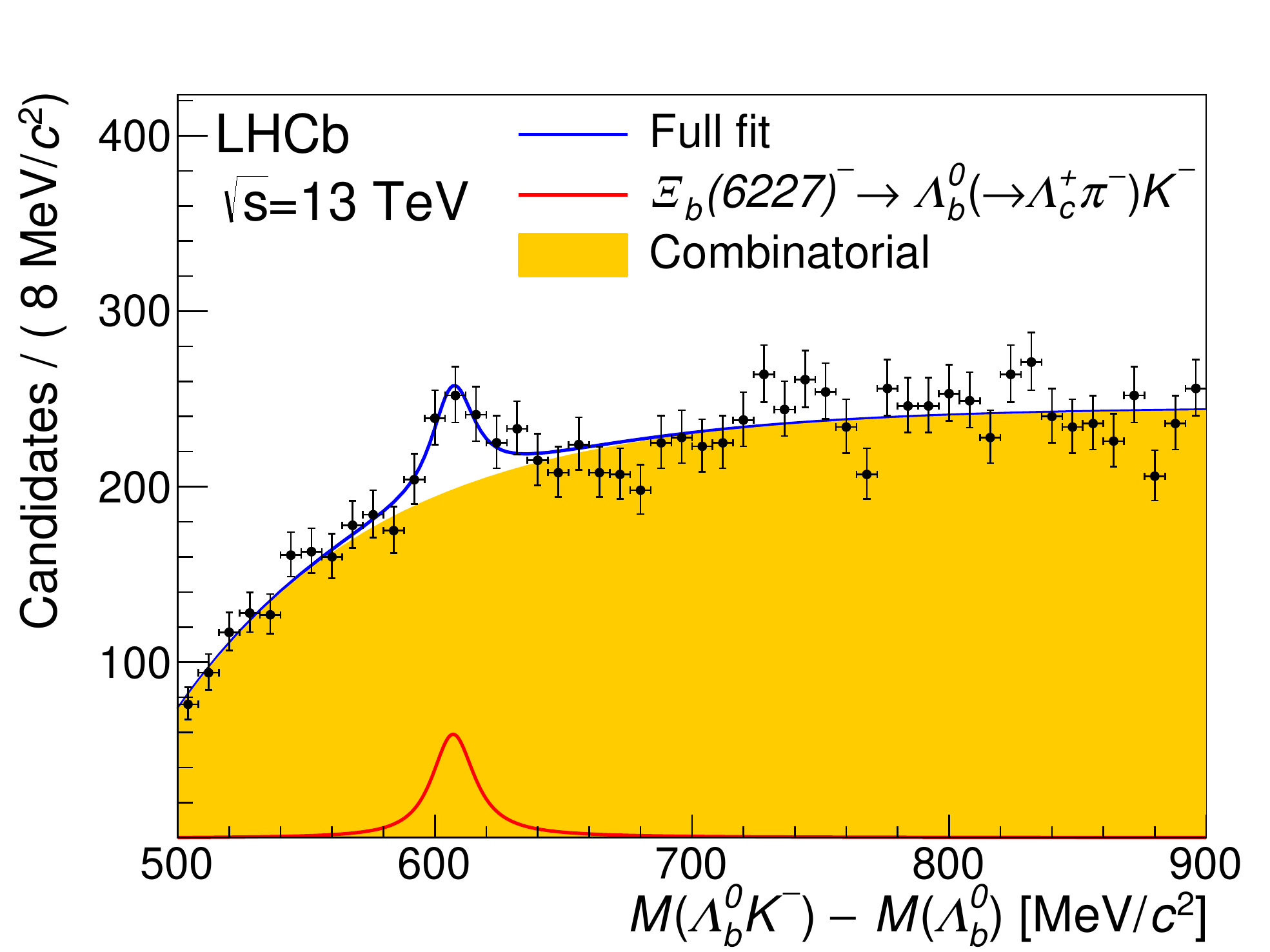}
    }
    \put( 50,0){ 
      \includegraphics*[width=50mm,height=40mm,%
      ]{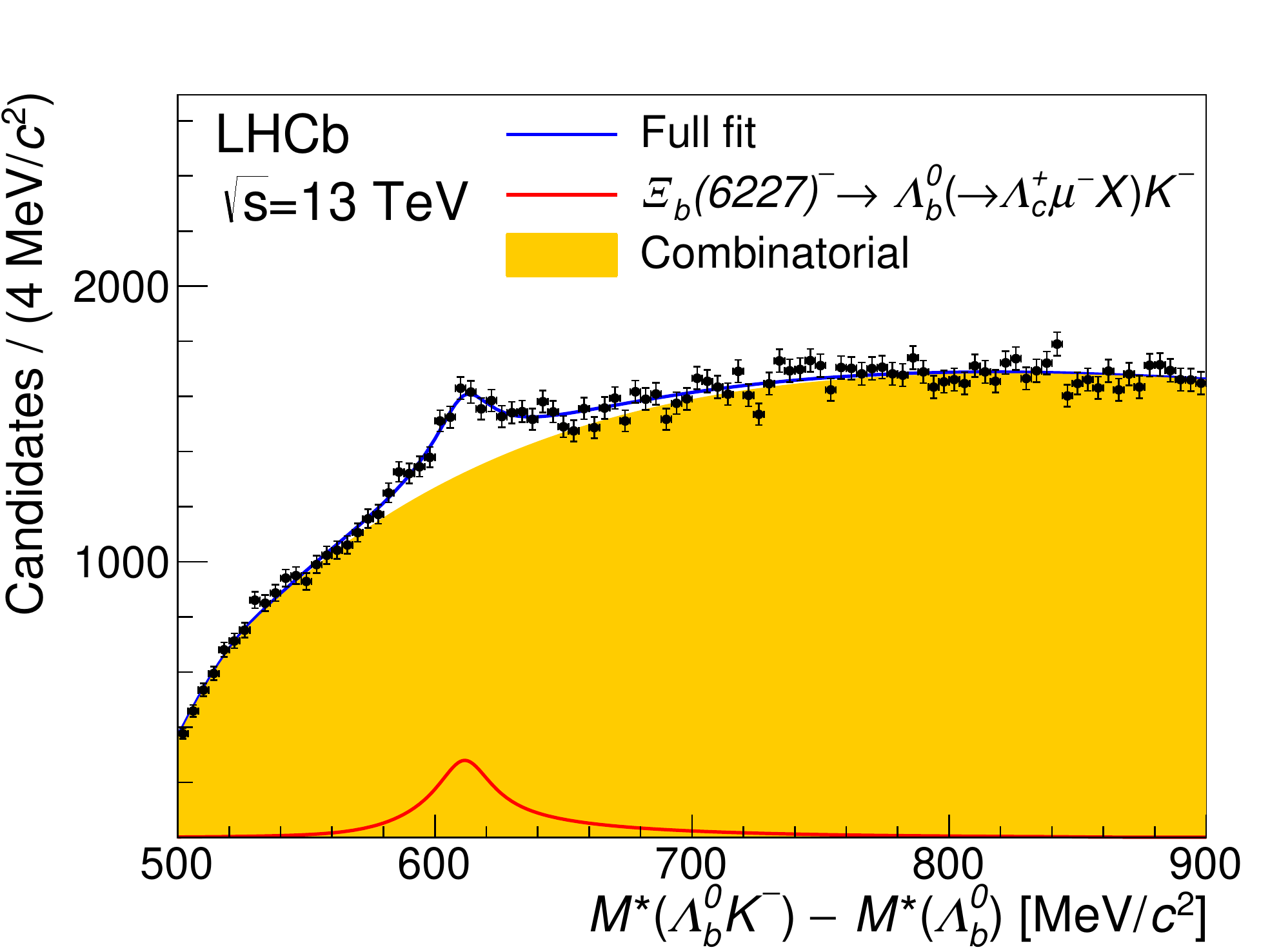}
    }
    \put(100,0){ 
      \includegraphics*[width=50mm,height=40mm,%
      ]{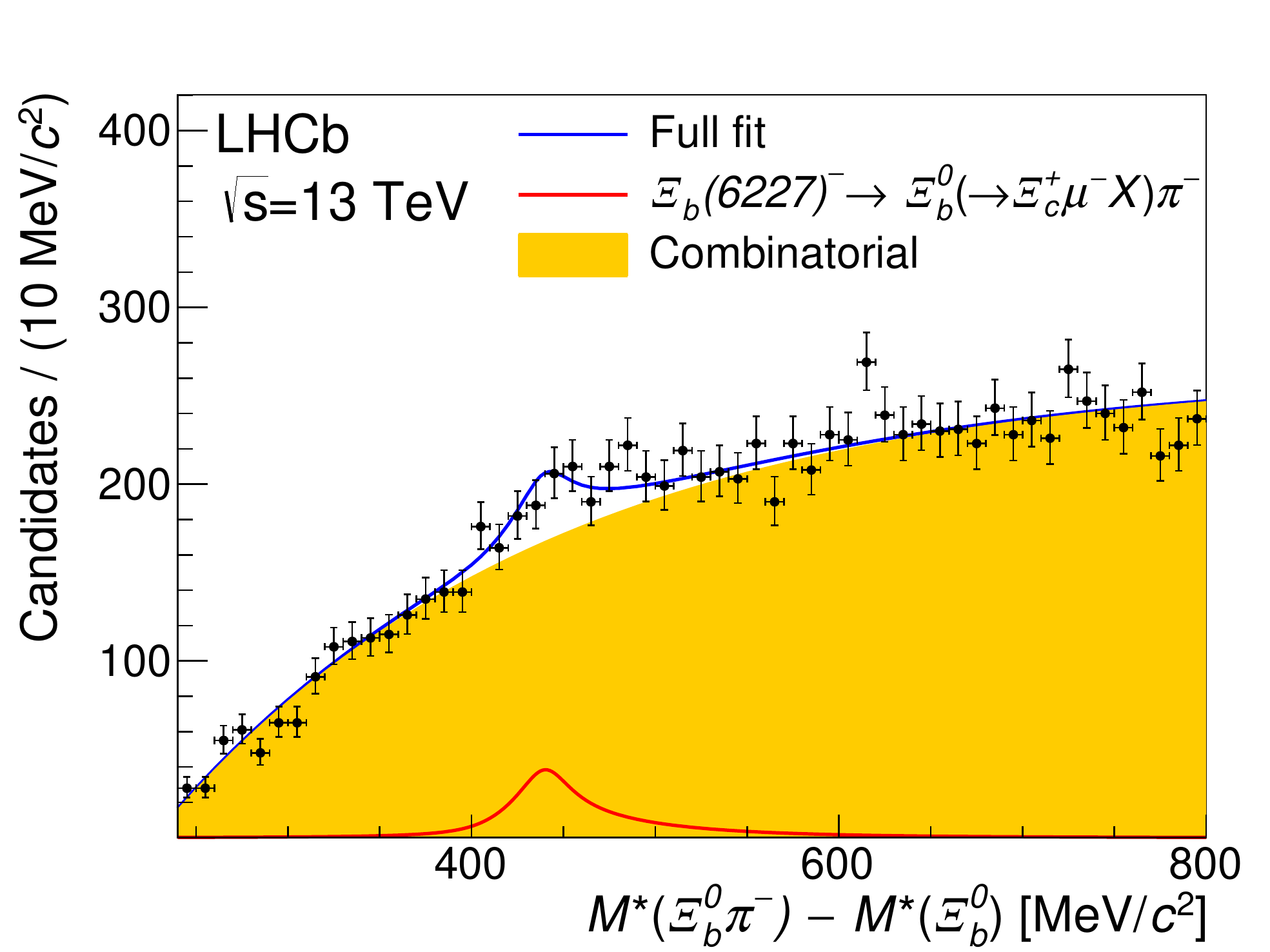}
    }
  \end{picture}
  \caption{\small
    Spectra of mass differences for $\PXi_{\bquark}^*$~candidates, reconstructed in the~final states
    (left)~$\Lb\Km$, with \mbox{$\Lb\to\Lc\pim$}, 
    (middle)~$\Lb\Km$, with \mbox{$\Lb\to\Lc\mun X$},
    and (right) $\Xibz\pim$, with \mbox{$\Xibz\to\Xicp\mun X$}, 
    along with fits to the data~\cite{LHCb-PAPER-2018-013}.
    The~top row is for 7 and 8~TeV data and 
    the~bottom is for 13~TeV. 
    The~symbol $M^*$ represents the~mass after 
    the~ 4-vector constraint~\mbox{$(p_{H_{\cquark}^+}+p_{\mun}+p_{\rm miss})^2=m_{H_{\bquark}^0}^2$} 
    is applied.
  }
  \label{fig:spectra:xibh}
\end{figure}

The~mass\nobreakdash-difference  spectra are shown in 
Fig.~\ref{fig:spectra:xibh}, where
the~peak locations for all three modes are seen to
agree well. 
The~statistical significances of the new excited baryon, 
dubbed the $\PXi_{\bquark}(6227)^-$, are found 
to be
$7.9\sigma$ 
for the~\mbox{$\decay{\PXi_{\bquark}(6227)^-}{\Lb\Km}$} followed by 
\mbox{$\decay{\Lb}{\Lc\pim}$}, 
$25\sigma$
for the~\mbox{$\decay{\PXi_{\bquark}(6227)^-}{\Lb\Km}$} followed by 
\mbox{$\decay{\Lb}{\Lc\mun X}$}
and 
$7.2\sigma$
for the~\mbox{$\decay{\PXi_{\bquark}(6227)^-}{\Xibz\pim}$} followed by 
\mbox{$\decay{\Xibz}{\Xicp\mun X}$}.

\begin{figure}[ht]
  \centering
  \setlength{\unitlength}{1mm}
  \begin{picture}(150,70)
    \put(0, 0){ 
      \includegraphics*[height=70mm,width=150mm%
      ]{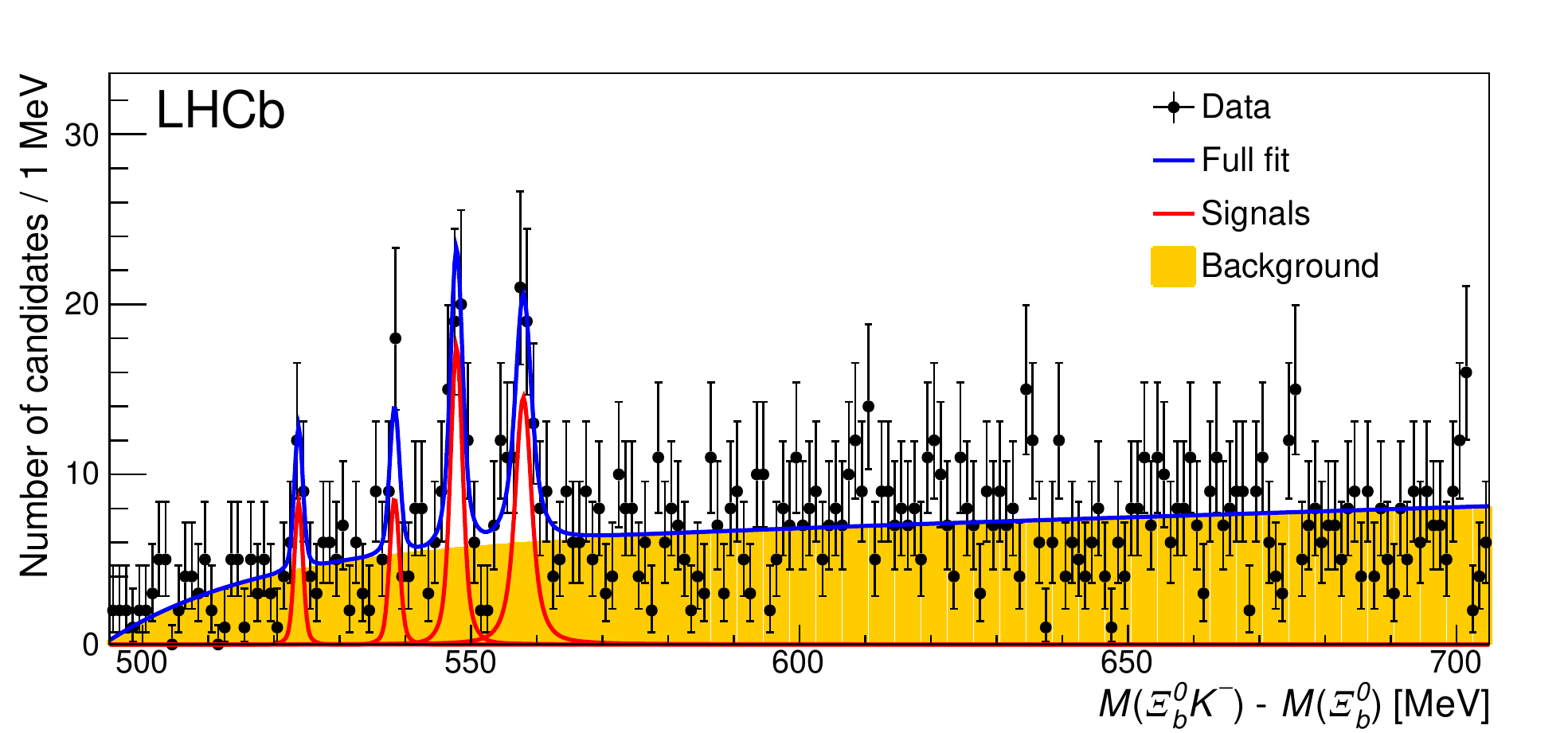}
    }
    \put( 15,52){\textcolor{white}{\rule{20mm}{10mm}}}
    \put( 20,55){\footnotesize{\lhcb~Run~1\&2}}
  \end{picture}
  \caption { \small
    The mass difference~$\delta m \equiv m(\Xibz\Km)-m(\Xibz)$   for selected 
    $\Xibz\Km$~candidates~\cite{LHCb-PAPER-2019-042}.
  }
  \label{fig:spectra:omegab}
\end{figure}

\paragraph{Four narrow excited $\POmega_{\bquark}$~states} 
have been observed in the $\Xibz\Km$~mass spectrum 
using the~full Run~1 and 2 \lhcb data-sets~\cite{LHCb-PAPER-2019-042}.
The~$\Xibz$~candidates were reconstructed in $\Xicp\pim$~final states
with \mbox{$\decay{\Xic}{\proton\Km\pip}$}. After multivariate selection, 
a low\nobreakdash-background sample of $(19.2\pm0.2)\times10^3$
\mbox{$\decay{\Xibz}{\Xicp\pim}$}~decays has been  obtained. 
The~mass-difference spectrum \mbox{$m(\Xibz\Km)-m(\Xibz)$} for 
$\Xibz\Km$~combinations exhibits four narrow peaks, shown in 
Fig.~\ref{fig:spectra:omegab}.
The~natural widths of the~three lower mass states are consistent 
with zero, while the width of the high\nobreakdash-mass state is found 
to be~\mbox{$1.4^{+1.0}_{-0.8}\pm0.1\mev$}.
The~peaks have local significances that range from~3.6 to 7.2~standard deviations. 
After~accounting for the~look\nobreakdash-elsewhere effect, 
the~significances of the two
low-mass peaks are reduced to~2.1$\sigma$ and 2.6$\sigma$,~respectively, 
whilst the~two higher-mass peaks exceed~5$\sigma$.
The~observed $\Xibz\Km$ peaks seen here are similar to those observed in 
the $\Xicp\Km$ invariant mass spectrum~\cite{LHCb-PAPER-2017-002}.
Arguably, the~simplest interpretation  is that the peaks correspond 
to excited $\Omegab$ states, in particular the~$L=1$ angular momentum excitation 
of the~ground state, or possibly an $n=2$ radial excitation.

Many~of the quark-model calculations predict $L=1$ states in this mass 
region, 
and at least some of the~states should be narrow. 
In~particular,  using the~$^3P_0$~model, five states in this mass region are predicted, 
with approximately 8\mev mass splittings; the~four lightest have partial widths, 
$\Gamma(\Xibz\Km)$, below 1\mev, whilst the~one with the~largest mass has
$\Gamma(\Xibz\Km)=1.49\mev$. 
Conversely, predictions using the~chiral quark model indicate that 
the $J^P=(3/2)^-$ and $(5/2)^-$ states are narrow, 
but the~$(1/2)^-$ states are wide~\cite{Wang:2017kfr}.
Quark\nobreakdash-diquark models have also predicted several excited $\Omegab$ states 
in the~region around 6.3\gev, 
with mass splittings  similar to that  observed here, however, 
there are no predictions for the~decay widths.  
Molecular models have also been employed, 
where two narrow $J^P=(1/2)^-$ states are predicted at 6405\mev and 6465\mev~\cite{Debastiani:2017ewu},
however do not match well with the~\lhcb measurements.

An~alternate interpretation for one or more of the~observed peaks is that they arise 
from the~decay of a~higher\nobreakdash-mass excited $\POmega_{\bquark}^{\ast\ast - }$~state 
to $\PXi_{\bquark}^{\prime 0}(\to\Xibz\piz)\Km$, where the~$\piz$ meson is undetected. 
If the~mass of a not\nobreakdash-yet\nobreakdash-observed $\PXi_{\bquark}^{\prime 0}$~state is in 
the~region \mbox{$m_{\Xibz}+m_{\piz}<m_{ \PXi_{\bquark}^{\prime 0}  }<m_{ \PXi_{\bquark}^{\prime-}}$},  
each of the~observed narrow peaks can be interpreted as having come from the~above decay, 
provided that the~corresponding excited $\POmega_{\bquark}^{\ast\ast - }$~state is narrow, 
$\Gamma_{\POmega_{\bquark}^{\ast\ast - }}\le 1\mev$. 
In~this case, their masses can be evaluated as 
$m_{\POmega_{\bquark}^{\ast\ast-}}=m_{\PXi_{\bquark}^{\prime 0}}+\delta m_{\rm peak}$, 
where $\delta m_{\rm peak}$ is a~measured position of the~peak in the~$m(\Xibz\Km)-m(\Xibz)$~spectrum.

\subsection{Conventional  charmonia and bottomonia}

\begin{figure}[htb]
  \centering
  \setlength{\unitlength}{1mm}
  \begin{picture}(150,65)
    \put(  0,0){ 
      \includegraphics*[width=75mm,height=65mm,%
      ]{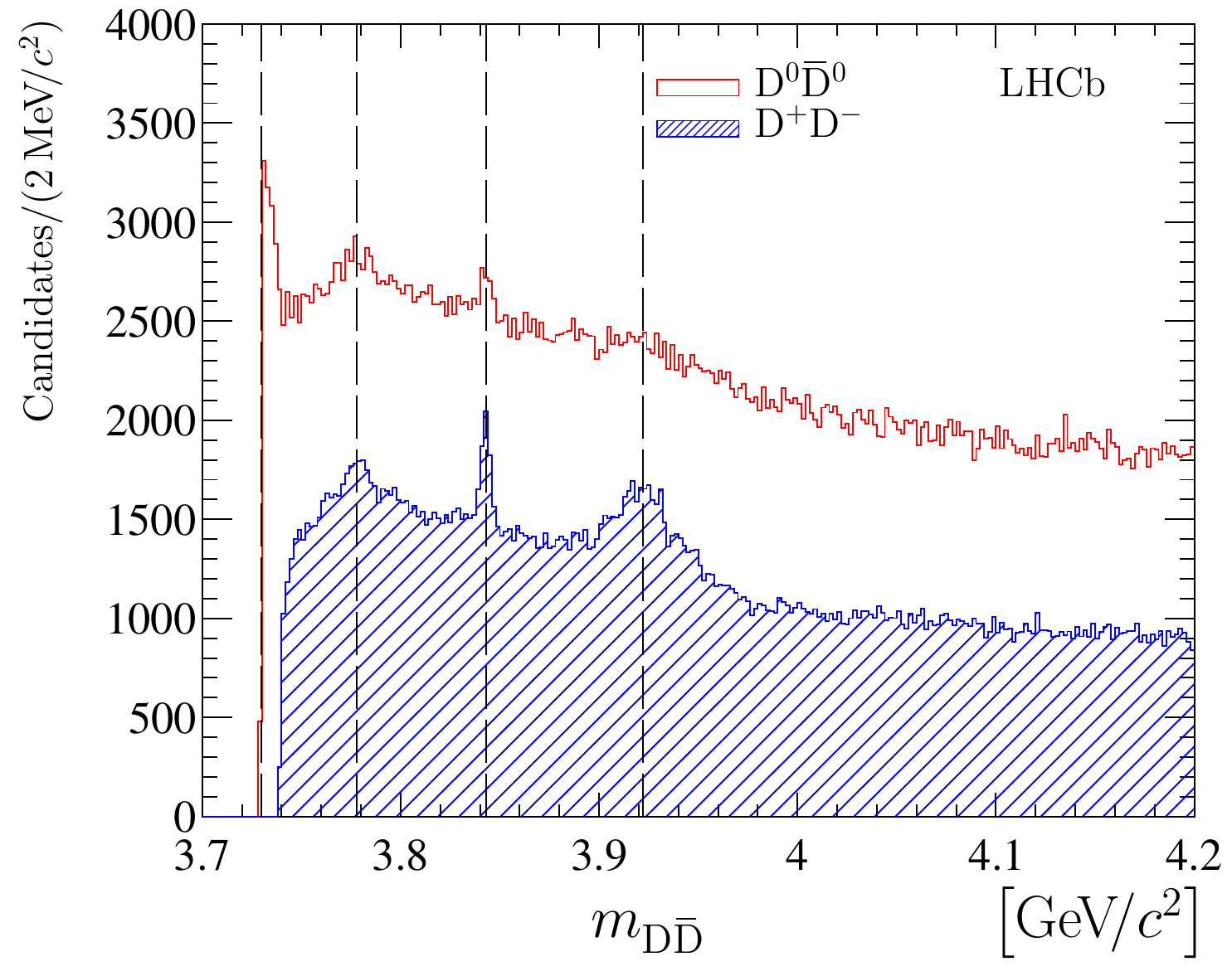}
    }
    \put( 75,3){ 
      \includegraphics*[width=75mm,height=62mm,%
      ]{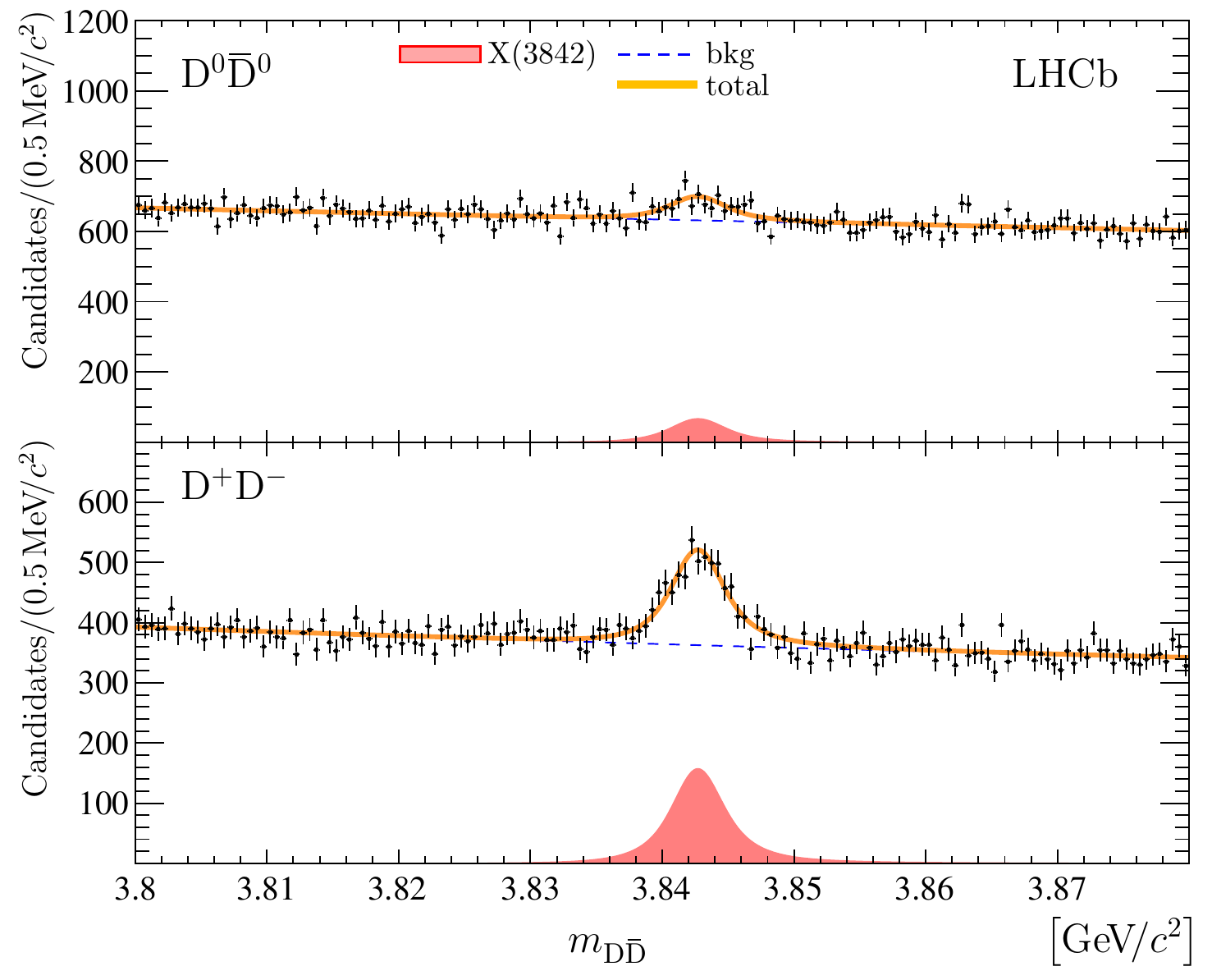}
    }
    \put( 55,58){\textcolor{white}{\rule{17mm}{4mm}}}
    \put( 54,56){\small{$\begin{array}{l}\lhcb \\ \mathrm{Run~1\&2} \end{array}$}}
    \put(130,58){\textcolor{white}{\rule{17mm}{4mm}}}
    \put(129,56){\small{$\begin{array}{l}\lhcb \\ \mathrm{Run~1\&2} \end{array}$}}
    \put( 95,1){\textcolor{white}{\rule{58mm}{6mm}}}
    \put(110,2){\small$m(\D\Dbar)$}
    \put(135,2){\small$\left[\!\gevcc\right]$}
    \put( 20,1){\textcolor{white}{\rule{58mm}{6mm}}}
    \put( 35,2){\small$m(\D\Dbar)$}
    \put( 60,2){\small$\left[\!\gevcc\right]$}
  \end{picture}
  \caption{\small
    (left)~The $\D\Dbar$ mass spectra. The open\,(red) histogram shows $\Dz\Dzb$ combinations
    whilst the dashed\,(blue) histogram shows $\Dp\Dm$.
    (right)~The~mass spectra of (top)~$\Dz\Dzb$ and (bottom)~$\Dp\Dm$~combinations 
    in the narrow mass region \mbox{$3.80\le m_{\D\Dbar} \le 3.88\gevcc$},  with  fits  superimposed.
    Different components employed in the fit are indicated in the~legend.}
  \label{fig:spectra:dd_one}
\end{figure}
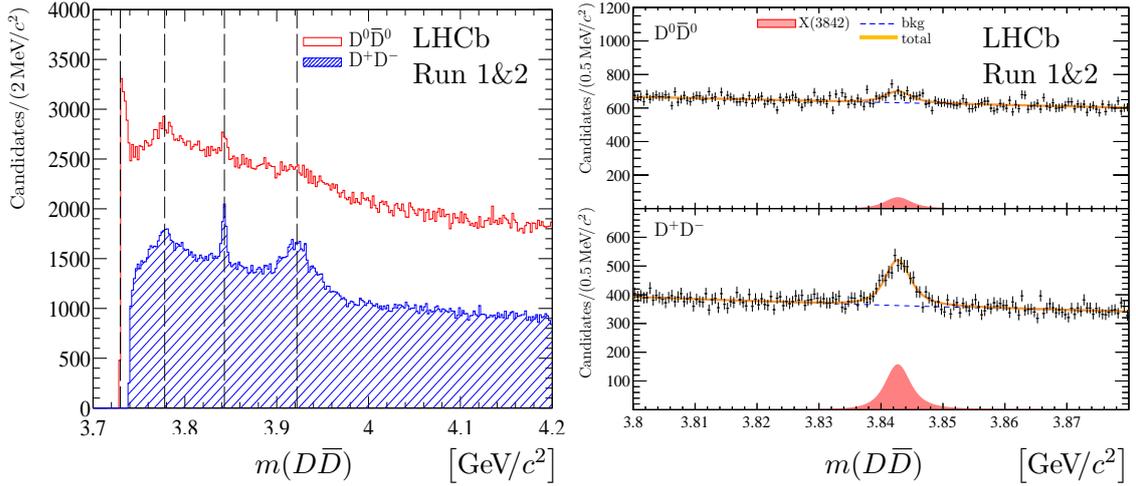

\paragraph{Charmonium states in $\D\bar{\D}$~mass 
spectra near threshold} have been studied
using  the~full \lhcb statistical sample collected 
in Runs~1 and 2~\cite{LHCb-PAPER-2019-005}.
$\Dz$ and $\Dp$~candidates were  reconstructed 
in the \mbox{$\decay{\Dz}{\Km\pip}$} 
and \mbox{$\decay{\Dp}{\Km\pip\pip}$} 
decay modes.
In~total, $3.6\times10^6$
$\Dz\Dz$ and $2.0\times10^6$
$\Dp\Dm$~pairs have been  selected. 
The~mass spectra for $\Dz\Dzb$ and $\Dp\Dm$~combinations are shown in
Fig.~\ref{fig:spectra:dd_one}(left), with the~zoomed-in  region between \mbox{$3.80\le m_{\D\Dbar} \le 3.88\gevcc$} 
presented in Fig.~\ref{fig:spectra:dd_one}(right).

Four peaking structures are observed in the spectra.
Two of the~peaks correspond to the known~$\Ppsi(3770)$ and~$\Pchi_{\cquark2}(3930)$~charmonium states.
A~narrow peak close to  threshold represents partially reconstructed \mbox{$\decay{\Pchi_{\cquark1}(3872)}{\Dstarz\Dzb}$} decays, subsequently 
with   \mbox{$\decay{\Dstarz}{\Dz\Pgamma}$} or \mbox{$\decay{\Dstarz}{\Dz\Ppi^0}$} with the $\Pgamma$ or $\Ppi^0$~meson missing. 
The~narrow peak with mass around $3840\,\mevcc$ is identified as a~new charmonium state.
Its~mass value and small natural width   suggest an interpretation 
as the~$\Ppsi_3(1^3\PD_3)$ charmonium state with quantum numbers 
$\PJ^{\PP\PC} = 3^{--}$~\cite{PhysRevD.72.054026}.
In~addition, prompt hadroproduction of the $\Pchi_{\cquark2}(3930)$ and $\Ppsi(3770)$~charmonium states have been observed for 
the~first time, and  precise measurements of their resonance parameters have been performed.

\paragraph{Observation of the~decays $\decay{\Pchi_{\cquark}}{\jpsi\mup\mun}$}
using a 4.9\invfb  data-set 
 collected at $\sqs=7$, 8 and 13\tev
enabled  the~most precise direct determination of 
the~masses of the~$\Pchi_{\cquark1}$ and 
$\Pchi_{\cquark2}$~states
and the~width of the $\Pchi_{\cquark2}$~to be performed with unprecedented precision~\cite{LHCb-PAPER-2017-036}.
The~observation of these decay modes provides 
opportunity for the~precise measurements of
the~$\Pchi_{\cquark1,\cquark2}$~production and polarization, 
that in turn is vital for tests of QCD models of charmonia production.

\begin{figure}[htb]
	\setlength{\unitlength}{1mm}
	\centering
	\begin{picture}(150,60)
	\definecolor{root8}{rgb}{0.35, 0.83, 0.33}
        \put( 0,0){\includegraphics*[width=75mm,height=60mm]{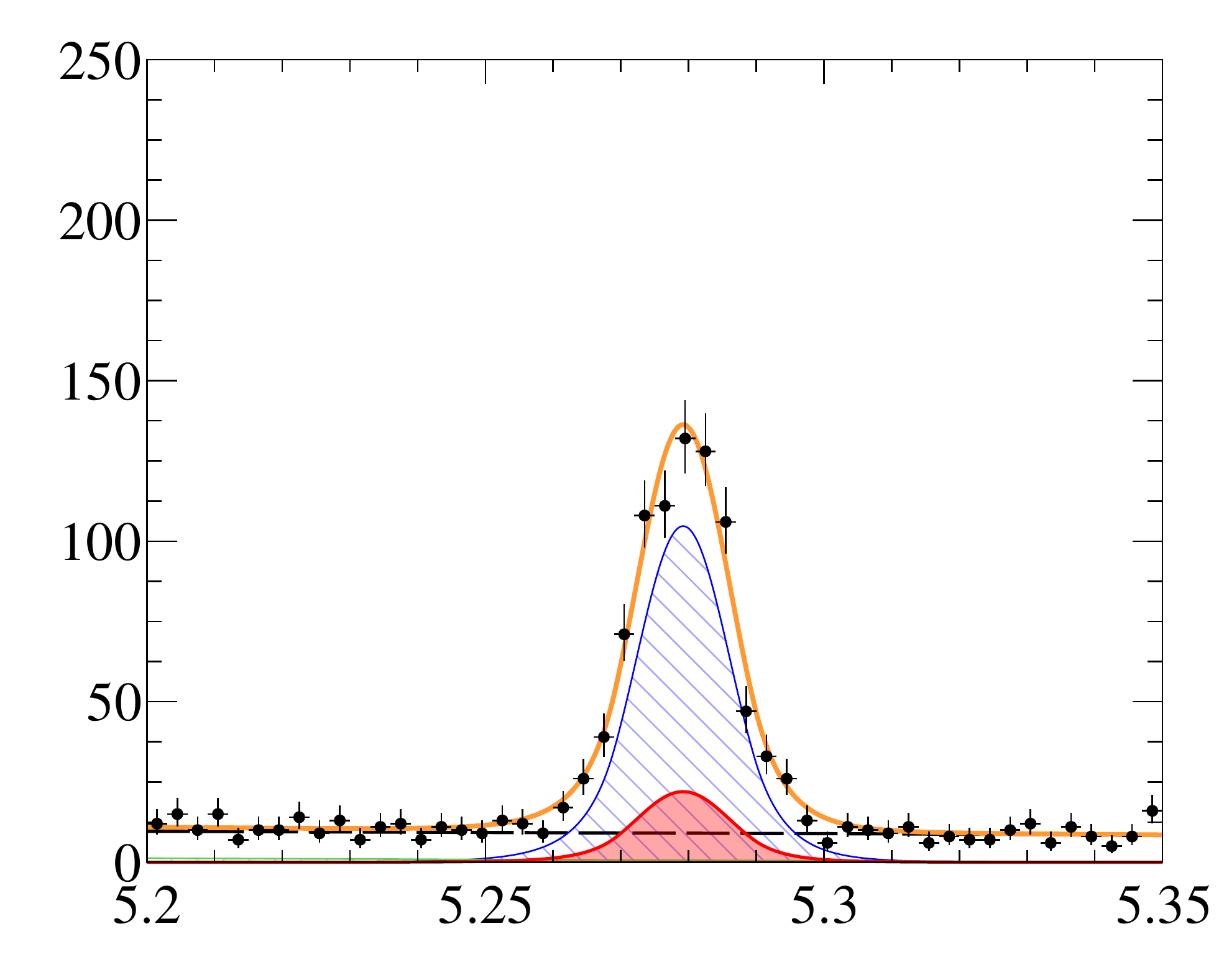}}
	\put(80,0){\includegraphics*[width=75mm,height=60mm]{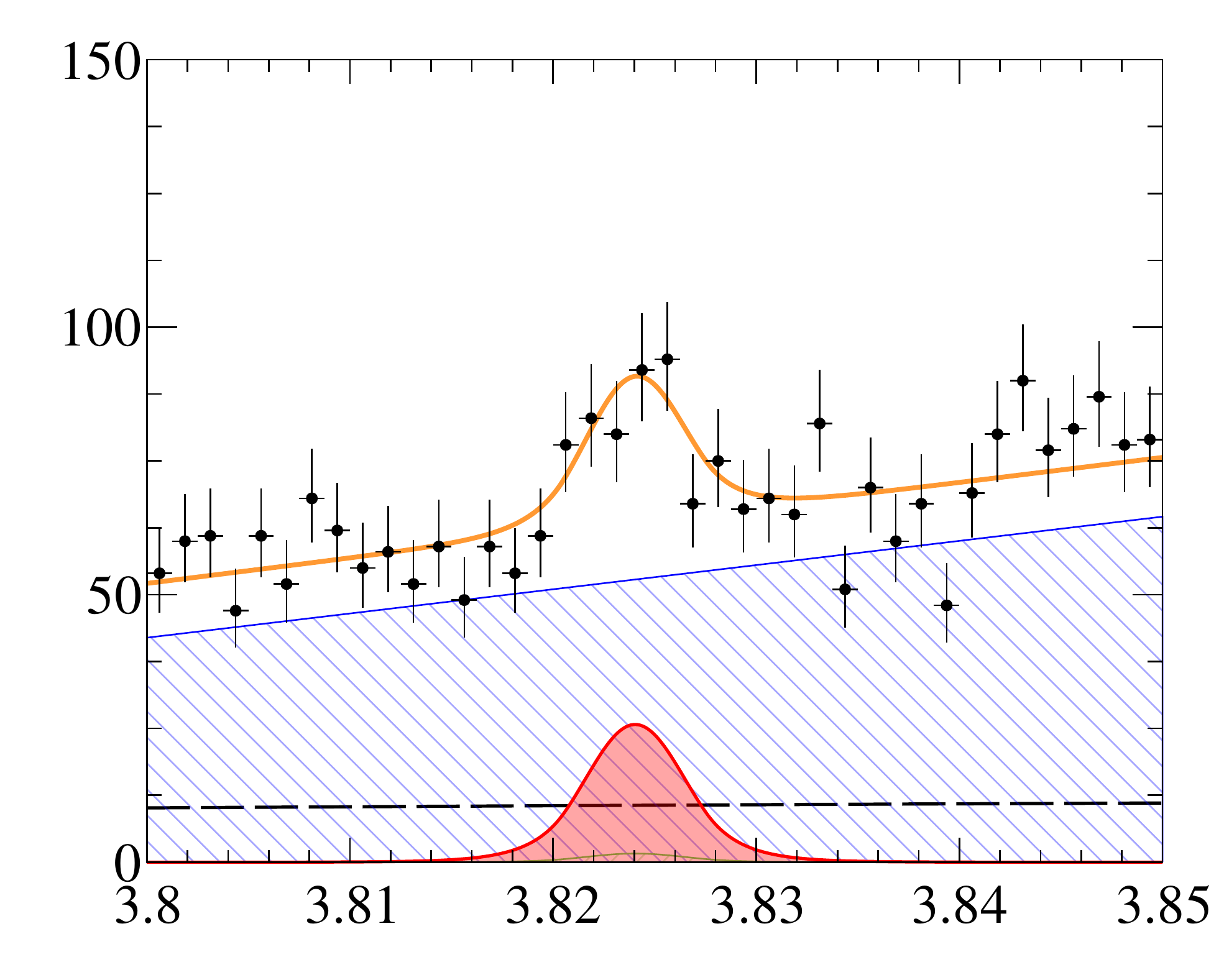}}
	\put(  -3,18){\begin{sideways}\small{Candidates/$(3\mevcc)$}\end{sideways}}	
	\put(  78,13){\begin{sideways}\small{Candidates/$(1.25\mevcc)$}\end{sideways}}
	\put(30 ,0){$m_{\jpsi\pip\pim\Kp}$}	
	\put(112 ,0){$m_{\jpsi\pip\pim}$}
	\put( 57,-1){$\left[\!\gevcc\right]$}
	\put( 137,-1){$\left[\!\gevcc\right]$}
	\put( 12, 51){\scriptsize$3.82<m_{\jpsi\pip\pim}<3.83\gevcc$}
	\put( 91, 51){\scriptsize$5.26<m_{\jpsi\pip\pim\Kp}<5.30\gevcc$}
        \put(  57,51){\tiny$\begin{array}{l}\lhcb\\ \text{Run~1\&2}\end{array}$}
        \put( 138,51){\tiny$\begin{array}{l}\lhcb\\ \text{Run~1\&2}\end{array}$}
        \put(12,45.5) {\begin{tikzpicture}[x=1mm,y=1mm]\filldraw[fill=red!35!white,draw=red,thick]  (0,0) rectangle (8,3);\end{tikzpicture} }
        \put(12,40.5){\begin{tikzpicture}[x=1mm,y=1mm]\draw[thin,blue,pattern=north west lines, pattern color=blue]  (0,0) rectangle (8,3);\end{tikzpicture} }
        \put(12,35.5){\begin{tikzpicture}[x=1mm,y=1mm]\draw[thin,root8,pattern=north east lines, pattern color=root8]  (0,0) rectangle (8,3);\end{tikzpicture} }
	\put(12,32.5){\color[RGB]{0,0,0}     {\hdashrule[0.0ex][x]{8mm}{1.0pt}{2.0mm 0.3mm} } }
	\put(12,28.5){\color[RGB]{255,153,51} {\rule{8mm}{2.0pt}}}
	\put( 22, 46.3){\scriptsize{\decay{\Bp}{\Ppsi_2(3823)\Kp}}}
	\put( 22, 41){\scriptsize{\decay{\Bp}{\left(\jpsi\pip\pim\right)_{\mathrm{NR}}\Kp}}}
	\put( 22,36.5){\scriptsize{comb. $\Ppsi_2(3823)\Kp$}}
	\put( 22,32){\scriptsize{comb. bkg.}}
	\put( 22,28){\scriptsize{total}}
	\end{picture}
	\caption {\small 
	Distributions of 
	the~(left)\,\jpsi\pip\pim\Kp and 
	(right)\,\mbox{$\jpsi\pip\pim$} mass
	for selected
        \mbox{$\decay{\Bu}{\left(\decay{\Ppsi_2(3823)}{\jpsi\pip\pim}\right)\Kp}$}~candidates~\cite{LHCb-PAPER-2020-009}.}
	\label{fig:psitwod}
\end{figure}

\paragraph{Observation of  
  \mbox{$\decay{\Ppsi_2(3823)}{\jpsi\pip\pim}$}}
in 
\mbox{$\decay{\Bu}{\left(\decay{\Ppsi_2(3823)}{\jpsi\pip\pim} \right)\Kp}$} decays
using the~full \lhcb Run~1 and 2 data-sets, has allowed the~most precise
determination of the~mass of the tensor
$\Ppsi_2(3823)$~state and the best constrained
upper limit of its width~\cite{LHCb-PAPER-2020-009}. The  observed mass distribution  is shown in Fig.~\ref{fig:psitwod}.
Within the~factorization approach, the~branching fraction
for the decay \mbox{$\decay{\Bu}{\Ppsi_2(3823)\Kp}$}
vanishes, and a~non\nobreakdash-zero value for 
this~branching fraction allows an evaluation of  the~contribution
of the~$\D_{\squark}^{(*)+}\bar{\D}{}^{(\ast)0}$~rescattering 
amplitudes in the~\decay{\Bu}{\ccbar\Kp}~decays. 

\subsection{Pentaquarks}

\begin{figure}[htb]
  \centering
  \setlength{\unitlength}{1mm}
  \begin{picture}(150,62)
    \put(  0, 0){ 
      \includegraphics*[width=75mm,height=60mm,%
      ]{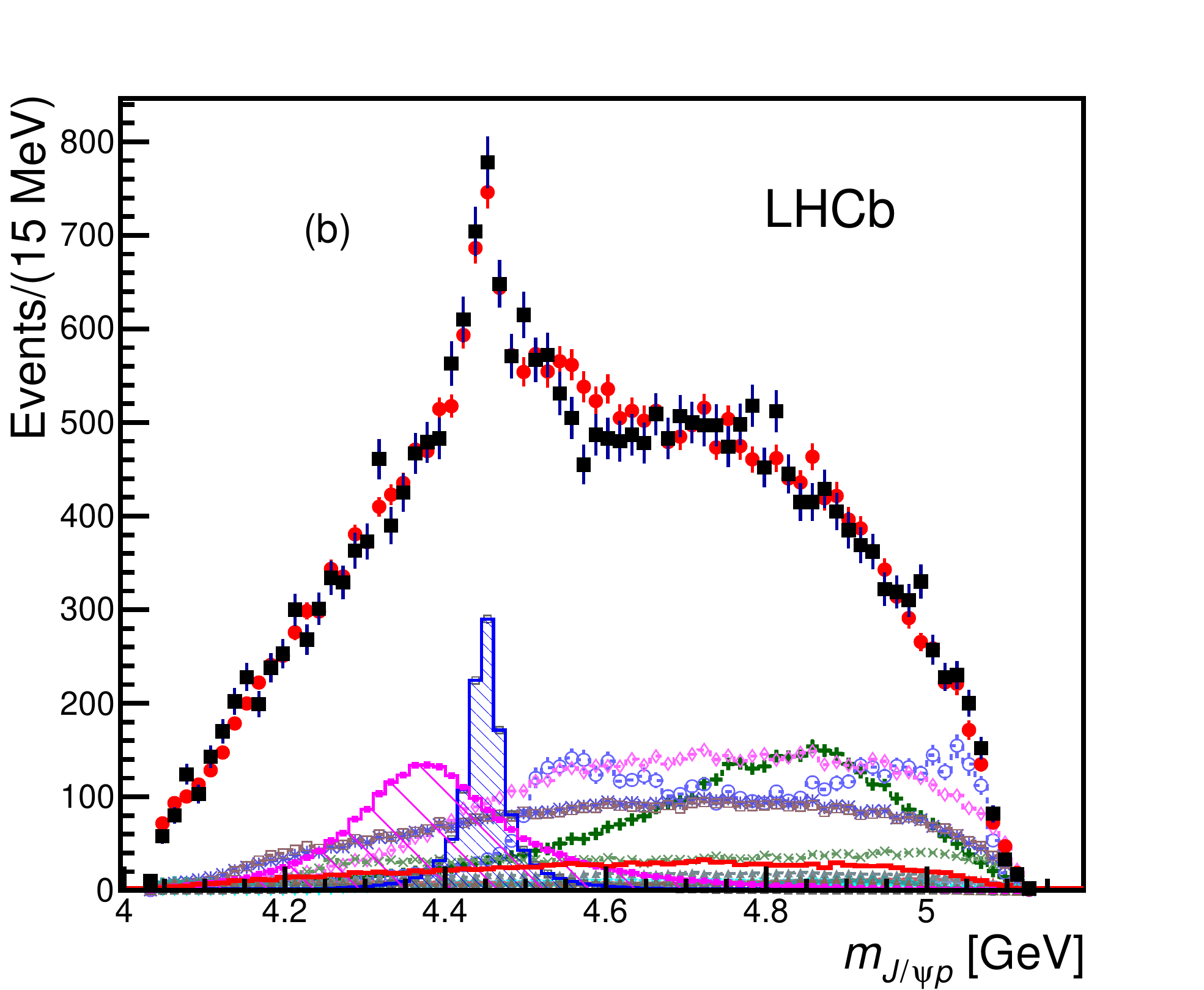}
    }
    \put( 75, 0){ 
      \includegraphics*[width=75mm,height=60mm,%
      ]{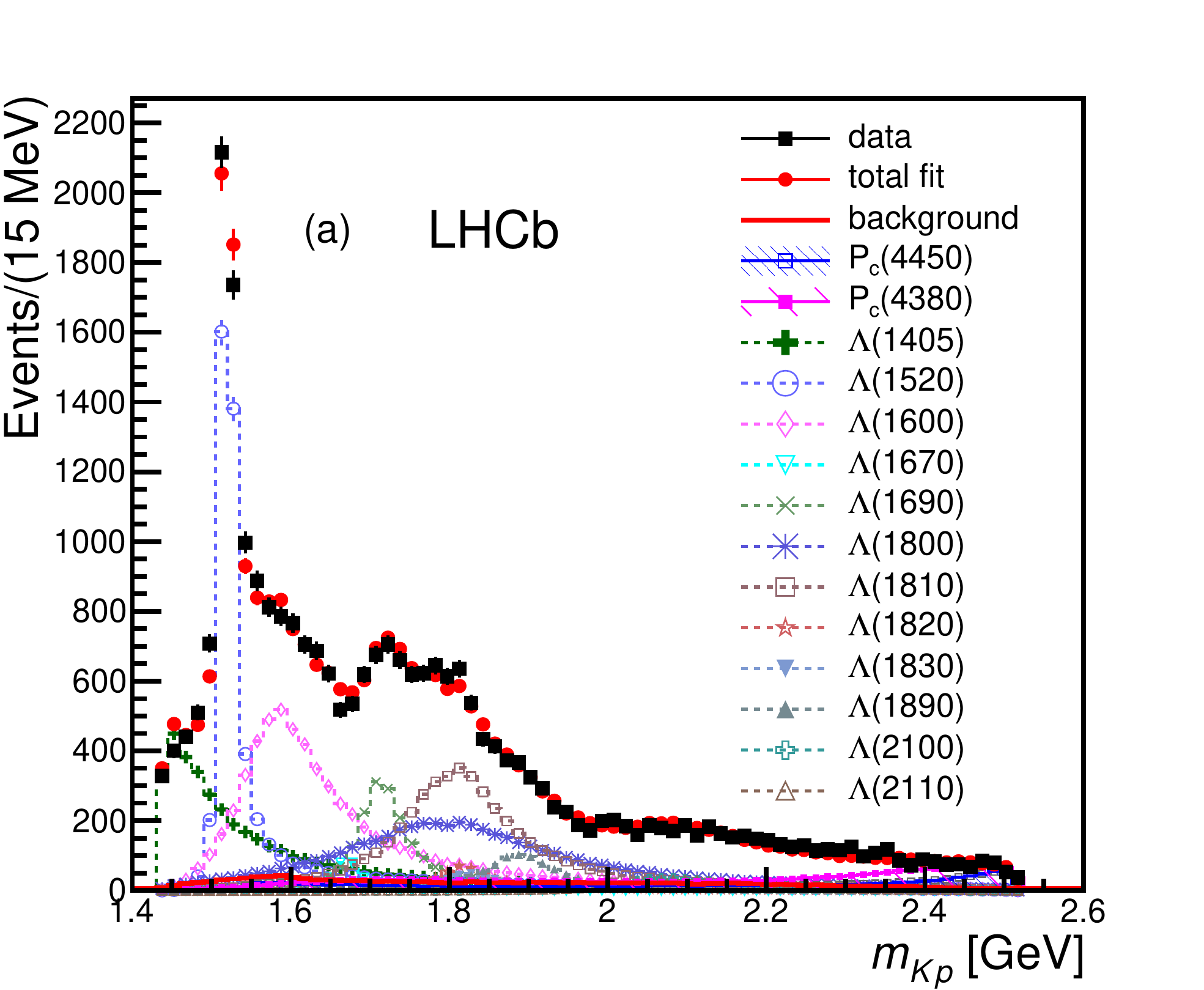}
    }
    \put( 15,40){\textcolor{white}{\rule{10mm}{10mm}}}
    \put( 95,40){\textcolor{white}{\rule{20mm}{10mm}}}
    \put( 48,40){\textcolor{white}{\rule{14mm}{10mm}}}
    \put( 48,45){\small{$\begin{array}{l}\lhcb\\ \text{Run~1}\end{array}$}}
    \put(100,45){\small{$\begin{array}{l}\lhcb\\ \text{Run~1}\end{array}$}}
  \end{picture}
  \caption{\small
    The
    (left)~$m_{\jpsi \proton}$ and 
    (right)~$m_{Kp}$  mass distributions, showing     superimposed   fit projections as solid\,(red) points.
    The~solid\,(red) histogram shows the~background distribution, 
    the blue 
    shaded histogram represents
    the~$P_c(4450)^+$ state, and the~shaded purple histogram 
    represents the~$P_c(4380)^+$. 
    Each~$\Lz^*$ component is also shown. 
  }
  \label{fig:penta_one}
\end{figure}
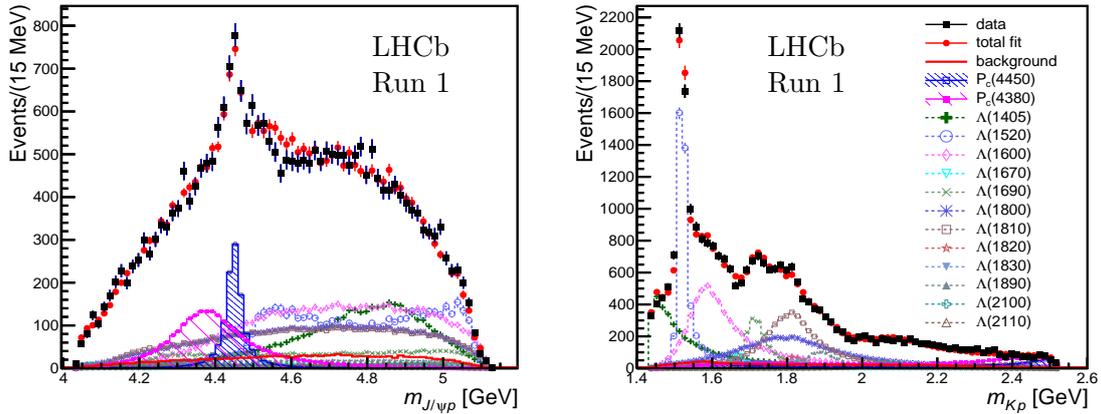

The~LHCb collaboration studied 
 $\decay{\Lb}{\jpsi\proton\Km}$ decays using the~Run~1 data-set~\cite{LHCb-PAPER-2015-029}. 
In~total $(26.0\pm0.2)\times 10^3$ signal \Lb~candidates were selected, and an anomalous
peak in the $\jpsi\proton$~mass spectrum is observed,  shown in Fig.~\ref{fig:penta_one}(left).
If~ the~peak structure represents  a~resonance which strongly decays into $\jpsi\proton$, 
the~minimal valence quarks would be $\cquark\cquarkbar\uquark\uquark\bquark$, 
a~charmonium pentaquark  state. A~full six\nobreakdash-dimensional amplitude fit 
with resonance invariant masses, three helicity angles and two differences
between decay planes has been applied to describe the~data. 
The~amplitude model in the~fit contains 14~well\nobreakdash-defined 
$\PLambda^{\ast}$~states and two pentaquark states, labelled as  
$\PP_{\cquark}(4380)^+$ and 
$\PP_{\cquark}(4450)^+$. 
The~projections of the fit are shown in Fig.~\ref{fig:penta_one}.
The~masses and widths of the~wide 
$\PP_{\cquark}(4380)^+$ and narrow $\PP_{\cquark}(4450)^+$~states have been measured.
The~preferred spin\nobreakdash-parity  assignments are 
$\left(\tfrac{3}{2}^-, \tfrac{5}{2}^+\right)$,
$\left(\tfrac{3}{2}^+, \tfrac{5}{2}^-\right)$ and 
$\left(\tfrac{5}{2}^+, \tfrac{3}{2}^-\right)$, where 
the~first value is the~$J^P$~assignment
given by the~best fit.

%


Following the first observation,  the~exotic hadronic character of the~$\jpsi\Pp$ structure around $4450\mevcc$ was confirmed 
in a~model\nobreakdash-independent way~\cite{LHCb-PAPER-2016-009}.
This analysis gave similar results and   excluded 
that the data could be described by the~$\proton\Km$ constributions alone. 
Further confirmation comes from  the~amplitude analysis of the
Cabibbo\nobreakdash-suppressed decay $\decay{\Lb}{\jpsi\proton\pim}$~\cite{LHCb-PAPER-2016-015},
where $1885\pm50$~signal candidates were investigated. 
There~are different theoretical interpretations suggested, including a tightly 
bound $\dquark\uquark\uquark\ccbar$~state
a~loosely bound molecular baryon\nobreakdash-meson state
or  a triangle\nobreakdash-diagram processes. 

\begin{figure}[htb]
  \centering
  \setlength{\unitlength}{1mm}
  \begin{picture}(150,62)
    \put(  0,-2){ 
      \includegraphics*[width=75mm,height=60mm,%
      ]{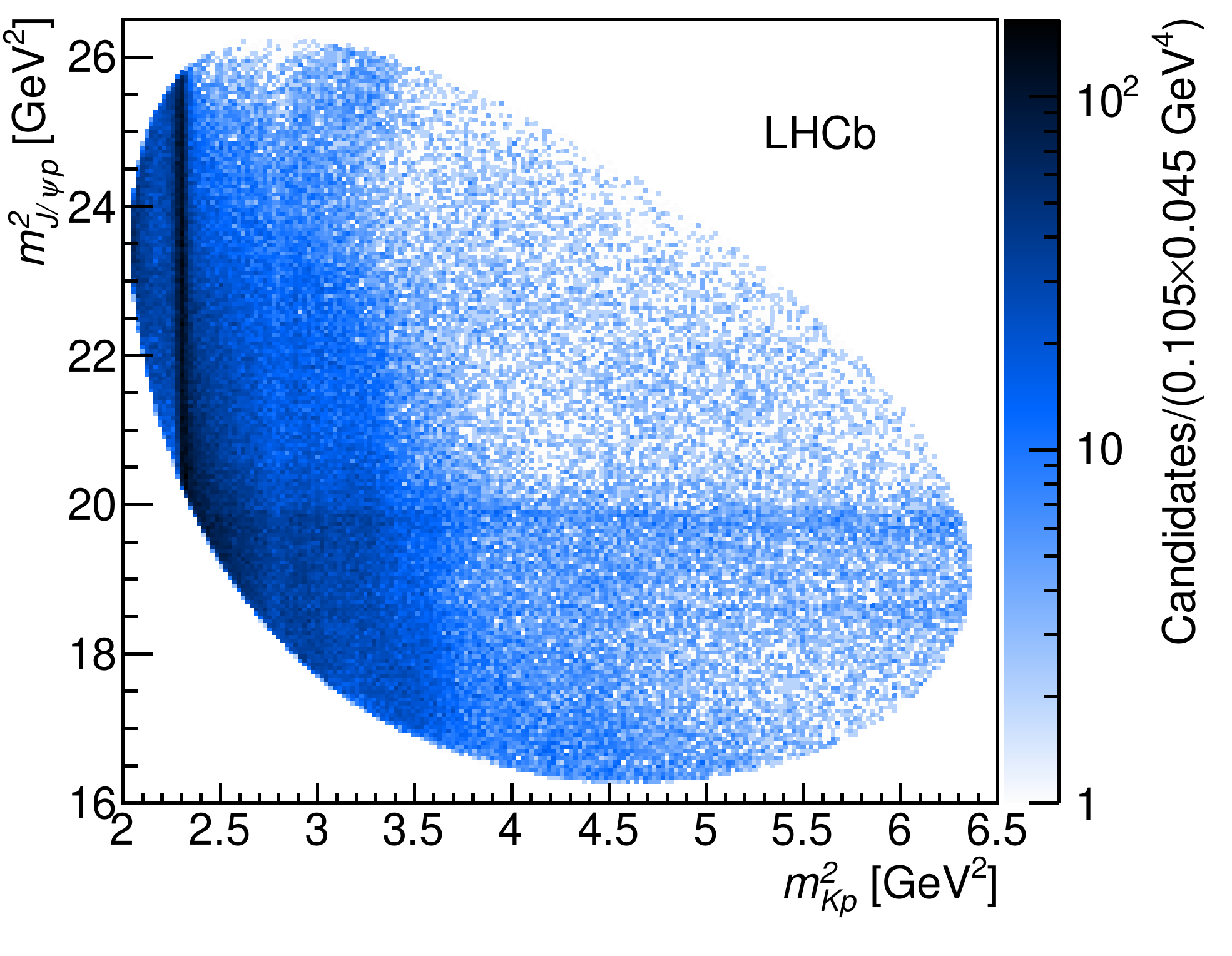}
    }
    \put( 75 ,2){ 
      \includegraphics*[width=75mm,height=60mm,%
      ]{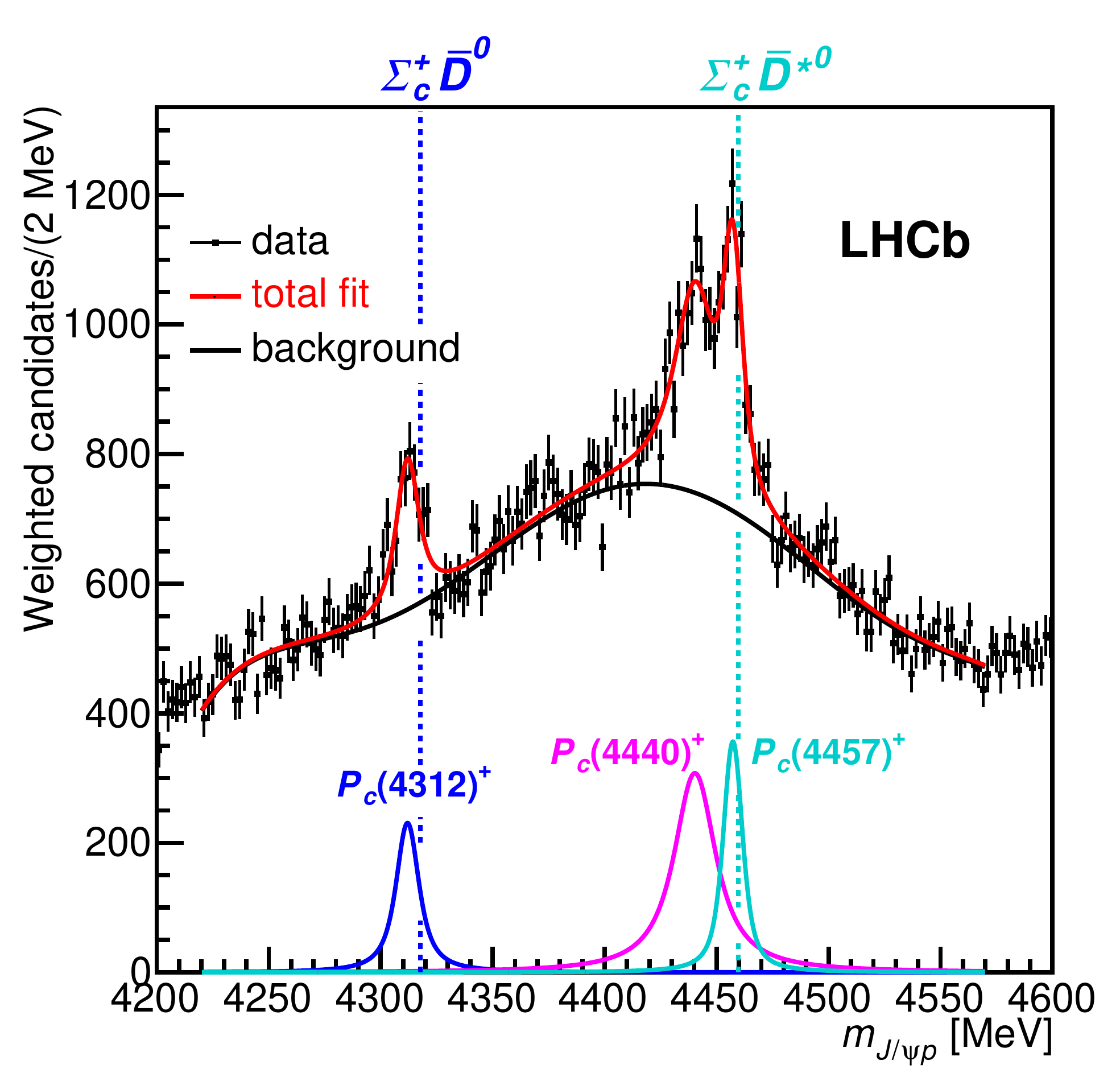}
    }
    \put( 45,44){\textcolor{white}{\rule{14mm}{10mm}}}
    \put(129,44){\textcolor{white}{\rule{14mm}{10mm}}}
    \put( 44,47){\small{$\begin{array}{l}\lhcb\\ \text{Run~1\&2}\end{array}$}}
    \put(129,47){\small{$\begin{array}{l}\lhcb\\ \text{Run~1\&2}\end{array}$}}
    \put( 95,-0.4){\textcolor{white}{\rule{58mm}{6mm}}}
    \put(110,+0.5){\small$m(\jpsi\proton)$}
    \put(132,+0.5){\small$\left[\!\mevcc\right]$}
  \end{picture}
  \caption{\small
    (left)~Dalitz plot of $\Lb\to\jpsi p\Km$ candidates. 
    The~vertical bands correspond to $\Lz^*$ resonances. 
    The~horizontal bands correspond to the~$P_c(4312)^+$, $P_c(4440)^+$, and $P_c(4457)^+$ structures 
    at $m_{\jpsi p}^2 = 18.6$, $19.7$, and $19.9\gev^2$, respectively.
    (right)~Projection of the {\mbox{$\PLambda^{\ast}$-suppressed}}   $m_{\jpsi\proton}$ distribution showing a fit  
     with three Breit$-$Wigner 
    amplitudes and a~sixth-order polynomial background. 
    The~mass thresholds for the~$\PSigma_{\cquark}\Dzb$  and $\PSigma_{\cquark}\Dstarzb$~final states are superimposed.
  }
  \label{fig:penta_two}
\end{figure}

A~partial update of the above analysis was made using the~full Run~1\&2 data  sample~\cite{LHCb-PAPER-2019-014}.
A~nine-fold increase of statistics is achieved due to the larger data sample, an~improved selection criteria, 
and increased \mbox{$\decay{\Pp\Pp}{\bbbar}$}~cross-section at $\sqrt{\rm s}=13\,\tev$ in Run~2.
For candidates with a mass consistent with the nominal $\Lb$~baryon mass, the~$\jpsi\Pp$ and $\Pp\PK^-$~mass spectra were 
investigated. In~the~distribution of $\jpsi\Pp$~mass, the previously reported peaking structure around $4450\mevcc$ 
was confirmed, and a~new narrow peak with mass around $4312\,\mevcc$ was found.
The~\mbox{$\decay{\PLambda^*}{\Pp\PK^-}$} contributions are clearly seen in the~Dalitz plot,   shown 
in Fig.~\ref{fig:penta_two}~(left).

Since the~newly observed peaks are narrow, 
the~full amplitude analysis faces computational challenges. This is 
because resolution effects should be included in 
the~formalism which complicates the fitting procedure.
Conversely, narrow peaks can not be due to reflections from
$\PLambda^*$~states, motivating the~validity of the~one\nobreakdash-dimensional 
fit approach to study the~$\jpsi\Pp$~invariant mass.
The $\jpsi\Pp$~mass in the narrow-resonance region together with the  result of the fit is shown in Fig.~\ref{fig:penta_two}~(right).
The previously reported peak around the $4450\mevcc$ mass is now resolved into a two\nobreakdash-peak 
structure of $\PP_\cquark(4440)^+$ and $\PP_\cquark(4457)^+$~states.
In~total, three narrow pentaquark states are observed.
The statistical significance of the~two\nobreakdash-peak interpretation of the 
previously-reported single $\PP_\cquark(4450)^+$ 
structure is $5.4\sigma$.
The~statistical significance of a~new $\PP_\cquark(4312)^+$~state is $7.3\sigma$.
The masses and widths of the pentaquark candidates are measured.
Taking into account  systematic uncertainties, 
the~widths are consistent with the~mass resolution.
Hence, upper limits on the natural widths at the $95\,\%$ confidence level~(CL) are obtained.

In summary, whilst the~existence  of pentaquark\nobreakdash-like resonances 
is certainly beyond doubt, their exact nature  is still unclear. 
They can be genuine five-quark bound states,
or \eg near\nobreakdash-threshold  meson\nobreakdash-baryon molecules.
More studies are  required to clarify this.

\subsection{Charmonium-like exotic states}

\paragraph{The enigmatic $X(3872)$~particle} was discovered in $B^+$ decays
by the~Belle bollaboration~\cite{Choi:2003ue}.
Subsequently, its existence has been confirmed by several
other experiments \cite{CDFPhysRevLett.93.072001,D0Abazov:2004kp,BaBarPhysRevD.71.071103}.
The~nature of this state is rather unclear. 
Among~the~open possibilities are 
conventional charmonium and exotic states such as 
$D^{*0}\bar{D}^0$ molecules~\cite{Tornqvist:2004qy}, 
tetra-quarks~\cite{Maiani:2004vq}, or 
their mixtures~\cite{Hanhart:2011jz}.
Determination of the $J^{PC}$ quantum numbers  
is important to shed light on this ambiguity.
The~$C$\nobreakdash-parity of the state is positive
since the~\mbox{$X(3872)\to\jpsi\g$} decay has been observed~\cite{Aubert:2006aj,Bhardwaj:2011dj}.
The CDF experiment analysed three\nobreakdash-dimensional angular correlations in 
a~relatively high\nobreakdash-background
sample of $2292\pm113$  inclusively-reconstructed \mbox{$X(3872)\to\jpsi\pip\pim$}, 
\mbox{$\jpsi\to\mumu$}~decays, 
dominated by prompt production in $p\bar p$ collisions.
The~unknown polarisation of the~$X(3872)$ limits 
the~sensitivity of the~measurement of $J^{PC}$~\cite{Abulencia:2006ma}.
A~$\chi^2$ fit of $J^{PC}$ hypotheses to the binned three-dimensional distribution
of the~$\jpsi$ and $\pip\pim$~helicity angles 
and the~angle between their decay planes~\cite{Jacob:1959at,Richman:1984gh,PhysRevD.57.431},
excluded all spin\nobreakdash-parity assignments except 
for~$1^{++}$ and~$2^{-+}$.

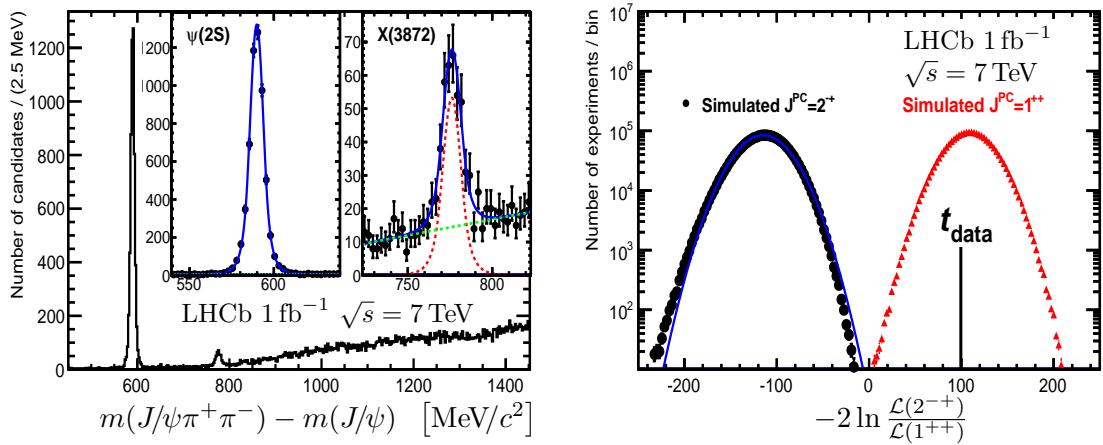
\begin{figure}[htb]
  \centering
  \setlength{\unitlength}{1mm}
  \begin{picture}(150,60)
    \put(0 , 0){ 
      \includegraphics*[width=75mm,height=60mm,%
      ]{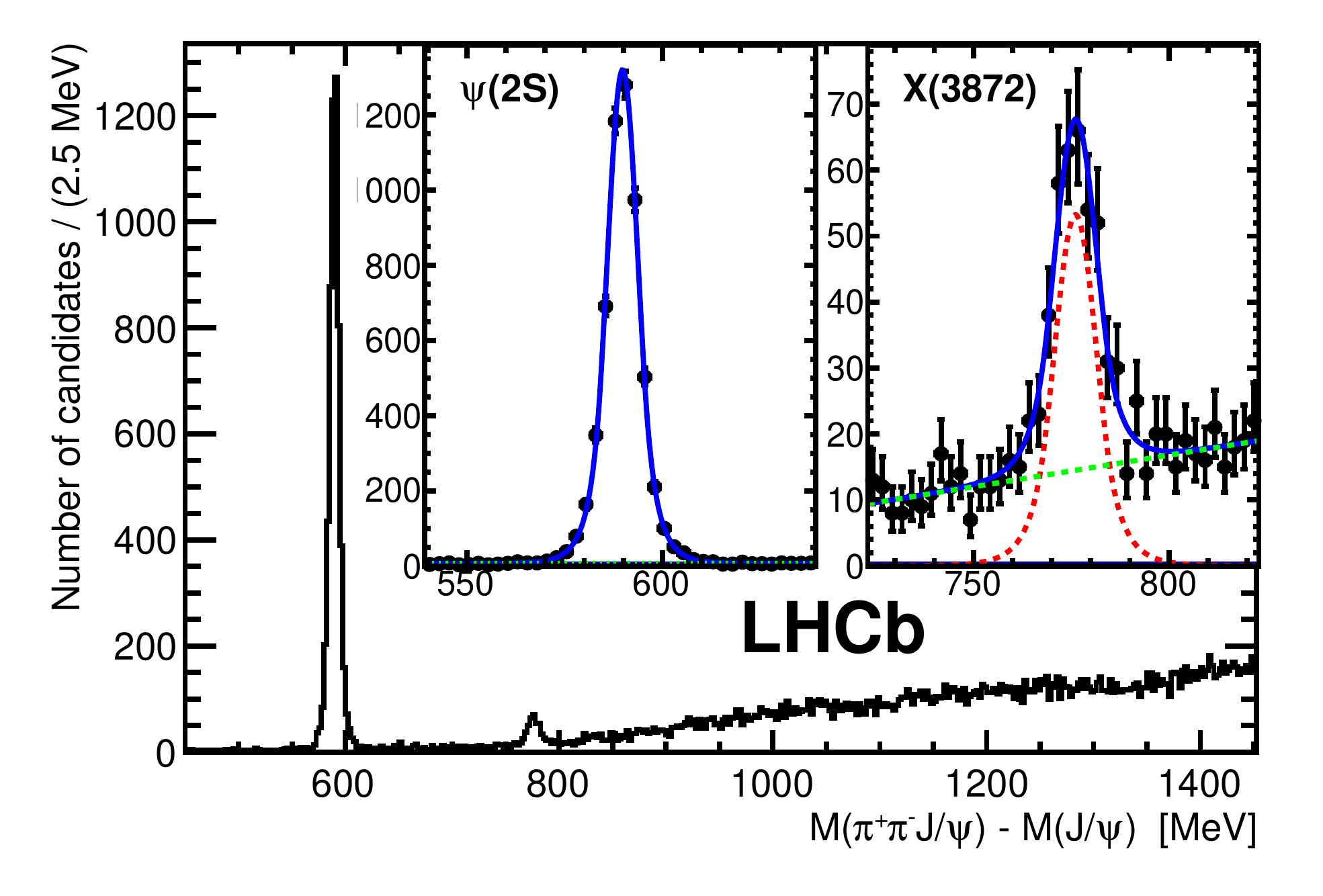}
    }
    \put(75, 0){ 
      \includegraphics*[width=75mm,height=60mm,%
      ]{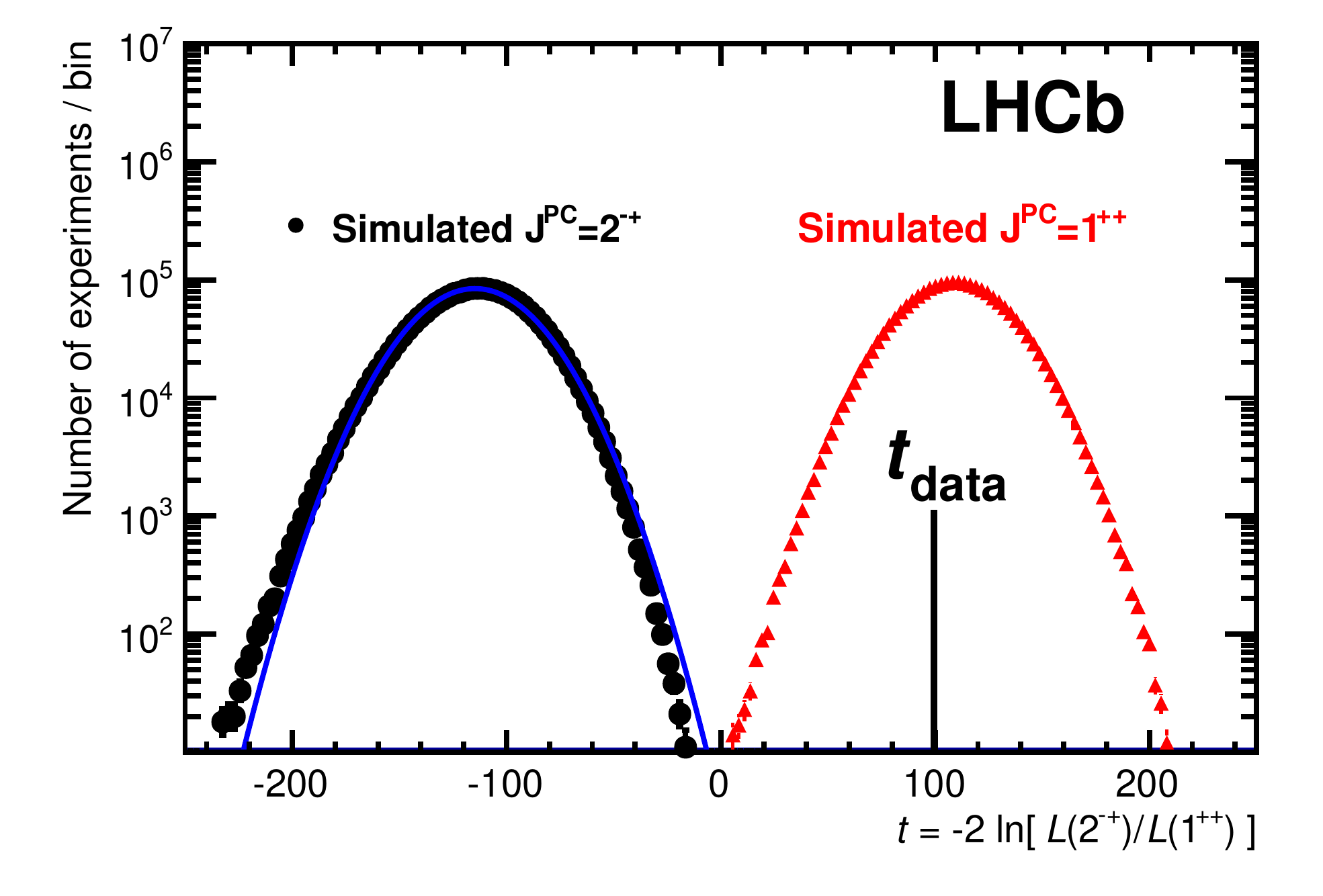}
    }
    \put( 40,15){\textcolor{white}{\rule{15mm}{5mm}}}
    \put( 25,16){\footnotesize{$\begin{array}{l}\lhcb~1\invfb~\sqs=7\tev\end{array}$}}
    \put(120,50){\textcolor{white}{\rule{25mm}{6mm}}}
    \put(120,50){\footnotesize{$\begin{array}{l}\lhcb~1\invfb\\ \sqs=7\tev\end{array}$}}
    \put( 20,0){\textcolor{white}{\rule{58mm}{6mm}}}
    \put( 16,2){\small$m(\jpsi\pip\pim)-m(\jpsi)$}
    \put( 58,2){\small$\left[\!\mevcc\right]$}
    \put( 95,0){\textcolor{white}{\rule{58mm}{6mm}}}
    \put(110,2){\small$-2\ln \tfrac{ \mathcal{L}(2^{-+})} {\mathcal{L}(1^{++})}$}
  \end{picture}
  \caption{\small  
    (left)~Distribution of $\Delta m \equiv m(\jpsi\pip\pim)-m(\jpsi)$ 
    for \mbox{$\Bu\to\jpsi\pip\pim\Km$}~candidates.
    The~fits of the $\psi(2S)$ and $X(3872)$~signals are displayed.
    The~solid blue, dashed red, and dotted green lines represent 
    the~total fit, signal component, and background component, 
    respectively.
    (right)~Distribution of the~test statistic 
    $t\equiv -2\ln\tfrac{\mathcal{L}(2^{-+})} {\mathcal{L}(1^{++})}$
    for the~simulated experiments 
    with $J^{PC}=2^{-+}$~(black circles on the~left)
    and with $J^{PC}=1^{++}$ (red triangles on the~right).
    The~value of the~test statistic for the~data, $t_{\rm data}$,
    is shown by the~solid vertical line.
  }
  \label{fig:spectra:xxxx}
\end{figure}

Using $\sqs=7\tev$  $\proton\proton$~collision data corresponding to 1\invfb
collected in 2011, the~LHCb collaboration performed the~first analysis of the~complete 
five\nobreakdash-dimensional angular correlations 
of the~\mbox{$\B^+\to X(3872) K^+$}, 
\mbox{$X(3872)\to\jpsi\pip\pim$}, 
\mbox{$\jpsi\to\mumu$}~decay chain~\cite{LHCb-PAPER-2013-001}.
About~$38\,000$~candidates  passed the multivariate selection 
in a~$\pm2\sigma$ range around the~$\B^+$ $m_{\jpsi\pip\pim\Km}$~mass distribution, 
with a~signal purity of~89\%.
The~\mbox{$\Delta m \equiv m_{\jpsi\pip\pim}-m_{\jpsi}$} distribution is 
shown in Fig.~\ref{fig:spectra:xxxx}(left).
The~fit yields 
$5642\pm76$ and 
$313\pm26$ candidates for 
\mbox{$\psi(2S)\to\jpsi\pip\pim$} and 
\mbox{$X(3872)\to\jpsi\pip\pim$}~signals, respectively.

The~angular correlations in the~$B^+$~decay  
carry information about the~$X(3872)$~quantum numbers.
To~discriminate between the~$1^{++}$ and $2^{-+}$~assignments,
a~likelihood\nobreakdash-ratio test is used. 
A~test statistic $t$~is  defined as 
$-2\ln[\mathcal{L}(2^{-+})/\mathcal{L}(1^{++})]$.
Positive\,(negative) values of the~test statistic for the~data,
$t_{\rm data}$, favour the~$1^{++}$\,($2^{-+}$) hypothesis.
The~value of the~test statistic observed in the~data is 
$t_{\rm data}=+99$, thus favouring the~$1^{++}$~hypothesis. 
A~rejection of the~$2^{-+}$ hypothesis 
with greater than $5\sigma$ significance is demonstrated using 
a~large  
sample of pseudo-experiments.
As~shown in Fig.~\ref{fig:spectra:xxxx}(right) 
the~distribution of $t$ is reasonably well approximated by a~Gaussian function.
Based on the~mean and r.m.s.\ spread of the~$t$ distribution for 
the~$2^{-+}$ experiments, this hypothesis is rejected with 
a~significance of~$8.4\sigma$. 
Hence the~obtained  results correspond to an
unambiguous assignment 
of the~$X(3872)$ state to be~$1^{++}$.
%

The above result rules out the~explanation of the~$X(3872)$ 
as a~conventional $\eta_{c2}(1^1D_2)$~state. 
Among~the~remaining possibilities are 
the $\chi_{c1}(2^3P_1)$ charmonium state,
and unconventional explanations such 
as a~$D^{*0}\bar{D}^0$ molecule 
tetraquark state 
charmonium\nobreakdash-molecule mixture. 

With a~larger $3\invfb$ data-set 
at $\sqs=7$ and 8\,TeV, the~analysis has been 
repeated in 
the~decay \mbox{$X(3872)\to\jpsi\Prho^0$} without an~assumption on the~orbital angular
momentum ~\cite{LHCb-PAPER-2015-015}.
The~analysis confirmed the~$J^{PC}=1^{++}$ assignment for 
the~$X(3872)$~state and also  set an~upper limit of 4\%
at 90\%\,C.L. on the~D\nobreakdash-wave contribution. 

A precise determination of the~mass and width of the~$X(3872)$~state was performed
using two minimally overlapping data-sets. The first was the  3\invfb ~Run~1 data-set 
in which the $X(3872)$~particles were now
selected from decays of hadrons containing \bquark~quarks~\cite{LHCb-PAPER-2020-008}.
The second was the~full Run~1 and 2 data-sets using a~sample of $(547.8\pm0.8)\times10^3$
$\decay{\Bu}{\jpsi\pip\pim\Kp}$~decays~\cite{LHCb-PAPER-2020-009}.
In~both cases the~$X(3872)$ 
was reconstructed in the~$\decay{X(3872)}{\jpsi\pip\pim}$~final state.
The mass and width were determined from a fit to the $\jpsi\pip\pim$~mass
distribution assuming a Breit--Wigner line shape for the~$X(3872)$~state, measured as follows:
\begin{equation*}
  \begin{array}{rclrcll}
    \m_{X(3872)} & = & 3871.70\pm0.07 \pm 0.07\mevcc\,, &  \Gamma_{\mathrm{BW}} & =  &  1.39\pm0.24\pm0.10\mev        \,,\\
    \m_{X(3872)} & = & 3871.60\pm0.06 \pm 0.03\mevcc\,, &  \Gamma_{\mathrm{BW}} & =  &  0.96^{+0.19}_{-0.18}\pm0.21\mev \,,  
  \end{array} 
\end{equation*}
where the~first and the~second lines correspond to Refs.~\cite{LHCb-PAPER-2020-008} and~\cite{LHCb-PAPER-2020-009},
respectively. The above measurements represent the~most precise determination  of the~mass of the~$X(3872)$~state and
the~first measurements of its width.  
The~measured mass corresponds to 
the~binding
energy $\delta E$, defined as \mbox{$m_{\Dz}c^2+ m_{\Dstarz}c^2 - m_{X(3872)}c^2$}, which is~\mbox{$70\pm120\kev$}.

Whilst the~proximity of the~measured mass of the
$X(3872)$  to 
the $\Dz\Dstarzb$~threshold~\mbox{\cite{
  Aubert:2008gu,
  Aaltonen:2009vj,
  LHCb-PAPER-2011-034,
  Choi:2011fc,
  Ablikim:2013dyn}} favours the~interpretation of 
this state as a $\Dz\Dstarzb$~molecule, 
the~large production   cross-section of the $X(3872)$ 
\mbox{\cite{
  CDFPhysRevLett.93.072001,
  D0Abazov:2004kp,
  LHCb-PAPER-2011-034,
  Aaboud:2016vzw,
  Chatrchyan:2013cld}} disfavours this.
The~pure molecular interpretation is further disfavoured 
by the observation of  the decay \mbox{$X(3872)\to\psi(2S)\g$}.
The~ratio
of the decay rates to
$\psi(2S)\g$ and $\jpsi\g$~final states is very sensitive to the nature of
the~$X(3872)$~state.
This  is predicted to be in the range $( 3 - 4 )\times10^{-3}$ 
for 
a~$\D\Dstarb$~molecule 
$1.2 - 15$ for a pure charmonium  state
and $0.5 - 5$ for a~molecule-charmonium mixture. 

\lhcb performed the search for \mbox{$X(3872)\to \psi(2S)\g$}~decays using 
the decay chain \mbox{$\Bu\to X(3872)\Kp$},\mbox{$X(3872)\to \psi(2S)\g$}.
The analysis was based on a 1\invfb ~data sample 
at $7\tev$ and 2\invfb at $8\tev$.
The~significance of the
\mbox{$\decay{\Bu}{\left(\decay{X(3872)}{\psi(2S)\g}\right)\Kp}$} signal  is
4.4~standard deviations.
The~branching fraction, normalised to that of the $X(3842)\to\jpsi\g$ 
decay mode, is measured to be 
\begin{equation*}
\dfrac{ \BR(X(3842)\to\psi(2S)\g)}{ \BR(X(3842)\to\jpsi\g )} = 2.46\pm0.64\pm0.29.
\end{equation*}
This result is compatible with, but more precise than, 
previous measurements~\cite{Aubert:2008ae,Bhardwaj:2011dj}, 
and strongly disfavours a pure molecular 
interpretation of the~$X(3872)$~state.

\begin{figure}[htb]
  \centering
  \setlength{\unitlength}{1mm}
  \begin{picture}(150,60)
   \put(  0, 5){ 
      \includegraphics*[width=75mm,height=55mm,%
      ]{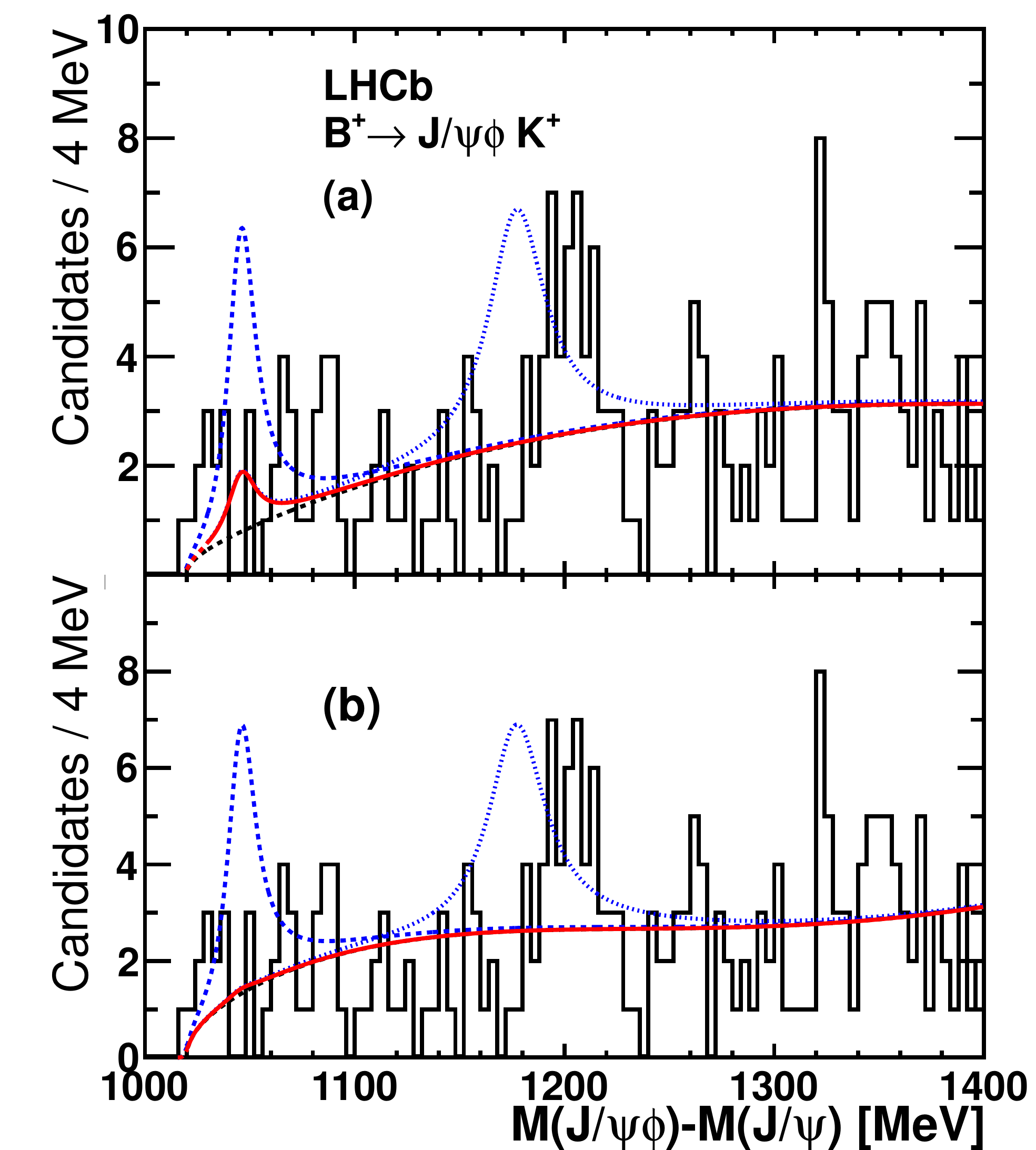}
    }
    \put( 75, 0){ 
      \includegraphics*[width=75mm,height=60mm,%
      ]{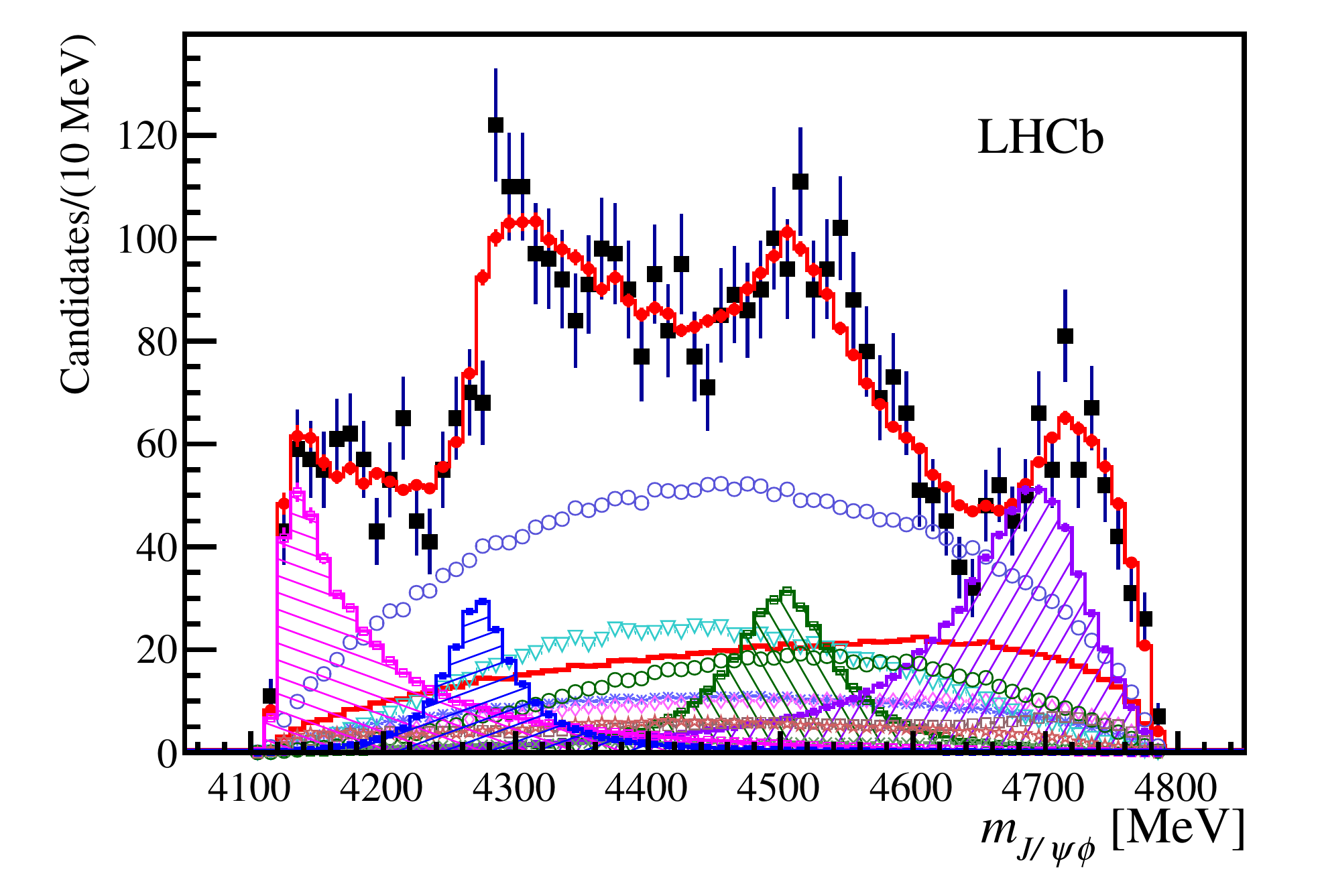}
    }
    \put( 23,52){\textcolor{white}{\rule{20mm}{6mm}}}
    \put( 23,47){\textcolor{white}{\rule{10mm}{6mm}}}
    \put( 19,53){\tiny{$\begin{array}{l}\lhcb~0.37\invfb\\ \sqs=7\tev\end{array}$}}
    \put( 23,25){\textcolor{white}{\rule{10mm}{6mm}}}
    \put(125,47){\textcolor{white}{\rule{20mm}{10mm}}}
    \put(130,50){\footnotesize{$\begin{array}{l}\lhcb\\ \text{Run~1}\end{array}$}}
    \put( 20,1){\textcolor{white}{\rule{58mm}{6mm}}}
    \put( 20,2){\small$m(\jpsi\Pphi)-m(\jpsi)$}
    \put( 58,2){\small$\left[\!\mevcc\right]$}
    \put( 95,0){\textcolor{white}{\rule{58mm}{6mm}}}
    \put(110,2){\small$m(\jpsi\Pphi)$}
    \put(133,2){\small$\left[\!\mevcc\right]$}
  \end{picture}
  \caption{\small  
    \small 
    (left)~The distribution of mass difference $m(\jpsi\phi)-m(\jpsi)$ in a 0.37\invfb data-set for 
    selected \mbox{$\decay{\Bu}{\jpsi\Pphi\Kp}$}~candidates~\cite{LHCb-PAPER-2011-033}.  
    A fit~of the $X(4140)$ signal on top of a~smooth background is superimposed\,(solid red line). 
    The~dashed blue\,(dotted blue) lines   illustrates 
    the~expected $X(4140)$\,($X(4274)$) signal yield from 
    the~CDF measurement~\cite{Aaltonen:2009tz}.
    The~top and bottom plots differ by the~background function
    used in the fit: (top)~an efficiency-corrected three-body phase-space;                      
    (bottom)~a~quadratic function multiplied by the~efficiency-corrected 
    three-body phase-space factor.
    (right) The distribution of $\jpsi\Pphi$  mass in the full Run~1 data-set for 
    $\decay{\Bu}{\jpsi\Pphi\Kp}$~candidates\,(black data points) 
    compared with the~results of the~amplitude fit 
    containing eight $K^{*+}\to\phi\Kp$\,contributions (shown  with open symbols) and 
    five $X\to\jpsi\Pphi$ resonance contributions  (shown as hatched histograms) 
    ~\cite{LHCb-PAPER-2016-018,LHCb-PAPER-2016-019}.
    The~total fit is given by the~red 
    histogram.
  }
  \label{fig:psiphi}
\end{figure}

\paragraph{Structures  in the $\jpsi\phi$~system} 
have  acquired great experimental and 
theoretical interest since the~CDF~collaboration reported  
$3.8\sigma$~evidence\,($14\pm5$ events) 
for a~narrow\,(\mbox{$\Gamma=11.7^{+8.3}_{-5.0}\pm3.7\mev$}) 
near\nobreakdash-threshold $X(4140)$~mass peak
in a  sample of $75\pm11$~reconstructed 
$\decay{\Bu}{\jpsi\Pphi\Kp}$~decays~\cite{Aaltonen:2009tz}.
Much larger widths are expected for charmonia states at this mass, 
therefore its~possible interpretations as a~molecular state, 
a~tetraquark state, a~hybrid state or a~rescattering effect have been discussed. 
The~$X(4140)$~structure was confirmed by  CMS~\cite{Chatrchyan:2013dma}
and  D0~\cite{Abazov:2013xda,Abazov:2015sxa}, 
however  searches 
in 
\mbox{$\decay{\Bu}{\jpsi\Pphi\Kp}$}~decays
were negative in
the~Belle~\cite{Brodzicka:2010zz,ChengPing:2009vu} 
and BaBar~\cite{Lees:2014lra}
experiments. 
Using a~0.37\invfb data-set 
at $\sqs=7\tev$\,($346\pm20$ signal 
\mbox{$\decay{\Bu}{\jpsi\Pphi\Kp}$}~decays)
\lhcb initially found no evidence for 
the~narrow~$X(4140)$~structure~\cite{LHCb-PAPER-2011-033}, in $2.4\sigma$~disagreement with the~measurement by~CDF,
seen in~Fig~\ref{fig:psiphi}(left).
However, using a significantly larger sample of
$4286\pm151$~\mbox{$\decay{\Bu}{\jpsi\Pphi\Kp}$}~decays\,(the~Run~1 data-set),
with roughly uniform efficiency across 
the~entire $\jpsi\Pphi$~mass region,
\lhcb performed a full amplitude analysis, including 
resonant contributions from $\kaon^{\ast}$~resonances decaying 
into~$\Pphi\Kp$ and possible resonances in the  $\jpsi\Pphi$~system~\cite{LHCb-PAPER-2016-018,LHCb-PAPER-2016-019}.
Four~resonance contributions 
labeled as 
$X(4140)$,
$X(4274)$,
$X(4500)$ and 
$X(4700)$ 
with quantum numbers 
$1^{++}$,
$1^{++}$,
$0^{++}$ and 
$0^{++}$, respectively, are observed, 
shown in Fig.~\ref{fig:psiphi}(right).
The~statistical significance varies from $5.6$ to~$8.4\sigma$.
The~widths of the~states are found to be  between 56 and 120\mev, 
significantly  exceeding the~narrow-width of the $X(4140)$ 
reported by CDF.

\paragraph{The charged charmonium-like~state $Z_{\cquark}(4430)^-$} was first observed by the~Belle collaboration
in the~$\Ppsi(2S)\pim$~mass spectrum of \mbox{$\decay{\Bd}{\Ppsi(2S)\Km\pim}$}~decays  
~\cite{Choi:2007wga}.
The state appeared as
as a~narrow\,($\Gamma=44^{+17+30}_{-13-11}\mev$) 
structure with a significance of $6.5\sigma$.
Later, the  collaboration performed a full amplitude analysis  
of \mbox{$2010\pm50\pm40$} 
 \mbox{$\decay{\Bd}{\Ppsi(2S)\Km\pim}$}~signal decays, 
determining the~quantum numbers 
as $J^P=1^+$, and finding a much 
broader width of $\Gamma=200^{+41+26}_{-46-35}\mev$~\cite{Chilikin:2013tch}.

The~\lhcb experiment has collected about $25,000$~signal \Bd decays\,(the~Run~1 data set)
and observed the~$Z_{\cquark}(4430)^-$ with a~significance
exceeding~$13.9\sigma$~\cite{LHCb-PAPER-2014-014}.
Model\nobreakdash-independent as well as full amplitude analyses were 
performed. 
The~spin\nobreakdash-parity is confirmed as~$1^+$, 
other hypotheses are excluded by at least $9.7\sigma$.

Exotic particles with quantum numbers
which can decay into 
$\Peta_{\cquark}\Ppi^-$
are predicted in several models~\cite{Voloshin:2013dpa}.
Using a $4.7\invfb$ data sample at 7,~8 and $13\,\tev$, 
the~\lhcb collaboration has performed a~Dalitz-plot analysis of 
\mbox{$\decay{\Bz}{\Peta_{\cquark}\PK^+\Ppi^-}$} decays, where the~$\Peta_{\cquark}$~meson is reconstructed in 
the~\mbox{$\decay{\Peta_{\cquark}}{\proton\antiproton}$}   final state~\cite{LHCb-PAPER-2018-034}.
Evidence was found for a~new exotic resonance 
in the~$\Peta_{\cquark}\pim$~system, 
later dubbed the~$\PX(4100)^{-}$ by the PDG.
The~significance of this new resonance 
exceeds three standard deviations.

\subsection{Structures in the $\jpsi\jpsi$~mass spectrum}

\begin{figure}[h]
  \centering
  \setlength{\unitlength}{1mm}
  \begin{picture}(150,90)
  \put(15.0,5.0) {\includegraphics[width=130mm,height=85mm]{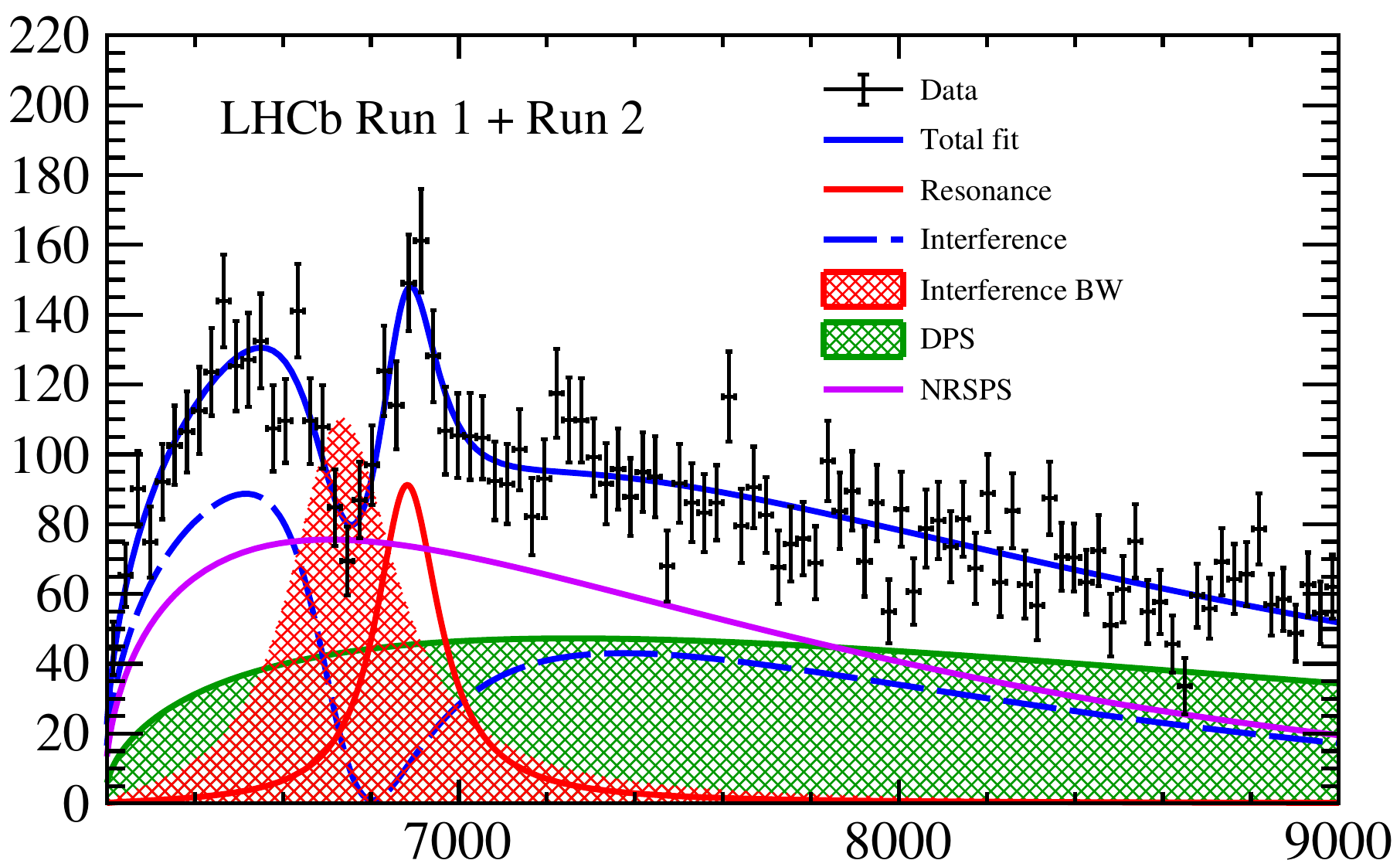}} 
  \put(8,35) {\large
      \begin{sideways}%
        Candidates/(28\mevcc) 
      \end{sideways}%
    } 
  \put(75,0) {\Large$m_{\jpsi\jpsi}$}
  \put(118,0){\Large$\left[\!\mevcc\right]$}
  \put(117,77){\footnotesize{$\begin{array}{l}\lhcb\\ \text{Run~1\&2}\end{array}$}}
  \put(30,73){\textcolor{white}{\rule{50mm}{10mm}}}
 \end{picture}
   \caption{\small The $\jpsi$-pair mass spectrum
   with 
   the fit result superimposed. The fit accounts  for 
   the~interference between 
   a~resonance and non-resonant SPS contribution 
  ~\cite{LHCb-PAPER-2020-011}.
   }\label{fig:fitXcut}
\end{figure}
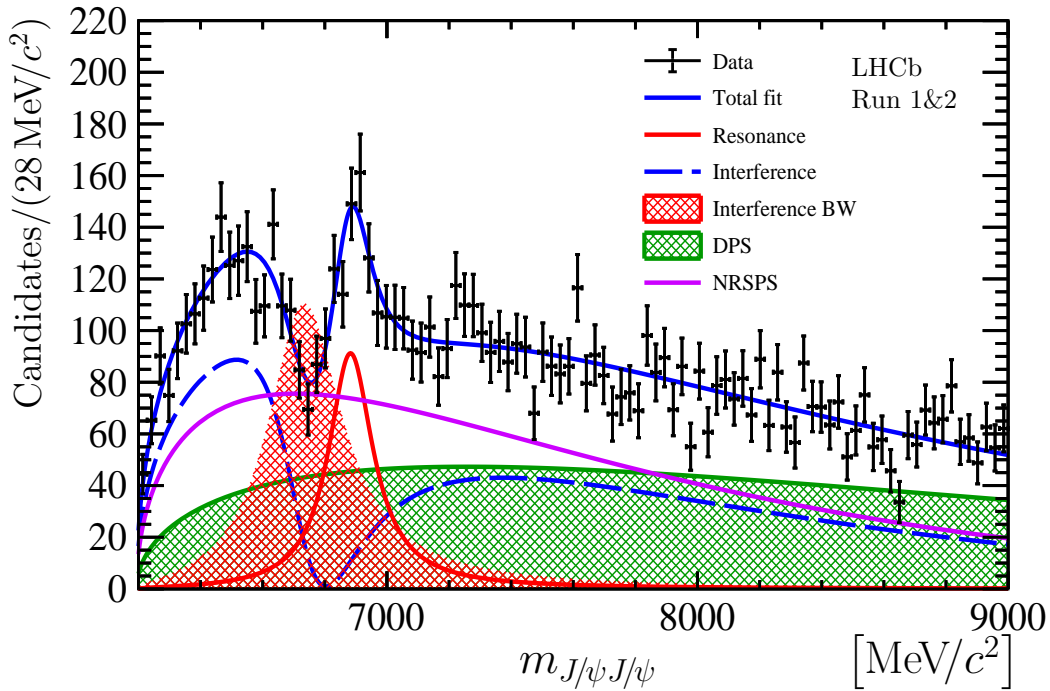

The~production of $\jpsi\jpsi$~pairs in 
high\nobreakdash-energy $\proton\proton$~collisions 
was observed for the first time by the~\lhcb experiment using
a~37.5\invpb data sample collected in 2010 at $\sqs=7\tev$
~\cite{LHCb-PAPER-2011-013}. 
The~$\jpsi\jpsi$ mass spectrum 
was studied in a ~sample of  \mbox{$116\pm16$}~signal 
 pairs,  and no structures was found.
The~subsequent analysis of  279\invpb of data  
collected in 2015 at $\sqs=13\tev$  ~\cite{LHCb-PAPER-2016-057} 
showed the~dominant role of 
the~double\nobreakdash-parton scattering\,(DPS) 
mechanism
for $\jpsi\jpsi$~production
over the~single-parton scattering mechanism\,(SPS).
This in turn includes both a non\nobreakdash-resonant 
SPS contribution
and $\cquark\cquark\cquarkbar\cquarkbar$~tetraquark
production.
Using the~full Run~1 and 2 data-set,
the~$\jpsi\jpsi$ mass spectrum was studied 
in more detail~\cite{LHCb-PAPER-2020-011}.
The data, shown in Fig.~\ref{fig:fitXcut},  were found to be inconsistent with the~hypothesis of 
non\nobreakdash-resonant SPS plus DPS 
in the~$6.2<m_{\jpsi\jpsi} <7.4\gevcc$ range,
where $\cquark\cquark\cquarkbar\cquarkbar$~tetraquarks
decaying into $\jpsi\jpsi$~pairs are expected.
A~narrow peaking structure 
at $m_{\jpsi\jpsi}\approx6.9\gevcc$
matching the~lineshape 
of a~resonance,
and a~broader structure near to threshold,
were found.

The~global significances of the~broader structure close to 
 threshold
or the~narrow peak around $6.9\gevcc$
(provided that the~other structure exists),
are determined to be larger than 5~standard deviations. 
The~structures are consistent with hadron states made up 
of four charm quarks, 
alternatively they may also result from near\nobreakdash-threshold rescattering effects,
as the~$\chiczero\chiczero$ and $\chicone\chiczero$ thresholds
sit at $6829.4\mevcc$ and $6925.4\mevcc$, respectively.

\subsection{Light hadron spectroscopy}

\paragraph{$\Peta-\Peta^{\prime}$~mixing} has been studied by \lhcb 
in $\decay{\BdorBs}{\jpsi\Peta^{(\prime)}}$~decays, resulting in
four observed decays modes, using 
the~Run~1 data-set~\cite{LHCb-PAPER-2014-056}. The $\Peta$ and $\Peta^{\prime}$ were identified in the decay modes
\mbox{$\decay{\Peta^{\prime}}{\Peta\pip\pim}$},
\mbox{$\decay{\Peta}{\pip\pim\piz}$} 
and \mbox{$\decay{\Peta^{\prime}}{\Prho^0\g}$}. 
For decays of  \Bs\,(\Bd)~mesons, the $\Peta^{(\prime)}$~mesons
are formed from initial $\squark\squarkbar$\,($\dquark\dquarkbar$) quark pairs, 
hence the~measurement of the~ratios of branching fractions 
of these decays allows  a precise measurement
of the~ $\Peta-\Peta^{\prime}$~mixing angle. It also
probes the~gluonium component in the~$\Peta^{\prime}$~meson. 


\paragraph{Excited strange mesons}~have been studied in the 
$\Pphi\Kp$~system from~a full amplitude fit of 
$\decay{\Bu}{\jpsi\Pphi\Kp}$ decays
using the \lhcb~Run~1 data-set~\cite{LHCb-PAPER-2016-018,LHCb-PAPER-2016-019}.
Even though no peaking structures are observed in the $\phi K^+$ mass distributions.
correlations in the~decay angles reveal a rich spectrum of $K^{*+}$ resonances.
In~addition to the~angular information contained in the~$K^{*+}$ and 
$\phi$ decays, the $\jpsi$ decay also helps to probe these resonances, 
as the~helicity states of the~$K^{*+}$ and $\jpsi$ mesons originating from 
the~$B^+$ decay  must be equal. 
Unlike the earlier scattering experiments
investigating $K^{*}\to\phi K$~decays, a~good sensitivity to states with 
both natural and unnatural spin\nobreakdash-parity combinations is achieved.

The~dominant $1^{+}$ partial wave 
has a~substantial non-resonant component, 
and at least one resonance that has a significance of  $7.6\sigma$. 
There is also $2\sigma$~evidence that this structure can be better 
described with two resonances 
matching  expectations for  two 
$2P_1$ excitations of the~kaon. 
Also prominent is the~$2^{-}$~partial wave
which contains at least one resonance at~$5.0\sigma$~significance. 
This structure is also better described with two
resonances at $3.0\sigma$~significance. 
Their masses and widths
are in good agreement with the well-established $K_2(1770)$ and $K_2(1820)$
states, matching the predictions for 
the~two~$1D_2$~kaon excitations.
The~$1^-$ partial wave
exhibits $8.5\sigma$ evidence for a~resonance
which matches the~$K^*(1680)$~state, which was well established 
in other decay modes, and matches  expectations for 
the~$1^3D_1$~kaon excitation.  
This~is the first observation of its decay to 
the $\phi K$~final state. 
The $2^+$ partial wave has a smaller intensity
but provides $5.4\sigma$~evidence for a~broad structure that 
is consistent with the~$K_2^*(1980)$~state, previously observed 
in other decay modes, and matches expectations for 
the~$2^3P_2$ state.
The~$K(1830)$ state\,($3^1S_0$ candidate),  
earlier observed in the $\phi K$ decay 
mode in~$K^-p$ scattering,
is also confirmed  at $3.5\sigma$~significance.
Its~mass and width  is now properly evaluated with uncertainties for the first
time. 


\clearpage

\section{Measurements not originally planned in LHCb }

While originally designed to study the production and decay of $\bquark$ and $\cquark$ hadrons, \lhcb has extended its physics programme to also include other areas, such as physics with jets, the production of \W and \Z bosons, searches for new particles in open mode, and nuclear collisions. Selected highlights are summarised below.


\subsection{Production of EW bosons \W and \Z}
\begin{figure}[ht]
\begin{center}
\includegraphics[width=0.48\linewidth]{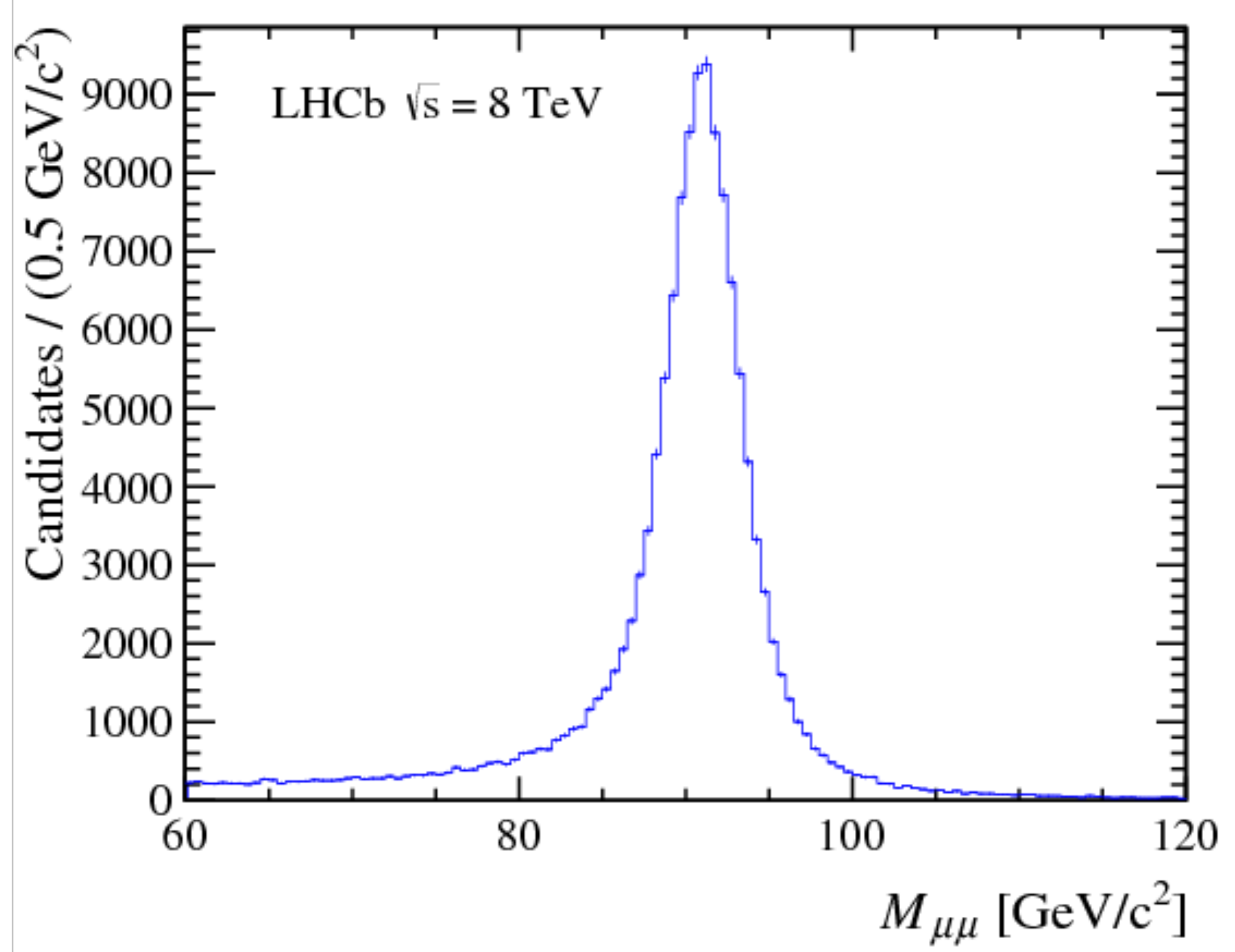}
\includegraphics[width=0.48\linewidth]{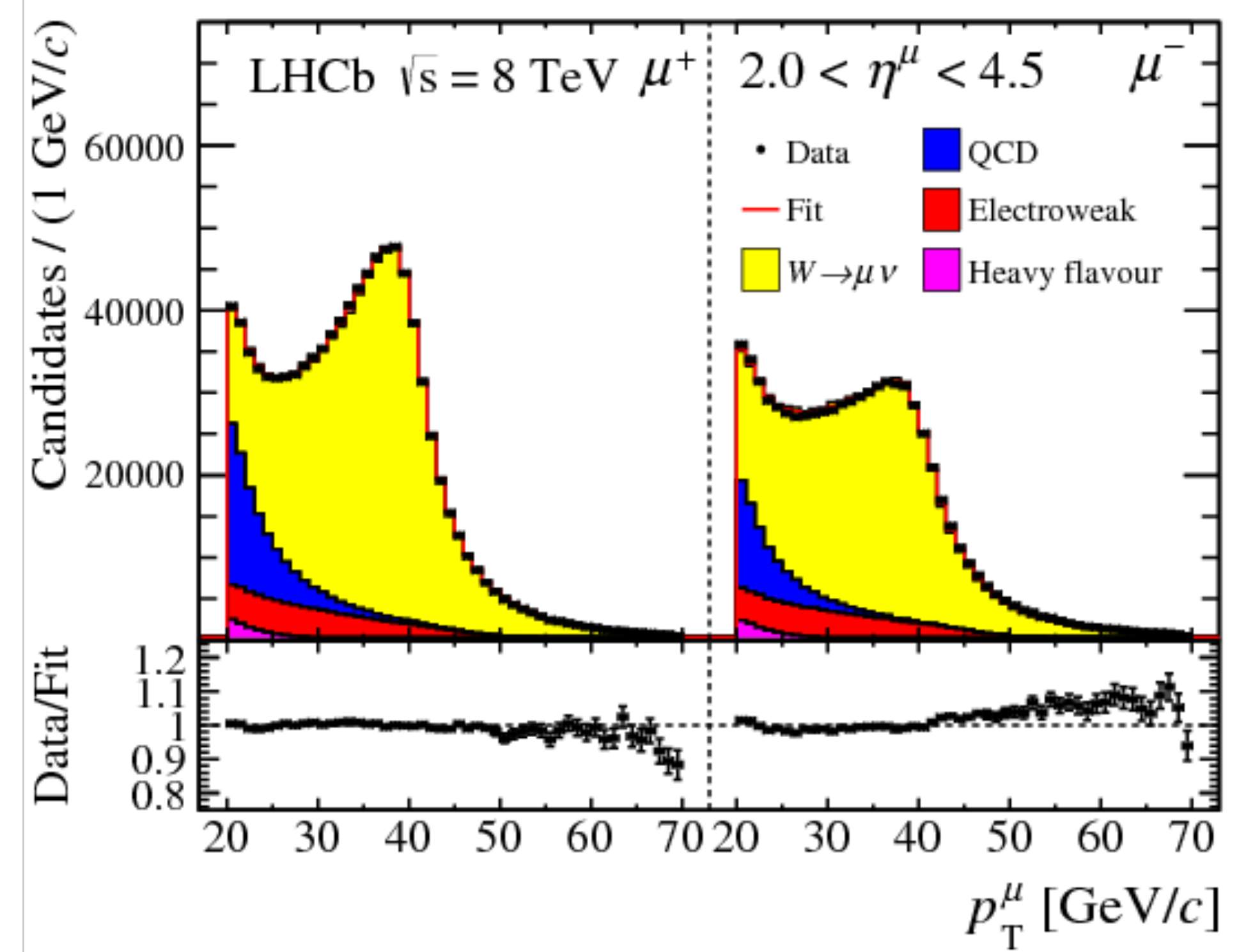}
\caption{(a)~Invariant mass distribution of dimuon pairs in the \Z-candidate sample.
(b) \lhcb data compared to QCD, electroweak and heavy flavour background, 
for positive (left) and negative (right) muon $\pt$ spectra of $W$ candidates~\cite{W-Z_prod_8tev}.
}
\label{Z-prod}
\end{center}
\end{figure}

\begin{figure}[ht]
 \begin{center}
\includegraphics[width=0.49\linewidth,height=65mm]{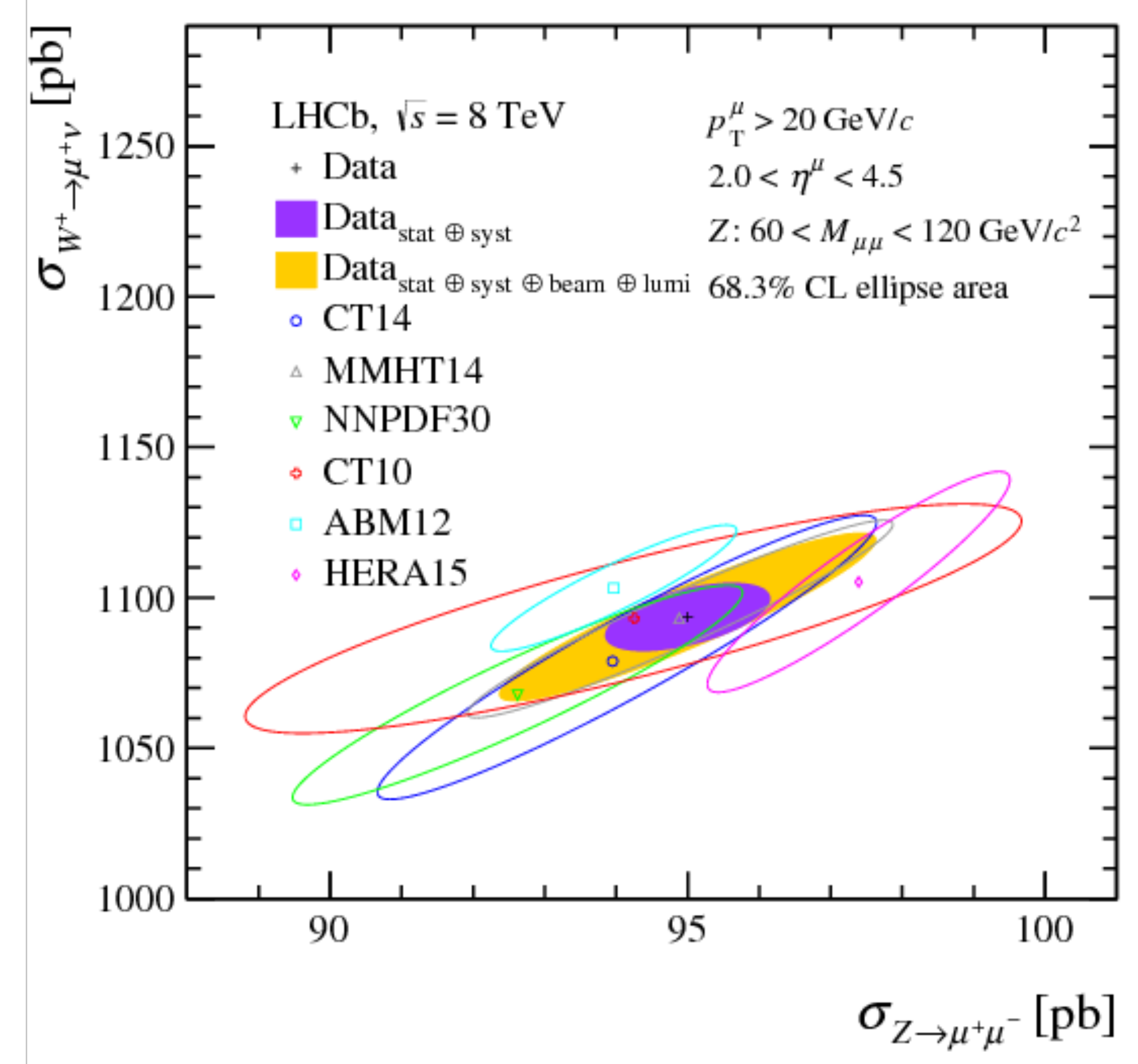}
\includegraphics[width=0.49\linewidth,height=65mm]{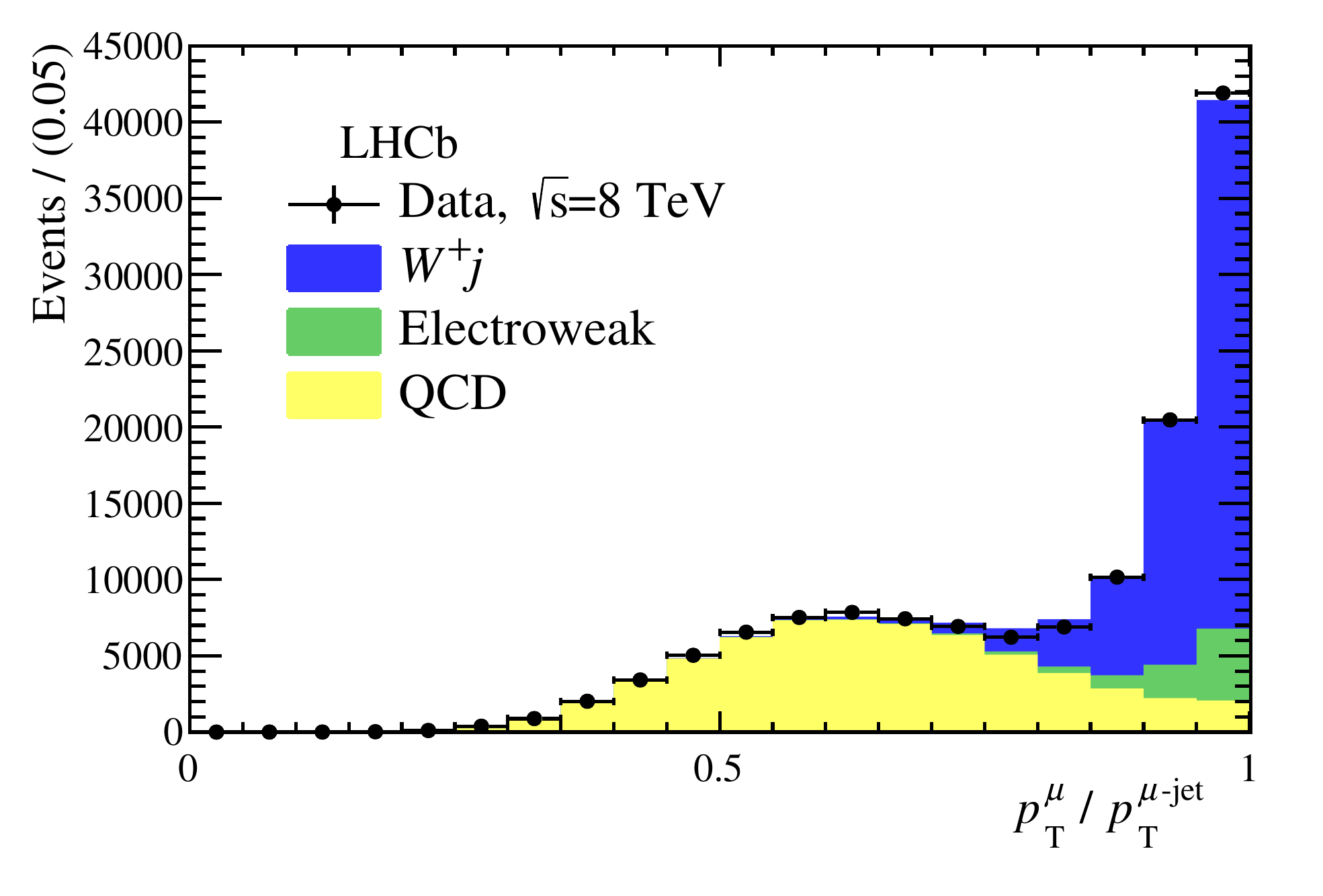}
\caption{(Left)~The  \W versus \Z measured production cross sections, showing comparison with theoretical predictions.
(Right) Contributions to the selected \W plus jet sample in the discriminating variable $p_T^\mu/p_T^{\mu-jet}$ (described in the text).
}
\label{W-vs-Z}
\end{center}
\end{figure}

\lhcb has measured the production of \Z and \W bosons 
inclusively~\cite{W-Z_prod_8tev}
and in association with jets,
 reconstructed in mainly muonic final states,
 using 
the data collected at $\sqrt{s}$ = 8 TeV~\cite{W-Z-jet}. 
Also decays to $e^+e^-$ \cite{Z-to-ee}, $\tau^+ \tau^-$ \cite{Z-to-tau} and $e\nu$ \cite{W-to-enu}, have been measured, 
however the muon channel is the most efficient 
due to the excellent performance of the muon system (see Sect.~\ref{sez:rivelatore_mu}). 
The~$Z\rightarrow \mu^+\mu^-$ decay shows  a spectacularly  clean signal, as shown in Fig.~\ref{Z-prod}(a)
 ~\cite{Z-prod_13tev}.
The $W \rightarrow \mu \nu$ channel also manifests in a clear signal, 
shown in Fig.~\ref{Z-prod}(b) 
~\cite{W-Z_prod_8tev}.
The absolute and differential cross sections, their ratios, 
and charge asymmetries have been measured and compared to 
theoretical predictions. 
Figure~\ref{W-vs-Z} (Left)  shows the comparison of
\W and \Z cross section measurements to   SM  predictions, showing good agreement.

\subsection{Jets in LHCb}

 Measurements of jets at \lhcb address several interesting areas of study: 
\begin{itemize}
\item Jet properties and heavy-quark jet tagging;
\item The constraining of proton 
parton density functions (PDFs) and to probe hard QCD in a 
unique kinematic range. 
Figure~\ref{kin-region} shows the domain 
in the (x,Q$^2$) plane  covered by the \lhcb detector, 
complementing the kinematic ranges of ATLAS and CMS;

\begin{figure}[ht]
\begin{center}
\includegraphics[width=0.70\linewidth]{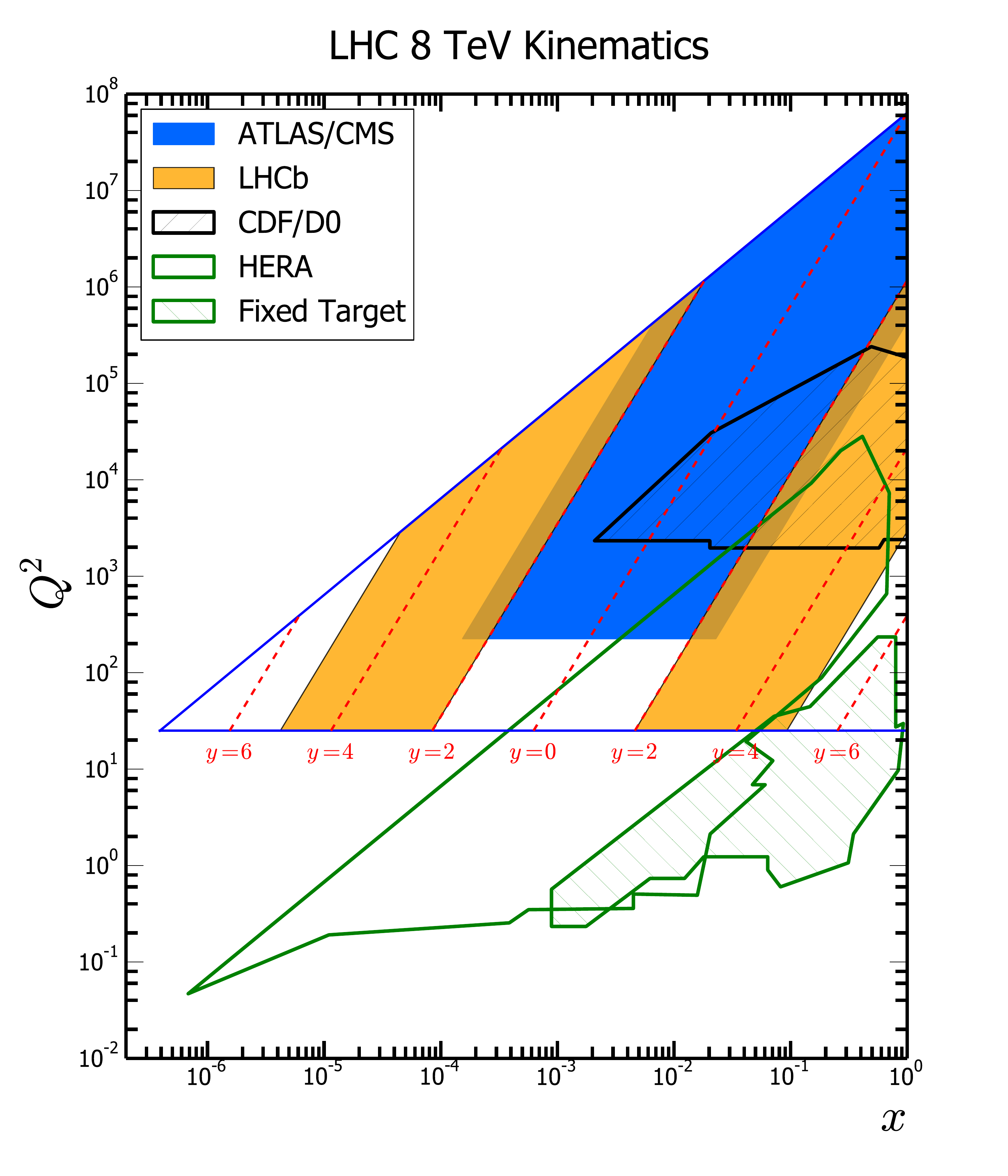}
\caption{The region in the $x-Q^2$ plane probed by \lhcb, compared to ATLAS, CMS and previous experiments. }
\label{kin-region}
\end{center}
\end{figure} 
\item Direct searches for the Higgs boson decaying to $b\bar{b}$ and $c\bar{c}$ final states;  
\item Direct searches for long-lived beyond-the-SM  particles decaying into jets. 

\end{itemize} 

Jets are reconstructed in \lhcb using a particle flow algorithm~\cite{jet-tag} 
 clustered using the anti-$kT$ algorithm with R = 0.5~\cite{kt-algo}.
The calibration of  
 jet reconstruction is performed in data using $\Z \to \mu^+\mu^-$ 
decays which also contain a jet, where the jet is
 reconstructed back-to-back with respect to the \Z. 
The efficiency for reconstructing and identifying jets is around 90$\%$ for jets
 with transverse momentum $\pt > 20$ \gevc. 
Furthermore, LHCb has developed
 a method to tag jets~\cite{jet-tag} and to determine whether 
 they correspond to a $\bquark$ or $\cquark$ quark or to a lighter quark. 
 Jets are tagged whenever a secondary 
vertex (SV) is reconstructed close enough to the jet in terms of 
 R =$\sqrt{(\Delta\phi^2 + \Delta\eta^2)}$. 
This provides a light-jet mistag rate below 1$\%$, with an 
efficiency for $\bquark$ ($\cquark$) jets of $\sim 65\%$ ($\sim 25 \%$).

\begin{figure}[ht]
\begin{center}
\includegraphics[width=0.65\linewidth]{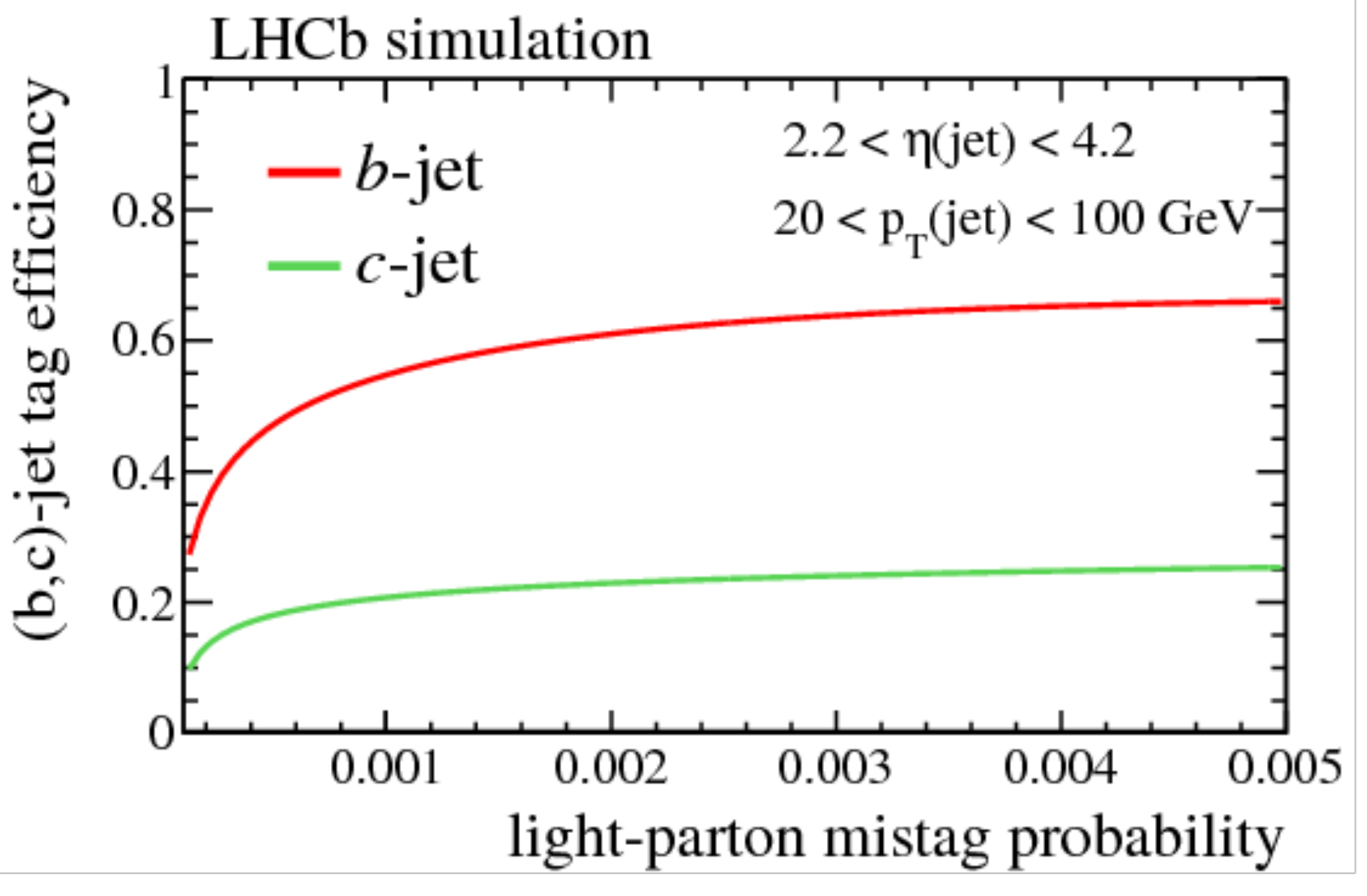}
\caption{Simulated efficiencies for SV-tagging a $\bquark, \cquark$-jet as a function of mistag probability for a light-quark jet.}
\label{eff-jetsID}
\end{center}
\end{figure}

Moreover, using the SV and jet 
properties, two boosted decision trees (BDTs)  have been developed, one to separate heavy from light 
jets 
and one to separate $\bquark$ from $\cquark$  jets.
 A summary of the obtained performance is shown in Fig.~\ref{eff-jetsID},
 where the efficiency of flavour identification is plotted as a function of  the misidentification of light jets. 


\begin{figure}[htb]
\begin{center}
\includegraphics[width=0.70\linewidth]{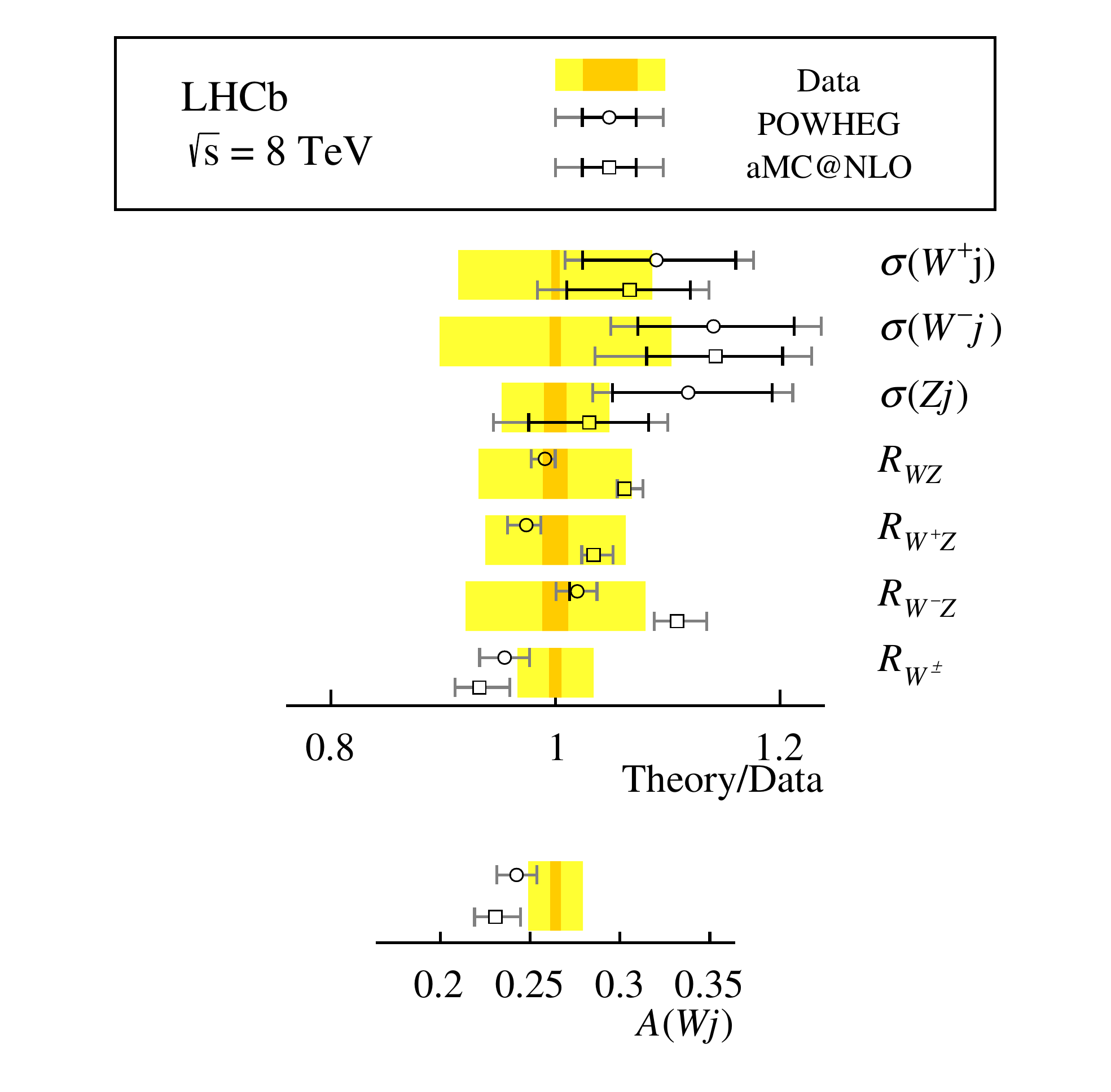}
\caption{Comparison with theoretical calculations of measured cross sections for \W and \Z production accompanied by jets.
The orange bands represent the statistical uncertainty only, the yellow bands are the quadratic sum of statistical and systematic uncertainties.
}
\label{WZ-with-jets}
\end{center}
\end{figure}

\paragraph{Production of \W and \Z with jets.}
\W and \Z production have also been studied in association with 
jets~\cite{W-Z-jet}, 
in  $W+j$, $Z+j$, $W+b\bar b$ and $W+c\bar c$.
Jets are reconstructed as described above, 
whilst \Z and \W bosons are reconstructed mainly in
muonic final states. 
The production of \W boson plus jets is discriminated from misidentified QCD
background processes  using a muon isolation variable, 
which is built as the ratio between the p$_T$ of the jet containing 
the muon and the $p_T$ of the muon alone. 
Figure~\ref{W-vs-Z}(b) 
shows the distribution of this variable,
with genuine muons from the \W boson peaking at 1. 
Figure~\ref{WZ-with-jets} shows the comparison of the measured cross-sections in \lhcb with theoretical expectations, showing very good agreement.

\lhcb has measured the $\W^{\pm}+b\bar b$, $\W^{\pm}+c\bar c$, 
production cross sections using a sample of $pp$ collisions taken at 
$\sqrt s$ = 8 TeV with a high-\pt 
isolated lepton from the \W  decay (electron or muon) and two heavy flavour 
($\bquark$ or $\cquark$) tagged jets
 in the final state. 
The heavy-quark tagging uses the method described above.
In this analysis, the  $W+c\bar{c}$ channel is studied for the first time. 

\begin{figure}[htb]
\begin{center}
\includegraphics[width=0.70\linewidth]{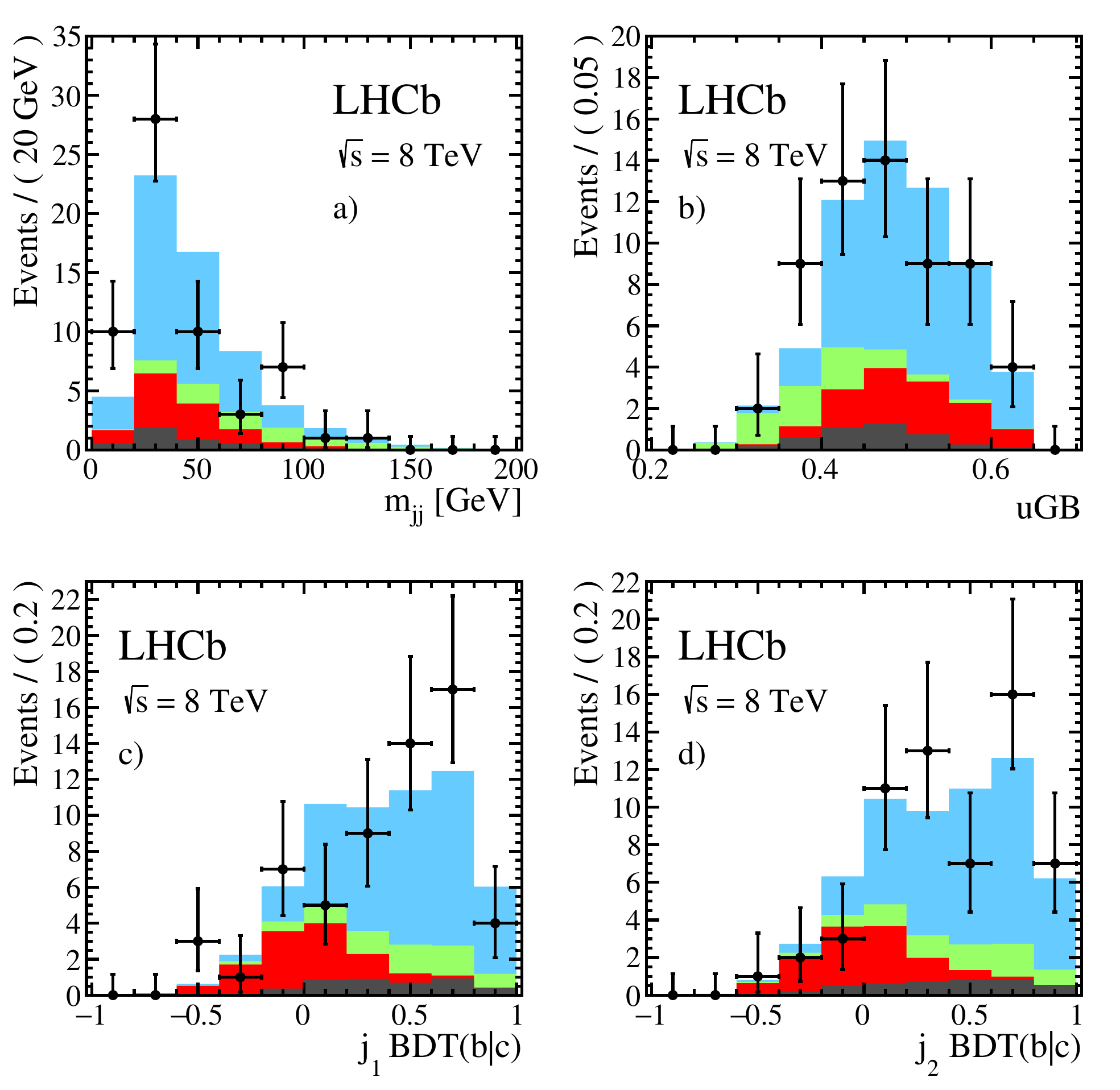}
\caption{
Projections of the simultaneous four-dimensional fit for the 
$\mu^+$ sample~\cite{Wjets} to: (a)~the dijet mass, 
(b)~the discriminator to separate 
$t\bar{t}$ from $W + b\bar{b}$ and $W + c\bar{c}$, and
(c)~the discriminator to separate $b$ and $c$ leading jets (d) sub-leading jets. In light blue is $W+b\bar{b}$, in green $t\bar{t}$, in red $W+c\bar{c}$ and in black the background.
}
\label{W-hq-jets}
\end{center}
\end{figure}

In order to extract the different signal components, 
a simultaneous four-dimensional fit is performed 
on the $\mu^+$, $\mu^-$, $e^+$ and $e^-$ samples.
Here the electron channels   
are used  to increase   statistics.
The four variables used in the fit are the dijet mass, 
a multivariate discriminator to separate $t\bar t$ 
from $W+b\bar b$ and $W+c\bar c$ events 
and a multivariate discriminator 
to separate  $b$- and $c$-jets, used for both  accompanying
jets~\cite{Wjets}. 
In this fit,  the background from QCD multi-jets is extrapolated from a 
control sample in data, while other background contributions are fixed to  
SM theoretical expectations. Only the signal components are then unconstrained.
 The projections of the resulting fit on four input variables for the $\mu^+$
 sample are illustrated in Fig.~\ref{W-hq-jets}. 
The statistical significance of the measured 
 $W^++ b\bar{b}$,  $W^++ c\bar{c}$, $W^-+ b\bar{b}$, 
 $W^-+ c\bar{c}$ and $t\bar{t}$ 
production cross sections are 7.1$\sigma$, 4.7$\sigma$, 
5.6$\sigma$, 2.5$\sigma$ and 4.9$\sigma$, respectively.
 The cross sections measured in the \lhcb fiducial 
 acceptance agree well with the Next-to-Leading-Order (NLO) theory predictions.

\paragraph{Search for long-lived new particles.}

The \lhcb detector has been designed to measure very rare decays of $\bquark$ quarks, with the aim of detecting the  presence of new  beyond-the-SM particles through their couplings in loops, which could change the expected SM branching ratios. 
This implies reaching mass scales higher than those explored in  $open$ new particle searches, where the particle is produced directly in $pp$ collisions. 
The traditional way to search for new particles is, as for ATLAS and CMS, reconstructing their decays through exclusive final states in their invariant masses, or with missing energy techniques. Here hermiticity is a mandatory feature of the detector.
In all cases, assumptions on their coupling, their production 
 mechanism, and on their decay modes must be made, and
 the search is therefore guided by theoretical models. 
 
\begin{figure}[htb]
\begin{center}
\includegraphics[width=0.9\linewidth]{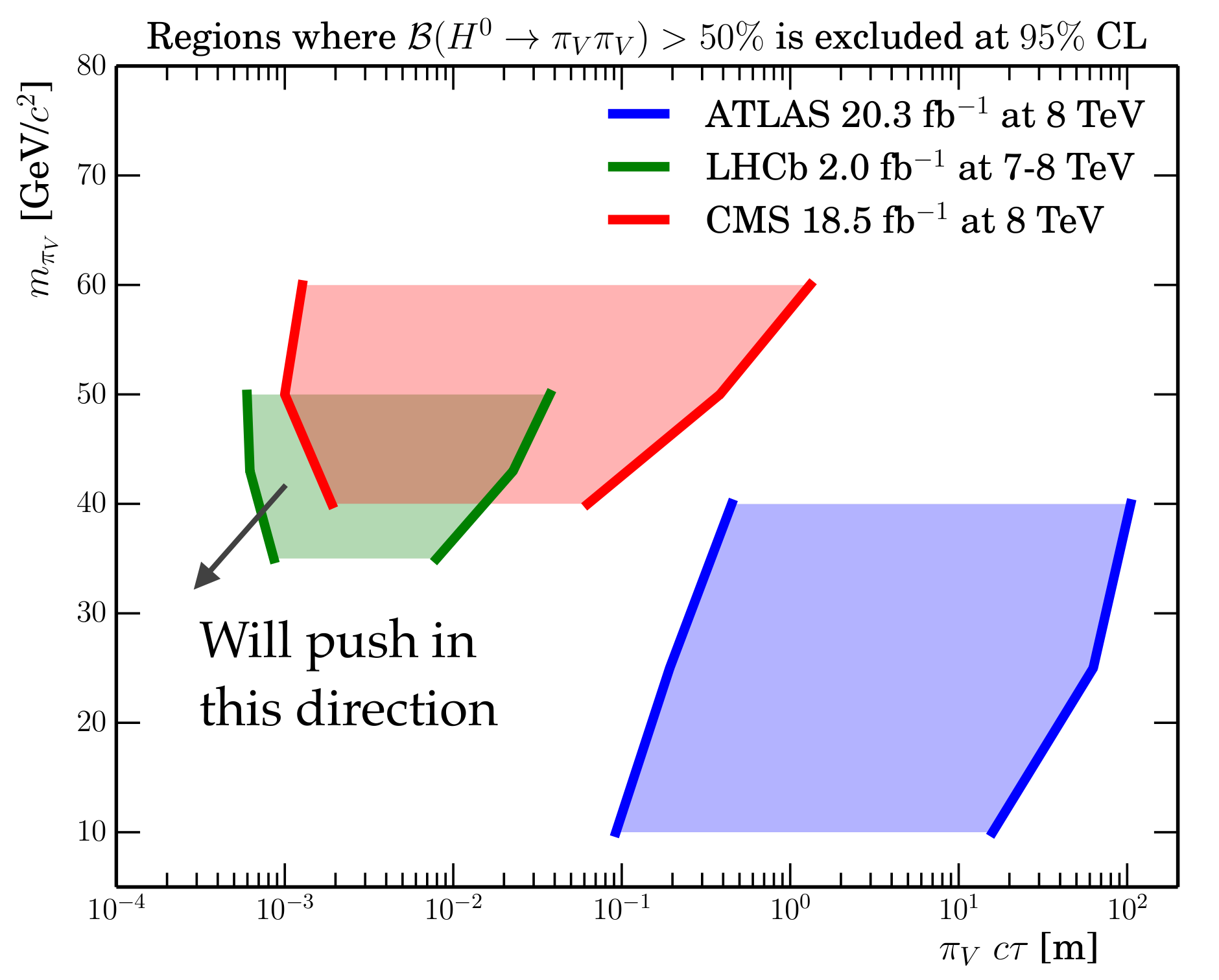}
\caption{
Comparison of the \lhcb exclusion region for the Branching Ratio of H $H^0 \to \pi _V \pi_V$, where $\pi_V$ is a long-lived particle 
decaying to jets. Exclusion regions for ATLAS and CMS
are also shown.
}
\label{exclusion_range_vs_GPD}
\end{center}
\end{figure}

For \lhcb, the most promising final states are  
 those decays which form a secondary vertex, 
 for which the LHCb VELO (see Sect.~\ref{sec:velo}) 
 is extremely efficient, i.e. the new particles are long lived.
 Several negative results have been published which
 are often less sensitive than the General-Purpose Detector (GPD) results. 
However in some cases, \lhcb can extend the exclusion region.
An example is given in Fig.~\ref{exclusion_range_vs_GPD}
 for a Higgs-like particle $H^0$ decaying into jets 
 forming a separate secondary vertex.
The \lhcb exclusion region is compared with that of 
ATLAS and CMS, which demonstrates the complementary of the \lhc experiments.
 The limits are, in this specific case,  competitive, 
 despite a factor 10 less luminosity.
 
\paragraph{Production of $t\bar{t}$ pairs.}

Top quark production 
is an excellent example where the 
forward acceptance of the \lhcb detector has several advantages with respect to
 the central region instrumented by ATLAS and CMS. 
The $t$ quark cross-section can provide important constraints on the 
large-$x$ gluon PDF, where
the forward kinematic region is particularly sensitive. 
In addition the forward region provides a greater fraction of events with 
quark-initiated production than in the central region, and  
enhances the size of $t\bar{t}$ asymmetries visible at \lhcb.
The challenge for \lhcb to measure $t\bar{t}$ production is the small acceptance 
and the impossibility of a missing energy measurement.
Also the fact that the luminosity is limited by the need to reduce multiple interactions for
measurements in the $\bquark$ sector, disfavours  $t\bar{t}$ statistics.

\begin{figure}[ht]
\begin{center}
\includegraphics[width=0.49\linewidth,height=65mm]{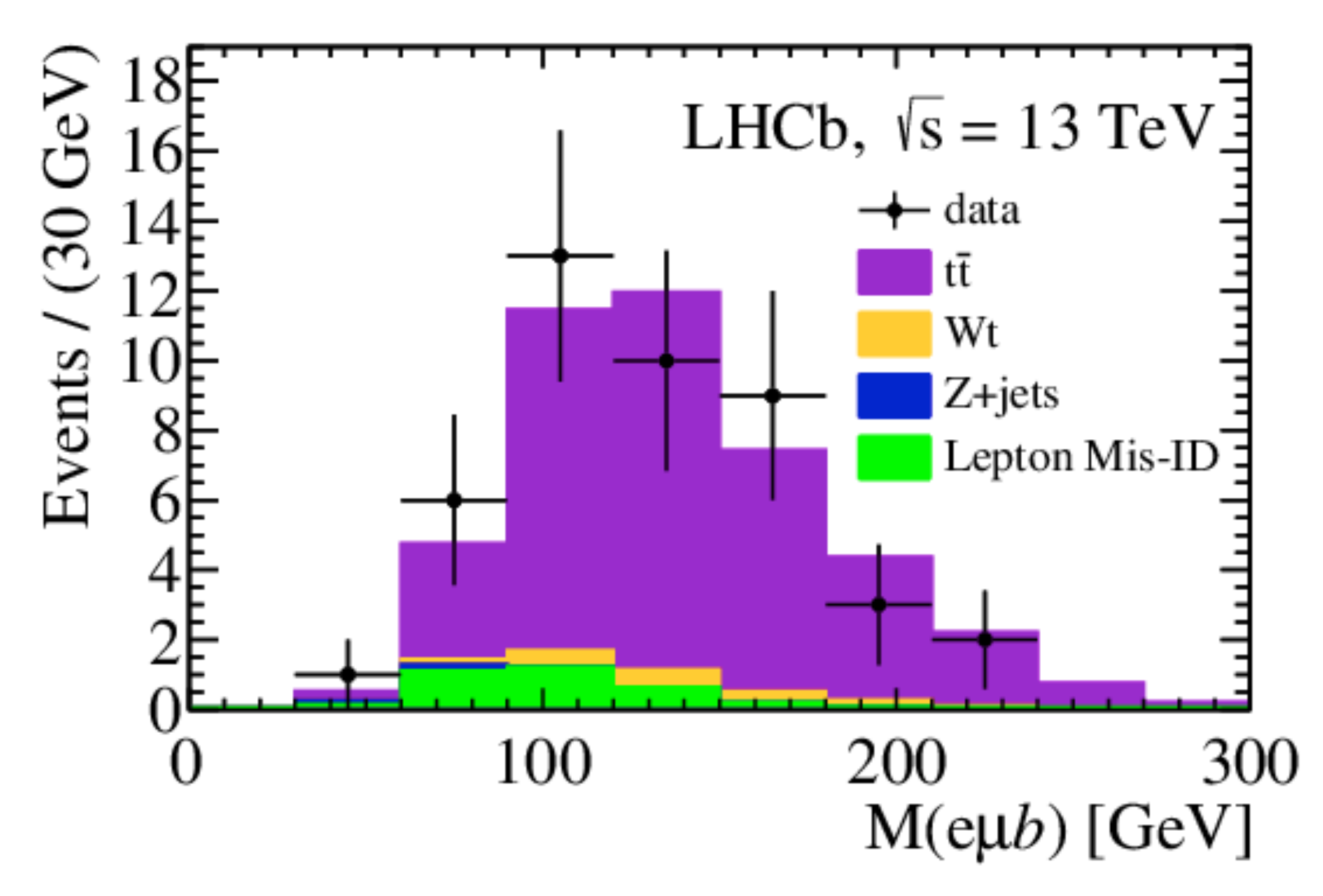}
\includegraphics[width=0.49\linewidth,height=65mm]{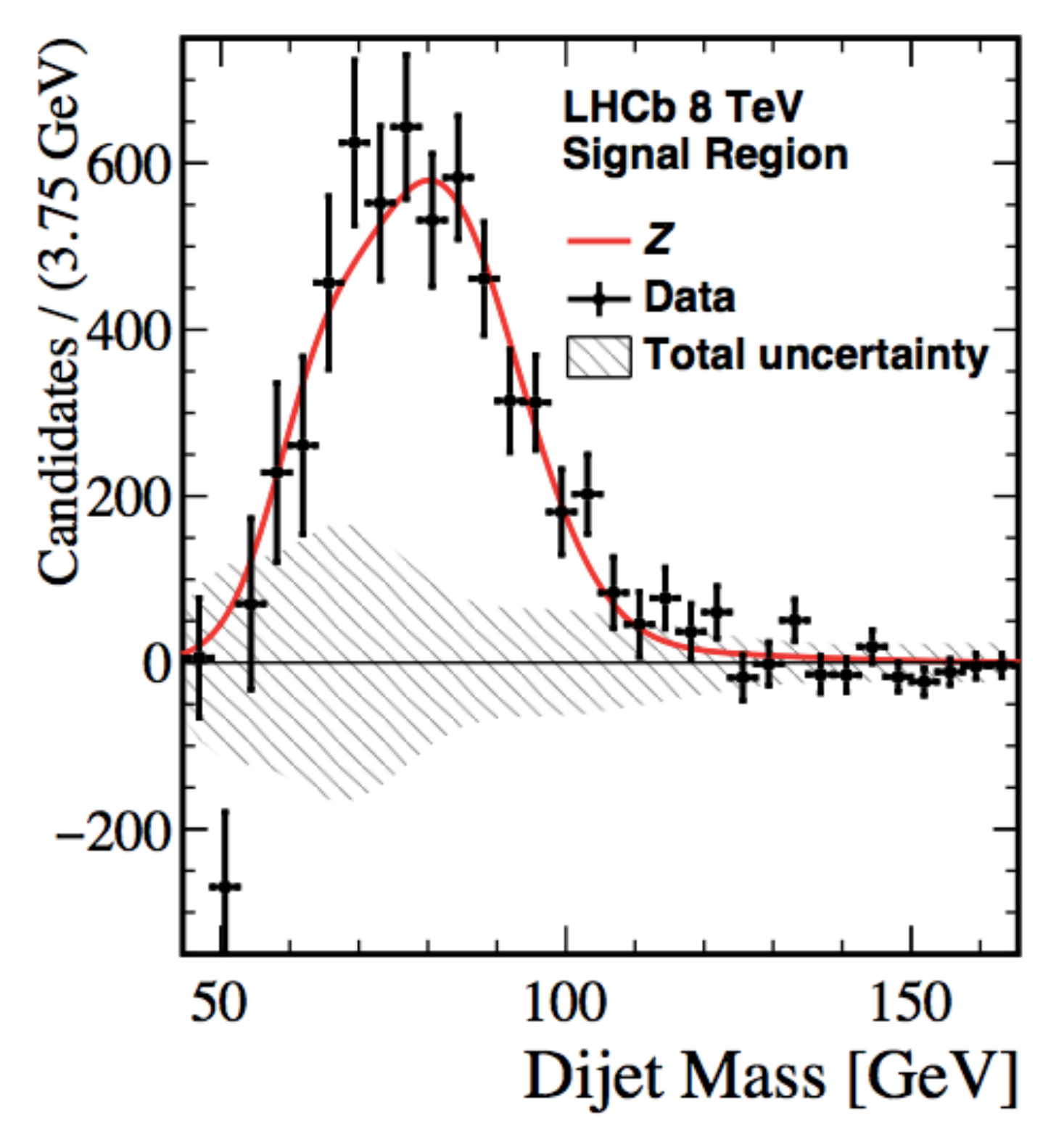}
\caption{ (Left)~The $e\mu b$ invariant mass for all 44 selected $t\bar{t}$
candidates, illustrates the excellent signal purity.
(Right)~The background-subtracted dijet mass spectrum showing the 
Z$\rightarrow b\bar{b}$ signal~\cite{LHCb-PAPER-2017-024}.
}
\label{tt_TeV}
\end{center}
\end{figure}

Top-quark production is presented here at $\sqrt s$=13 TeV, which gives an increase in the  production rate of an order of magnitude with respect to 8 TeV, and which 
brings these  new channels into statistical reach. 
The  $t\bar{t}$ analysis is based on 
 an integrated 
luminosity of 2~fb$^{-1}$, and 
with $e\mu b$ measured in the final state.
Hence the final state is the decay chain
$t\bar{t} \rightarrow bW^+  bW^- \rightarrow e^+ \mu^- bb$, 
where at least one $\bquark$-jet is reconstructed.  
This is a very pure final state, as the second lepton 
suppresses $W+b\bar{b}$ production and the different flavoured leptons suppress $Z+b\bar{b}$.
The signal purity is illustrated in Fig.~\ref{tt_TeV}(a),
and the observable cross section is measured to be 
$\sigma_{t\bar{t}} = 126\pm19(\stat)\pm16(\syst)\pm5$(lumi)  fb,
which is compatible with SM predictions.


\paragraph{$Z \rightarrow b\bar{b}$ decay.}
This measurement is an important validation of the \lhcb jet reconstruction and
 $\bquark$-tagging performance. 
Two $\bquark$-tagged jets are reconstructed, with a third balancing jet also 
reconstructed to help control the QCD background and define signal and
control regions using a multivariate technique. 
The background-subtracted signal distribution is shown in 
Fig.~\ref{tt_TeV}(b)~\cite{LHCb-PAPER-2017-024}.
The signal is observed with a statistical significance 
of 6$\sigma$ and the measured cross section is found to 
be compatible with SM predictions at next-to-leading order.

\subsection{Dark Photons}
The possibility that dark matter particles may interact via unknown forces, 
almost not felt by SM particles, has motivated substantial 
effort to search for dark-sector forces (see \cite{DM-review} for a review). 
A dark-force scenario involves a massive dark photon, $A^{\prime}$.
In the minimal model, the dark photon 
does not couple directly to charged SM particles, but it can gain a weak 
coupling to the SM electromagnetic current via kinetic mixing. 
The strength of this coupling is suppressed by a factor $\epsilon$ with respect to the SM photon. 
If the kinetic mixing arises from processes whose amplitudes involve one or 
two loops containing high-mass particles, perhaps even at the Planck scale, 
then   $10^{-12} \leq   \epsilon^2 \leq 10^{-4}$ is expected 
\cite{DM-review}. 

Constraints have been placed on visible $A^{\prime}$~decays by previous 
beam-dump, 
 fixed-target, 
collider 
and rare meson decay 
experiments; 
the few-loop region is ruled out for dark photon masses 
$m(A^{\prime})\sim10$ \mevcc.
 Additionally, the region
 $\epsilon^2 <  5\times 10^{-7}$
 is excluded for   $m(A^{\prime})< 10.2$ \gevcc,
 along with about half of the remaining few-loop 
region below the dimuon threshold. 
Many ideas have been proposed to further explore
the [$m(A^{\prime}),\epsilon^2$] 
parameter space, 
including an inclusive search for
$A^{\prime}\rightarrow \mu^-\mu^+$ decays with the 
\lhcb experiment.
A dark photon produced in proton-proton collisions via 
$\gamma^*–A^{\prime}$~mixing 
inherits the production mechanisms of an off-shell photon with
$m(\gamma^\ast)=m(A^{\prime})$, 
therefore both the production and decay kinematics of the 
$A^{\prime}\rightarrow \mu^+\mu^-$
and $\gamma^\ast \rightarrow \mu^+\mu^-$ 
processes are identical.

\begin{figure}[htb]
\begin{center}
\includegraphics[width=1.0\linewidth]{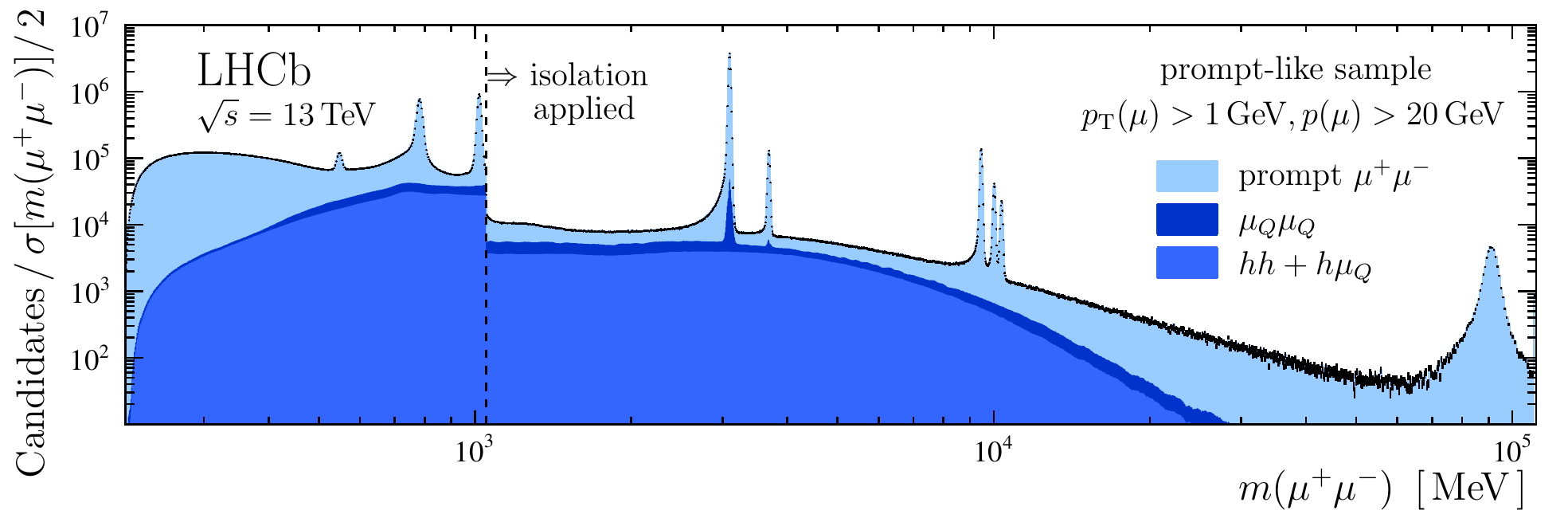}
\caption{
The prompt-like $\mu^+\mu^-$ mass spectrum.
}
\label{mu-mu-mass_full}
\end{center}
\end{figure}

\lhcb has performed searches for both prompt-like and long-lived dark photons \cite{DM-photon}
produced in $pp$ collisions at a centre-of-mass energy of 13~TeV, using 
$A^{\prime} \rightarrow \mu^+\mu^-$
 decays and a data sample corresponding to an integrated luminosity of
1.6 fb$^{-1}$ collected during 2016. 
The prompt-like $A^{\prime}$~search is 
performed from near the dimuon threshold up to 70~\gev, above which
the  $m(\mu^+\mu^-)$ spectrum is dominated by the $\Z$ boson.
The prompt-like dimuon spectrum is shown in Fig.~\ref{mu-mu-mass_full}. 

\noindent Three main types of background contribute to the prompt-like $A^{\prime}$~search:
prompt off-shell $\gamma^\ast\rightarrow \mu^+\mu^-$, which is irreducible; 
resonant decays to $\mu^+\mu^-$, whose mass 
peak regions are excluded in the search (see Fig.~\ref{mu-mu-mass_full}), 
and various types of misidentification,
which are highly suppressed by the stringent muon-identification 
and prompt-like requirements applied in the trigger. 

For the long-lived dark photon search, i.e. with displaced dimuon vertices,
the stringent criteria applied in the 
trigger make contamination from prompt muon candidates negligible. 
The long-lived $A^{\prime}$~search is restricted 
to the mass range
$214\leq m(A^{\prime})\leq 350$ \mevcc,
where the data sample potentially provides sensitivity.
In this case the background composition is 
dominated by
photon conversions to $\mu^+\mu^-$ 
in the VELO,
 $b$-hadron decays 
where two muons are produced in the decay chain, 
and the low-mass tail from 
\KS $\rightarrow \pi^+\pi^-$ 
decays where both pions are misidentified as muons.


\begin{figure}[htb]
\begin{center}
\includegraphics[width=1.0\linewidth]{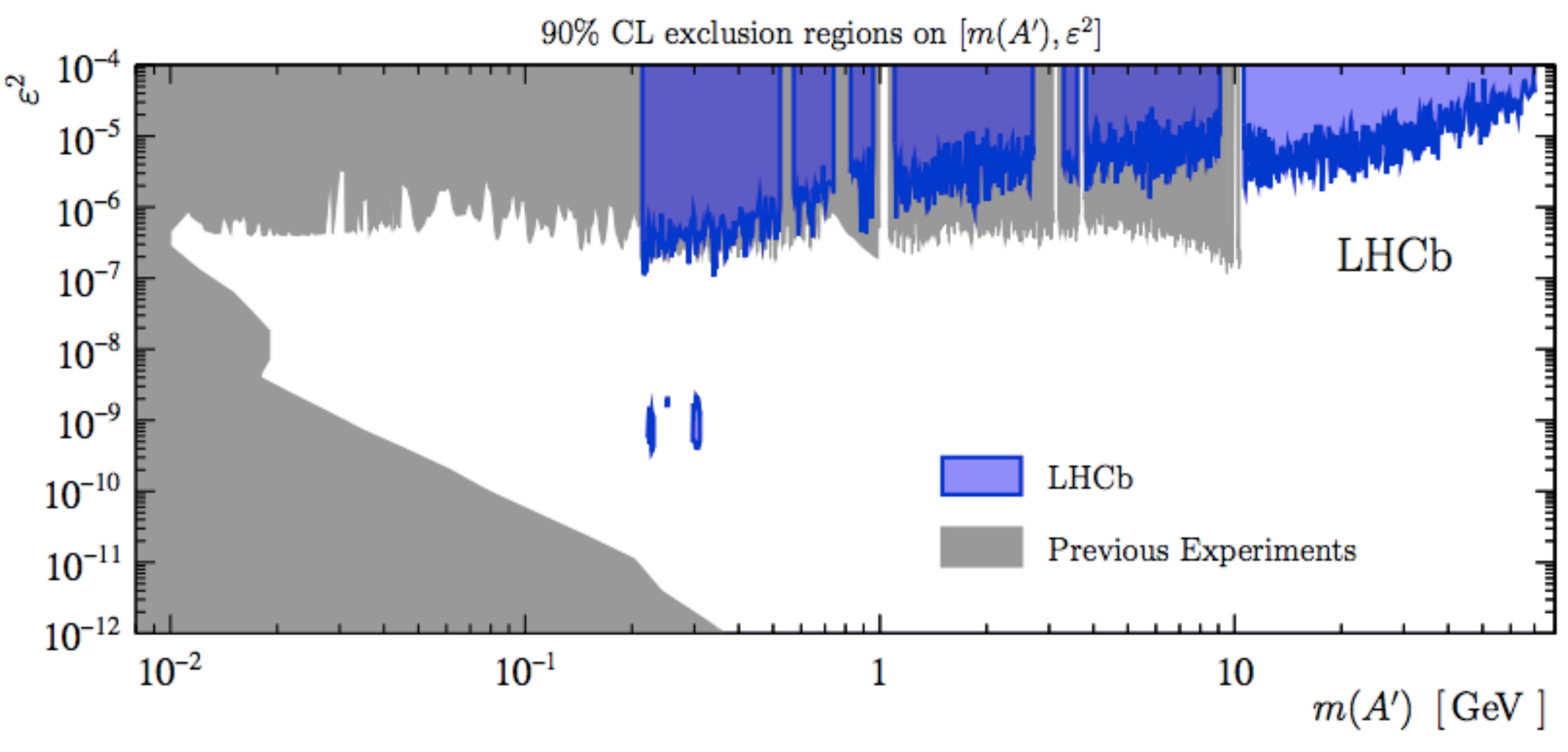}
\caption{
Results of the dark photon search. Both prompt-like (top) and displaced 
(centre) exclusions are shown.
}
\label{fig:exclusion-dark-photon}
\end{center}
\end{figure} 
In the dark-photon searches, no evidence for a signal is found, and 90\% CL exclusion 
regions are set on the $\gamma-A^{\prime}$~kinetic-mixing strength,  shown in Fig.~\ref{fig:exclusion-dark-photon}.  
The constraints placed on prompt-like dark photons are the most stringent to 
date for the mass range   $10.6\leq m(A^{\prime})\leq 70$ \gevcc,
and are comparable to the best existing limits for  $m(A^{\prime})\leq0.5$ \gevcc. 
The search for long-lived dark photons is the first to achieve sensitivity 
using a displaced-vertex signature.
These results demonstrate the unique sensitivity of the \lhcb experiment to dark
 photons, even using a data sample collected with a trigger that is inefficient
 for low-mass $A^{\prime}\rightarrow \mu^+ \mu^-$ decays. 
Using knowledge gained from this analysis, the
 software-trigger efficiency for low-mass dark photons has been significantly 
improved for 2017 data taking. 

In Run 3 to come, the planned increase in luminosity and removal of the
 hardware- trigger stage should increase the number of expected
 $A^{\prime}\rightarrow \mu^+ \mu^-$ decays in the low-mass region by 
 $\order(100–-1000)$ compared to the 2016 data sample.

\subsection{Nuclear Collisions}
Ultra-relativistic heavy-ion collisions allow the study of 
the so-called Quark-Gluon Plasma (QGP) state of matter, 
a hot and dense medium of deconfined quarks and gluons
 where heavy quarks are crucial probes. 
 Produced via hard interactions at the early stage of the nucleus-nucleus 
collision, before the QGP formation, heavy quarks experience the entire 
evolution of the QGP. 
A correct interpretation of these probes requires a full understanding of   
Cold Nuclear Matter (CNM) effects, which are present regardless of the 
formation of the deconfined medium. 
To disentangle the CNM   from   genuine QGP effects, heavy-flavour 
production in proton-nucleus collisions is studied.

The \lhcb experiment has collected data of 
proton-lead ($p$Pb) and lead-lead (PbPb) collisions. 
Since the LHCb detector covers only one direction of the full acceptance, there
 are two distinctive beam configurations for the pPb collisions. 
In the forward (backward) configuration, the proton (lead) beam enters the \lhcb
 detector from the interaction point. 
The proton beam and the lead beam have different energies per nucleon in the 
laboratory frame, hence the nucleon-nucleon centre-of-mass frame is boosted in 
the proton direction with a rapidity, $y$, shift. 
This results in the \lhcb acceptance for the forward configuration as 
$1.5~<~y~<~4$, and for the backward 
configuration $-5~<~y~<~-2.5$.

In addition, \lhcb provides the unique capability at the \lhc to collect 
fixed-target collisions utilising the System for Measuring the Overlap with Gas (SMOG) system\cite{SMOG}.
 Originally designed for precise luminosity measurements, SMOG provides the injection of a noble gas such as argon or helium 
inside the primary \lhc vacuum around the 
VELO detector with pressure $\mathcal{O}$$(10^{-7})$ mbar,
allowing measurements of $p$-gas and ion-gas collisions, 
and operating \lhcb as a fixed target experiment. 
Since 2015, \lhcb has exploited SMOG in physics runs using 
special fills not devoted to $pp$ physics, with a variety of  beam ($p$ or Pb) and 
target configurations. This allows unique production studies which are relevant 
to cosmic ray and heavy-ion physics.

 The heavy-ion results on heavy-flavour production in $p$Pb, 
PbPb and fixed-target collisions collected by \lhcb 
bring  yet more diversity and complementarity into the field.
Also in this context, the excellent momentum resolution and particle 
identification provided by \lhcb are especially suited for measuring heavy quark production.
The \lhcb collaboration joined the other participants into the \lhc heavy-ion collider
programme with a $p$Pb run at 5 TeV in 2013 and with a PbPb run in 2015. 
Following these pioneering data runs, significantly larger data-sets have been
successfully recorded.
\paragraph{Fixed target collisions.} 



\begin{figure}[htb]
\begin{center}
\includegraphics[width=0.90\linewidth]{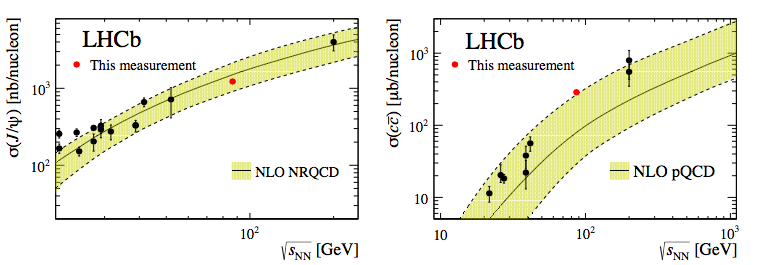}
\caption{
\jpsi (left) and $c\bar{c}$ (right) cross-section measurements as a 
function of the centre-of-mass energy, compared with other experimental data 
(black points). The bands correspond to   fits based on NLO 
NRQCD calculations for \jpsi and NLO pQCD calculations for $c\bar{c}$, respectively. 
More details are given in  
\cite{cc-sigma}.
}
\label{nuclear_cc_prod}
\end{center}
\end{figure}

\lhcb has reported  first measurements of heavy-flavour production with the 
fixed-target mode~\cite{cc-sigma}. 
\jpsi~production cross-sections and D$^0$ yields have been 
 measured in $p$He collisions at $\sqrt{s_{NN}} = 86.6$ GeV and $p$Ar 
collisions at $\sqrt{s_{NN}} = 110.4$ GeV, over the rapidity range $2~ < ~y ~< ~4.6$. The cross-section measurements are made for $p$He data only, since the 
luminosity determination is only available for this sample. 
After correction for  acceptances, efficiencies and  branching fractions, 
the cross-sections are extrapolated to the full phase space. The $D^0$
measurement is used to extract the $c\bar{c}$ cross-section. 
The \jpsi~ and $\ccbar$ measurements are compared in
Fig.~\ref{nuclear_cc_prod} 
with other experiments
at different centre-of-mass energies and with theoretical predictions.

With  $p$He data, \lhcb also measured the antiproton production cross section
\cite{LHCb-PAPER-2018-031}, a very interesting direct determination, helping the 
interpretation of the antiproton cosmic-ray flux detected by space experiments \cite{cosmic-antip}.

\paragraph{Collider mode.}
In collider mode, the LHCb experiment has collected proton-lead collision data   
 at $\sqrt{s_{NN}}$ = 5 TeV in 2013 and at 8.16 TeV in 2016. 
 The 2013 data sample corresponds to an integrated luminosity of
 $1.06\pm0.02$\,nb$^{-1}$  for the forward
 and $0.52\pm0.01$\,nb$^{-1}$ for the 
backward regions, whilst the 2016 data corresponds to 13.6$\pm$~0.3~nb$^{-1}$ 
 for the forward   and 20.8 $\pm$ 0.5 nb$^{-1}$ for the backward. 
These data samples are used to measure   quarkonium and open charm or beauty 
production.

$\Upsilon$(nS)-meson production is studied
in the  decay to two opposite-sign muons~\cite{pPb-upsilon-prod}. 
The measurements include the differential production cross-sections of 
$\Upsilon$(1S), $\Upsilon$(2S)  
states and nuclear modification factors, 
performed as a function of   transverse momentum and rapidity in the 
nucleon-nucleon centre-of-mass frame of the $\Upsilon$(nS) state. Also the
production cross-sections for the $\Upsilon$(3S) is measured, integrated over phase space, and the production ratios between all three $\Upsilon$(nS) states are determined. 

\begin{figure}[htb]
\begin{center}
\includegraphics[width=0.90\linewidth]{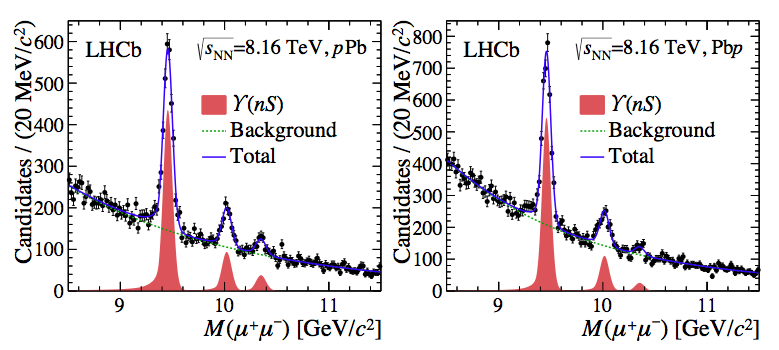}
\caption{
Invariant-mass distribution of $\mu^+\mu^-$ pairs from the (left) $p$Pb and 
(right) Pb$p$ samples after   trigger and offline selections.
}
\label{nuclear_upsilon_prod}
\end{center}
\end{figure}

The three states are well identified in both $p$Pb and Pb$p$ configurations as shown in Fig.~\ref{nuclear_upsilon_prod}. 
The nuclear modification factors are compared with theoretical predictions, 
and suppressions for bottomium in $p$Pb collisions are observed.
The \lhcb measurements 
improve the understanding of cold nuclear matter 
effects down to low \pt. 

\section{ Future prospects }

Over the years 2011 - 2018, both the \lhc machine and the \lhcb detector performed
 extremely well, providing great improvements
with respect to the \bfactory measurements, in particular the pioneering \CP violation measurements (Sect. 3), the observation of 
the rarest beauty meson decays (Sect.4)
 and the discovery of pentaquarks (Sect. 5).
 \lhcb also observed and reported a number of interesting hints of anomalies related 
to the flavour sector, which has generated much theoretical attention, 
especially relating to rare decays and lepton flavour universality. The precision achieved by the experiment is in line with prior expectations, as documented in \cite{BigReport}, and which demonstrates the remarkable understanding of all aspects of the detector. 

To further pursue these exciting results and fully exploit the 
flavour physics potential of the \lhc, the \lhcb detector required an upgrade,
to increase the rate and efficiency of data taking beyond the Long Shutdown 2 (LS2). 
Consequently the \lhcb detector is now undergoing a major upgrade that is well 
underway, and will allow the experiment to pursue its superb performance into the future.

At present, the hardware-based trigger limits the amount of data taken each 
year to a maximum of about 2 fb$^{-1}$. 
In addition, most of the detector sub-systems would not  cope with higher 
luminosity due to either their outdated readout electronics or radiation-induced damage
 sustained during Run 1 and Run 2 data taking. 
The initial ideas regarding the upgrade were  formulated in 2011 
\cite{upgrade_LoI},
and further solidified in 2012 when the Technical Design Report was released 
\cite{LHCb-TDR-012}.
Many of the subdetector components are largely unchanged in the upgrade, with the exception of a new pixel vertex detector replacing the current VELO, the TT stations being replaced by a new silicon micro-strip upstream tracker (UT), and the straw outer chambers replaced by a scintillating fibre detector. 
Details of each subdetector upgrade can be found in refs.
\cite{LHCb-TDR-012,LHCb-TDR-015,LHCb-TDR-013,LHCb-TDR-014,LHCb-TDR-018}.

The crucial point of the upgrade project is to build a reliable and robust 
detector capable of operating at higher luminosity without compromising the 
excellent physics performance of the current detector. 
This, in turn, cannot be achieved by redesigning the hardware components alone,
but has to be augmented by a new innovative and flexible trigger system. 
A critical part of the upgrade strategy is  the 
design of a so-called ``trigger-less'' 
front-end electronics system capable of reading out the full detector at 40 MHz, i.e.
 at the \lhc clock frequency. 
Completely new and novel chips have been designed and tested for the pixel sensors 
\cite{LHCb-TDR-015}
the UT 
\cite{LHCb-TDR-013}
and RICH detectors
\cite{LHCb-TDR-014}.

\begin{figure}[h]
\begin{center}
\includegraphics[width=1.\linewidth]{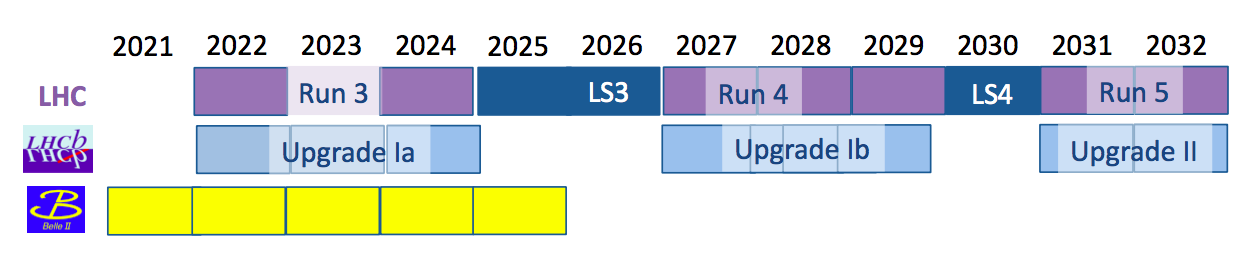}
\caption{
A time-line showing the operations of the \lhc and the HL-\lhc over the next decade, including long shutdown (LS) 
periods, as  can be estimated today. 
The  operational periods of \lhcb and  Belle II are shown
\cite{BELLE_operation}.
}
\label{LHC-time-schedule}
\end{center}
\end{figure}

The upgraded detector will operate at an instantaneous luminosity of 
$2 \cdot 10^{33}$ cm$^{2}$s$^{-1}$ 
which allows collection of around 10 fb$^{-1}$ of data per year as a target, also keeping pace with
 Belle II \cite{BELLE_operation},
 the other major flavour-physics experiment.  
 Figure \ref{LHC-time-schedule} shows the corresponding time-line for \lhcb operations over the next decade. 

To efficiently run at increased luminosity, the present 
hardware-based trigger will be replaced, and events will be selected by the 
software-based HLT alone. 
To cope with the much higher event rate (typically five proton-proton interactions per 
beam crossing), a flexible software trigger will be employed and coupled 
with a re-optimized network capable of handling a multi-terabyte data stream. 
The upgraded trigger will process every event (the visible rate at   \lhcb is 
estimated to reach 30 MHz) using information from every sub-detector to enhance
 its decision and maximize signal efficiencies, especially for the hadronic 
channels.
The precision of particle identification and track-quality information will be identical than "offline" and able to reduce the rate down to 20-100 kHz.
The new trigger strategy will increase the triggering efficiency for the 
hadronic channels by a factor 2 to 4 with respect to Run 1 
\cite{LHCb-TDR-016},
corresponding to an increase of a factor 10 to 20 for the hadronic yields.

Finally, plans for a further future upgrade (called Upgrade II) to use the full potential of flavour physics during the HL-LHC operation
have now started \cite{LHCb-PII-EoI, LHCb-PII-Physics}. This upgrade would require a complete redesign of the detector able to take data at instantaneous luminosities of $2 \cdot 10^{34}$ cm$^{2}$s$^{-1}$, and collect $\sim 50$~fb$^{-1}$ of data per year,  
guaranteeing \lhcb operation beyond 2030. 

The \lhcb Upgrades I and II will significantly improve  the reach of key physics measurements. By way of example, the precision quoted in Section~\ref{Sec:gamma} on today's measurement of the CKM angle  $\gamma,~\mathcal{O} (5^\circ)$, will be improved to $1^\circ$ with the new Upgrade I hadronic trigger and luminosity increase. This further improves to $0.35^\circ$ with the large statistics accumulated with \lhcb Upgrade II. As discussed in Section~\ref{Sec:bsmm}, currently there is not   enough sensitivity to measure the rare decay $\BR(\Bdmm)$. With \lhcb Upgrade I, a first observation should be possible, but it will require  Upgrade II to reach a $\sim 10\%$ precision on its branching ratio. This sensitivity will  allow  a meaningful measurement of the ratio between \Bd and \Bs  into  the $\mu^+\mu^-$ final state, and will constitute a clean and powerful test of extensions beyond the SM. 


\section{Summary and Conclusions}


In  ten years of operation at the \lhc,
the \lhcb experiment has delivered a remarkably rich programme of physics measurements.
In this paper the 25-year evolution of the experiment since its inception has been described, and its  successes and achievements have been summarised. The diversity of the physics output has truly shown \lhcb to be a ``general-purpose detector in the forward region''.

Over the last ten years, \lhcb has measured the CKM quark mixing matrix elements and \CP violation parameters to world-leading precision in the
\bquark- and \cquark-quark systems. The experiment has  measured very rare
decays of \bquark~and \cquark  mesons and baryons, some  with branching ratios down to order 10$^{-9}$, testing Standard Model predictions to unprecedented levels. Hints of new physics in rare-decay angular distributions and through tests of lepton universality in electron-muon decay modes have generated considerable theoretical interest.
The global knowledge of $\bquark$ and $\cquark$ quark states has  improved significantly, through discoveries of many new resonances already anticipated in the quark model, and also by the observation of new exotic tetraquark and pentaquark states.
In addition, many interesting measurements have been made
that were not anticipated in the original \lhcb proposal,    such as electroweak physics, jet measurements,  
new long-lived particle searches and heavy-ion physics.
\textcolor{red}{ 
An incredibly rich harvest of fundamental  results has been produced, many of these will remain in textbooks for years to come.  }

\lhcb has recently been upgraded and will start data-taking  early in 2022 at a factor 5 higher luminosity, incorporating new subdetectors and a software-based trigger.   Statistics in hadronic modes  will be improved by at a factor 10--20, allowing much more precise measurements, especially of very rare $\bquark$- and $\cquark$-hadron decays.  
In addition, the future planned Upgrade II at the (HL)-LHC early in 2030 will ensure that \lhcb maintains its lead in flavour physics for at least the next two decades. 


 
 
 

\section{Acknowledgments}

LHCb is at present a collaboration of about 1000 authors.
The rich variety of outstanding results has been made possible by the dedicated work of many colleagues: detector builders and operators, data verifiers and analysts.

We would like to acknowledge the important roles played by T. Nakada as first spokesman and by the late H.J.Hilke as Technical Coordinator, who successfully managed the realisation of this very complex detector.

 Finally, we would also like to thank our colleagues 
 P. Koppenburg and G. Passaleva 
 who made helpful and insightful comments to this paper.

\clearpage 
\addcontentsline{toc}{section}{References}
\bibliographystyle{LHCb}
\bibliography{Introduction,CKM,RD,SP,EW_NP,EW_NP_bis,main,standard,LHCb-PAPER,LHCb-CONF,LHCb-DP,LHCb-TDR,nuclear,chapter_8}
 
\end{document}